\documentclass[twocolumn]{aa}

\usepackage{graphicx}
\usepackage{amsmath,amsfonts,amssymb}
\usepackage[varg]{txfonts}
\usepackage{color}
\usepackage{natbib}
\usepackage{float}
\usepackage{dblfloatfix}
\usepackage{afterpage}
\usepackage{ifthen}
\usepackage[morefloats=12]{morefloats}
\usepackage{placeins}
\usepackage{multicol}
\usepackage{multirow}
\bibpunct{(}{)}{;}{a}{}{,}
\usepackage[switch]{lineno}
\definecolor{linkcolor}{rgb}{0.6,0,0}
\definecolor{citecolor}{rgb}{0,0,0.75}
\definecolor{urlcolor}{rgb}{0.12,0.46,0.7}
\usepackage[breaklinks, colorlinks, urlcolor=urlcolor,
    linkcolor=linkcolor,citecolor=citecolor,pdfencoding=auto]{hyperref}
\hypersetup{linktocpage}
\usepackage{bold-extra}
\usepackage{xcolor}
\usepackage{capt-of}

\usepackage{lipsum}

\def\setsymbol#1#2{\expandafter\def\csname #1\endcsname{#2}}
\def\getsymbol#1{\csname #1\endcsname}

\def\Planck{\textit{Planck}}

\newbox\tablebox    \newdimen\tablewidth
\def\leaderfil{\leaders\hbox to 5pt{\hss.\hss}\hfil}
\def\endPlancktablewide{\tablewidth=\textwidth 
    $$\hss\copy\tablebox\hss$$
    \vskip-\lastskip\vskip -2pt}
\def\tablenote#1 #2\par{\begingroup \parindent=0.8em
    \abovedisplayshortskip=0pt\belowdisplayshortskip=0pt
    \noindent
    $$\hss\vbox{\hsize\tablewidth \hangindent=\parindent \hangafter=1 \noindent
    \hbox to \parindent{$^#1$\hss}\strut#2\strut\par}\hss$$
    \endgroup}
\def\doubleline{\vskip 3pt\hrule \vskip 1.5pt \hrule \vskip 5pt}

\def\L2{\ifmmode L_2\else $L_2$\fi}

\def\DeltaT{\ifmmode \Delta T\else $\Delta T$\fi}
\def\deltat{\ifmmode \Delta t\else $\Delta t$\fi}
\def\fknee{\ifmmode f_{\rm knee}\else $f_{\rm knee}$\fi}
\def\Fmax{\ifmmode F_{\rm max}\else $F_{\rm max}$\fi}
\def\solar{\ifmmode{\rm M}_{\mathord\odot}\else${\rm M}_{\mathord\odot}$\fi}
\def\Msolar{\ifmmode{\rm M}_{\mathord\odot}\else${\rm M}_{\mathord\odot}$\fi}
\def\Lsolar{\ifmmode{\rm L}_{\mathord\odot}\else${\rm L}_{\mathord\odot}$\fi}
\def\inv{\ifmmode^{-1}\else$^{-1}$\fi}
\def\mo{\ifmmode^{-1}\else$^{-1}$\fi}
\def\sup#1{\ifmmode ^{\rm #1}\else $^{\rm #1}$\fi}
\def\expo#1{\ifmmode \times 10^{#1}\else $\times 10^{#1}$\fi}
\def\,{\thinspace}
\def\lsim{\mathrel{\raise .4ex\hbox{\rlap{$<$}\lower 1.2ex\hbox{$\sim$}}}}
\def\gsim{\mathrel{\raise .4ex\hbox{\rlap{$>$}\lower 1.2ex\hbox{$\sim$}}}}

\def\simprop{\mathrel{\raise .4ex\hbox{\rlap{$\propto$}\lower 1.2ex\hbox{$\sim$}}}}
\def\deg{\ifmmode^\circ\else$^\circ$\fi}
\def\pdeg{\ifmmode $\setbox0=\hbox{$^{\circ}$}\rlap{\hskip.11\wd0 .}$^{\circ}
          \else \setbox0=\hbox{$^{\circ}$}\rlap{\hskip.11\wd0 .}$^{\circ}$\fi}
\def\arcs{\ifmmode {^{\scriptstyle\prime\prime}}
          \else $^{\scriptstyle\prime\prime}$\fi}
\def\arcm{\ifmmode {^{\scriptstyle\prime}}
          \else $^{\scriptstyle\prime}$\fi}
\newdimen\sa  \newdimen\sb
\def\parcs{\sa=.07em \sb=.03em
     \ifmmode \hbox{\rlap{.}}^{\scriptstyle\prime\kern -\sb\prime}\hbox{\kern -\sa}
     \else \rlap{.}$^{\scriptstyle\prime\kern -\sb\prime}$\kern -\sa\fi}
\def\parcm{\sa=.08em \sb=.03em
     \ifmmode \hbox{\rlap{.}\kern\sa}^{\scriptstyle\prime}\hbox{\kern-\sb}
     \else \rlap{.}\kern\sa$^{\scriptstyle\prime}$\kern-\sb\fi}
\def\ra[#1 #2 #3.#4]{#1\sup{h}#2\sup{m}#3\sup{s}\llap.#4}
\def\dec[#1 #2 #3.#4]{#1\deg#2\arcm#3\arcs\llap.#4}
\def\deco[#1 #2 #3]{#1\deg#2\arcm#3\arcs}
\def\rra[#1 #2]{#1\sup{h}#2\sup{m}}

\def\dots{\relax\ifmmode \ldots\else $\ldots$\fi}
\def\WHzsr{\ifmmode $W\,Hz\mo\,sr\mo$\else W\,Hz\mo\,sr\mo\fi}
\def\mHz{\ifmmode $\,mHz$\else \,mHz\fi}
\def\GHz{\ifmmode $\,GHz$\else \,GHz\fi}
\def\mKs{\ifmmode $\,mK\,s$^{1/2}\else \,mK\,s$^{1/2}$\fi}
\def\muKs{\ifmmode \,\mu$K\,s$^{1/2}\else \,$\mu$K\,s$^{1/2}$\fi}
\def\muKRJs{\ifmmode \,\mu$K$_{\rm RJ}$\,s$^{1/2}\else \,$\mu$K$_{\rm RJ}$\,s$^{1/2}$\fi}
\def\muKHz{\ifmmode \,\mu$K\,Hz$^{-1/2}\else \,$\mu$K\,Hz$^{-1/2}$\fi}
\def\MJysr{\ifmmode \,$MJy\,sr\mo$\else \,MJy\,sr\mo\fi}
\def\MJysrmK{\ifmmode \,$MJy\,sr\mo$\,mK$_{\rm CMB}\mo\else \,MJy\,sr\mo\,mK$_{\rm CMB}\mo$\fi}
\def\microns{\ifmmode \,\mu$m$\else \,$\mu$m\fi}

\def\muK{\ifmmode \,\mu$K$\else \,$\mu$\hbox{K}\fi}
\def\microK{\ifmmode \,\mu$K$\else \,$\mu$\hbox{K}\fi}
\def\muW{\ifmmode \,\mu$W$\else \,$\mu$\hbox{W}\fi}
\def\kms{\ifmmode $\,km\,s$^{-1}\else \,km\,s$^{-1}$\fi}
\def\kmsMpc{\ifmmode $\,\kms\,Mpc\mo$\else \,\kms\,Mpc\mo\fi}
\providecommand{\sorthelp}[1]{}

\def\WMAP{\emph{WMAP}}
\def\WMAPnine{\emph{WMAP9}}
\def\COBE{\emph{COBE}}
\def\wmap{\emph{WMAP}}
\def\planck{\emph{Planck}}
\def\Planck{\emph{Planck}}
\def\LCDM{$\Lambda$CDM}

\def\commander{\texttt{Commander}}

\def\commanderthree{\texttt{Commander3}}

\def\sroll2{\texttt{SRoll2}}

\newcommand{\dv}[0]{\vec{d}}

\newcommand{\A}[0]{\mathrm{A}}
\newcommand{\B}[0]{\mathrm{B}}

\newcommand{\n}[0]{\vec{n}}

\newcommand{\s}[0]{\vec{s}}
\renewcommand{\a}[0]{\vec{a}}

\newcommand{\bv}[0]{\vec{b}}

\renewcommand{\L}[0]{\tens{L}}
\newcommand{\g}[0]{\vec{g}}
\newcommand{\N}[0]{\tens{N}}
\newcommand{\M}[0]{\tens{M}}

\renewcommand{\P}[0]{\tens{P}}

\newcommand{\Te}[0]{T_{\rm e}}

\newcommand{\BP}{\textsc{BeyondPlanck}}
\newcommand{\bp}{\textsc{BeyondPlanck}}
\newcommand{\cosmoglobe}{\textsc{Cosmoglobe}}
\newcommand{\Cosmoglobe}{\textsc{Cosmoglobe}}

\newcommand{\K}[0]{\textit K}
\newcommand{\Ka}[0]{\textit{Ka}}
\newcommand{\Q}[0]{\textit Q}
\newcommand{\V}[0]{\textit V}
\newcommand{\W}[0]{\textit W}
\newcommand{\e}{\mathrm e}

\def\Tcmb{\ifmmode T_\mathrm{CMB}\else $T_{\mathrm{CMB}}$\fi}
\def\Tdust{\ifmmode T_\mathrm{d}\else $T_{\mathrm{d}}$\fi}
\def\scmb{\ifmmode s_\mathrm{CMB}\else $s_{\mathrm{CMB}}$\fi}
\def\squad{\ifmmode s_\mathrm{quad}\else $s_{\mathrm{quad}}$\fi}
\def\ssynch{\ifmmode s_\mathrm{s}\else $s_\mathrm{s}$\fi}
\def\sdust{\ifmmode s_\mathrm{d}\else $s_{\mathrm{d}}$\fi}
\def\ssdust{\ifmmode s_\mathrm{sd}\else $s_{\mathrm{sd}}$\fi}
\def\same{\ifmmode s_\mathrm{AME}\else $s_{\mathrm{AME}}$\fi}
\def\ssrc{\ifmmode s_\mathrm{src}\else $s_{\mathrm{src}}$\fi}
\def\sco{\ifmmode s_\mathrm{CO}\else $s_{\mathrm{CO}}$\fi}
\def\sff{\ifmmode s_\mathrm{ff}\else $s_{\mathrm{ff}}$\fi}
\def\gff{\ifmmode g_\mathrm{ff}\else $g_{\mathrm{ff}}$\fi}
\def\fsynch{\ifmmode f_\mathrm{s}\else $f_{\mathrm{s}}$\fi}
\def\fsd{\ifmmode f_\mathrm{sd}\else $f_{\mathrm{sd}}$\fi}
\def\fame{\ifmmode f_\mathrm{AME}\else $f_{\mathrm{AME}}$\fi}
\def\alphasrc{\ifmmode \alpha_\mathrm{src}\else $\alpha_{\mathrm{src}}$\fi}
\def\bdust{\ifmmode \beta_\mathrm{d}\else $\beta_{\mathrm{d}}$\fi}
\def\bsynch{\ifmmode \beta_\mathrm{s}\else $\beta_{\mathrm{s}}$\fi} 
\def\bsun{\ifmmode \beta_\mathrm{sun}\else $\beta_{\mathrm{sun}}$\fi} 
\def\nuzeros{\ifmmode \nu_{0,\mathrm{s}}\else $\nu_{0,\mathrm{s}}$\fi} 
\def\nuzeroff{\ifmmode \nu_{0,\mathrm{ff}}\else $\nu_{0,\mathrm{ff}}$\fi} 
\def\nuzerod{\ifmmode \nu_{0,\mathrm{d}}\else $\nu_{0,\mathrm{d}}$\fi} 
\def\nuzeroame{\ifmmode \nu_{0,\mathrm{AME}}\else $\nu_{0,\mathrm{AME}}$\fi} 
\def\nuzerosd{\ifmmode \nu_{0,\mathrm{}}\else $\nu_{0,\mathrm{sd}}$\fi} 
\def\nuzerosrc{\ifmmode \nu_{0,\mathrm{src}}\else $\nu_{0,\mathrm{src}}$\fi} 
\def\nup{\ifmmode \nu_{\mathrm{p}}\else $\nu_{\mathrm{p}}$\fi} 
\def\alphasd{\ifmmode \alpha_{\mathrm{sd}}\else $\alpha_{\mathrm{sd}}$\fi} 
\def\Te{\ifmmode T_{\mathrm{e}}\else $T_{\mathrm{e}}$\fi} 
\def\kB{\ifmmode k_\mathrm{B}\else $k_{\mathrm{B}}$\fi}

\usepackage[T1]{fontenc}

\newcommand{\ncorr}{\vec n^\mathrm{corr}}
\newcommand{\data}{\vec d}

\def\inv{^{-1}}

\begin{document}

\title{\bfseries{\Cosmoglobe\ DR1 results. I. Improved \emph{Wilkinson Microwave Anisotropy Probe} maps through Bayesian end-to-end analysis}}
\newcommand{\oslo}[0]{1}
\newcommand{\iiabangalore}[0]{2}

\author{\small
D.~J.~Watts\inst{\ref{uio}}\thanks{Corresponding author: D.~J.~Watts; \url{duncanwa@astro.uio.no}}
\and
A.~Basyrov\inst{\ref{uio}}
\and
J.~R.~Eskilt\inst{\ref{uio},\ref{imperial}}
\and
M.~Galloway\inst{\ref{uio}}
\and
L.~T.~Hergt\inst{\ref{ubc}}
\and
D.~Herman\inst{\ref{uio}}
\and
H.~T.~Ihle\inst{\ref{uio}}
\and
S.~Paradiso\inst{\ref{waterloo}}
\and
F.~Rahman\inst{\ref{iiabangalore}}
\and
H.~Thommesen\inst{\ref{uio}}
\and
R.~Aurlien\inst{\ref{uio}}
\and
M.~Bersanelli\inst{\ref{milan}}
\and
L.~A.~Bianchi\inst{\ref{milan}}
\and
M.~Brilenkov\inst{\ref{uio}}
\and
L.~P.~L.~Colombo\inst{\ref{milan}}
\and
H.~K.~Eriksen\inst{\ref{uio}}
\and
C.~Franceschet\inst{\ref{milan}}
\and
U.~Fuskeland\inst{\ref{uio}}
\and
E.~Gjerl\o w\inst{\ref{uio}}
\and
B.~Hensley\inst{\ref{princeton}}
\and
G.~A.~Hoerning\inst{\ref{saopaulo}}
\and
K.~Lee\inst{\ref{uio}}
\and
J.~G.~S.~Lunde\inst{\ref{uio}}
\and
A.~Marins\inst{\ref{saopaulo}}
\and
S.~K.~Nerval\inst{\ref{dunlap1},\ref{dunlap2}}
\and
S.~K.~Patel\inst{\ref{iit_bhu}}
\and
M.~Regnier\inst{\ref{apc}}
\and
M.~San\inst{\ref{uio}}
\and
S.~Sanyal\inst{\ref{iit_bhu}}
\and
N.-O.~Stutzer\inst{\ref{uio}}
\and
A.~Verma\inst{\ref{iit_bhu}}
\and
I.~K.~Wehus\inst{\ref{uio}}
\and
Y.~Zhou\inst{\ref{berkeley}}
}
\institute{\small
Institute of Theoretical Astrophysics, University of Oslo, Blindern, Oslo, Norway\label{uio}
\and
Imperial Centre for Inference and Cosmology, Department of Physics, Imperial College London, Blackett Laboratory, Prince Consort Road, London SW7 2AZ, United Kingdom\label{imperial}
\and
Department of Physics and Astronomy, University of British Columbia, 6224 Agricultural Road, Vancouver BC, V6T1Z1, Canada\label{ubc}
\and
Waterloo Centre for Astrophysics, University of Waterloo, Waterloo, ON N2L 3G1, Canada\label{waterloo}
\and
Indian Institute of Astrophysics, Koramangala II Block, Bangalore, 560034, India\label{iiabangalore}
\and
Dipartimento di Fisica, Università degli Studi di Milano, Via Celoria, 16, Milano, Italy\label{milan}
\and
Instituto de Física, Universidade de São Paulo - C.P. 66318, CEP: 05315-970, São Paulo, Brazil\label{saopaulo}
\and
Department of Astrophysical Sciences, Princeton University, 4 Ivy Lane, Princeton, NJ 08540\label{princeton}
\and
David A. Dunlap Department of Astronomy \& Astrophysics, University of Toronto, 50 St. George Street, Toronto, ON M5S 3H4, Canada\label{dunlap1}
\and
Dunlap Institute for Astronomy \& Astrophysics, University of Toronto, 50 St. George Street, Toronto, ON M5S 3H4, Canada\label{dunlap2}
\and
Department of Physics, Indian Institute of Technology (BHU), Varanasi - 221005, India\label{iit_bhu}
\and
Laboratoire Astroparticule et Cosmologie (APC), Université Paris-Cité, Paris, France\label{apc}
\and
Department of Physics, University of California, Berkeley, Berkeley, CA 94720, USA\label{berkeley}
}

\authorrunning{Watts et al.}
\titlerunning{\cosmoglobe\ \wmap{}  analysis}

\abstract{
  We present \cosmoglobe\ Data Release 1, which implements the first joint analysis of \textit{WMAP} and \textit{Planck} LFI time-ordered data, processed within a single Bayesian end-to-end framework. This framework builds directly on a similar analysis of the LFI measurements by the \BP\ collaboration, and approaches the CMB analysis challenge through Gibbs sampling of a global posterior distribution, simultaneously accounting for calibration, mapmaking, and component separation. The computational cost of producing one complete \WMAP+LFI Gibbs sample is 812\,CPU-hr, of which 603\,CPU-hrs are spent on \WMAP\ low-level processing; this demonstrates that end-to-end Bayesian analysis of the \WMAP\ data is computationally feasible. We find that our \WMAP\ posterior mean temperature sky maps and CMB temperature power spectrum are largely consistent with the official \textit{WMAP9} results. Perhaps the most notable difference is that our CMB dipole amplitude is $3366.2\pm1.4\,\mathrm{\mu K}$, which is 11\muK\ higher than the \WMAPnine\ estimate and $2.5\,\sigma$ higher than \bp; however, it is in perfect agreement with the HFI-dominated \Planck\ PR4 result. In contrast, our \WMAP\ polarization maps differ more notably from the \WMAPnine\ results, and in general exhibit significantly lower large-scale residuals. We attribute this to a better constrained gain and transmission imbalance model. It is particularly noteworthy that the \W-band polarization sky map, which was excluded from the official \WMAP\ cosmological analysis, for the first time appears visually consistent with the \V-band sky map. Similarly, the long standing discrepancy between the \WMAP\ \K-band and LFI 30\,GHz maps is finally resolved, and the difference between the two maps appears consistent with instrumental noise at high Galactic latitudes. Relatedly, these updated maps allow us for the first time to combine \WMAP\ and LFI polarization data into a single coherent model of large-scale polarized synchrotron emission. Still, we identify a few issues that require additional work, including 1) low-level noise modeling, 2) large-scale temperature residuals at the 1--2\muK\ level; and 3) a strong degeneracy between the absolute $K$-band calibration and the dipole of the anomalous microwave emission component. We conclude that leveraging the complementary strengths of \WMAP\ and LFI has allowed the mitigation of both experiments' weaknesses, and resulted in new state-of-the-art \WMAP\ sky maps. All maps and the associated code are made publicly available through the \cosmoglobe\ web page. 
}


\maketitle

\tableofcontents

\section{Introduction}
\label{sec:introduction}

The discovery of the cosmic microwave background (CMB) by \citet{penzias:1965} marked a paradigm shift in the field of cosmology, providing direct evidence that the Universe was once much hotter than it is today, effectively ruling out the steady-state theory of the universe \citep{dicke:1965}. This discovery spurred a series of ground-breaking cosmological experiments, including the Nobel Prize-winning measurements by \COBE-FIRAS that confirmed the blackbody nature of the CMB \citep{mather:1994} and \COBE-DMR that measured temperature variations from the primordial gravitational field \citep{smoot:1992}.

The NASA-funded \textit{Wilkinson Microwave Anisotropy Probe} (\WMAP; \citealp{bennett2003:MAP}) mission was launched a decade after \COBE-DMR, and mapped the microwave sky with 45 times higher sensitivity and 33 times higher angular resolution, thereby revolutionizing our understanding of early universe physics \citep{bennett2003:MAP}.  In addition, the 3-year measurements presented by \citet{page2007} included the first ever detection of large-scale polarization in the CMB, opening a new window into the process of cosmic reionization. As quantified by \citet{bennett2012}, the permissible parameter space volume for a standard \LCDM\ model was decreased by a factor of 68,000 by \WMAP, and the best pre-\WMAP\ determination of the age of the universe was $t_0<14\,\mathrm{Gyr}$ from Boomerang \citep{lange:2001}, with best-fit values of 9--11\,Gyr; the latter values in apparent contradiction with direct measurements of the oldest globular clusters \citep{hu:2001}. 

The ESA-led \Planck\ satellite \citep{planck2016-l01} was developed concurrently with \WMAP, and their operation lifetimes partially overlapped, with \WMAP\ observing from 2001--2011 and \Planck\ from 2009--2013. \Planck's stated goal was to fully characterize the primary CMB temperature fluctuations from recombination, as well as to characterize the polarized microwave sky on large angular scales.  Overall, \Planck's raw CMB sensitivity was an order of magnitude higher than \WMAP's, and its angular resolution more than twice higher. Today, \Planck\ represents the state-of-the-art in terms of full-sky microwave sky measurements.

\Planck\ comprised two independent experiments, namely the Low Frequency Instrument (LFI; \citealp{planck2016-l02}) and High Frequency Instrument (HFI; \citealp{planck2016-l03}), respectively. The LFI detectors were based on HEMT (high electron mobility transistor) amplifiers, spanning three frequency channels between 30 and 70\,GHz, while the HFI detectors were based on spider-web and polarization sensitive bolometers, and spanned six frequency channels between 100 and 857\,GHz. For comparison, \WMAP\ was also HEMT-based, with comparable sensitivity to LFI alone, and spanned five frequencies between 23 and 94\,GHz. At the same time, the two experiments implemented very different scanning strategies, and as a result they are highly complementary and synergistic; together they provide a clearer view of the low-frequency microwave sky than either can alone.

Towards the end of the \Planck\ analysis phase it became clear that the interplay between instrument calibration and astrophysical component separation was a main limiting factor in terms of systematic effects for high signal-to-noise measurements \citep{planck2016-l02}. Specifically, in order to calibrate the instrument to sufficient precision, it became apparent that it was necessary to know the true sky to a comparably high precision -- but to know the sky, it was also necessary to know the instrumental calibration. The data analysis was thus fundamentally circular and global in nature. The final official \Planck\ LFI analysis performed four complete iterations between calibration and component separation \citep{planck2016-l02}, aiming to probe this degeneracy. However, it was recognized that this was not sufficient to reach full convergence, and this sub-optimality led to the \BP\ project \citep{bp01}, which aimed to perform thousands of complete analysis cycles, as opposed to just a handful. This framework was implemented using the \commanderthree\ \citep{bp03} code, a CMB Gibbs sampler that performs integrated high-level and low-level parameter estimation in a single integrated framework. This analysis demonstrated the feasibility of end-to-end CMB analysis through Gibbs sampling analysis, while at the same time it provided the highest-quality LFI maps to date.

Rather than simply probing the degeneracy between instrument calibration and component separation, a better solution is to actually break it. The optimal approach to do so is by jointly analyzing complementary datasets, each of which provide key information regarding the full system. This insight led to the \cosmoglobe\footnote{\url{https://cosmoglobe.uio.no}} initiative, which is an Open Source and community-wide effort that aims to derive a single joint model of the radio, microwave, and sub-millimeter sky by combining all available state-of-the-art experiments. An obvious first extension of the LFI-oriented \BP\ project is to analyze the \WMAP\ measurements in the same framework. Indeed, already as part of the \BP\ suite of papers, \citet{bp17} integrated \WMAP\ \Q-band time-ordered data (TOD) into the \commanderthree\ framework, calibrated off of the \BP\ sky model.

In this paper, we present the first end-to-end Bayesian analysis of the full \WMAP\ TOD, processed within the \commander\ framework. As such, this paper also presents the first ever joint analysis of two major CMB experiments (LFI and \WMAP) at the lowest possible level, and it therefore constitutes a major milestone of the \cosmoglobe\ initiative. We refer to the current products as \cosmoglobe\ Data Release 1 (CG1), and the scientific results from this are described in a series of four papers. The current paper gives a detailed discussion of data processing methods, instrumental parameters, frequency maps, and preliminary astrophysical results, while updated constraints on anomalous microwave emission and polarized synchrotron emission are presented by \citet{watts2023_ame} and \citet{fuskeland:2023}, respectively. \citet{eskilt:2023} use these new products to provide new constraints on cosmic birefringence. In the future, many more datasets and astrophysical components will be added to this framework, gradually providing stronger and stronger constraints on both the true astrophysical sky and the instrumental calibration of all previous experiments. 

The rest of this paper is organized as follows. In Sect.~\ref{sec:methods}, we provide a brief review of the Bayesian end-to-end statistical framework used in this work, before describing the underlying data and computational expenses in Sect.~\ref{sec:data}. The main results, as expressed by the global posterior distribution, are described in Sects.~\ref{sec:instrument}--\ref{sec:astrophysics}, summarizing instrumental parameters, frequency sky maps, and preliminary astrophysical results, respectively. In Sect.~\ref{sec:issues} we address unresolved issues that should be further analyzed in future work. We conclude in Sect.~\ref{sec:conclusions}, and lay a path forward for the \cosmoglobe\ project.

\section{End-to-end Bayesian CMB analysis}
\label{sec:methods}

The general computational analysis framework used in this work has been described in detail by \citet{bp01} and \citet{bp17} and references therein. In this section, we give a brief summary of the main points, and emphasize in particular the differences with respect to earlier work. 

\subsection{Official \WMAP\ instrument model and analysis pipeline}
\label{sec:wmap_instmodel}

The main goal of the current paper is to perform a similar analysis as \citet{bp01} did for \Planck\ LFI, but this time including the \WMAP\ in terms of time-ordered data, and thereby solve some of the long-standing unresolved issues with the official maps, in particular related to poorly constrained large-scale polarization modes. Before presenting our algorithm, however, it is useful to briefly review the official \WMAP\ instrument model and analysis pipeline, which improved gradually over a total of five data releases, often referred to as the 1-, 3-, 5-, 7-, and 9-year data releases, respectively. Unless otherwise noted, we will refer to the final 9-year results \citep{bennett2012}, and denote these as \WMAPnine. A concise summary of the \WMAP\ mission, data processing, and results is available in \citet{komatsu2014}. The full data archive can be found on LAMBDA.\footnote{\url{https://lambda.gsfc.nasa.gov/product/wmap/dr5/m_products.html}}

The \WMAP\ satellite carried twenty differential polarization-sensitive
radiometers, grouped into ten differencing assemblies (DAs), where one 
was sensitive to the difference in signal at one polarization
orientation and the other sensitive to the orthogonal
polarization. In total, of the ten DAs there were:
one \K-band (23\,GHz), one \Ka-band (33\,GHz), two \Q-bands (41\,GHz),
two \V-bands (61\,GHz), and four \W-bands (94\,GHz). Each radiometer comprised two detector diodes, which each recorded a science sample every $1.536/N_\mathrm{obs}$ seconds, where $N_\mathrm{obs}$ is 12, 12, 15, 20 and 30 for \K, \Ka, \Q, \V, and \W, respectively. The raw data are recorded as 16-bit integers with units du (digital unit). 

The \WMAP\ bandpasses were measured pre-launch on the ground, sweeping a signal source through 201 frequencies and recording the output \citep{jarosik2003:MAP}. The bandpass responses available on LAMBDA have not been updated since the initial data release. However, as noted by \citet{bennett2012}, there has been an observed drift in the center frequency of \K, \Ka, \Q, and \V-band corresponding to a $\sim$$\,0.1\,\%$ decrease over time. In practice, this did not affect the \WMAP\ data processing because each year was mapped separately and co-added afterwards. An effective frequency calculator was delivered in the DR5 release as part of the IDL library to mitigate this effect during astrophysical analyses.\footnote{\url{https://lambda.gsfc.nasa.gov/product/wmap/dr5/m_sw.html}}

The beams were characterized in the form of maps, with separate products for the central portion of the beam pattern and the far sidelobes. The main beam and near sidelobes were characterized using a combination of physical optics codes and observations of Jupiter for each horn separately. The maps of Jupiter were then combined with the best-fit parameters from physical optics codes to create a map of the beam response \citep{hill2009,weiland2010,bennett2012}.

Far sidelobes were estimated using a combination of laboratory measurements and Moon data taken during the mission \citep{barnes2003}, as well as a physical optics model described by \citet{hinshaw2009}. To remove the far sidelobe  in the TOD, an estimate was calculated by convolving the intensity map and the orbital dipole signal with the measured sidelobe signal \citep{jarosik2007}. Although the sidelobe pickup was modeled by \citet{barnes2003}, it was determined that the results were small enough to be neglected and have not been explicitly reported in any of the subsequent \WMAP\ data releases.

The \WMAP\ pointing solution was determined using the boresight vectors of individual feedhorns in spacecraft coordinates, in combination with on-board star trackers. Thermal flexure of the tracking structure introduced small pointing errors, as discussed by \citet{jarosik2007}. Using the temperature variation measured by onboard thermistors, the pointing solution was corrected using a model that returns angular deviation per kelvin. The residual pointing errors were computed using observations of Jupiter and Saturn, and the reported upper limit was estimated to be 10\arcsec\ \citep{wmapexsupp,bennett2012}.

The \WMAP\ data were calibrated by jointly estimating the time-dependent gains, $\g$, and baselines, $\bv$, as described by \citet{hinshaw2007}, \citet{hinshaw2009}, and \citet{jarosik2010}.
The TOD were initially modeled as having
constant gain and baseline for a 1--24\,hour period, with parameters that were fit to the orbital
dipole assuming $T_0$ from \citet{mather:1999} and a map made from a previous
iteration of the mapmaking procedure. Once the gain and baseline solution had
converged, the data were fit to a parametric form of the radiometer response
as a function of housekeeping data, given in Appendix~A of \citet{wmapexsupp}.

\WMAP\ had two primary mirrors positioned on opposite sides of the vertical satellite axis, tilted approximately $19.5^\circ$ towards the Solar shield. Essentially, when horn A was pointed at pixel $p_\A$, horn B was pointed at a pixel $p_\B$ approximately $141^\circ$ away \citep{page2003:MAP}. The incoming radiation was differenced in the electronics before being deposited on the detectors, recording radiation proportional to the observed maps $\boldsymbol m$ at their respective pixels, $\boldsymbol m_{p_\A}-\boldsymbol m_{p_\B}$ and $\boldsymbol m_{p_\B}-\boldsymbol m_{p_\A}$ \citep{jarosik2003:MAP}. Each radiometer had a partner that observed the same pixels with sensitivity to the orthogonal polarization direction. Taking all these effects into account, the total data model for a single radiometer is given by
\begin{align}
	d_t^\mathrm{imbal}&\propto (1+x_\mathrm{im})T_{p_\A}-(1-x_\mathrm{im})T_{p_\B}
	\\
	&=(T_{p_\A}-T_{p_\B})+x_\mathrm{im}(T_{p_\A}+T_{p_\B}),
\end{align}
where $T_{p_\A}$ and $T_{p_\B}$ are the A- and B-side antenna temperatures, and $x_\mathrm{im}$ is the differential optical pickup between horns A and B. This effect is taken into account during mapmaking. However, inaccuracies in the determination of $x_\mathrm{im}$ will yield a spurious polarization component, and create artificial imbalance modes due to coupling with the sky signal, in particular with the bright Solar CMB dipole \citep{jarosik2007}. The \WMAP\ transmission imbalance factors were fit to the Solar dipole in TOD space, accounting for both common and differential modes \citep{jarosik2003a,jarosik2007}.

Data were flagged and masked before the final mapmaking step. In particular, station-keeping maneuvers, solar flares, and unscheduled events caused certain data to be unusable -- the full catalog of these events is listed in Table~1.8 of \citet{wmapexsupp}. In addition, data were masked depending on the channel frequency and the planet itself, with the full list of exclusion radii enumerated in Table~4 of \citet{bennett2012}.

To create the sky maps $\boldsymbol m$, the calibrated data were put into the asymmetric mapmaking equation,
\begin{equation}
	\label{eq:asymmetric_mapmaking}
	\mathsf P_\mathrm{am}^T\mathsf N^{-1}\mathsf P\boldsymbol m=\mathsf P^T_\mathrm{am}\mathsf N^{-1}\boldsymbol d,
\end{equation}
where $\mathsf N$ is the noise covariance matrix, and the pointing matrix $\mathsf P$ is implicitly defined for each datastream, $\boldsymbol d_1$ and $\boldsymbol d_2$ sensitive to different polarization orientations. 
The asymmetric mapmaking matrix, $\mathsf P_\mathrm{am}$, was used because, as noted by \citet{jarosik2010}, large signals observed in one beam could leak into the solution for the pixel observed by the other beam, leading to incorrect signals in the final map. The asymmetric mapmaking solution is defined by only updating the matrix multiplication for beam A when beam A is in a high emission region and beam B is not, and vice versa. \citet{bennett2012} also identified that these effects are pronounced when one horn is crossing a large temperature gradient, leading to excesses $140^\circ$ away from the Galactic center if an appropriate processing mask is not used.
For each side $\A/\B$, the maps are defined as a function of the Stokes parameters $T_{\A/\B}$, $Q_{\A/\B}$, and $U_{\A/\B}$, with polarization angle $\gamma_{\A/\B}$, such that
\begin{align}
	\boldsymbol d_1&=\mathsf P_1\boldsymbol m
	\nonumber
	\\
	&=(1+x_\mathrm{im})[T_\A+Q_\A\cos2\gamma_\A+U_\A\sin2\gamma_\A+S_\A]
	\nonumber
	\\
	&+(1-x_\mathrm{im})[-T_\B-Q_\B\cos2\gamma_\B-U_\B\sin2\gamma_\B-S_\B],
	\label{eq:mapeq1}
\end{align}
and
\begin{align}
	\boldsymbol d_2&=\mathsf P_2\boldsymbol m
	\nonumber
	\\
	&=(1+x_\mathrm{im})[T_\A-Q_\A\cos2\gamma_\A-U_\A\sin2\gamma_\A-S_\A]
	\nonumber
	\\
	&+(1-x_\mathrm{im})[-T_\B+Q_\B\cos2\gamma_\B+U_\B\sin2\gamma_\B+S_\B].
	\label{eq:mapeq2}
\end{align}
In this formalism, $S_{\A/\B}$ acts as an extra Stokes parameter that absorbs the effects of differing bandpass responses between radiometers $\boldsymbol d_1$ and $\boldsymbol d_2$ \citep{jarosik2007}.

An accurate noise model was necessary both to perform the maximum likelihood mapmaking and for the evaluation of the dense time-space inverse noise covariance matrix  $\mathsf N^{-1}$. The \WMAP\ team defined this in the form of a time domain autocorrelation function that was estimated separately for each year of data. This was then Fourier transformed, inverted, and inverse Fourier transformed to create an effective inverse noise operator $N_{tt'}^{-1}$. 
Finally, to create the sky maps themselves, the \WMAP\ team processed the data one year at a time, producing maps by solving Eq.~\eqref{eq:asymmetric_mapmaking} using the iterative Bi-Conjugate Gradient Stabilized Method \citep[BiCG-STAB,][]{bicgstab,bicgstab_template}.

\subsection{\cosmoglobe\ instrument model}
\label{sec:cosmoglobe_instmodel}

\begin{figure*}
	\includegraphics[width=\textwidth]{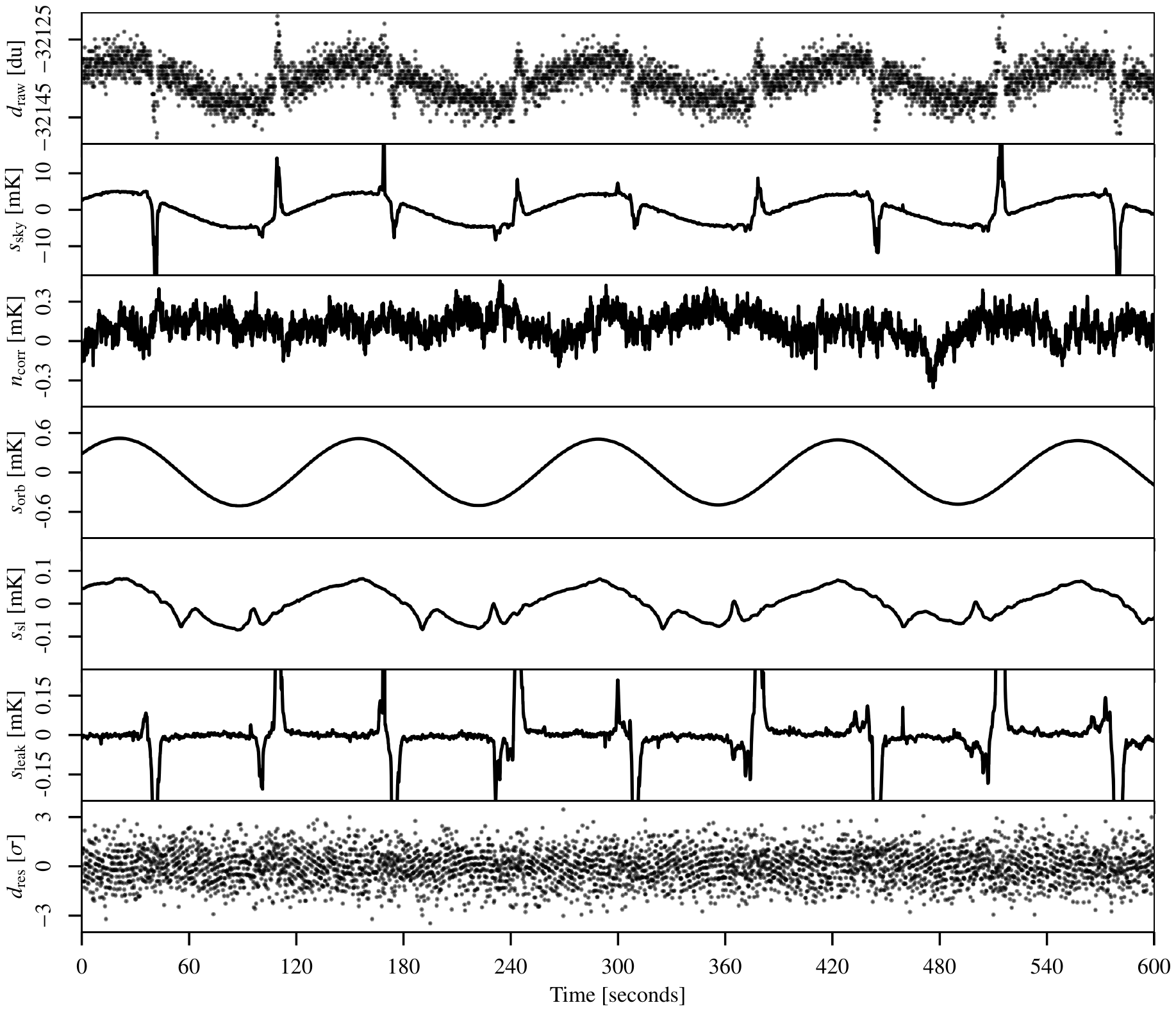}
	\caption{Time-ordered data segment for the \K113 diode. From top to bottom, the panels show 1) raw uncalibrated TOD $\boldsymbol d$; 2) sky signal $\boldsymbol s_\mathrm{sky}$; 3) calibrated correlated noise $\boldsymbol n_\mathrm{corr}$; 4) orbital CMB dipole signal $\boldsymbol s_\mathrm{orb}$; 5) sidelobe correction $\boldsymbol s_\mathrm{sl}$; 6) bandpass leakage correction $\boldsymbol s_\mathrm{leak}$; and 7) residual TOD, $\boldsymbol d_\mathrm{res}=(\boldsymbol d -\boldsymbol n_\mathrm{corr}-\boldsymbol b)/g-\boldsymbol s_\mathrm{sky}-\boldsymbol s_\mathrm{orb}-\boldsymbol s_\mathrm{leak} -\boldsymbol s_\mathrm{sl}$, in units of $\sigma_0[\mathrm{du}]$ for this TOD segment. Note that the vertical range varies and units vary from panel to panel.
		}
	\label{fig:timestreams}
\end{figure*}

A fundamental difference between the \cosmoglobe\ and \WMAP\ analysis pipelines (and those of most other CMB experiments) is that while the \WMAP\ pipeline models each channel in isolation, the \cosmoglobe\ framework simultaneously considers all data, both internally within \WMAP, and also from all other sources, and most notably from \Planck\ LFI. The main advantage of such a global approach is significantly reduced parameter degeneracies, as data from observations with different frequency coverages and instrumental designs break the same degeneracies. For this approach to be computationally tractable, one must establish a global parametric model that simultaneously accounts for both the astrophysical sky and all relevant instruments. For the current \WMAP+LFI oriented analysis, we adopt the following expression \citep{bp01},
\begin{equation}
	\label{eq:model}
	\boldsymbol d =\mathsf G\mathsf P[\mathsf B^\mathrm{symm}\mathsf M\boldsymbol a+\mathsf B^\mathrm{4\pi}(\boldsymbol s^\mathrm{orb}
	+\boldsymbol s^\mathrm{fsl})] + \boldsymbol s^\mathrm{inst}+ \boldsymbol n^\mathrm{corr}+\boldsymbol n^\mathrm w,
\end{equation}
where $\mathsf G$ is the time-dependent gain in the form of the matrix $\mathrm{diag}(g_t)$; $\mathsf P$ is the $n_p\times n_\mathrm{TOD}$ pointing matrix, where $n_p$ is the number of pixels and $n_\mathrm{TOD}$ length of the TOD;
$\mathsf B^\mathrm{symm}$ and $\mathsf B^{4\pi}$ are the symmetrized and full asymmetric beam, respectively; $\mathsf M$ is the mixing matrix between a given sky component $c$ with spectral energy distribution $f_c(\nu/\nu_{0,c})$ and a detector $i$ with bandpass $\tau_i(\nu)$, given by
\begin{equation}
	\mathsf M_{ci}=\int d\nu\,\tau_i(\nu)f_c(\nu/\nu_{c,0}).
\end{equation}
(In practice, $\M$ also accounts for unit conversion, but this is suppressed for readability in this expression; see \citealp{bp09} for further details.)
The maps $\boldsymbol a$ represent the Stokes parameters for each astrophysical component, while $\boldsymbol s^\mathrm{orb}$ is the orbital dipole induced by the motion of the telescope with respect to the Sun, and $\boldsymbol s^\mathrm{fsl}$ is the time-dependent far sidelobe signal. Following \citet{bp06}, we model the correlated noise component $\boldsymbol n^\mathrm{corr}$ in terms of a $1/f$ power spectral density (PSD), which explicitly takes the form $P_n(f) = \sigma^2_0 (1 + (f/f_\mathrm{knee})^\alpha)$, where $\sigma_0$ denotes the white noise amplitude, $f_\mathrm{knee}$ is the so-called $1/f$ knee frequency, and $\alpha$ is a free power law slope. For notational purposes, we denote the set of all correlated noise parameters by $\boldsymbol\xi_n = \{\sigma_0, f_{\mathrm{knee}}, \alpha\}$. We note that this model represents a significant approximation compared to the more flexible \WMAP\ autocorrelation model, as the actual \WMAP\ noise is known to be colored  at high temporal frequencies \citep{jarosik2007}. The main impact of this approximation is a worse-than-expected $\chi^2$ goodness of fit statistic. However, measured in absolute noise levels the effect is very small, and has very little if any impact on the final science results; for further discussion of this approximation, see Sect.~\ref{sec:noisemodel}. 

The term $\boldsymbol s^\mathrm{inst}$ denotes any instrument-specific terms that might be required for a given experiment. For instance, for LFI it is used to model the 1\,Hz spike contribution due to electronic cross-talk. For \WMAP, we use it for first-order baseline corrections, and set $s^\mathit{WMAP}_t = b_0 + b_1\,t$, where $b_0$ and $b_1$ represent the mean and slope of the baselines over the data segment in question. We note that while the \WMAP\ team fitted a single constant baseline over either 1- or 24-hour periods, our data segments are typically several days long (corresponding to a number of samples chosen to optimize Fourier transforms). A natural question is therefore whether nonlinear baseline variations could induce artifacts. In this regard, it is important to note that the correlated noise component effectively acts as a single-sample baseline correction that can absorb by far most such nonlinearities, as long as their total effect on the power spectrum does not exceed that imposed by the $1/f$ model. In practice, this is a very mild constraint. At the same time, visual inspection of $\n^{\mathrm{corr}}$ projected into sky maps provides a very powerful check on any potential baseline residuals, which will appear as correlated stripes aligned with the \WMAP\ scanning path; for the full set of correlated noise maps derived for all ten \WMAP\ DAs, see Fig.~\ref{fig:ncorr} in Appendix~\ref{sec:map_survey}. Such maps have been used to identify and mitigate modeling errors several times in the course of this analysis. In sum, it is important to note that the \cosmoglobe\ model allows for a more flexible baseline behaviour than the \WMAP\ pipeline, even though the dedicated baseline parameters themselves apply to relatively long timescales.

A third notable difference between the \WMAP\ and \cosmoglobe\ data models concerns bandpass mismatch. While the \WMAP\ pipeline simply projects out any bandpass difference from the polarization maps by solving for the spurious $S$ maps, we model it explicitly through the use of the global astrophysical sky model \citep{bp09}. Explicitly, the expected calibrated sky signal for diode $i$ is given by
\begin{equation}
	m_{p,i}=\mathsf B_{p,p'}\sum_c\mathsf M_{c,i}a^c_{p'}+n_{i,p}^\mathrm w.
\end{equation}
Since $\mathsf M_{c,i}$ encodes the bandpass response of every detector $i$ to every sky component $c$, the detector-specific maps, $\boldsymbol m_i$, will each be slightly different depending on their bandpass $\tau_i$. Therefore, before averaging different detectors together, we estimate the average over all detectors in a given frequency channel $\boldsymbol m\equiv \langle \boldsymbol m_i\rangle$, and subtract it directly in the timestream;
\begin{equation}
	\delta s_{t,i}^\mathrm{leak}=\mathsf P_{t,p}^i\mathsf B_{p,p'}^i\left(\boldsymbol m_{i,p'}-\boldsymbol m_{p'}\right).
\end{equation}
This leakage term uses the expected bandpass response to remove the expected component that deviates from the mean in the timestream, directly reducing polarization contamination.%

To build intuition regarding this model, we plot in Fig.~\ref{fig:timestreams} both the TOD and the individual model components for an arbitrarily selected ten-minute segment for the \WMAP's \K113 diode. The uncalibrated data, $\boldsymbol d_\mathrm{raw}$, are displayed in the top panel, with the sky signal $\boldsymbol s_\mathrm{sky}=\mathsf P\mathsf B^\mathrm{symm}\mathsf M\boldsymbol a$ plotted directly underneath. The next four panels show the correlated noise realization $\boldsymbol n_\mathrm{corr}$, the orbital dipole $\boldsymbol s_\mathrm{orb}$, the far sidelobe contribution $\boldsymbol s_\mathrm{sl}$, and the bandpass leakage $\boldsymbol s_\mathrm{leak}$. Finally, we also plot the time-ordered residual for this segment of data, obtained by subtracting the model from the raw data, in units of the estimated white noise level.

\subsection{Sky model}\label{subsec:sky_model}

Following \citet{bp01}, we assume that the sky across the frequency range of interest can be modeled as a linear combination of CMB fluctuations ($\vec{a}_{\mathrm{CMB}}$), synchrotron ($\vec{a}_{\mathrm{s}}$), free-free emission ($\vec{a}_{\mathrm{ff}}$), anomalous microwave emission (AME; $\vec{a}_{\mathrm{AME}}$), thermal dust ($\vec{a}_{\mathrm{d}}$), and radio point sources ($\vec{a}_{j,\mathrm{src}}$). Explicitly, we assume that the astrophysical sky (in units of brightness temperature) may be modeled as follows,
\begin{align}
  \vec{s}_{\mathrm{RJ}} &= \left(\vec{a}_{\mathrm{CMB}}+\vec{a}_{\mathrm{quad}}(\nu)\right) \frac{x^2 \e^x}{(\e^x -1)^2}+\label{eq:cmb_astsky}\\
  &+ \vec{a}_{\mathrm{s}} \left(\frac{\nu}{\nuzeros}\right)^{\bsynch} + \label{eq:synch_astsky}\\
  &+ \vec{a}_{\mathrm{ff}} \left(\frac{\nuzeroff}{\nu}\right)^2 \frac{g_{\mathrm{ff}}(\nu;\Te) }{g_{\mathrm{ff}}(\nuzeroff;\Te)} +\label{eq:ff_astsky}\\
  &+ \vec{a}_{\mathrm{AME}} \e^{\beta_{\mathrm{AME}}(\nu-\nu_{0,\mathrm{AME}})}+\label{eq:ame_astsky}  \\
  &+ \vec{a}_{\mathrm{d}} \left(\frac{\nu}{\nuzerod}\right)^{\bdust+1} \frac{\e^{h\nuzerod/\kB\Tdust}-1}{\e^{h\nu/\kB\Tdust}-1}+ \label{eq:dust_astsky}\\
  &+ U_{\mathrm{mJy}} \sum_{j=1}^{N_{\mathrm{src}}} \vec{a}_{j,\mathrm{src}} \left(\frac{\nu}{\nuzerosrc}\right)^{\alpha_{j,\mathrm{src}}-2}, \label{eq:sum_ptsrc}
\end{align}
where $x=h\nu/kT_{\mathrm{CMB}}$; $\nu_{0,c}$ is a reference frequency for component $c$; $\beta_{\mathrm{s}}$ is a power-law index for synchrotron emission (which may take different values for temperature and polarization); $T_e$ is the electron temperature, and $g_{\mathrm{ff}}$ is the so-called Gaunt factor \citep{dickinson2003}; $\beta_{\mathrm{AME}}$ is an exponential scale factor for AME emission (see below); $\beta_{\mathrm{d}}$ and $T_{\mathrm{d}}$ are the emissivity and temperature parameters for a single modified blackbody thermal dust model; $\alpha_{j,\mathrm{src}}$ is the spectral index of point source $j$ relative to the same source catalog as used by \citet{planck2016-l04}; and $U_{\mathrm{mJy}}$ is the conversion factor between flux density (in millijansky) and brightness temperature (in $\mathrm{K_{RJ}}$) for the channel in question. Finally, $\a_{\mathrm{quad}}$ accounts for a relativistic quadrupole correction due to the Sun's motion through space \citep{Notari:2015}.

In general, this model is nearly identical to the one adopted by \citet{bp01}. However, there is one notable exception, namely the spectral energy density (SED) for the AME component, $\boldsymbol s_0^\mathrm{sd}(\nu)$. In this work, we adopt a simple exponential function for this component, as for instance proposed by \citet{hensley:2015}, and this is notably different from the \texttt{SpDust2} model \citep{ali-haimoud:2009, ali-haimoud:2010, silsbee:2011} that was used in the \bp\ analysis. The motivation for this modification is discussed in detail by \citet{watts2023_ame}. First and foremost, the current combination of \WMAP\ and LFI data appears to prefer a higher AME amplitude at frequencies between 40 and 60\,GHz than can easily be supported by \texttt{SpDust2}. This was first noted by \citet{planck2014-a12}, who solved this issue by introducing a second independent AME component. For the original \bp\ analysis, on the other hand, this excess was not statistically significant, simply because that analysis did not include the powerful \WMAP\ \K-band data. In the current analysis, the excess is obvious. The observation that a simple one-parameter exponential model fits the data as well as the complicated multi-parameter model of \citet{planck2014-a12} is a novel result from the current work. Indeed, it performs about as well as the commonly used log-normal model derived by \citet{Stevenson_2014}, which also has one extra parameter. By virtue of having fewer degrees of freedom than any of the previous models, we adopt the exponential model.

\subsection{Priors and poorly measured modes}
\label{sec:priors}

\begin{figure}
	\centering
	\includegraphics[width=\columnwidth]{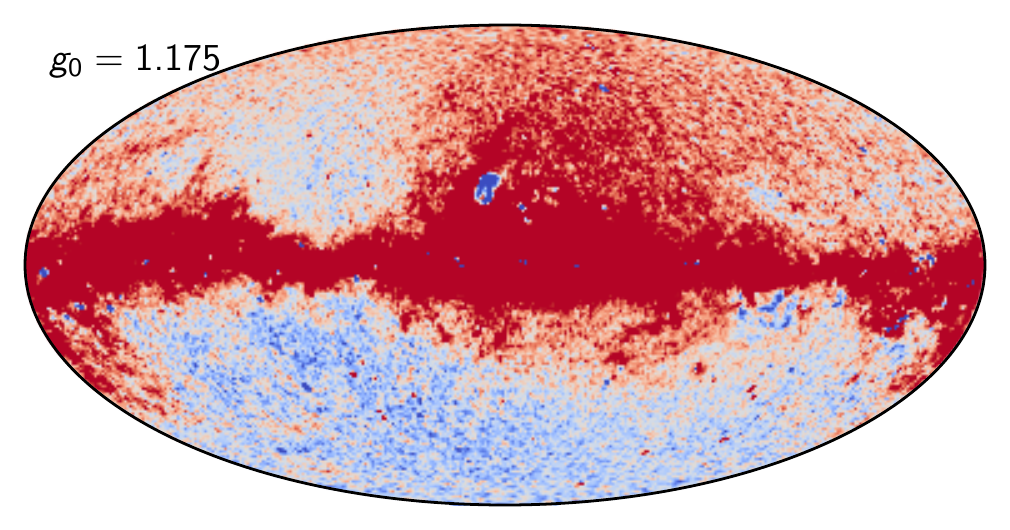}
	\includegraphics[width=\columnwidth]{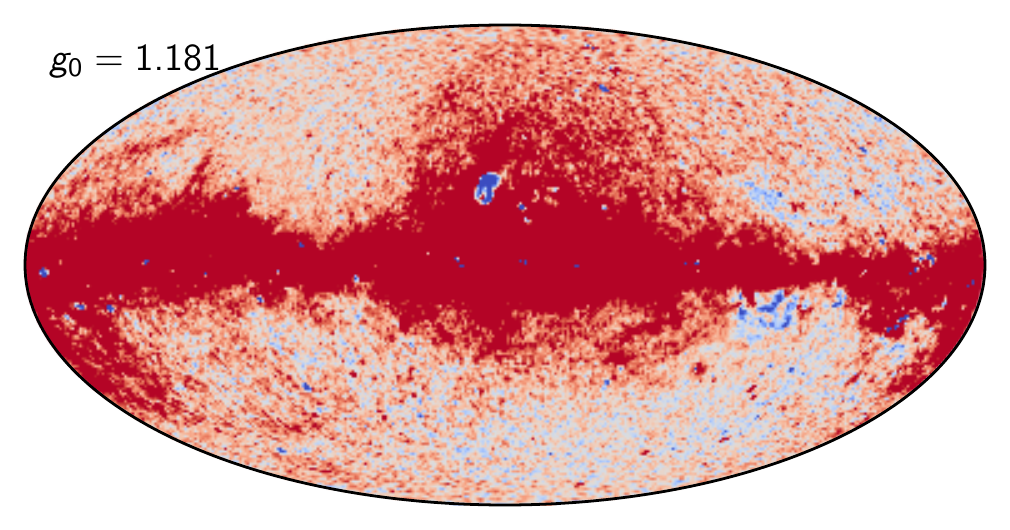}
	\includegraphics[width=\columnwidth]{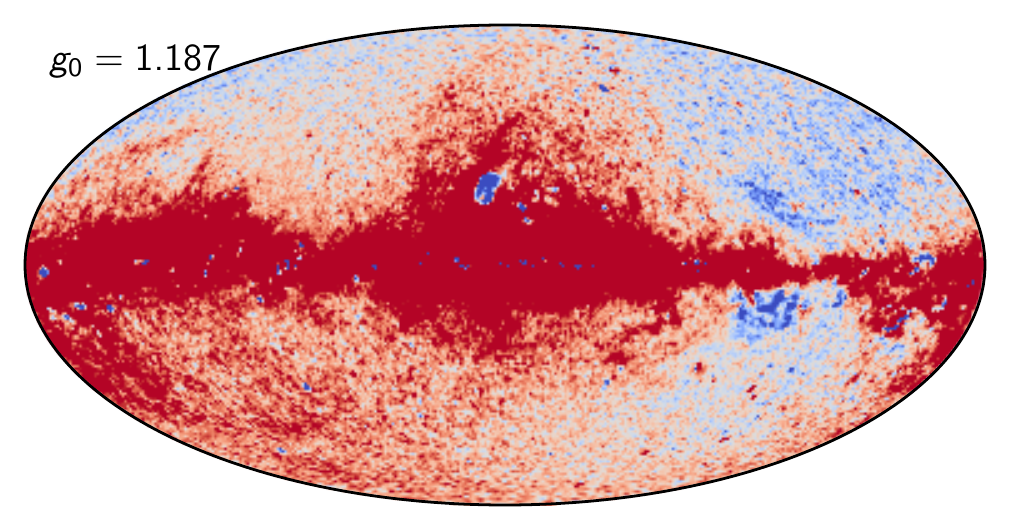}
	\includegraphics[width=0.5\columnwidth]{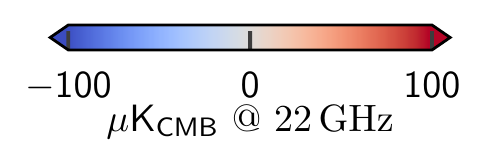}
	\caption{Dependence of AME amplitude evaluated at 22\,GHz on the absolute calibration. Each map comes from the fifth iteration of a dedicated \commanderthree\ run that fixed $g_0$ while letting all other TOD parameters be fit. The values of $g_0=1.175$ and $g_0=1.187$ represent $6\sigma$ draws from the prior distribution with mean $1.181$ and standard deviation $0.001$. Extreme outliers were chosen to illustrate this effect. The dipole visible in the top and bottom panels is aligned perfectly with the Solar dipole, and is directly due to variations in the \K-band absolute calibration.}
	\label{fig:g0_ame}
\end{figure}

The model described in Sects.~\ref{sec:cosmoglobe_instmodel} and \ref{subsec:sky_model} is prone to several degeneracies, allowing for unphysical solutions to be explored in the Gibbs chain. Such unphysical degeneracies are highly undesirable for two main reasons. First, they increase the statistical uncertainties on most (if not all) other important parameters in the model -- sometimes to the point that the target quantity is rendered entirely unmeasurable. Secondly, and perhaps even more importantly, the data model described above is known to be a (sometimes crude) approximation to the real observations, and there will invariably be modeling errors. Degeneracies tend to amplify their impact, in the sense that any unconstrained parameters will typically be used to fit such small modeling errors. For both these reasons, it is preferable to impose either informative or algorithmic priors on the unconstrained parameters, rather than to leave them entirely unconstrained in the model.

An important example of an algorithmic prior is the foreground smoothing prior used by \citet{planck2016-l04} and \citet{bp13}, which dictates that astrophysical foregrounds must be smooth on small angular scales. This is justified by noting that the angular spectrum on large and intermediate scales typically falls as a power-law in multipole space; extrapolating this into the noise dominated regime prevents the overall foreground model from becoming degenerate at small scales.

Correspondingly, important examples of informative priors are the use of HFI constraints on the thermal dust SED parameters, $\beta_{\mathrm{d}}$ and $T_{\mathrm{d}}$ in \bp. Because that analysis only included the highest HFI frequency channel, they had very little constraining power on the thermal dust SED. Rather than trying to fit these directly from LFI \WMAP\ alone, they instead imposed informative Gaussian priors on each of these parameters, as derived from the HFI observations \citep{planck2016-l04}.

Unless otherwise noted, we adopt the same algorithmic and informative priors as \citet{bp01}. However, there are three notable exceptions, as detailed below. All of these are dictated either by the fact that we include the \WMAP\ \K-band channel (which has a strong impact on the low-frequency foreground model), or by the fact that we now process the \WMAP\ data in the time domain, and therefore are subject to the same degeneracies as the official \WMAP\ low-level pipeline; degeneracies that were solved with either implicit or explicit priors in the original analysis.

First and foremost, and discussed further in Sect.~\ref{sec:ame_Kband}, we observe a very strong degeneracy between the absolute calibration of the \K-band channel and the dipole of the AME map. This is not unexpected, considering \K-band is by far the strongest channel in terms of AME signal-to-noise ratio, exceeding that of LFI 30\,GHz by about 50\,\% (see Sect.~\ref{sec:s2n} for details). Effectively, a small variation in the absolute gain may be countered by subtracting the corresponding CMB Solar dipole variation from the AME map, and end up with a nearly identical total $\chi^2$; the orbital CMB dipole is not bright enough at 23\,GHz relative to AME emission to break this degeneracy on its own.

This is illustrated in Fig.~\ref{fig:g0_ame}, which shows the AME amplitude map as derived for three different values of the mean \K-band gain, $g_0$, namely 1.175, 1.181, and $1.184\,\mathrm{du\,mK^{-1}}$; the extreme values differ only by 0.5\,\%. All of these three values appear equally acceptable from a pure $\chi^2$ point-of-view, relative to the noise level and modeling errors of these data. At the same time, it is clear from visual inspection that only the middle value actually makes physical sense, given what we know about the structure of the Milky Way. For this reason, we apply a Gaussian prior on the absolute \K-band gain of $g_0 \sim \mathcal N(1.181, 0.001^2)$ to regularize this issue. Thus, the extreme panels in Fig.~\ref{fig:g0_ame} represent $\pm6\,\sigma$ outliers, respectively, and will appear in our Markov chains with a frequency of about 1-in-$10^9$.

It is reasonable to ask why the \WMAP\ pipeline produced sensible results without applying such a prior during their calibration procedure. We posit that the answer is due to the main difference between the two approaches. While \cosmoglobe\ attempts to fit a single overall parametric model to all data at once, the \WMAP\ pipeline calibrated each channel independently by co-adding data from one channel into a map, subtracting that map from the TOD, fitting the gain to the orbital dipole, and iterating until the solution became stable. An advantage of the single-channel approach is that the solution is independent of the assumed sky model. However, a disadvantage is that it is impossible to break any potential inherent degeneracies; it cannot be combined with external observations in any meaningful way. One important example of this regarding the \WMAP\ data is a strong degeneracy between the transmission imbalance factors and the polarized sky signal; it is exceedingly difficult to break this degeneracy using data from only one DA alone, and the resulting errors will propagate to most other aspects in the analysis. In the global approach, on the other hand, the polarization modes that are poorly measured by \WMAP\ alone are well measured by \Planck\ and vice-versa, resulting in an overall better constrained fit. 

Second, as reported by \citet{bp14} for the \BP\ analysis, another important degeneracy in the current global model concerns the spectral index of polarized synchrotron emission versus the time-variable detector gain; when fitting both the polarized synchrotron amplitude and calibration freely without priors, the synchrotron spectral index at high Galactic latitudes tend to be biased toward unreasonably flat values, $\beta_\mathrm s \lesssim -2.5$, which was likely due to a low level of unmodeled systematics, for instance temperature-to-polarization leakage, rather than true polarized synchrotron emission. In turn, this resulted in a contaminated CMB sky map with a strong synchrotron morphology. To break this degeneracy, \citet{bp14} chose to marginalize the high-latitude synchrotron spectral index over a Gaussian prior of $\mathcal N(-3.30,0.1^2)$, informed by \citet{planck2016-l05}, rather than estimate it from the data themselves. We observe the same degeneracy, and the introduction of the \K-band data is not sufficient to break it on its own. For this reason, we choose to apply the same informative prior.

Third and finally, we also marginalize over the AME scale index with a prior of $\beta_{\mathrm{AME}}\sim\mathcal N(3.56, 0.1^2)$. The parameters of these priors were determined by running a grid over $\beta_{\mathrm{AME}}$, and identifying the range that resulted in reasonable residuals near the Galactic plane, similar to that shown in Fig.~\ref{fig:g0_ame} for the absolute calibration of \K-band. We note that this prior should in principle be replaced with direct $\chi^2$-based posterior optimization, combined with a properly tailored analysis mask. However, the recent release of the QUIJOTE data \citep{QUIJOTE_IV}, which covers the 11--19\,GHz frequency range, suggests that the entire AME model should be revisited in a future joint \WMAP+LFI+QUIJOTE analysis. We therefore leave detailed prior and SED optimization to that work. For further information regarding AME modeling with the current dataset, we refer the interested reader to \citet{watts2023_ame}.

In sum, we impose strong priors on all foreground spectral indices, with parameters that are informed by the requirement of obtaining physically meaningful component maps. These strong priors imply that the foreground spectral parameters sampled within the chain do not carry independent significance in the traditional posterior sense, but are in practice only nuisance parameters used to marginalize over externally defined uncertainties. 

\subsection{Posterior distribution and Gibbs sampling}

As shown by \citet{bp01}, this joint parametric description of the instrumental effects and sky allows us to write down a total model for the data, $\boldsymbol d=\boldsymbol s^\mathrm{tot}(\boldsymbol\omega)+\boldsymbol n^\mathrm w$, where $\boldsymbol s^\mathrm{tot}$ encompasses all of the terms in Eq.~\eqref{eq:model} except for the white noise term. Assuming that all instrumental effects have been modeled adequately, and that the white noise is Gaussian distributed, the data should then also be Gaussian distributed with a mean of $\boldsymbol s^\mathrm{tot}(\boldsymbol\omega)$ and variance $\boldsymbol \sigma_0^2$. In general, the likelihood reads
\begin{equation}
	P(\boldsymbol d\mid\boldsymbol\omega)\propto\exp\left(-\frac12\sum_t\frac{(d_t-s^\mathrm{tot}_t(\boldsymbol\omega))^2}{\sigma_0^2}
	\right).
\end{equation}
If $\boldsymbol d\sim\mathcal N(\boldsymbol s^\mathrm{tot},\boldsymbol\sigma_0^2)$ is the correct model for the data, the argument of the exponent is proportional to a $\chi^2$-distribution with $n_\mathrm{TOD}$ degrees of freedom, where $n_\mathrm{TOD}$ number of datapoints within a given datastream. In the limit of large $n$, a $\chi^2$ distribution is well-approximated by a Gaussian with mean $n$ and variance $2n$. Therefore we define and use the reduced normalized $\chi^2$ statistic,
\begin{equation}
	\chi^2\equiv \frac{\sum_t((d_t-s_t^\mathrm{tot})/\sigma_0)^2 - n_\mathrm{TOD}}{\sqrt{2n_\mathrm{TOD}}},
  \label{eq:chisq}
\end{equation}
which is approximately drawn from the standard normal distribution $\mathcal N(0,1)$.

Following \citet{bp01}, the \cosmoglobe\ Gibbs chain for this analysis is given by
\begin{alignat}{10}
\label{eq:samp_inst}\boldsymbol s^\mathrm{inst} &\,\leftarrow                                 P(\boldsymbol s^\mathrm{inst}&\,             \mid \data, &\,\phantom{\boldsymbol s^{\mathrm{inst}},} &\,\g, &\,\boldsymbol x_{\mathrm{im}}, &\,\ncorr, &\,\boldsymbol\xi_n,      &\,\boldsymbol\beta, &\,\boldsymbol a, &\,C_{\ell})\\
\label{eq:gain_samp_dist}\g &\,\leftarrow          P(\g&\,               \mid \data, &\,\boldsymbol s^{\mathrm{inst}}, &\,\phantom{\g,} &\,\boldsymbol x_{\mathrm{im}}, &\,\ncorr, &\,\boldsymbol\xi_n,      &\,\boldsymbol\beta, &\,\boldsymbol a, &\,C_{\ell})\\
\label{eq:xim} \boldsymbol x_{\mathrm{im}} &\,\leftarrow        P(\boldsymbol x_{\mathrm{im}}&\, \mid \data, &\,\boldsymbol s^{\mathrm{inst}}, &\,\g, &\,\phantom{\boldsymbol x_{\mathrm{im}},} &\,\ncorr, &\,\boldsymbol\xi_n,      &\,\boldsymbol\beta, &\,\boldsymbol a, &\,C_{\ell})\\
\label{eq:ncorr_samp_dist} \ncorr &\,\leftarrow    P(\ncorr&\,        \mid \data, &\,\boldsymbol s^{\mathrm{inst}}, &\,\g, &\,\boldsymbol x_{\mathrm{im}}, &\,\phantom{\ncorr,} &\,\boldsymbol\xi_n,      &\,\boldsymbol\beta, &\,\boldsymbol a, &\,C_{\ell})\\
\label{eq:xi_samp_dist} \boldsymbol\xi_n &\,\leftarrow        P(\boldsymbol\xi_n&\,            \mid \data, &\,\boldsymbol s^{\mathrm{inst}}, &\,\g, &\,\boldsymbol x_{\mathrm{im}}, &\,\ncorr, &\,\phantom{\boldsymbol\xi_n,}      &\,\boldsymbol\beta, &\,\boldsymbol a, &\,C_{\ell})\\
\label{eq:beta_samp}\boldsymbol\beta &\,\leftarrow                     P(\boldsymbol\beta)\\
\boldsymbol a &\,\leftarrow                                   P(\boldsymbol a&\,            \mid \data, &\,\boldsymbol s^{\mathrm{inst}}, &\,\g, &\,\boldsymbol x_{\mathrm{im}}, &\,\ncorr, &\,\boldsymbol\xi_n,      &\,\boldsymbol\beta, &\,\phantom{\boldsymbol a,} &\,C_{\ell})\\
C_{\ell} &\,\leftarrow                             P(C_{\ell}&\,         \mid \boldsymbol a)\label{eq:cl_sampling}
\end{alignat}
with each step requiring its own dedicated sampling algorithm. The \commanderthree\ pipeline is designed so that results of each Gibbs sample can be easily passed to each other, and that the internal calculations of each step do not directly depend on the inner workings of each other, which greatly increases modularity of the code.

\subsection{Sampling algorithms}
\label{sec:algorithms}

Before we discuss the results of this Gibbs chain as applied to the \Planck\ LFI and \WMAP\ data, we summarize the TOD processing steps in this section. Each step of the Gibbs chain requires its own conditional distribution sampling algorithm. In Sect.~\ref{ssec:oldsamplers} we review the sampling algorithms implemented in the \bp\ suite of papers, while Sects.~\ref{ssec:mapmaking}--\ref{ssec:baseline} provide an overview of the \WMAP-specific processing steps.

\subsubsection{Review of sampling algorithms}
\label{ssec:oldsamplers}

Most of the techniques required for \WMAP\ data analysis have already been  described in the \bp\ project and implemented in \commanderthree. This section includes a summary of the algorithms that were used previously for the analysis of LFI data. In each of these cases, every part of the model not explicitly mentioned is held fixed unless specified otherwise.

Noise estimation and calibration are described by \citet{bp06} and \citet{bp07}, respectively. As noted in those works, these two steps are strongly correlated, simply because the timestream
\begin{equation}
	\boldsymbol d_i=\boldsymbol g_i\boldsymbol s_{i}^\mathrm{tot}+\boldsymbol n_{i}^\mathrm{corr}+\boldsymbol n_{i}^\mathrm{w}
\end{equation}
may be almost equally well fit by two solutions defined schematically by
\begin{equation}
	\boldsymbol g'=\boldsymbol g\boldsymbol s^\mathrm{tot}/(\boldsymbol s^\mathrm{tot})'
\end{equation}
or 
\begin{equation}
	{(\boldsymbol n^\mathrm{corr})'=\boldsymbol n^\mathrm{corr}+\boldsymbol g\boldsymbol s^\mathrm{tot}+\boldsymbol g'(\boldsymbol s^\mathrm{tot})'};
\end{equation} the only thing that breaks this degeneracy is the noise PSD, which is a relatively loose constraint. A Gibbs sampler is not very effective for nearly degenerate distributions, and we therefore instead define a joint sampling step for the correlated noise and gain. In practice, this is done by first drawing the calibration from its marginal distribution with respect to $\boldsymbol n^\mathrm{corr}$, and then drawing $\boldsymbol n^\mathrm{corr}$ from its conditional distribution with respect to $\boldsymbol g$,
\begin{alignat}{10}
	\g&\,\leftarrow P(\g&\,\mid\data, &\, &\,&\,\boldsymbol\xi_n)
	\label{eq:gmarg}
	\\
	\ncorr&\,\leftarrow P(\ncorr&\,\mid\data, &\,\g, &\,&\,\boldsymbol\xi_n).
	\label{eq:ncorr_only}
\end{alignat}
One can see that this is a valid sample from the joint distribution from the definition of a conditional distribution, $P(\g,\ncorr\mid\boldsymbol\omega)=P(\ncorr\mid\g,\boldsymbol\omega)P(\g\mid\boldsymbol\omega)$. In practice, this simply means that when sampling for $\g$, the covariance matrix $\mathsf N=\mathsf N_\textrm{w}+\mathsf N_\textrm{corr}$ must be used, rather than just $\mathsf N_\mathrm{w}$.

\commanderthree\ models the gain at each timestream $t$ for a detector $i$ as
\begin{equation}
	g_{t,i}=g_0+\Delta g_i+\delta g_{q,i},
\end{equation}
where $q$ labels the time interval for which we assume the gain is constant, typically a single scan. In order to sample the gain, we write down a generative model for the TOD,
\begin{equation}
	\dv_i=\g_{i}\s_{i}^\mathrm{tot} +\n_{i}^\mathrm{tot}\sim\mathcal N(\g_i\s_{i}^\mathrm{tot},\mathsf N_i).
\end{equation}
Since $\dv_i$ is given as a linear combination of the fixed signal and the gains, a random sample of the gain can be drawn by solving\footnote{See, e.g., Appendix A.2 of \citet{bp01} for a derivation of this result.}
\begin{equation}
	\label{eq:gain_gauss_samp}
	[(\s_i^\mathrm{tot})^T\mathsf N_i^{-1}\s_i^\mathrm{tot}]\g_i
	=(\s_i^\mathrm{tot})^T\N^{-1}_i\dv_i
	+(\s_i^\mathrm{tot})^T\N^{-1/2}_i\boldsymbol\eta,
\end{equation}
where $\boldsymbol\eta\sim\mathcal N(0,1)$ is a vector of standard normal variables.
Note that $\mathsf N_i$ depends implicitly on the noise PSD $\boldsymbol \xi_n$, while the fluctuations corresponding to $\ncorr$ are properly downweighted by $\mathsf N_{\mathrm{corr},i}$.
As detailed by \citet{bp07}, \commanderthree\ samples $g_0$, $\Delta g_i$, and $\delta g_{q,i}$ in separate steps. Specifically, the absolute calibration $g_0$, for the CMB-dominated channels, is fitted using only the orbital dipole, while the relative calibrations, $\Delta g_i$, exploits the full sky signal. The same is true for the time-dependent gain fluctuations, $\delta g_{q,i}$, and in this case an additional smoothness prior is applied through an effective Wiener filter. The Gibbs chain is formally broken by fitting the absolute gain $g_0$ to the orbital dipole alone, as opposed to the full sky signal. However, this makes the sampling more robust with respect to unmodeled systematic effects, somewhat analogous to applying a confidence mask when estimating the CMB power spectrum.

The correlated noise sampling, described by \citet{bp06}, follows a similar procedure, except this now conditions upon the previous gain estimate, which is sampled immediately before the correlated noise component in the code. Similar to the gain case, we can write a generative model for the data,
\begin{equation}
	\dv_i=\g_i\s_i^\mathrm{tot}+\n_i^\mathrm{corr}+\n^\mathrm{w}_i
	\sim\mathcal N(\g_i\s_i^\mathrm{tot},\N_{\mathrm{corr},i}+\N_{\mathrm{w},i}).
\end{equation}
Given fixed $\boldsymbol r_i=\dv_i-\g_i\s_i^\mathrm{tot}$, we can again write a sampling equation, 
\begin{equation}
	\label{eq:ncorr_gauss_samp}
	(\N_{\mathrm{corr},i}^{-1}+\N_{\mathrm{w},i}^{-1})
	\n^\mathrm{corr}_i
	=\N^{-1}_{\mathrm{w},i}\boldsymbol r_i
	+\N^{-1/2}_{\mathrm{w},i}\boldsymbol\eta_1
	+\N^{-1/2}_{\mathrm{corr},i}\boldsymbol\eta_2.
\end{equation}
This gives a sample of the underlying correlated noise. %

To sample the correlated noise parameters, we assume that the correlated noise is drawn from a correlated Gaussian and from the conditional posterior distribution,
\begin{equation}
	P(\boldsymbol\xi_n\mid\n^\mathrm{corr})\propto\frac{\exp{[-\frac12(\n^\mathrm{corr})^T\N_\mathrm{corr}^{-1}\n^\mathrm{corr}]}}
	{\sqrt{|\N_\mathrm{corr}|}}P(\boldsymbol\xi_n).
\end{equation}
where $P(\boldsymbol\xi _n)$ is a flat prior over the PSD parameters.
The simplest and most commonly used parametrization for correlated noise is given by
\begin{equation}
	\mathsf N_\mathrm{corr}(f)=\sigma_0^2\left(\frac f{f_\mathrm{knee}}\right)^\alpha.
\end{equation}
This can in principal be modified, and for \Planck\ LFI a Gaussian log-normal bump was added at a late stage in the \bp\ analysis. Rather than sampling for $\sigma_0$, we effectively fix the white noise level to the noise level at the highest frequency, e.g.,
\begin{equation}
	\sigma_0^2\equiv\frac{\mathrm{Var}(r_{t+1}-r_t)}2,
\end{equation}
where $t$ and $t+1$ are consecutive time samples, and ${\boldsymbol r\equiv\boldsymbol d-\boldsymbol g\boldsymbol s^\mathrm{tot}
-\boldsymbol n^\mathrm{corr}}$. In practice, this makes $\sigma_0$ a deterministic function of the sampled sky and gain parameters. The parameters $\alpha$ and $f_\mathrm{knee}$ are not linear in the data, but they can be sampled efficiently using a standard inversion sampler (see, e.g., Appendix~A.3 of \citealt{bp01} or Chapter~7.3.2 of \citealt{numerical_recipes} for further details). In practice, this requires computing the posterior over a linear grid one parameter at a time.

Once the instrumental parameters have been sampled, \commanderthree\ computes the calibrated TOD for each band,
\begin{equation}
	r_{t,i}=\frac{d_{t,i}-n_{t,i}^\mathrm{corr}}{g_{t,i}}-\left(s_{t,i}^\mathrm{orb}
	+s_{t,i}^\mathrm{fsl}+\delta s_{t,i}^\mathrm{leak}+{s}_{t,i}^\mathrm{inst}\right)
	\label{eq:caltod}
\end{equation}
where $\s^\mathrm{orb}$ is the orbital dipole \citep{bp07}, $\s^\mathrm{fsl}$ is the far sidelobe timestream \citep{bp08}, $\delta\s^\mathrm{leak}$ is the bandpass leakage \citep{bp09}, and $\s^\mathrm{inst}$ is some instrumental-specific contribution, e.g., the 1\,Hz electronic spike for LFI. With a correlated noise realization removed, one can perform simple binned mapmaking, weighting each pixel by the white noise amplitude.

\subsubsection{Differential mapmaking}
\label{ssec:mapmaking}

The first additional algorithm that needs to be added to \commanderthree\ in order to process \WMAP\ TOD data is support for differential mapmaking \citep{bp17}. After calibration and correction for instrumental effects, the TOD can be modeled as
\begin{equation}
	\boldsymbol d=\mathsf P\boldsymbol m+\boldsymbol n^\mathrm{w},
\end{equation}
where
\begin{equation}
	\boldsymbol m=\mathsf B^\mathrm{symm}\mathsf M\boldsymbol a
\end{equation}
is the expected map for each detector after removing the orbital dipole, far sidelobe, baseline, and a realization of correlated noise. The differential pointing strategy can be represented in matrix form as 
\begin{align}
	\label{eq:wmap_pointing}
	\mathsf P_{tp}&=
	(1+x_\mathrm{im})(\delta_{p^I_{\phantom\A}p_\A^I}
	+\delta_{p^Q_{\phantom\A}p^Q_\A}\cos2\psi_\A
	+\delta_{p^U_{\phantom\A}p^U_\A}\sin2\psi_\A)
	\\
	&-(1-x_\mathrm{im})(\delta_{p^I_{\phantom\B}p_\B^I}
	-\delta_{p^Q_{\phantom\A}p^Q_\B}\cos2\psi_\B
	-\delta_{p^U_{\phantom\A}p^U_\B}\sin2\psi_\B)\nonumber
\end{align}
where $p_\A$ and $p_\B$ are the time-dependent pointings for each DA. Note that this is equivalent to Eqs.~\eqref{eq:mapeq1} and \eqref{eq:mapeq2}, except without the spurious component $S_{\A/\B}$ as the bandpass mismatch is explicitly subtracted beforehand in Eq.~\eqref{eq:caltod}. The maximum likelihood map can now in principle be derived using the usual mapmaking equation,
\begin{equation}
	\label{eq:mapmapking_eqn1}
	\mathsf P^T\mathsf N^{-1}\mathsf P\boldsymbol m=\mathsf P^T\mathsf N^{-1}\boldsymbol d.
\end{equation}
For a single-horn experiment, i.e., \Planck\ LFI, this reduces to a $3\times3$ matrix that can be inverted for each pixel independently. For the pointing matrix in Eq.~\eqref{eq:wmap_pointing}, this is no longer possible, as there is inherently coupling between horns A and B in the timestreams. The $3N_\mathrm{pix}\times3N_\mathrm{pix}$ matrix can in principle be solved using an iterative algorithm, e.g., preconditioned conjugate gradients \citep{shewchuk:1994}.

\citet{jarosik2010} identified an issue where a large difference in the sky temperature values at pixel A versus pixel B induced artifacts in the mapmaking procedure. We adopt the procedure first described by \citet{hinshaw2003a} where only the pixel in a bright region, defined by a small processing mask \citep{bennett2012} is accumulated, thus modifying the mapmaking equation to
\begin{equation}
	\mathsf P^T_\mathrm{am}\mathsf N^{-1}\mathsf P\boldsymbol m
	=\mathsf P^T_\mathrm{am}\mathsf N^{-1}\boldsymbol d.
	\label{eq:bicg_stab}
\end{equation}
This equation can be solved using the BiCG-STAB algorithm for a non-symmetric matrix $\mathsf A$ where $\mathsf A\boldsymbol x=\boldsymbol b$. We apply a preconditioner $\mathsf M$ by numerically inverting the same problem with $N_\mathrm{side}=16$ maps and applying a diagonal noise matrix. Numerically, we define convergence as when the residual $\boldsymbol r\equiv\boldsymbol b-\mathsf A\boldsymbol x$ satisfies $\boldsymbol r^T\mathsf M^{-1}\boldsymbol r/\boldsymbol b^T\mathsf M^{-1}\boldsymbol b<10^{-10}$, which typically takes about 20 iterations for producing frequency maps.

\begin{figure}
	\centering
	\includegraphics[width=\columnwidth]{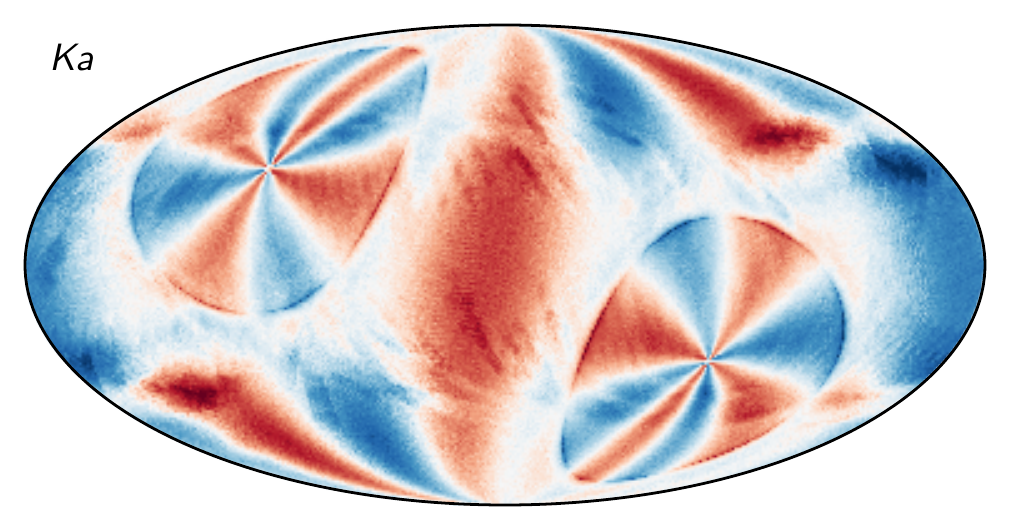}\\
	\includegraphics[width=\columnwidth]{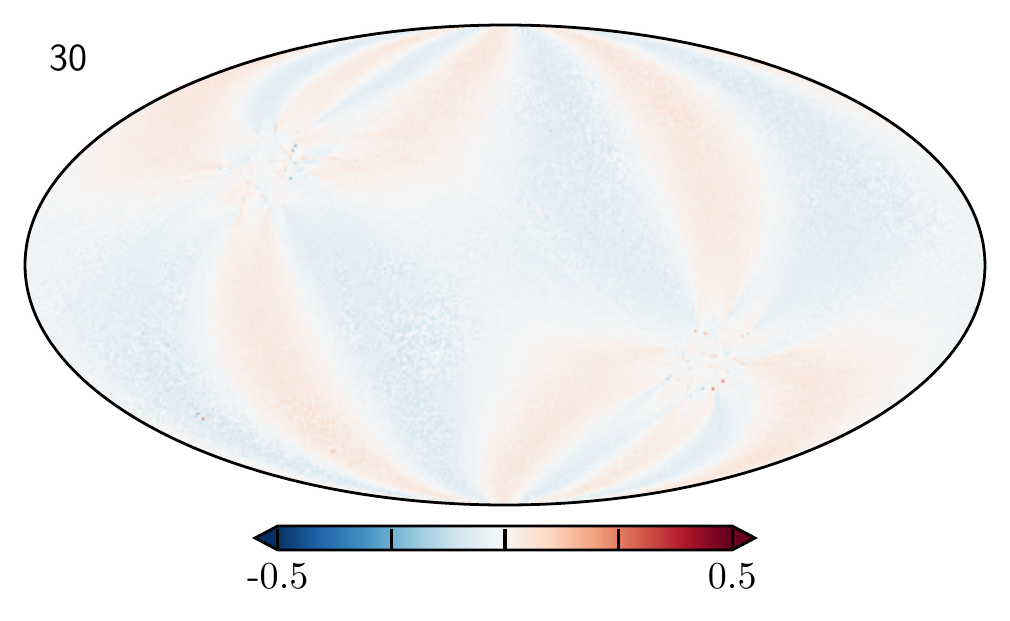}
	\caption{Cross-correlation $\rho_{QU}$ per pixel for \textit{(top:)} \Ka\ and \textit{(bottom:)} LFI 30\,GHz.}
	\label{fig:cross_cor}
\end{figure}

The full noise covariance matrix $\mathsf N_{pp'}$ is given by the inverse of $\mathsf P^T_\mathrm{am}\mathsf N^{-1}\mathsf P$, where the diagonals $\mathsf N_{pp}$ are the white noise variance for each Stokes parameter. An additional quantity that was computed in \bp\ but not delivered in the final products is the covariance of the Stokes parameters within a single pixel, $\sigma_{QU,p}$. We find that the correlation between Stokes parameters, $\rho_{QU}\equiv\sigma_{QU}/\sqrt{\sigma_{QQ}\sigma_{UU}}$, is of order 0.5 for the \WMAP\ DAs, as shown in Fig.~\ref{fig:cross_cor}. For \Planck\ LFI, the 30 and 70\,GHz channels have $|\rho_{QU}|\sim0.1$, while the 44\,GHz correlations are notably higher with $|\rho_{QU}|\sim0.5$. The reason for this difference is that the 44\,GHz channel has three horns. Two of those are aligned with the scanning direction in the focal plane, and have polarization angles that are rotated by $45^{\circ}$ with respect to each other. Together those two horns therefore disentangle polarization information very efficiently. The third horn, however, does not have a corresponding partner, and relies only on satellite precession to recover individual Stokes parameters. For comparison, all 30 and 70\,GHz horns have partners aligned with the scanning direction. In the current work, we have implemented support for the full $3\times 3$ noise matrices, including $\sigma_{QU},$ for component separation and map-based $\chi^2$ calculations for both \WMAP\ and LFI.

\subsubsection{Baseline sampling}
\label{ssec:baseline}

The data model adopted by \citet{hinshaw2003a} can be written in raw du as
\begin{equation}
	\data = \mathsf{GPBM}\boldsymbol a+\boldsymbol n+\boldsymbol b,
\end{equation}
where $\boldsymbol b$ is the instrumental baseline and $\boldsymbol n$ is the
total instrumental noise. As noted above, \commanderthree\ divides
the noise into $\n=\n^\mathrm w+\n^\mathrm{corr}$, a white noise term and a
correlated noise term. Because the white noise is by definition uncorrelated in time, it does not have any
correlations between adjacent pixels, so that any pixel-pixel covariance should
be fully described by realizations of the $\n^\mathrm{corr}$ timestream.

\commanderthree\ estimates the baseline using the full estimate of the current sky
model, $\boldsymbol r=\boldsymbol d-\boldsymbol g\boldsymbol s^\mathrm{tot}=\boldsymbol
b+\boldsymbol n$. Modeling $\boldsymbol b=b_0+b_1\Delta t$, we solve for $b_0$
and $b_1$ using linear regression in each timestream while masking out samples
that lie within the processing mask. Strictly speaking, this is breaking the
Gibbs chain, as we are not formally sampling $b_0$ and $b_1$ for each TOD
chunk. In practice, baseline estimation uncertainty propagates to correlated
noise realizations and PSD parameters, as discussed below.

The approach detailed by \citet{hinshaw2003a} and the
\commanderthree\ implementation differ mainly in two ways. First, the
assumed stable timescales are different -- the initial
\WMAP\ baseline is estimated over one hour timescales, and assumed to
be an actual constant, whereas \commanderthree\ assumes constant values
through the entire time chunk, which is 3--7 days depending on the
band in question, but allows a linear term in the baseline. Second,
the two methods differ in how they treat nonlinear residuals in the
first-order baseline model.  As noted by \citet{hinshaw2003a},
residual baseline variations manifest as correlated noise stripes in
the final maps, and \WMAPnine\ solves this using a time domain filter,
downweighting the data based on the noise characterization. This
is similar to the \commanderthree\ approach, which
accounts for this as part of the correlated noise component. The main
advantages of the latter is that it allows for proper error
propagation at all angular scales without the use of a dense
pixel-pixel noise covariance, and provides a convenient
means for inspecting the residuals visually by binning the correlated
noise into a sky map.

\subsubsection{Transmission imbalance estimation}
\label{ssec:imbalance}

Transmission imbalance, the differential power transmission of the optics and waveguide components between horns A and B, can be parameterized as
\begin{equation}
	d_{t,i}=g_{t,i}[(1+x_{\mathrm{im},i})s_{t,i}^\mathrm{tot,A}-(1-x_{\mathrm{im},i})s_{t,i}^\mathrm{tot,B}]+n_t.
\end{equation}
This can be decomposed into a differential (d) and common-mode (c) signal such that
\begin{equation}
	d_{t,i}=g_{t,i}[s_{t,i}^\mathrm d+x_{\mathrm{im},i}s_{t,i}^\mathrm c]+n_t.
\end{equation}
In this form, the imbalance parameters can be estimated by drawing Gaussian samples from the standard mean and standard deviation over the entire mission. To draw samples for $x_{\mathrm{im},i}$, we construct a sampling routine analogous to the gain estimation of Eq.~\eqref{eq:gain_gauss_samp} and correlated noise estimation of  Eq.~\eqref{eq:ncorr_gauss_samp}, with $\boldsymbol r=\boldsymbol d-\boldsymbol g\boldsymbol s^\mathrm d$,
\begin{equation}
	[(\boldsymbol g\boldsymbol s^\mathrm c)^T\mathsf N^{-1}\boldsymbol g\boldsymbol s^\mathrm c]x_\mathrm{im}
	=(\boldsymbol g\boldsymbol s^\mathrm c)^T\mathsf N^{-1}\boldsymbol r+(\boldsymbol g\boldsymbol s^\mathrm c)^T\mathsf N^{-1/2}\boldsymbol\eta,
\end{equation}
cross-correlating the common-mode signal with $\boldsymbol r$ with appropriate weights and adding a Gaussian random variable with the correct weighting. Note that we are marginalizing over the correlated noise here by using $\N=\N_\mathrm{w}+\N_\mathrm{corr}$. This mitigates any baseline drifts being erroneously attributed to the common-mode signal and biasing the estimate of $x_\mathrm{im}$.

The \WMAP\ procedure, described by \citet{jarosik2003a}, fit for common-mode and differential coefficients along with a cubic baseline over 10 precession periods at a time, corresponding to 10 hours of observation. The mean and uncertainty were then calculated by averaging and taking the standard deviation of these values. This approach has the benefit of allowing for the tracking of possible transmission imbalance variation throughout the mission. However, none of the \WMAP\ suite of papers have found evidence for this, and it has not arisen in our analysis, so we model this as an effect whose value is constant throughout the mission.

\section{Data and data processing}
\label{sec:data}

\begin{table}
\caption{\cosmoglobe\ flagging statistics for each DA. The second column indicates the fraction of data that are removed by the official \WMAP\ flags, while the third column indicates the fraction that is additionally discarded in the current processing for computational reasons. The fourth column indicates the total fraction of data actually used to generate the final maps. }              %
\label{table:flagged_data}      %
\centering                                      %
\begin{tabular}{c c c c}          %
\hline\hline                        %
	Band & Flagged (\%) & Discarded (\%) & Used (\%) \\    %
\hline                                   %
	\K  &  1.72 & 0.87 & 97.4\\
	\Ka &  1.64 & 0.88 & 97.5\\      %
	\Q1 &  1.84 & 0.84 & 96.5\\
	\Q2 &  1.62 & 0.81 & 97.6\\
	\V1 &  1.62 & 1.10 & 97.3\\
	\V2 &  1.61 & 1.01 & 97.4\\
	\W1 &  1.76 & 1.03 & 97.2\\
	\W2 &  1.60 & 0.81 & 97.6\\
	\W3 &  1.61 & 0.87 & 97.5\\
	\W4 &  1.60 & 0.81 & 97.6\\
\hline                                             %
\end{tabular}
\end{table}

\begin{table*}[t]
  \begingroup
  \newdimen\tblskip \tblskip=5pt
  \caption{Computational resources required for end-to-end
    \cosmoglobe\ processing. All times correspond to CPU hours, and all data volumes are reported in GB. Reported times are
    averaged over more than 100 samples, and vary by $\lesssim\,5\,\%$ from sample to
    sample. Note that the average cost per sample takes into account the undersampling of 70, \V, and \W.
	}
  \label{tab:resources}
  \nointerlineskip
  \vskip -3mm
  \footnotesize
  \setbox\tablebox=\vbox{
    \newdimen\digitwidth
    \setbox0=\hbox{\rm 0}
    \digitwidth=\wd0
    \catcode`*=\active
    \def*{\kern\digitwidth}
    \newdimen\signwidth
    \setbox0=\hbox{-}
    \signwidth=\wd0
    \catcode`!=\active
    \def!{\kern\signwidth}
 \halign{
      \hbox to 5.0cm{#\leaderfil}\tabskip 1em&
      \hfil#\tabskip 1em&
      \hfil#\tabskip 1em&
      \hfil#\tabskip 1em&
      \hfil#\tabskip 1em&
      \hfil#\tabskip 1em&
      \hfil#\tabskip 1em&
      \hfil#\tabskip 1em&
      \hfil#\tabskip 1em&
      \hfil#\tabskip 1em&
      \hfil#\tabskip 1em&
      \hfil#\tabskip 1em&
      \hfil#\tabskip 1em&
      \hfil#\tabskip 1em&
      #\tabskip 0em\hfil\cr
    \noalign{\doubleline}
      \omit\textsc{Item}\hfil&
      \omit\hfil\textsc{30}\hfil&
      \omit\hfil\textsc{44}\hfil&
      \omit\hfil\textsc{70}\hfil&
      \omit\hfil\K\hfil&
      \omit\hfil\Ka\hfil&
      \omit\hfil\Q1\hfil&
      \omit\hfil\Q2\hfil&
      \omit\hfil\V1\hfil&
      \omit\hfil\V2\hfil&
      \omit\hfil\W1\hfil&
      \omit\hfil\W2\hfil&
      \omit\hfil\W3\hfil&
      \omit\hfil\W4\hfil&
      \omit\hfil\textsc{Sum}\hfil\cr
      \noalign{\vskip 4pt\hrule\vskip 4pt}
      \noalign{\vskip 5pt\hrule\vskip 5pt}
        \multispan5\textit{Data volume}\hfil\cr
        \noalign{\vskip 2pt}
	\hskip 10pt Compressed TOD volume & **86& *178& **597& **13& **12& **15& **15& **19& **18& **26&**26& **26& **26& 1\,053\cr
        \noalign{\vskip 2pt}      
      \multispan5\textit{Processing time (cost per run)}\hfil\cr
      \hskip 10pt TOD initialization/IO time                    &1.8& 2.5& 7.8& 0.7& 0.6& 0.8& 0.7& 0.9& 0.8& 1.3& 1.3& 1.0& 0.9& *21.1\cr
      \hskip 10pt Other initialization                                &  &  &  &  & & & & & & & & & & *14.6\cr
      \hskip 10pt {\bf Total initialization}                          &  &  &  &  & & & & & & & & & & {\bf *35.7}\cr
      \noalign{\vskip 2pt}      
      \multispan5\textit{Gibbs sampling steps (cost per sample)}\hfil\cr
      \hskip 10pt Huffman decompression                            & 1.2& 2.2& 23.2& 0.8& 0.9& 1.1& 1.1& 1.5& 1.4& 2.0& 2.0& 2.0& 2.0& *41.4\cr
      \hskip 10pt Array allocation                                 & 0.4& 0.9& 51.6& 1.3& 1.3& 1.5& 1.5& 3.1& 3.3& 4.0& 3.8& 4.0& 4.0& *80.7\cr
      \hskip 10pt TOD projection ($\P$ operation)                  & 0.9& 2.0&  12.3& 6.1& 7.1& 8.7& 8.9& 11.4& 11.3& 15.9& 15.8& 15.7& 15.8& 131.9\cr
      \hskip 10pt Sidelobe evaluation                              & 1.2& 2.6&  9.5& 3.0& 3.5& 4.1& 4.2& 5.5& 5.4& 7.8& 7.7& 7.7& 7.5& *69.7\cr
      \hskip 10pt Orbital dipole                                   & 0.9& 2.0& 9.0& 1.2& 1.5& 1.8& 1.9& 2.6& 2.5& 3.8& 3.8& 3.8& 3.8& *38.6\cr
      \hskip 10pt Gain sampling                                    & 0.6& 0.9& 2.2& 1.3& 1.3& 0.8& 0.8& 1.3& 1.3& 1.2& 1.2& 1.2& 1.2& *15.3\cr
      \hskip 10pt 1\,Hz spike sampling                             & 0.3& 0.4& 1.9& & & & & & & & & & & **2.7\cr      
      \hskip 10pt Correlated noise sampling                        & 2.1& 4.3& 24.8& 2.7& 2.9& 3.7& 3.8& 6.2& 5.4& 7.7& 7.4& 6.9& 8.3& *86.4\cr
      \hskip 10pt Correlated noise PSD sampling                    & 5.0& 6.2& 1.6& 0.3& 0.3& 0.3& 0.3& 0.5& 0.5& 0.7& 0.6& 0.6& 0.7& *17.6\cr
      \hskip 10pt TOD binning ($\P^t$ operation)                   & 0.1& 0.1& 10.5& 0.8& 0.8& 1.0& 1.0& 1.7& 1.6& 2.4& 2.4& 2.4& 2.4& *27.2\cr
      \hskip 10pt Mapmaking                                        & & & & 9.2& 9.7& 13.1& 12.7& 21.7& 20.2& 35.4& 34.9& 36.1& 39.3& 232.3\cr
      \hskip 10pt MPI load-balancing                      & 1.2& 1.7& 9.2& 2.2& 2.0& 2.2& 2.1& 3.6& 3.3& 4.8& 4.6& 4.5& 4.6&*46.0\cr
      \hskip 10pt Sum of other TOD processing                      & 0.7& 1.6& 13.1& 0.1& 0.2& 0.5& 0.4& 0.7& 0.8& 0.9& 1.0& 0.9& 1.2&*22.1\cr
      \hskip 10pt {\bf TOD processing cost per sample}             & {\bf 14.6}& {\bf 24.9}& {\bf 169.7}&  {\bf 28.8}& {\bf 31.5}& {\bf 38.7}& {\bf 38.7}& {\bf 59.8}& {\bf 57.0}& {\bf 86.6}& {\bf 85.2}& {\bf 85.8}& {\bf 90.8}& {\bf 812.1}\cr
      \noalign{\vskip 2pt}
      \hskip 10pt Amplitude sampling  &&&&&&&&&&&   &  &  & *16.2\cr
      \hskip 10pt Spectral index sampling  &&&&&&&&&&&   &  &  & *32.1\cr
      \noalign{\vskip 2pt}
      \hskip 10pt {\bf Average cost per sample}                  &&&&&&&&&& &   &  &  &  {\bf 418.9}\cr
      \noalign{\vskip 4pt\hrule\vskip 5pt} } }
  \endPlancktablewide \endgroup
\end{table*}

We describe the public \WMAPnine\ data products in Sect.~\ref{sec:products}, then describe the treatment we apply to make them compatible with \commanderthree\ in Sect.~\ref{sec:preprocessing}. Finally, we describe the computational requirements in Sect.~\ref{sec:resources}.

\subsection{Publicly available \WMAP\ products}
\label{sec:products}

The full \WMAP\ dataset is hosted at the Legacy Archive for Microwave Background Data Analysis (LAMBDA).\footnote{\url{https://lambda.gsfc.nasa.gov/product/wmap/dr5/m_products.html}} In addition to the primary scientific products, e.g., cosmological parameters, CMB power spectra and anisotropy maps and frequency maps, the time-ordered data (TOD) can be downloaded, both in uncalibrated and calibrated form.\footnote{\url{https://lambda.gsfc.nasa.gov/product/wmap/dr5/tod_info.html}} In principle, thanks to these data and the explanatory supplements \citep{wmapexsupp}, the entire data analysis pipeline can be reproduced from uncalibrated TOD to frequency maps.

For this analysis, we keep certain instrumental parameters fixed to the reported values. For example, we have made no attempts to rederive the pointing solutions, re-estimate the main beam response and far sidelobe pickup, or recover data that were flagged in the \WMAP\ event log. These and other analyses, such as estimating the bandpass shift over the course of the mission, are certainly possible within the larger Gibbs sampling framework. However, in this work we limit ourselves to recalibrating the TOD, estimating their noise properties, and applying bandpass corrections to the data before mapmaking.

\subsection{TOD preprocessing and data selection}
\label{sec:preprocessing}

The full nine-year \WMAP\ archive spans from 10 August 2001 to 10 August 2010, with the raw uncalibrated data comprising 626\,GB. A little over 1\,\% of the data were lost or rejected due to incomplete satellite telemetry, thermal disturbances, spacecraft anomalies, and station-keeping maneuvers, with an extra 0.1\,\% rejected due to planet flagging \citep{bennett2003a,hinshaw2007,hinshaw2009,bennett2012}. 
The final results reported by \citet{bennett2012} included roughly 98.4\,\% of the total data volume.
A full accounting of all data cuts can be found in Table~1.8 of \citet{wmapexsupp}. In this analysis we flag the same data indicated in the fiducial \WMAP\ analysis, and use the same planet exclusion radii.

As shown by \citet{bp03}, a large fraction of \commanderthree's computational time is spent performing Fast Fourier Transforms (FFTs) on individual scans. Rather than truncating datastreams to have lengths equal to ``magic numbers'' for which \texttt{FFTW} \citep{FFTW05} performs efficiently, as was done in the \bp\ analysis, 
we redistribute the data into scans of length $2^N$, where $N=22$ for \K--\Q, $N=23$ for \V--\W. This yields scans with lengths of 6.21 days for \K- and \Ka-band, 4.97 days for \Q-band, 7.46 days for \V-band, and 4.97 days for \W-band.\footnote{Note that scans with equal $n_\mathrm{TOD}$ cover different lengths of time due to the different sampling rate for each frequency.}
These datastream lengths are short enough to be processed quickly and distributed efficiently across multiple processors, while being long enough to properly characterize the noise properties of the timestreams, whose $f_\mathrm{knee}$ values are on the order $1\,\mathrm{mHz}$. Most importantly, \texttt{FFTW} performs fastest when the datastream is of length $2^N$. 

When redistributing the data, timestreams of length $2^N$ were interrupted by events logged in Table~1.8 of \citet{wmapexsupp}.
When we encountered these events, interrupted TOD segments were appended to the previous TOD, in most cases creating TODs with lengths $>2^N$. We found that events of length $<2^N$ were too short to accurately estimate the noise PSD parameters. This criterion led us to discard these otherwise useful data. In addition, when $>10\,\%$ of the TOD are flagged, the large number of gaps in the data makes the solution of Eq.~\eqref{eq:ncorr_gauss_samp} for $\ncorr$ computationally more expensive. Given that data near many large gaps are more likely to have unmodeled effects than stable data, and they are more expensive to process, we chose to remove these from the analysis. Together, these two effects led to $\simeq1\%$ of the data to be discarded. We summarize the full flagging statistics for our maps in Table~\ref{table:flagged_data}. In total, the \Cosmoglobe\ maps use about 1\,\% less data than the \textit{WMAP9} official products. The total difference in data volume can be entirely accounted for by the cuts described in this paragraph.

\subsection{Computational resources and future plans}
\label{sec:resources}

A key motivation of the current analysis is to evaluate whether it is feasible to perform a joint analysis of two datasets simultaneously, each with its own particular processing requirements and algorithmic treatment. One of the results from \citet{bp17} was that most of the data processing procedures for  \WMAP\ and \Planck\ LFI overlapped, with the notable exception of mapmaking. While the algorithmic requirements have been discussed in Sect.~\ref{sec:methods}, we have not yet quantified the requirements in terms of RAM and CPU hours. In Table~\ref{tab:resources}, we enumerate the RAM requirements and CPU time for each sampling step using a single AMD EPYC 7H12, 2.6\,GHz cluster node with 128~cores and 2\,TB of memory. As such, approximate wall runtimes can be obtained by dividing all numbers in Table~\ref{tab:resources} by 128.

Despite the relatively small data volume spanned by \WMAP, e.g., 86\,GB for 30\,GHz vs. 13\,GB for \K-band, the CPU time is comparable to each of the LFI channels.  The single largest reason for this is the mapmaking step, Eq.~\eqref{eq:bicg_stab}, which requires looping over the entire dataset for each matrix multiplication, a process which must be repeated $\sim20$ times. As discussed in Sec. \ref{ssec:mapmaking}, this is vastly sped up by the use of a low resolution preconditioner, reducing the number of iterations by an order of magnitude.

Additionally, operations that require the creation of  timestreams for each detector, i.e., TOD projection, sidelobe evaluation, and orbital dipole projection, take much longer than expected from a pure data volume scaling. Part of this is due to \commanderthree\ evaluating the sky in two pixels simultaneously, doubling the expected workload, but the other issue is that we are unable to benefit from the ring-clustering based TOD distribution scheme used for LFI. Due to \WMAP's more complex scan strategy and detector geometry, it is impossible to cluster scans with similar pixel coverage onto a single core, which makes pixel-space lookup operations less efficient in this case.

Gain sampling and correlated noise sampling include multiple FFTs. Typical LFI TODs are of length $\sim200\,000$, an order of magnitude smaller than the \WMAP\ TODs of length $\sim5\,000\,000$. Despite the TOD lengths being pre-determined to be $2^N$, this extra length still results in longer run times for equivalent data volumes, but does yield noise information on much longer time scales than we have for LFI. Note that \WMAP\ was had typical $f_\mathrm{knee}$'s over 100 times smaller than LFI's, so TODs that were over 100 times longer are necessary for characterizing its noise PSD properties.

\begin{figure}[t]
	\centering
	\includegraphics[width=\linewidth]{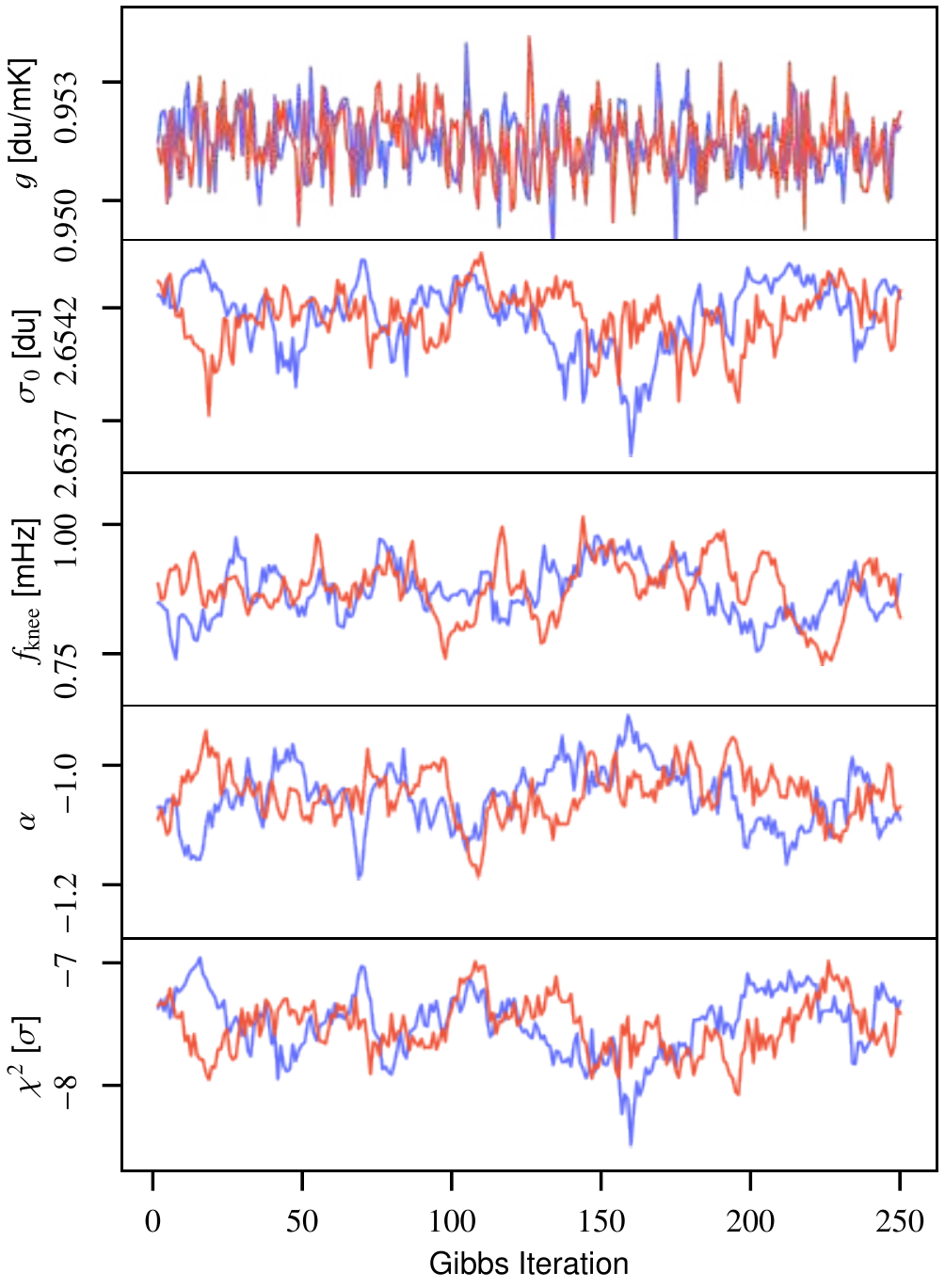}
	\caption{Trace plots of the \K113 gain and noise parameters for a single scan starting on MJD 52285.2. The two colors correspond to the two independent Markov chains produced in this analysis.}
	\label{fig:inst_K113_gibbs}
\end{figure}

\begin{figure*}[t]
	\centering
	\includegraphics[width=\textwidth]{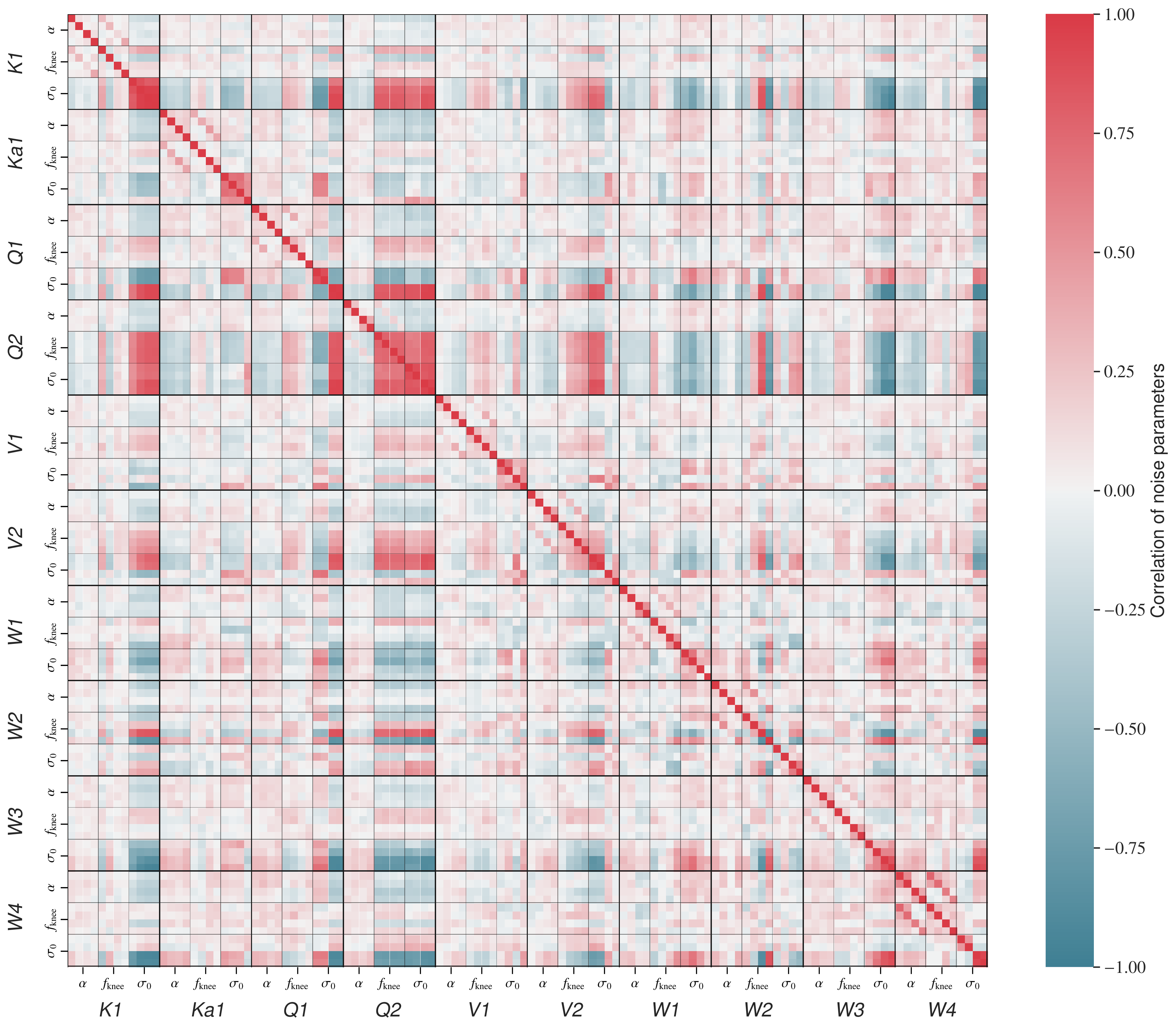}
	\caption{Noise parameter correlation matrix. We average over all Gibbs samples of the noise parameters $\boldsymbol\xi_n=\{\alpha,f_\mathrm{knee},\sigma_0\}$ for each PID. We then find the correlation in time between these averages for the different bands and detector. The results here are for the calibrated white noise level, $\sigma_0[\mathrm{mK}]$. The values for each detector are ordered 13, 14, 23, and 24.}
	\label{fig:correlation}
\end{figure*}

For the current analysis, which aims primarily to derive posterior-based \WMAP\ frequency maps, we produce a total of 500 main Gibbs samples, divided into two chains. Noting that the computational cost of the \W-channel carries almost half of the total expense of the \WMAP\ TOD processing, while being of less scientific importance than, say, the \K-band, we choose to only reprocess this channel every fourth main sample. Likewise, we only reprocess the \V-band every other main sample, and the LFI 70\,GHz sample every fourth sample. The total cost for producing 500 \WMAP\ \K, \Ka, \Q, \Planck\ 30, and 44\,GHz samples, 250 \V-band samples, and 125 \W-band and 70\,GHz samples is 210k CPU-hrs, and the total wall time is 33 days. Noting that the \bp\ analysis required 4000 samples to reach full convergence in terms of the optical depth of reionization \citep{bp12}, a corresponding complete LFI+\WMAP\ analysis will cost about 1.7M CPU-hrs, and take about 9 months of continuous runtime on two cluster nodes. While entirely feasible, this is sufficiently expensive that we choose to perform the analysis in two stages; first we present preliminary frequency maps in the current paper, and use these to identify potential outstanding issues, either in terms of data model or Markov chain stability. An important goal of this phase is also to invite the larger community to study these preliminary maps, and thereby identify additional problems that we may have missed. Then, when all issues appear to have been resolved, we will restart the process, and generate sufficient samples to achieve full convergence.

\section{Instrumental parameters}
\label{sec:instrument}

In this section and Sect.~\ref{sec:maps} we present the main results from the \cosmoglobe\ DR1 analysis, which may be summarized in terms of the joint posterior distribution. For organizational purposes, we will discuss instrumental parameters, frequency maps, and astrophysical results separately in this and the following two sections. It is important to remember that these results are all derived from one single highly multivariate posterior distribution, and every parameter is in principle correlated with all others. In this section, we focus on instrumental parameters, starting with visual inspection of the basic Markov chains and posterior means, before considering each instrumental parameter in turn.

\subsection{Markov chains, correlations and posterior mean statistics}
\label{sec:summary_stats}

To build intuition regarding the general Markov chain properties, we show in Fig.~\ref{fig:inst_K113_gibbs} the Markov chains for the gain and noise parameters for one arbitrary diode (\K113) and scan. Each panel corresponds to one single parameter, and the observed variation quantify the uncertainty in that single parameter due to the combination of white noise and correlations with other parameters. Here we see that the different parameters have quite different correlation lengths; the gain (in the top panel) has a very short autocorrelation length, as in just a few samples, while the noise parameters have typical correlation lengths of a few tens of samples. Even for these parameters, however, the full set of 500 samples provides a fairly robust estimate of the full marginal mean and uncertainty.

The bottom panel shows the reduced normalized $\chi^2$ for the same scan in units of $\sigma\equiv\sqrt{2n_\mathrm{TOD}}$, and we see that this also shows similar correlation lengths as the noise parameters. This makes sense since the TOD residual at the level of a single sample is strongly noise dominated. In contrast, small variations in either the sky signal or gain have relatively small impacts on this particular $\chi^2$; the goodness of fit of such global parameters is better measured through map-level residuals and $\chi^2$'s. In this respect, we also note that the absolute value of the TOD-level $\chi^2$ is for this particular scan about $-7.5\,\sigma$, which at first sight appears as a major goodness of fit failure. However, it is important to recall that a typical scan contains about five million data points, and this statistic is therefore extremely sensitive to any deviation in the noise model. Specifically, the reduced $\chi^2$ for this particular scan is $\chi_{\mathrm{raw}}^2/n_{\mathrm{TOD}}=0.993$, which corresponds to an overestimation of the white noise level of only 0.3\,\%. Furthermore, as discussed in Sect.~\ref{ssec:oldsamplers}, we currently assume a strict $1/f$ noise model for the \WMAP\ noise, while the true \WMAP\ noise is known to exhibit a very slight non-white noise excess at high frequencies \citep{bp17}. Properly modeling such non-white high-frequency noise is therefore an important goal for the next \cosmoglobe\ data release.  Such work is also a vital step in preparing for integration of other types of experiments with non-white noise into the framework, such as \Planck\ HFI. However, in absolute terms, the impact of this model failure is very limited, and not likely to significantly affect any astrophysical results; it is primarily a limitation for TOD-level goodness of fit testing.

For a survey of the entire experiment's noise properties, Fig.~\ref{fig:correlation} shows pairwise correlations between the various noise parameters for all DAs, averaged over all Gibbs samples and scans. It is important to note that a non-zero correlation in this plot does not indicate that the specific noise realization is correlated between DAs, but only that the noise PSD parameters are correlated. This is expected due to the \WMAP\ satellite motion around the Sun, which induces an annual variation in the system temperature. This correlation plot therefore primarily quantifies the sensitivity to this common-mode signal for each diode. Most notably, we see that the \Q2 DA exhibits particularly strong correlations, and we also note that the calibrated white noise $\sigma_0[\mathrm{mK}]=\sigma_0[\mathrm{du}]/g$ is generally more susceptible to these variations than $f_{\mathrm{knee}}$ and $\alpha$.

\begin{figure}[t]
	\centering
	\includegraphics[width=\linewidth]{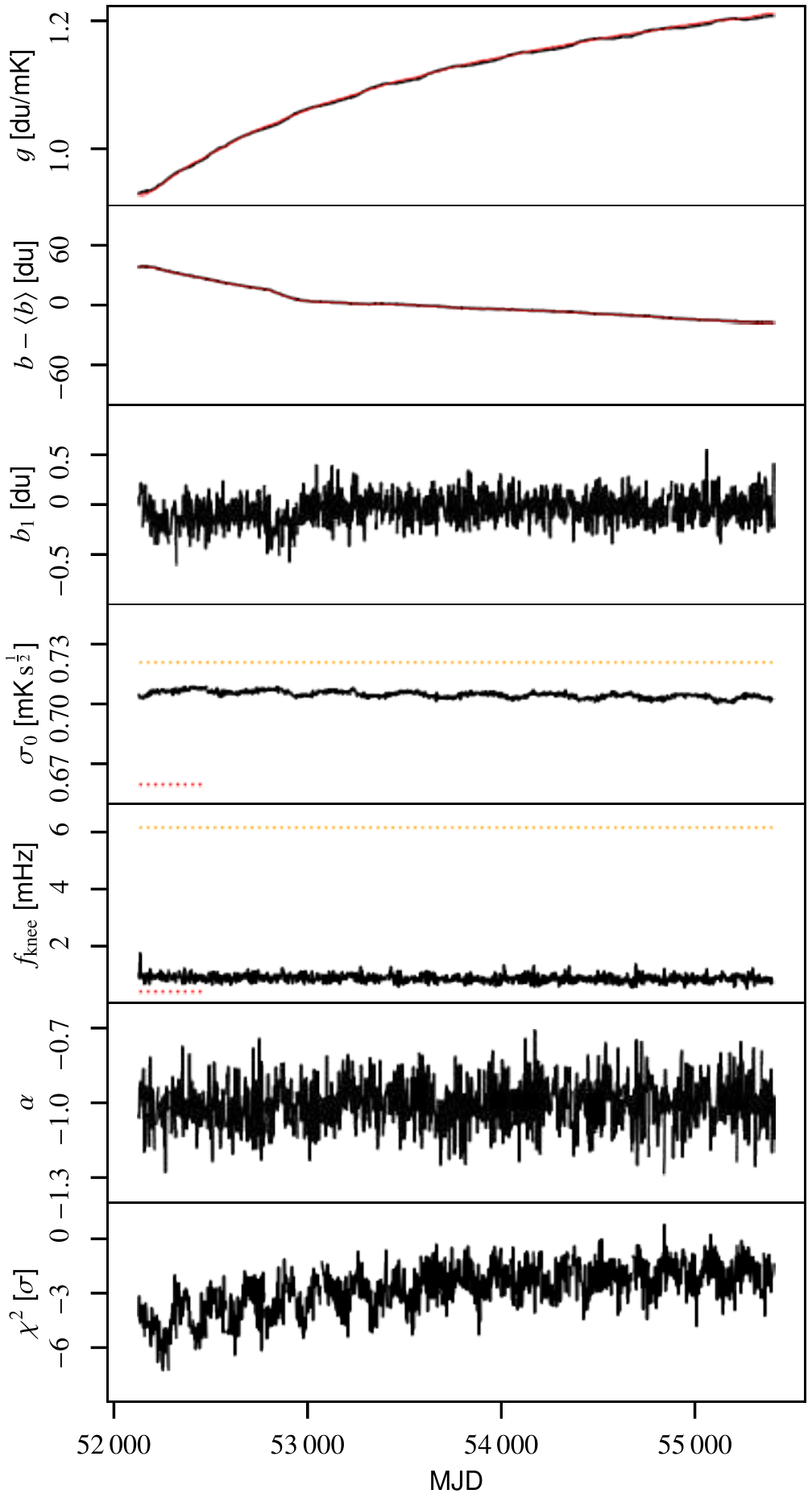}
	\caption{Overview of \K113. The red solid lines in the first and second panel  are the regressed gain and baseline from \WMAPnine, while the black lines in all panels are posterior means from the \cosmoglobe\ Gibbs chain. The red dashed and yellow dashed lines are reported $\sigma_0$ and $f_\mathrm{knee}$ values from the first-year \WMAP\ data analysis and GSFC measurements, respectively.}
	\label{fig:inst_K113}
\end{figure}

Next, in Fig.~\ref{fig:inst_K113} we show posterior mean values for each instrumental parameter for the same \K113 diode, in this case plotted as a function of time throughout the entire mission. The panels show, from top to bottom, 1) gain; 2) the difference between the baseline mean and its full-mission average; 3) the baseline slope; 4) the white noise level; 5) the correlated noise knee frequency; 6) the correlated noise slope; and 7) the TOD-level $\chi^2$. The \cosmoglobe\ results are shown as black curves, while the \WMAP\ results are (for the gain and baseline) shown as red curves; dotted red and orange line corresponds to the first-year \WMAP\ and Goddard Space Flight Center (GSFC) laboratory measurements, respectively. Note that the gain and baseline are nearly indistinguishable -- their differences will be discussed in more detail in Sect.~\ref{sec:gain}. For brevity, we have only shown the results for one single diode here. However, a complete survey of all instrumental parameter posterior means for all 40~diodes is provided in Appendix~\ref{sec:survey}, and all individual samples are also available in a digital format as part of the \cosmoglobe\ DR1.

\subsection{Gain and baselines}
\label{sec:gain}

We now consider the gain and baseline parameters in greater detail, and aim to compare our estimates with the \WMAPnine\ products. The \WMAPnine\ gain and baseline estimates are not directly available in terms of public data products, but only in terms of the general parametric models. For instance, the \WMAP\ gain model reads \citep{wmapexsupp}
\begin{equation}
	\label{eq:wmap_gain}
	g=\alpha\frac{\overline V-V_\circ-\beta(T_\mathrm{RXB}-290\,\mathrm K)}
	{T_\mathrm{FPA}-T_\circ}+(m\Delta t+c),
\end{equation}
where $\overline V$ represents the radio frequency bias powers per detector; $T_\mathrm{RXB}$ and $T_\mathrm{FPA}$ are the receiver box and focal plane assembly temperatures, which are recorded every 23.04\,s; $\alpha$, $\V_\circ$, $\beta$, $T_\circ$, $m$, and $c$ are all free parameters that are fit to a constant value across the mission for each diode. Evaluating this model as a function of $T_\mathrm{RXB}$ and $T_\mathrm{FPA}$ requires the housekeeping data for the thermistor that was physically closest to the relevant radiometer's focal plane on the satellite. The free parameters are fully tabulated in the \WMAP\ Explanatory Supplement \citep{wmapexsupp}, but the physical layout of the thermistors in the focal plane and receiver box is not readily available. We therefore do not attempt to reproduce the gain model given in Eq.~\eqref{eq:wmap_gain}.

\begin{figure}
	\includegraphics[width=\columnwidth]{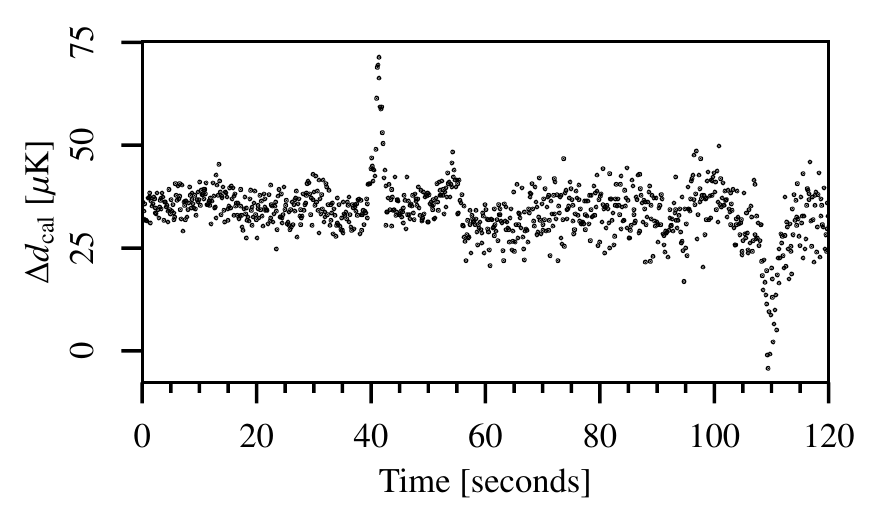}
	\caption{Difference between the \cosmoglobe\ $\boldsymbol d_\mathrm{cal}=\boldsymbol d/g-\boldsymbol b - \boldsymbol s_\mathrm{sl}$ and the delivered calibrated TOD from \WMAP. Note that this is for a small segment of the \K113 TOD displayed in Fig.~\ref{fig:timestreams}.}
	\label{fig:cal_comp_10min}
\end{figure}

Rather, we estimate the gains and baselines by comparing the uncalibrated \WMAP\ data with the calibrated \WMAP\ data, after subtracting a far sidelobe contribution convolved with the delivered \WMAPnine\ DA maps plus the Solar dipole.  We find that the calibrated and uncalibrated data can be related by
\begin{equation}
	d^\mathrm{raw}_t=g(d^\mathrm{cal}_t+s^\mathrm{sl}_t)+\sum_{i=0}^3c_i(t-t_0)^i,
\end{equation}
where the second term is a cubic polynomial with coefficients $c_i$ referenced to the time at the beginning of the scan $t_0$. The red curves in the top two panels of Fig.~\ref{fig:inst_K113} correspond to these estimates. The agreement between the \WMAPnine\ and \cosmoglobe\ gain and baseline models appear reasonable at this level and for this diode.

A complete comparison between the  \WMAP\ and \cosmoglobe\ gain and baseline models for all diodes is provided in Appendix~\ref{sec:survey}. In particular, Fig.~\ref{fig:baseline} shows the baseline differences as a function of time, and here we see that most diode differences scatter around a constant value that is close to zero; the precise constant value is of limited importance, since that only corresponds to a difference in the overall monopole of the maps, which for \WMAP\ is determined through post-processing. However, there are a few notable features. First, we see that the two \Q11 diodes exhibit large variations at the very beginning of the mission, with typical values of a few du's. Individual scans show notable spikes for many diodes. These are all relatively isolated in time, and will therefore have relatively minor impact on the final maps. Far more significant are the \W-band differences, for which one sees both slow drifts and abrupt changes. Furthermore, in many cases they vary notably between diodes within the same DA, which translates into differences in the large-scale polarization maps derived from the two pipelines.

Similarly, Fig.~\ref{fig:gain} compares the gain solutions directly, while Fig.~\ref{fig:dgain} shows the fractional differences in units of percent. Overall, we see that the two gain models agree to typically about 0.5\,\% in an absolute sense, and better than typically 0.1\,\% in terms of relative agreement between neighboring scans. The most striking feature in this plot is an annual variation that traces the \WMAP\ satellite's motion around the Sun. In general, such an oscillatory gain behaviour is entirely expected, because of known temperature variations in the satellite. However, the difficulty lies in estimating the magnitude of the oscillations, as different radiometers can respond differently to these temperature variations. In this respect, it is useful to recall that the \WMAP\ and \cosmoglobe\ gain estimation algorithms differ at a fundamental level. The \WMAP\ analysis considers each DA in isolation, fitting seven instrumental parameters, defined by Eq.~\eqref{eq:wmap_gain}, to the orbital dipole seen by each DA. The \cosmoglobe\ analysis considers the problem globally, and attempts to fit all gain parameters to the full sky signal (including both the Solar and orbital CMB dipole) simultaneously, without the use of a strong instrumental model prior. Returning to the absolute gains shown in Fig.~\ref{fig:gain}, it is difficult to determine visually which approach is better at this level alone, as the two models are quite similar; in some cases, such as \Ka124 and \Q214, the \WMAP\ model oscillates more strongly than the \cosmoglobe\ model, while in others, such as \K113 and \K114, the opposite is true. We also see the impact of the strong instrumental priors in the \WMAP\ solution particularly well in \W-band, where the \cosmoglobe\ gains are far more noisy than the \WMAP\ gains.

The impact of these differences at the TOD level is illustrated in Fig.~\ref{fig:cal_comp_10min}, which shows the calibrated \cosmoglobe\ timestream $\boldsymbol d/g-\boldsymbol s_\mathrm{sl}-\boldsymbol b$ minus the \WMAP\ calibrated signal in units of microkelvin. The most prominent feature is a $\sim25\,\mathrm{\mu K}$ offset, which is unsurprising, given the different treatment of baselines in our two pipelines. The second obvious difference is a series of spikes associated with Galactic plane crossings. Differences of order $50\,\mathrm{\mu K}$ are seen where the absolute sky brightness is about $10\,\mathrm{mK}$, equivalent to $\sim0.5\,\%$ deviations in the gain solution. This is twice as large as the 0.2\,\% uncertainty estimated in \citet{bennett2012} based on end-to-end simulations.

Another interesting feature in Fig.~\ref{fig:cal_comp_10min} is slow correlated variations at a timescale of $\sim$\,20\,sec timescale. There is nothing in the \cosmoglobe\ instrument model that varies on such short timescales, and this must therefore come from \WMAP. The most likely explanation is the fact that the \WMAP\ gain model depends directly on housekeeping data that are recorded with a 23.04\,sec sample rate, and these values appear to have been applied without any smoothing, resulting in sharp jumps in the final \WMAP\ gain model, as well as increases in the observed level of scatter. At the same time, it is also important to note that the \cosmoglobe\ gain model does not include any time-varying structure within a single scan, and if any artifacts resulting from this are identified in the current products, it may be worth incorporating housekeeping data in a future  \cosmoglobe\ data release.

\subsection{Transmission imbalance}
\label{sec:xim}

\begin{table}
\newdimen\tblskip \tblskip=5pt
\caption{Transmission imbalance parameters for each \WMAP\ radiometer as estimated in the current analysis (\emph{second column}) and in the official 9-year \WMAP\ analysis (\emph{third column}). Our uncertainties indicate $1\,\sigma$ marginal posterior standard deviations. }
\label{tab:xim}
\vskip -8mm
\footnotesize
\setbox\tablebox=\vbox{
 \newdimen\digitwidth
 \setbox0=\hbox{\rm 0}
 \digitwidth=\wd0
 \catcode`*=\active
 \def*{\kern\digitwidth}
  \newdimen\dpwidth
  \setbox0=\hbox{.}
  \dpwidth=\wd0
  \catcode`!=\active
  \def!{\kern\dpwidth}
  \halign{\hbox to 1.8cm{#\leaderfil}\tabskip 2em&
    \hfil$#$\hfil \tabskip 2em&
    \hfil$#$\hfil \tabskip 0em\cr
\noalign{\doubleline}
\omit\hfil\sc Radiometer \hfil& x_{\mathrm{im}}^{\mathrm{CG}}& x_{\mathrm{im}}^{\mathit{WMAP}}\cr
\noalign{\vskip 3pt\hrule\vskip 5pt}
\K11 &   0.00028 \pm  0.00019  &  -0.00067 \pm  0.00017\phantom{-} \cr
\K12 &   0.00442 \pm  0.00054  &   0.00536 \pm  0.00014 \cr
\Ka11 &   0.00338 \pm  0.00027  &   0.00353 \pm  0.00017 \cr
\Ka12 &   0.00138 \pm  0.00022  &   0.00154 \pm  0.00008 \cr
\Q11 &   0.00076 \pm  0.00043  &  -0.00013 \pm  0.00046\phantom{-} \cr
\Q12 &   0.00530 \pm  0.00041  &   0.00414 \pm  0.00025 \cr
\Q21 &   0.00993 \pm  0.00057  &   0.00756 \pm  0.00052 \cr
\Q22 &   0.01314 \pm  0.00071  &   0.00986 \pm  0.00115 \cr
\V11 &  -0.00016 \pm  0.00066\phantom{-}  &   0.00053 \pm  0.00020 \cr
\V12 &   0.00221 \pm  0.00063  &   0.00250 \pm  0.00057 \cr
\V21 &   0.00281 \pm  0.00059  &   0.00352 \pm  0.00033 \cr
\V22 &   0.00411 \pm  0.00087  &   0.00245 \pm  0.00098 \cr
\W11 &   0.01145 \pm  0.00146  &   0.01134 \pm  0.00199 \cr
\W12 &   0.00338 \pm  0.00131  &   0.00173 \pm  0.00036 \cr
\W21 &   0.01534 \pm  0.00180  &   0.01017 \pm  0.00216 \cr
\W22 &   0.01618 \pm  0.00155  &   0.01142 \pm  0.00121 \cr
\W31 &  -0.00105 \pm  0.00107\phantom{-}  &  -0.00122 \pm  0.00062\phantom{-} \cr
\W32 &   0.00375 \pm  0.00133  &   0.00463 \pm  0.00041 \cr
\W41 &   0.02596 \pm  0.00252  &   0.02311 \pm  0.00380 \cr
\W42 &   0.01886 \pm  0.00203  &   0.02054 \pm  0.00202 \cr
\noalign{\vskip 5pt\hrule\vskip 5pt}}}
\endPlancktablewide
\end{table}

\begin{figure}[t]
	\centering
	\includegraphics[width=\linewidth]{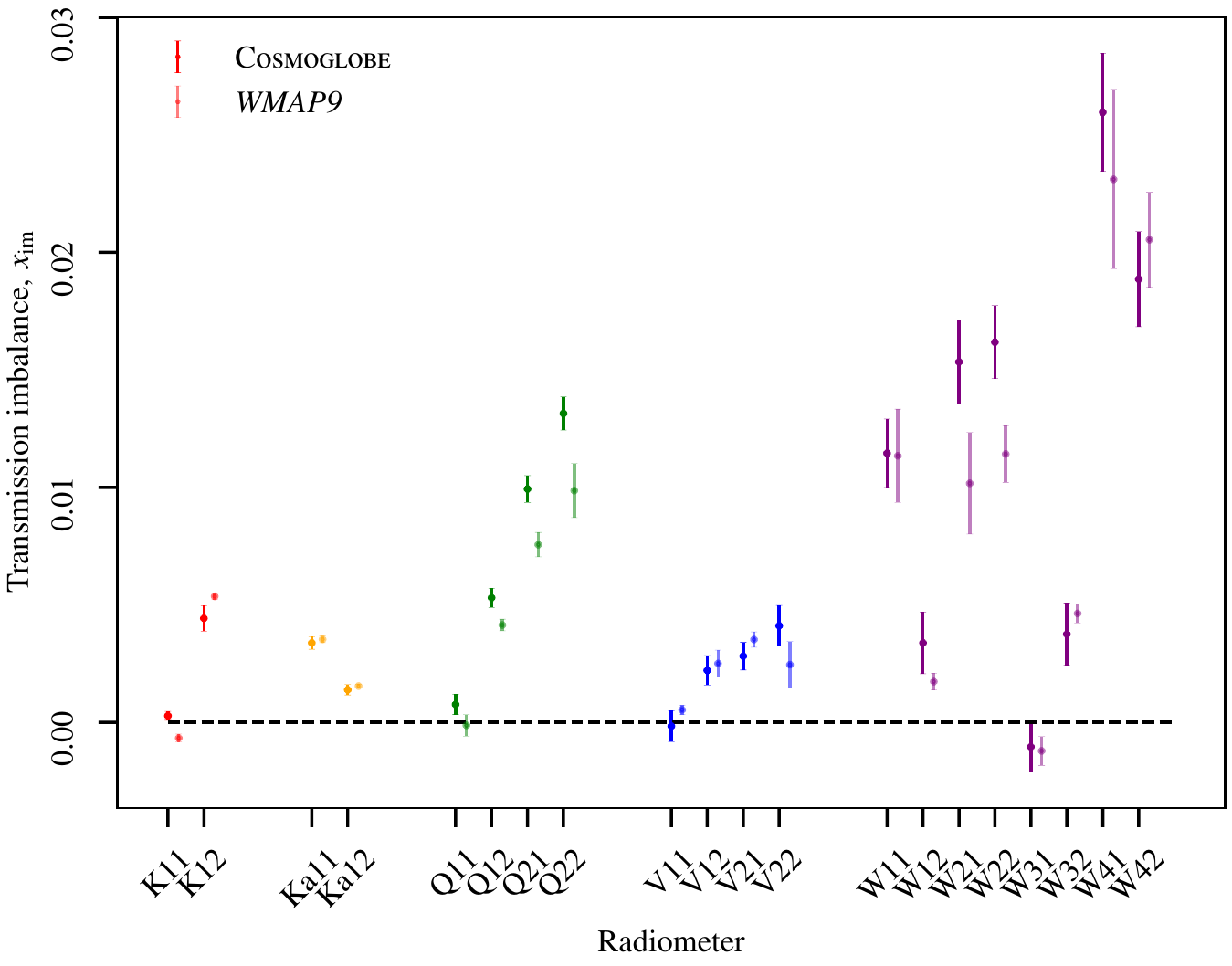}
	\caption{Comparison of the transmission imbalance factors, $x_{\mathrm{im}}$, estimated by \cosmoglobe\ (dark colors) and \WMAPnine\ (light colors) for each radiometer.}
	\label{fig:x_im}
\end{figure}

Closely related to the gain model is the transmission imbalance factor,
$x_{\mathrm{im}}$, quantifying the difference between responsivity in the two
horns, as described in Sect.~\ref{sec:wmap_instmodel}. These are listed for
each radiometer in Table~\ref{tab:xim} for both \cosmoglobe\ and \WMAPnine; for
\cosmoglobe\ the reported values correspond to marginal posterior means and
standard deviations. The same information is plotted in Fig.~\ref{fig:x_im}.
Overall, there is a reasonable agreement between the \cosmoglobe\ and
\WMAPnine\ estimates, with the same overall magnitude for each individual
radimoeter.  At the same time, we do observe several notable differences.
There are nearly $4\,\sigma$ differences between the derived \K11 radiometer,
while for \Q-band and \W2, the $x_\mathrm{im}$ factors are consistently larger
by about $1\,\sigma$ for all radiometers. The comparison between \W-band and
\V-band is especially noteworthy, as these are the bands with the most
prominent transmission imbalance modes in the \WMAPnine\ maps, especially \W4.
The uncertainties in \cosmoglobe\ are also larger for the lower frequency
channels than in \WMAPnine, which can be explained by the dependence of
$x_\mathrm{im}$ on the varying sky model within \commanderthree. At the same
time, the ratio of uncertainties in \W-band varies across radiometers, somewhat
depending on the knee frequency of the radiometer in question. Overall, the
variation in the noise levels and values is expected because of the larger
number of degrees of freedom in the \cosmoglobe\ model, while the amplitude of
the relative agreement shows the robustness of the data to the specific
pipeline choices.

While these differences may appear to be subtle, they could still be important both in terms of final sky map biases and error propagation. The reason for this is that $x_{\mathrm{im}}$ couples directly to the astrophysical sky signal, and in particular to the bright 3\,mK Solar CMB dipole. Even an inaccuracy at the $\mathcal{O}(10^{-3})$ level can therefore excite correlated large scale artifacts at the microkelvin level, which are comparable to (or larger than) the expected cosmological reionization of about 0.5\muK.

It is also important to note that even though the error bars reported in Table~\ref{tab:xim} are of similar order of magnitude, the detailed manners in which these uncertainties are propagated into higher-level astrophysical results are very different in the two pipelines. Specifically, while the \cosmoglobe\ sampling approach accounts for all couplings between the specific value of $x_{\mathrm{im}}$ and all other parameters (gain, baselines, correlated noise, CMB dipole, large scale polarization, etc.) at every single step of the Markov chain, the \WMAP\ approach only marginalizes over two linear templates in the low-resolution covariance matrix. These two templates are derived by changing $x_{\mathrm{im}}$ for one diode pair in a DA by +10\,\% and the other diode pair by $\pm10$\,\% with respect to their mean values. This linear low resolution approach can obviously only capture a limited subvolume of the full nonlinear effect of transmission imbalance uncertainties. Even cases for which the mean estimates formally agree within $1\,\sigma$ may therefore in practice result in significantly different sky maps and error estimates. We will return to the impact of transmission imbalance in Sect.~\ref{subsec:imbalance_template}.

\begin{table*}
\newdimen\tblskip \tblskip=5pt
\caption{Summary of noise properties. }
\label{tab:noise}
\vskip -4mm
\footnotesize
\setbox\tablebox=\vbox{
 \newdimen\digitwidth
 \setbox0=\hbox{\rm 0}
 \digitwidth=\wd0
 \catcode`*=\active
 \def*{\kern\digitwidth}
  \newdimen\dpwidth
  \setbox0=\hbox{.}
  \dpwidth=\wd0
  \catcode`!=\active
  \def!{\kern\dpwidth}
  \halign{\hbox to 2.cm{#\leaderfil}\tabskip 2em&
    \hfil#\hfil \tabskip 1em&
    \hfil$#$\hfil \tabskip 0.5em&
    \hfil$#$\hfil \tabskip 0.5em&
    \hfil$#$\hfil \tabskip 2em&    
    \hfil$#$\hfil \tabskip 0.5em&
    \hfil$#$\hfil \tabskip 0.5em&
    \hfil$#$\hfil \tabskip 2em&
    \hfil$#$\hfil \tabskip 0em\cr
\noalign{\doubleline}
\omit&&\multispan3\hfil Sensitivity, $\sigma_0$ (mK\,$\sqrt{\mathrm{s}}$) \hfil&
\multispan3\hfil Knee frequency, $f_{\mathrm{knee}}$ (mHz) \hfil& \cr
\noalign{\vskip -3pt}
\omit&&\multispan3\hrulefill&\multispan3\hrulefill&\omit\cr
\noalign{\vskip 3pt} 
Radiometer & Diode & \textrm{GSFC} & \textit{WMAP} & \textrm{CG}/\sqrt{2} & \textrm{GSFC} & \textit{WMAP} & \textrm{CG}/\sqrt{2} & \mathrm{CG\,Slope}, \alpha \cr
\noalign{\vskip 3pt\hrule\vskip 5pt}
\K11  &  1  &  0.72  &  0.66  &   0.704 \pm  0.002  &  6.13  &  0.4  &    0.82 \pm   0.20  &   -1.01 \pm   0.10 \cr
\omit &  2  &  &  &   0.708 \pm  0.003  &  &  &    0.63 \pm   0.14  &   -0.95 \pm   0.10 \cr
\K12  &  1  &  0.87  &  0.75  &   0.796 \pm  0.004  &  5.37  &  0.51  &    0.42 \pm   0.19  &   -0.93 \pm   0.12 \cr
\omit &  2  &  &  &   0.780 \pm  0.005  &  &  &    0.71 \pm   0.15  &   -1.02 \pm   0.10 \cr
\Ka11  &  1  &  0.75  &  0.71  &   0.788 \pm  0.001  &  1.66  &  0.71  &    1.20 \pm   0.22  &   -1.02 \pm   0.09 \cr
\omit &  2  &  &  &   0.777 \pm  0.001  &  &  &    1.19 \pm   0.22  &   -1.02 \pm   0.09 \cr
\Ka12  &  1  &  0.77  &  0.72  &   0.788 \pm  0.003  &  1.29  &  0.32  &    0.62 \pm   0.16  &   -0.99 \pm   0.11 \cr
\omit &  2  &  &  &   0.784 \pm  0.001  &  &  &    0.63 \pm   0.13  &   -1.01 \pm   0.11 \cr
\Q11  &  1  &  0.99  &  0.92  &   0.998 \pm  0.002  &  3.21  &  1.09  &    1.06 \pm   0.16  &   -1.09 \pm   0.09 \cr
\omit &  2  &  &  &   0.992 \pm  0.002  &  &  &    1.06 \pm   0.16  &   -1.10 \pm   0.09 \cr
\Q12  &  1  &  0.95  &  1.02  &   1.159 \pm  0.007  &  3.13  &  0.35  &    0.45 \pm   0.47  &   -0.98 \pm   0.11 \cr
\omit &  2  &  &  &   1.146 \pm  0.007  &  &  &    0.83 \pm   0.14  &   -1.00 \pm   0.09 \cr
\Q21  &  1  &  0.89  &  0.85  &   0.908 \pm  0.002  &  1.92  &  5.76  &    2.88 \pm   0.37  &   -1.10 \pm   0.07 \cr
\omit &  2  &  &  &   0.906 \pm  0.002  &  &  &    3.22 \pm   0.56  &   -1.10 \pm   0.06 \cr
\Q22  &  1  &  1.04  &  0.99  &   1.074 \pm  0.004  &  4.61  &  8.62  &    3.95 \pm   0.54  &   -1.11 \pm   0.06 \cr
\omit &  2  &  &  &   1.064 \pm  0.003  &  &  &    4.05 \pm   0.64  &   -1.11 \pm   0.06 \cr
\V11  &  1  &  1.25  &  1.22  &   1.551 \pm  0.003  &  2.56  &  0.09  &    1.27 \pm   0.15  &   -0.90 \pm   0.06 \cr
\omit &  2  &  &  &   1.539 \pm  0.003  &  &  &    1.19 \pm   0.14  &   -0.89 \pm   0.06 \cr
\V12  &  1  &  1.07  &  1.11  &   1.398 \pm  0.002  &  4.49  &  1.41  &    2.11 \pm   0.20  &   -0.97 \pm   0.05 \cr
\omit &  2  &  &  &   1.432 \pm  0.002  &  &  &    1.88 \pm   0.17  &   -0.96 \pm   0.05 \cr
\V21  &  1  &  1.01  &  0.97  &   1.241 \pm  0.298  &  2.43  &  0.88  &    1.50 \pm   0.24  &   -0.95 \pm   0.07 \cr
\omit &  2  &  &  &   1.217 \pm  0.294  &  &  &    1.60 \pm   0.26  &   -0.97 \pm   0.06 \cr
\V22  &  1  &  1.13  &  1.1  &   1.443 \pm  0.300  &  3.06  &  8.35  &    4.01 \pm   0.85  &   -1.00 \pm   0.08 \cr
\omit &  2  &  &  &   1.415 \pm  0.316  &  &  &    3.08 \pm   0.65  &   -1.01 \pm   0.08 \cr
\W11  &  1  &  1.18  &  1.35  &   1.938 \pm  0.005  &  16.2  &  7.88  &    5.59 \pm   0.53  &   -0.94 \pm   0.05 \cr
\omit &  2  &  &  &   1.895 \pm  0.005  &  &  &    8.99 \pm   0.85  &   -0.95 \pm   0.04 \cr
\W12  &  1  &  1.41  &  1.61  &   2.301 \pm  0.005  &  15.1  &  0.66  &    3.91 \pm   0.42  &   -0.89 \pm   0.05 \cr
\omit &  2  &  &  &   2.345 \pm  0.006  &  &  &    4.81 \pm   0.53  &   -0.89 \pm   0.05 \cr
\W21  &  1  &  1.38  &  1.61  &   2.225 \pm  0.007  &  1.76  &  9.02  &   13.57 \pm   1.47  &   -0.89 \pm   0.03 \cr
\omit &  2  &  &  &   2.292 \pm  0.006  &  &  &    5.06 \pm   0.95  &   -0.93 \pm   0.05 \cr
\W22  &  1  &  1.44  &  1.72  &   2.291 \pm  0.006  &  0.77  &  7.47  &    3.02 \pm   0.53  &   -0.98 \pm   0.05 \cr
\omit &  2  &  &  &   2.232 \pm  0.007  &  &  &    7.26 \pm   1.05  &   -0.95 \pm   0.04 \cr
\W31  &  1  &  1.47  &  1.65  &   2.328 \pm  0.005  &  1.84  &  0.93  &    1.30 \pm   0.46  &   -0.99 \pm   0.07 \cr
\omit &  2  &  &  &   2.322 \pm  0.006  &  &  &    1.97 \pm   0.28  &   -0.98 \pm   0.06 \cr
\W32  &  1  &  1.69  &  1.86  &   2.707 \pm  0.015  &  2.39  &  0.28  &    1.59 \pm   0.29  &   -0.98 \pm   0.07 \cr
\omit &  2  &  &  &   2.579 \pm  0.015  &  &  &    1.40 \pm   0.39  &   -1.00 \pm   0.07 \cr
\W41  &  1  &  1.6  &  1.71  &   2.519 \pm  0.010  &  8.46  &  46.5  &   26.81 \pm   1.83  &   -0.92 \pm   0.04 \cr
\omit &  2  &  &  &   2.479 \pm  0.009  &  &  &   24.75 \pm   1.63  &   -0.92 \pm   0.04 \cr
\W42  &  1  &  1.43  &  1.65  &   2.221 \pm  0.017  &  5.31  &  26.0  &   16.10 \pm   1.09  &   -0.94 \pm   0.04 \cr
\omit &  2  &  &  &   2.202 \pm  0.015  &  &  &   17.11 \pm   1.19  &   -0.94 \pm   0.04 \cr
\noalign{\vskip 5pt\hrule\vskip 5pt}}}
\endPlancktablewide
\end{table*}

\subsection{Instrumental noise}
\label{sec:noise}

Next, we consider the instrumental noise parameters, $\boldsymbol\xi_n=\{\sigma_0, \fknee, \alpha\}$. In this case, we recall three major differences between the \cosmoglobe\ and \WMAP\ analysis. First, while we model the noise explicitly with a $1/f$ noise profile in Fourier domain, the \WMAP\ analysis adopts a model independent approach by simply measuring the autocorrelation function directly. A notable advantage of the latter approach is that it naturally accounts for the non-white noise at high frequency without algorithmic modifications, while this has to be added manually in the parametric \cosmoglobe\ approach. A second difference is the fact that while \WMAP\ uses 1- or 24-hour segments to estimate the noise model, we use 3--5 days, and are therefore able to trace noise correlations to longer timescales. Thirdly, while \WMAP\ assumed the noise filter to be constant within each year of operation, we allow it to vary between scans, that is, on a timescale of days.

With these differences in mind, Figs.~\ref{fig:sigma0}--\ref{fig:alpha} provides a complete overview of the noise parameters for all 40 \WMAP\ diodes. As in Fig.~\ref{fig:inst_K113}, the solid black lines show \cosmoglobe\ results, while the dotted red and orange lines show the corresponding 1-year and GSFC measurements, where available. Starting with the white noise level, we see that these are overall relatively constant in time, although with slight traces for annual variations in some channels (e.g., \K113); slight instabilities near the beginning and/or end of the mission in other channels (e.g., \Ka); and slight drifts in yet others (e.g., \Q12 and \W32).

When comparing the \cosmoglobe\ values with the \WMAP\ values, it is worth noting that \WMAP\ only published results for each diode-pair, not for individual diodes. All \WMAP\ values are therefore the same for each diode pair. Still, from the \cosmoglobe\ results, which are reported individually for each diode, we see that diode pairs generally have quite similar white noise levels and vary at most by a percent.

To facilitate a more quantitative comparison, Table~\ref{tab:noise} compares the \cosmoglobe\ posterior mean results with the reported \WMAP\ results. Note that for $\sigma_0$, the \cosmoglobe\ values have been scaled by a factor of $\sqrt{2}$, in order to account for the fact that these apply to individual diodes, as opposed to diode-pairs. Both in Table~\ref{tab:noise} and Fig.~\ref{fig:sigma0}, we see that about half of the \cosmoglobe\ values lie between the two \WMAP\ results, while the other half are higher. In particular the \W-band noise levels are much higher in the \cosmoglobe\ solution, sometimes by as much as 50\,\%.

In this respect, it is worth recalling from Sect.~\ref{ssec:oldsamplers} that the white noise level in raw du is in \cosmoglobe\ not strictly sampled from the full posterior distribution, but rather estimated deterministically from the highest frequencies. This makes our estimate more sensitive to possible colored noise at high frequencies \citep{bp17}. At the same time, the calibrated white noise level $\sigma_0[\mathrm K]=\sigma_0[\mathrm{du}]/g$ depends on the gain, and this  allows us to test the effects of the calibration on the instrument sensitivity itself. The calibrated white noise level follows a biannual trend indicative of a system temperature variation, which is to be expected given the radiometer equation
\begin{equation}
	\sigma_0[\mathrm V]\propto gT_\mathrm{sys}.
\end{equation}
Aside from an overall amplitude shift due to the absolute calibration variation, the shape of the white noise level is stable throughout the Gibbs chain.

Another issue worth pointing out is the fact that we are not yet accounting for correlations between the white noise in diode pairs. However, the correlation coefficient between residuals is relatively small, with e.g., values of roughly 5\,\% for \K-band. Diode pairs in \V\ and \W\ have higher correlation of $\sim25\,\%$, but have similar orders of correlation with diodes from the other pair, indicating that the correlation is driven by unsubtracted sky signal.

In summary, we have not yet been able to identify the cause of the major difference in reported white noise levels in the \W-band; while we do detect goodness of fit failures of as much as 5--10\,$\sigma$ for many of these diodes at the TOD level (see Sect.~\ref{sec:summary_stats}), such significances correspond to sub-percent errors in the white noise level. For comparison, a white noise misestimation of 50\,\% would translate into an 800\,$\sigma$ $\chi^2$ failure. This is left to be understood through future work, but we do not expect it to indicate a real failure in either analysis, but it is more likely just a matter of different conventions.

Turning our attention to the low frequency parameters, we see in Table~\ref{tab:noise} and Fig.~\ref{fig:fknee} that our knee frequencies lie between the \WMAP\ ground and laboratory measurements, almost without exception, which on the one hand indicates generally good agreement between the two analyses. However, our values are in fact closer to the \WMAP\ laboratory measurements than the \WMAP\ flight measurements. This may be due to the longer time-scales used in the \cosmoglobe\ analysis.

\begin{figure}
	\includegraphics[width=\columnwidth]{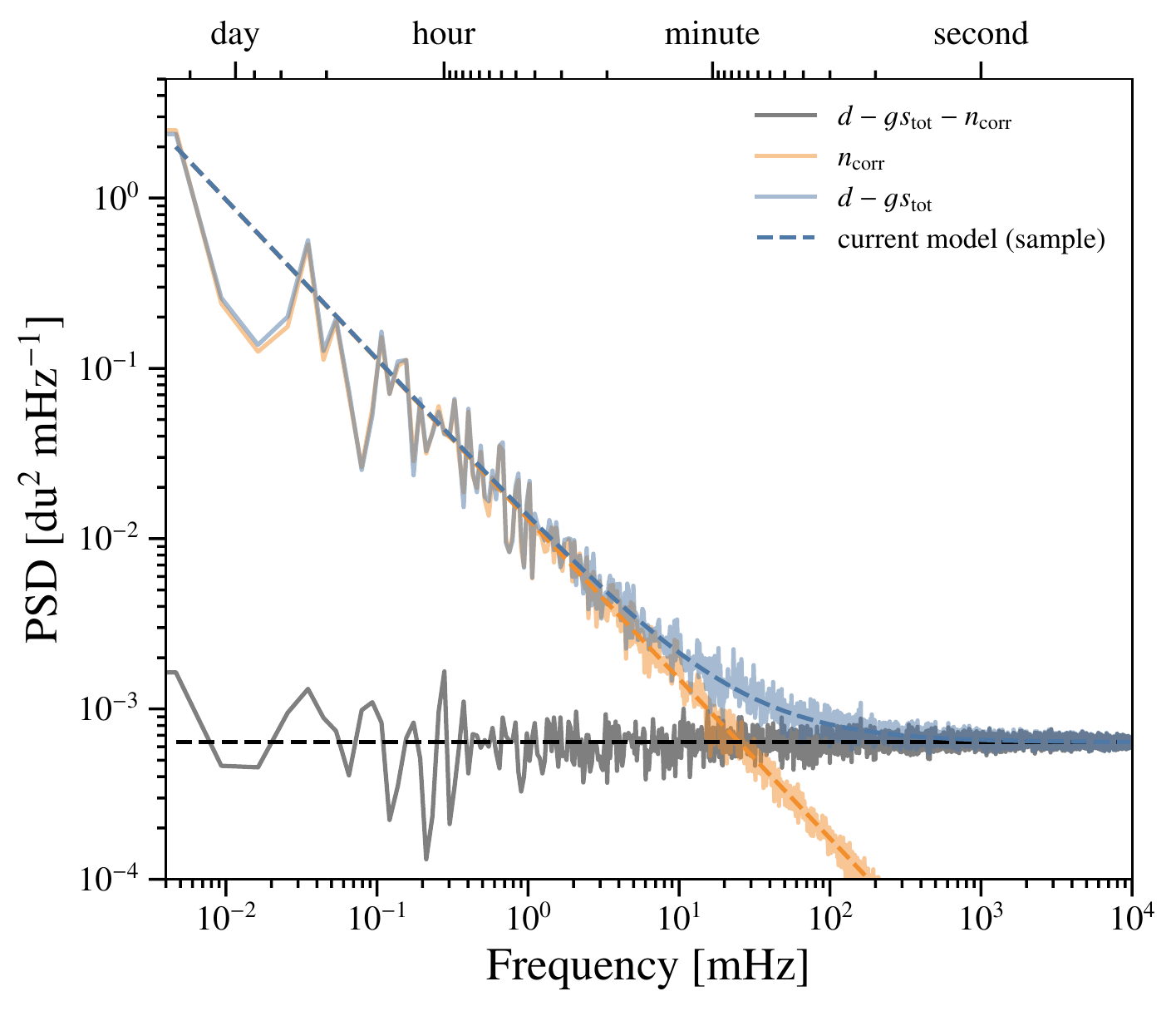}
	\caption{PSD for diode \W413 that spans MJDs 52252.3--52254.8. The power spectrum of the blue line corresponds to the residual, while the gray line is the residual with a correlated noise realization removed.}
	\label{fig:W413_psd}
\end{figure}

\begin{figure}
	\includegraphics[width=\columnwidth]{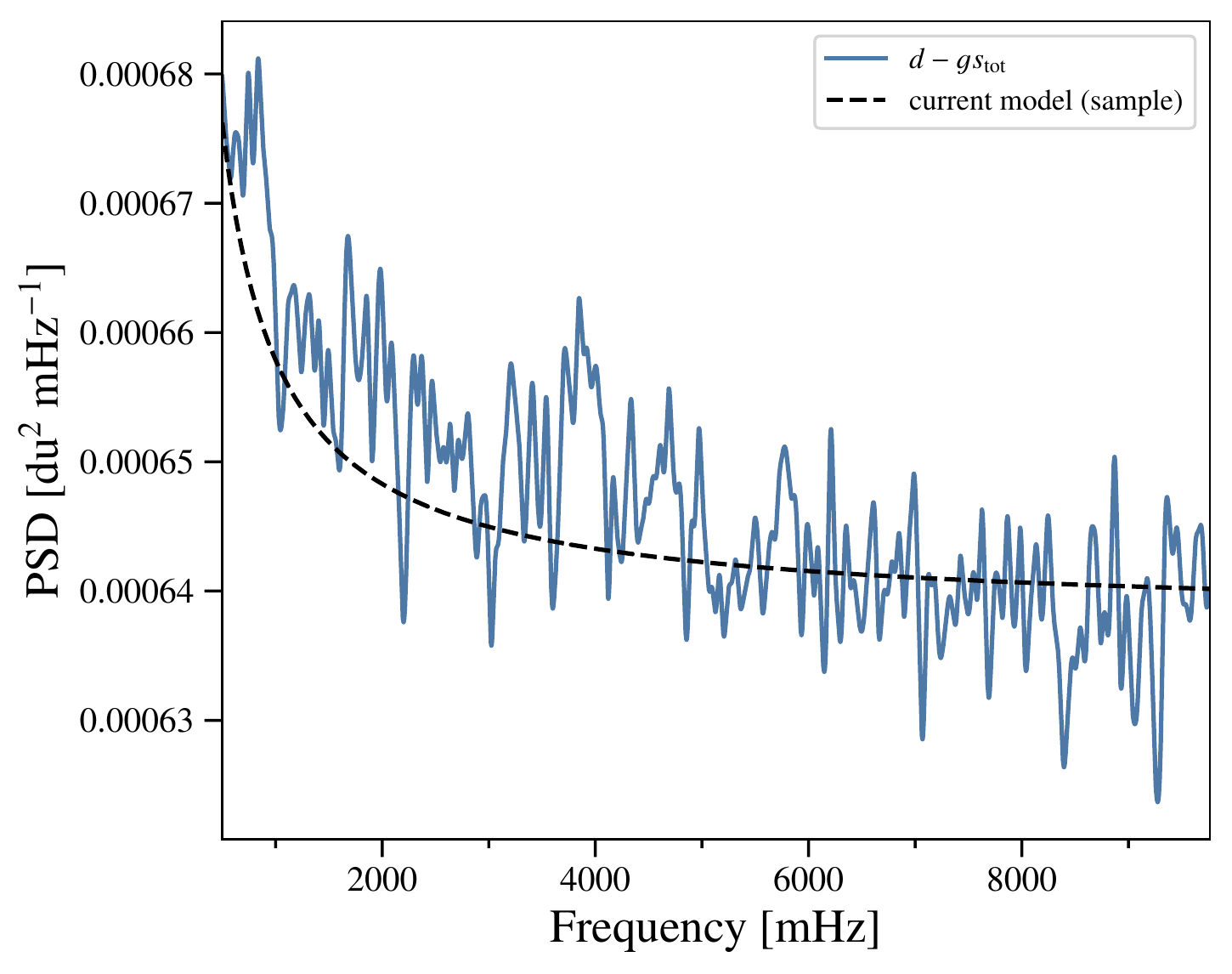}
	\caption{PSD for diode \W413 that spans MJDs 52252.3--52254.8. 
	The black dashed line is a sample of the theoretical PSD, while the blue line is the smoothed residual power spectrum.
	}
	\label{fig:W413_psd_zoom}
\end{figure}

Most diodes have constant $f_\mathrm{knee}$ throughout the mission, with a few notable exceptions. First, all \W-band channels display some amount of temporal variation that does not seem to be associated with any sinusoidal features. Second, all \Q2 channels, \V223, and \V224 all display a similar asymptotic drift in time. We have not found any instrumental effects that share this feature.
The PSD slope $\alpha$ is around $-1$ for each diode, albeit with high scatter for the lower frequencies. As expected, the uncertainty in $\alpha$ decreases as $f_\mathrm{knee}$ increases, since there are more datapoints to fit below $f_\mathrm{knee}$ where the constraining power on $\alpha$ is the strongest.

\begin{figure}
	\includegraphics[width=\columnwidth]{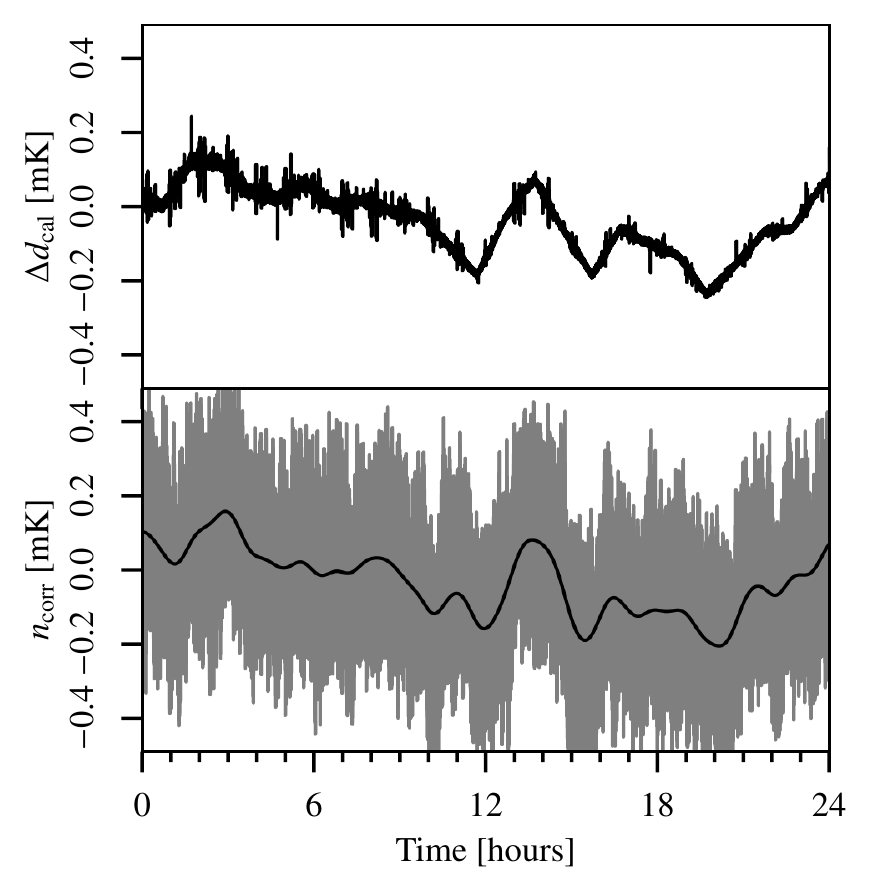}
	\caption{\textit{(top):} Difference between the \cosmoglobe\ $\boldsymbol d_\mathrm{cal}=\boldsymbol d/g-\boldsymbol b - \boldsymbol s_\mathrm{sl}$ and the delivered calibrated TOD from \WMAP. \textit{(Bottom):} Raw correlated noise (gray) and smoothed data with Gaussian kernel (black). This shows the hourly baseline subtraction from the \WMAP\ treatment.}
	\label{fig:cal_comp}
\end{figure}

For completeness, Fig.~\ref{fig:chisq} shows a summary of the reduced
normalized $\chi^2$ for all diodes. The most striking features in
these figures are the amplitude and semiannual periodicity.  Given
the noise model and data residual, we can evaluate the goodness of fit
in the form of the relative $\chi^2$. Here, we find that approximately
half of the diodes have a $\chi^2$ value at least $6~\sigma$ above
or below the expected value.  Given perfect Gaussian residuals, we
would expect these values to be within $\pm1~\sigma$ 68\,\% of the time. For
a typical \W-band scan of length $n_\mathrm{TOD}=2^{22}$, a $10~\sigma$
model failure corresponds to $\chi^2/n_\mathrm{TOD}=1.003$. It is
therefore exceedingly difficult to look at any given \WMAP\ scan in
the time domain and identify a model failure. To illustrate this,
Fig.~\ref{fig:W413_psd} compares the observed noise PSD with the
best-fit model for the \W413 diode. This is a $7~\sigma$ outlier;
despite this, the $1/f$ model appears to perform exceedingly well over
seven decades in frequency.

\begin{figure*}[t]
	\centering
	\includegraphics[width=0.32\textwidth]{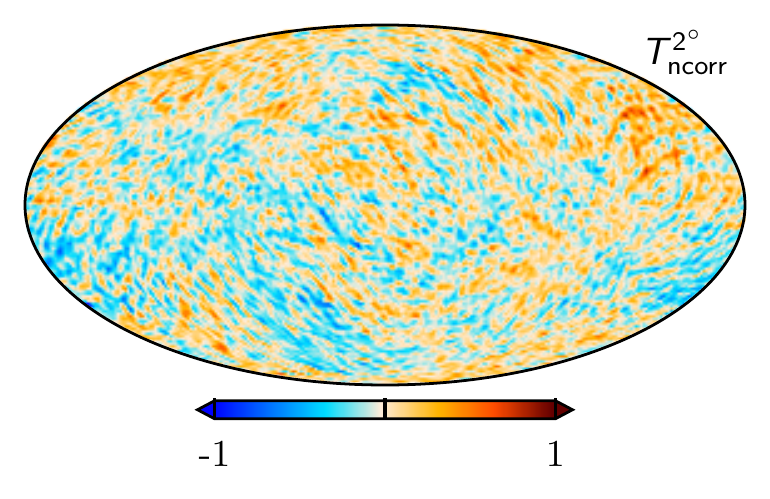}
	\includegraphics[width=0.32\textwidth]{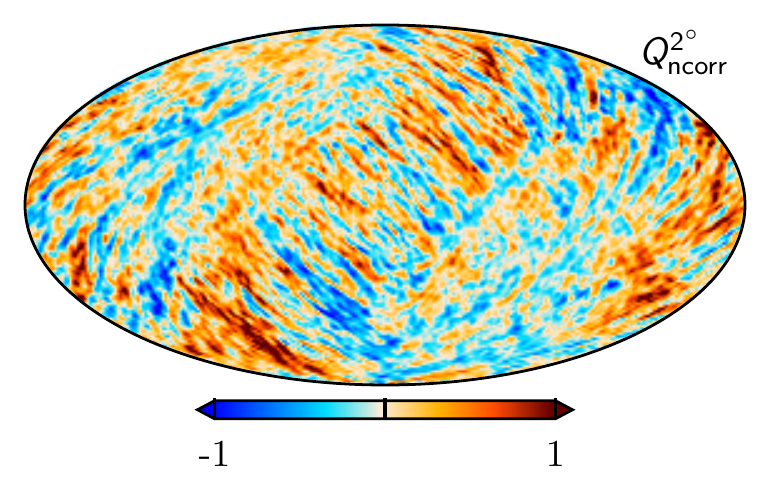}
	\includegraphics[width=0.32\textwidth]{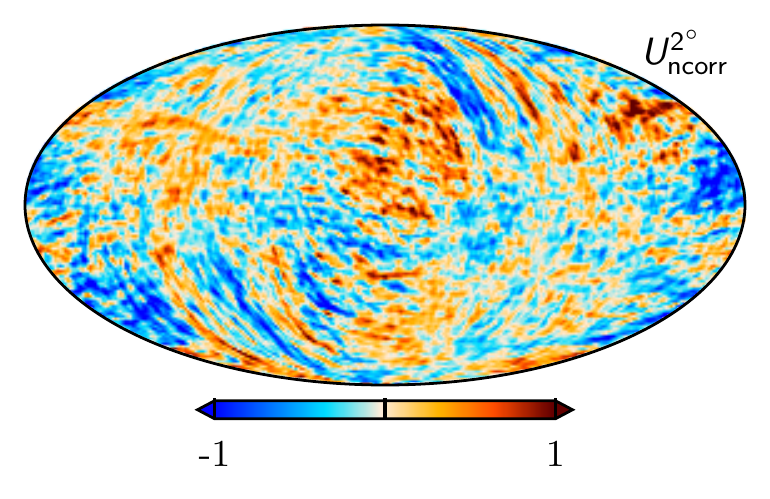}\\
	\includegraphics[width=0.32\textwidth]{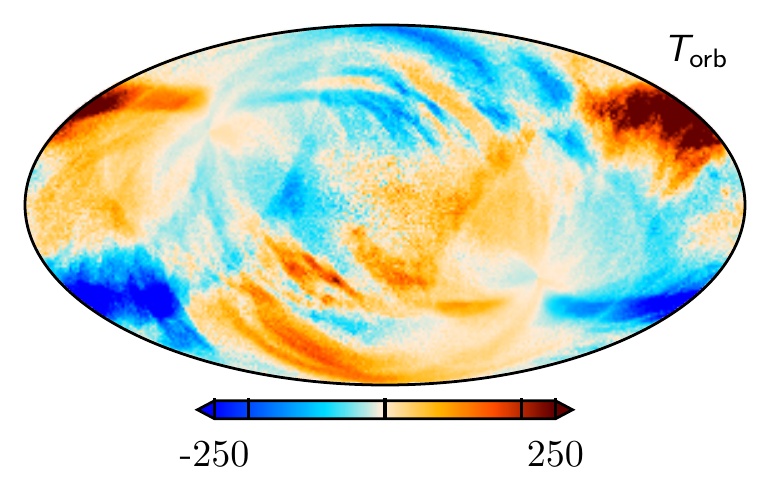}
	\includegraphics[width=0.32\textwidth]{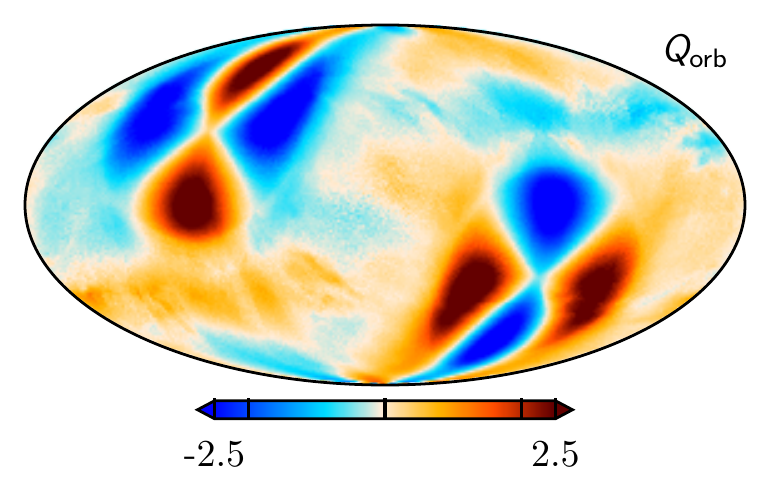}
	\includegraphics[width=0.32\textwidth]{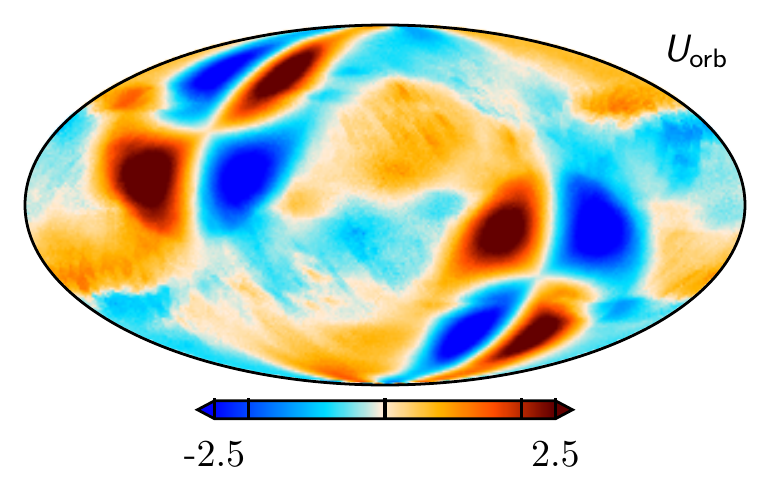}\\
	\includegraphics[width=0.32\textwidth]{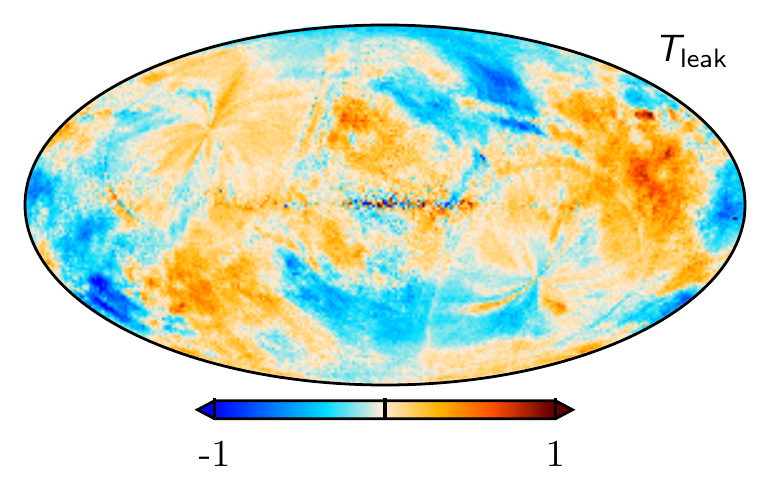}
	\includegraphics[width=0.32\textwidth]{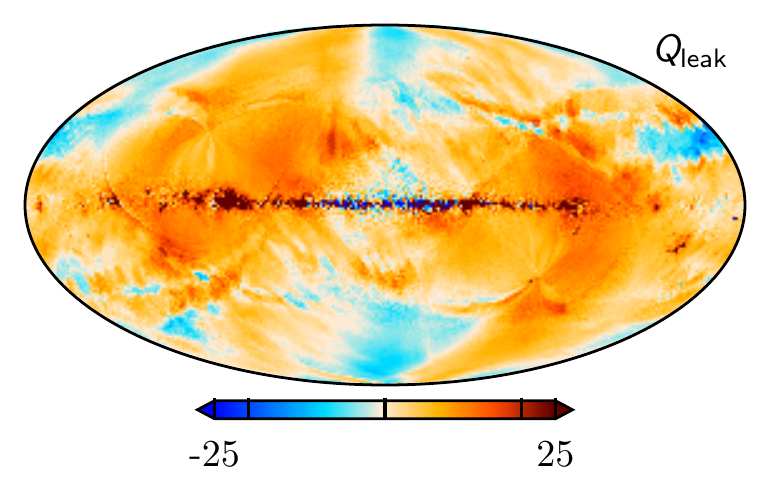}
	\includegraphics[width=0.32\textwidth]{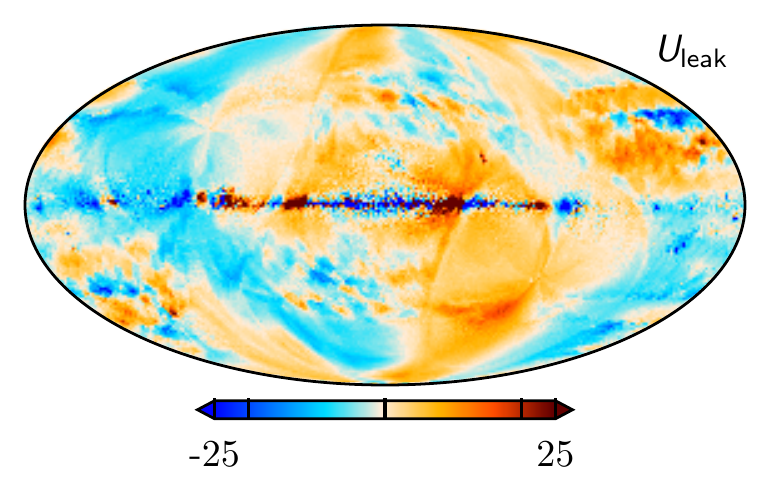}\\
	\includegraphics[width=0.32\textwidth]{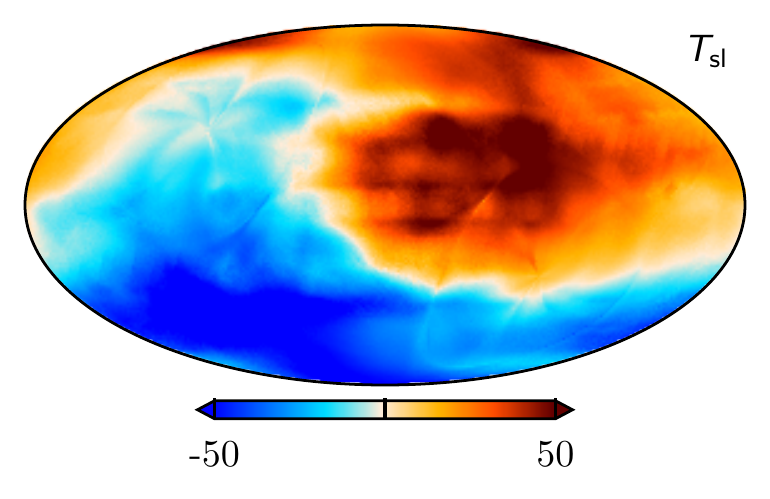}
	\includegraphics[width=0.32\textwidth]{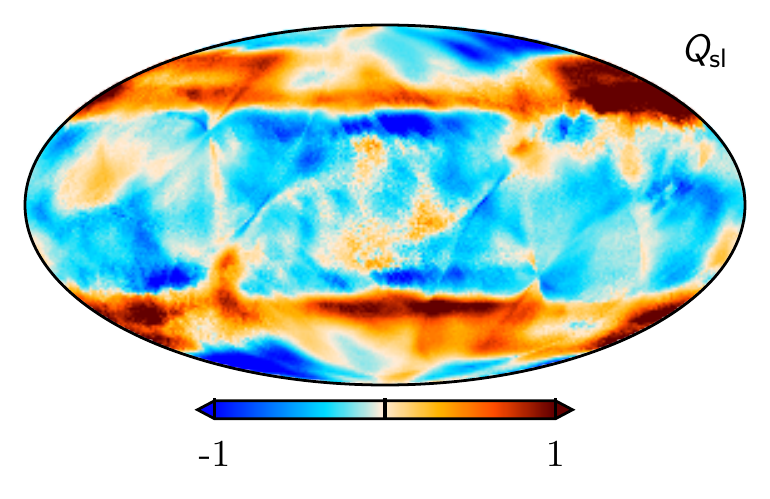}
	\includegraphics[width=0.32\textwidth]{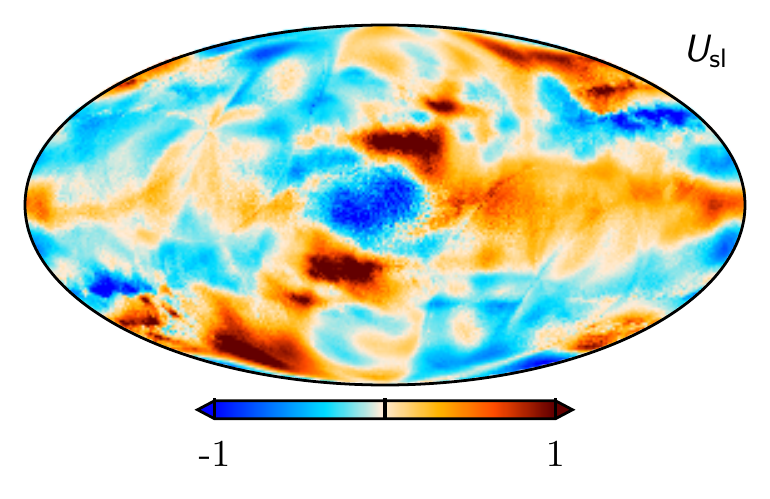}\\
	\includegraphics[width=0.32\textwidth]{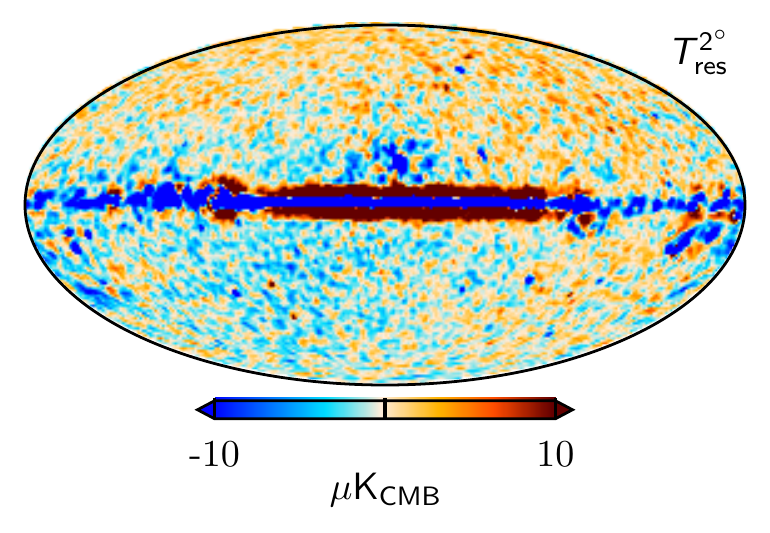}
	\includegraphics[width=0.32\textwidth]{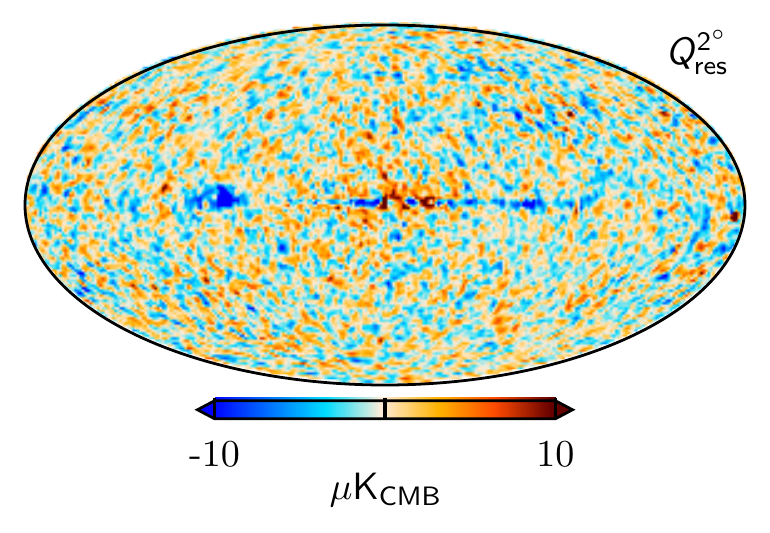}
	\includegraphics[width=0.32\textwidth]{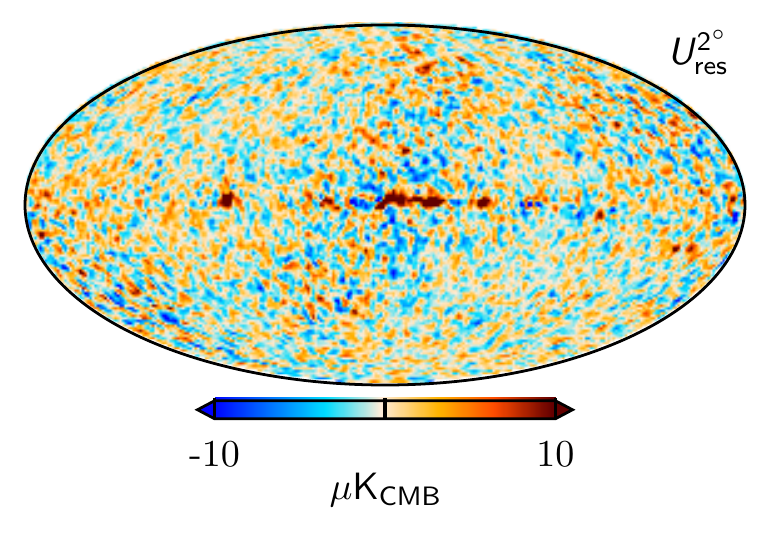}\\
	\caption{TOD corrections for \K-band for a single Gibbs sample, projected into maps. Columns show Stokes $T$, $Q$, and $U$ parameters. Rows show, from top to bottom, 1) correlated noise; 2) the orbital dipole; 3); bandpass mismatch leakage; and 4) sidelobe corrections. The bottom row shows the residual obtained when binning the sky and systematics-subtracted TOD into a sky map. Note that the correlated noise and residual have been smoothed by a $2^\circ$ Gaussian beam.}
	\label{fig:tod_corrections}
\end{figure*}

Only with aggressive smoothing does the model failure become apparent
at frequencies 1--10\,Hz. This is illustrated in
Fig.~\ref{fig:W413_psd_zoom}, which shows exactly the same underlying
data as in Fig.~\ref{fig:W413_psd}, but heavily smoothed. Here, it is
clear that despite fitting the data well at the highest and lowest
frequencies, it is in the intermediate range of 1--5\,Hz where the
$1/f$ model is a less accurate fit to the residual power
spectrum. Part of the cause of this failure is that the white noise
level is fixed by the value of the power spectrum at the
Nyquist frequency, as it was computed by differencing adjacent
samples. The power spectrum has a downward trend above 1\,Hz,
indicating that the data would be better fit by a polynomial in powers of
$f^\alpha$. This is phenomenologically similar to the
\WMAP\ collaboration's approach of describing the time-space
autocorrelation as a cubic polynomial in $\log\Delta t$
\citep{jarosik2007}.

In practice, the failure of the $1/f$ model has a small effect on the final data products, and
was not visible in noise models when we modeled the data in one day scans
rather than the longer 3--7 day scans due to the lower $n_\mathrm{TOD}$ giving
a higher uncertainty on the relative $\chi^2$.  Therefore, although this
strictly constitutes a deficiency in the model, it is in practice too small to
affect the results of the rest of the chain. The downturn of the noise PSD at
high frequencies is also present in, e.g., the \Planck\ HFI data
\citep[Fig.~1]{planck2014-a10}, so improved modeling of this form will be a
necessity in future \cosmoglobe\ endeavors, and will be used to improve the
\WMAP\ data processing.

\begin{figure*}
  \center	
   \includegraphics[width=0.95\linewidth]{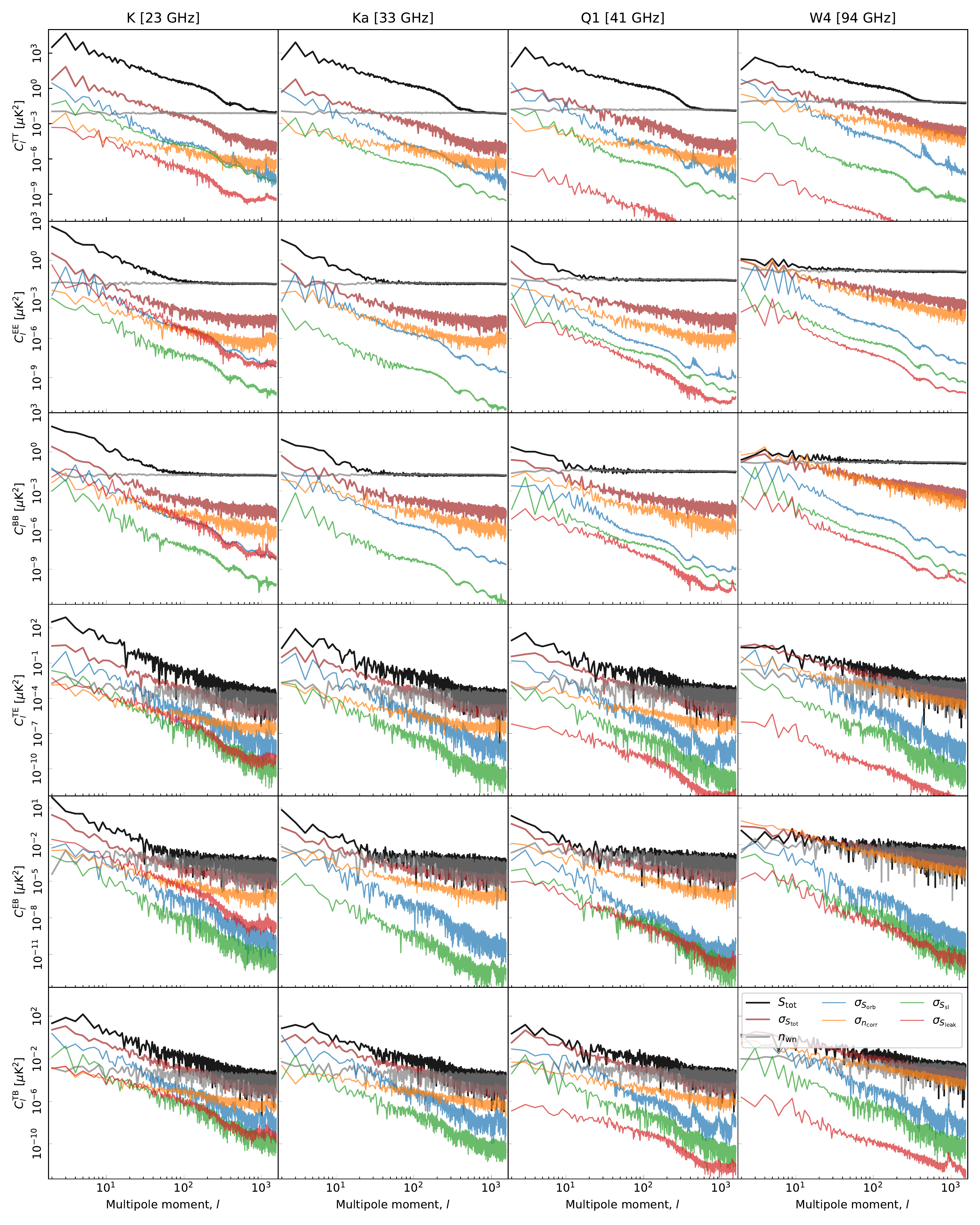}
  \caption{Pseudo-spectrum standard deviation for each instrumental
    systematic correction shown in
    Fig.~\ref{fig:tod_corrections} (\emph{thin
      colored lines}). For comparison, thick black lines show spectra
    for the mean of the full frequency map; thick red lines show their standard deviations (i.e., the full systematic uncertainty); and gray lines show white noise realizations.
    Columns show results for \K,
    \Ka, \Q1, and \W4, respectively, while rows show results for each of
    the six polarization states ($TT$, $EE$, $BB$, $TE$, $TB$, and
    $EB$). All spectra have been derived outside the CMB confidence
    mask presented by \citet{bp13} using the HEALPix \texttt{anafast}
    utility, correcting only for sky fraction and not for mask mode
    coupling. }
  \label{fig:corrmap_stddev}
\end{figure*}

Before concluding this section, we recall the close relationship between the correlated noise component and the baseline model. This is illustrated in Fig.~\ref{fig:cal_comp}, which shows the difference between the calibrated \cosmoglobe\ and \WMAP\ TOD data for \K113, i.e., the same as Fig.~\ref{fig:cal_comp_10min}, but plotted for 24 hours instead of 10 minutes. The bottom panel shows the \cosmoglobe\ correlated noise realization for the same period, both raw and smoothed. The most prominent feature in this figure is a varying signal of amplitude $0.2\,\mathrm{mK}$. This is due to the hourly baseline subtraction mentioned above, which contrasts with the \cosmoglobe\ approach of assigning a linear baseline solution for the entire scan, and then accounting for the nonlinearity through $\n^{\mathrm{corr}}$. The variations are commensurate with the correlated noise correlation length, which for \K113 has ${f_\mathrm{knee}\sim0.5\,\mathrm{mHz}}$, corresponding to a little over half an hour. Therefore, the hour-long baseline subtraction acts as a destriper, removing an estimate of the correlated noise.

\subsection{Instrumental corrections in map domain}

Returning to the global parametric data model in Eq.~\eqref{eq:model}, it is useful for intuition purposes to project each of the \WMAP\ TOD-level correction terms into sky maps. This is done for \K-band in Fig.~\ref{fig:tod_corrections}, in the same form as was done for LFI by \citet{bp10}. Columns correspond to Stokes $T$, $Q$, and $U$ parameters, while rows show, from top to bottom, 1) correlated noise; 2) the orbital CMB dipole; 3) bandpass leakage; and 4) sidelobe corrections. The bottom row shows the residual obtained after subtracting all modeled terms from the raw TOD. All maps correspond to one single and randomly selected Gibbs sample.

Starting with the correlated noise in the top row and the residual in the bottom row, we note that these are the only terms that are fundamentally stochastic in nature; all the instrumental terms are either primarily deterministic in nature, as they rely only on the sky model coupled to a small number of instrumental parameters, such as the gain, bandpass, or beam, whereas the random realizations of the sky model are highly constrained by the multifrequency component separation. As such, the correlated noise and residual maps act as the ``trash cans'' of Bayesian CMB analysis; together they show anything in the TOD that is not explicitly modeled. For the \K-band channel, we see that the correlated noise is limited to less than 1\,\muK\ over almost the full sky, while the residual appears like random noise over most of the sky. The main exceptions to this are strong residuals near the Galactic plane in temperature, which indicates the presence of unmodeled foreground features; typical candidates to explain this would be curvature in the synchrotron spectral index or a more complicated AME SED than that assumed here. Secondly, at high latitudes we see traces of a small dipole with an amplitude of 1--2\muK\ aligned with the CMB dipole. This indicates the presence of an absolute calibration deviation of about 0.03--0.06\,\%; this is within the uncertainty of the absolute \K-band calibration prior of 0.1\,\% discussed in Sect.~\ref{sec:priors}, and when inspecting different Markov chain samples of the same type, one can see that the amplitude and sign of this residual scatter is around zero as expected.

Next, the amplitude of the orbital dipole is about 250\muK\ in temperature and 2.5\muK\ in polarization. This component by itself is exceedingly well known, as it depends only on the satellite velocity and the CMB monopole temperature. However, when actually fitting this to the real data, it obviously also depends on both the gain, and sample-to-sample variations in this map, and  is therefore a good tracer of gain uncertainties. In addition, its physical appearance also in principle depends on the sidelobe model, but we do not yet propagate sidelobe uncertainties anywhere in the analysis.

The third row shows the corrections for bandpass leakage.\footnote{In principle, this also accounts for leakage due to beam differences, as discussed by \citet{bp09}, but for \WMAP\ this effect is much smaller than the bandpass effect.} As for \Planck\ LFI, this term is by far the largest correction in polarization, with an amplitude that is almost an order of magnitude larger than any of the others. It is highly nontrivial to compare this component to the \WMAPnine\ analysis, since this effect was accounted for by solving for a spurious $S$ map as part of their mapmaking. However, it is important to note that whether one models this effect explicitly, as we do, or project it out during mapmaking, as the \WMAP\ pipeline did, the accuracy of the bandpass corrections depends directly on the accuracy of the gain and transmission imbalance calibration.

The fourth row shows the sidelobe contribution. Here we see that the temperature amplitude reaches 50\,\muK, which corresponds to 1.5\,\% of the CMB Solar dipole amplitude. If it should turn out that the sidelobe model were incorrect by, say, 30\,\%, this translates directly into an error in the absolute \K-band calibration of about 0.5\,\%, which is significantly greater than the statistical error shown above. Given the degeneracies discussed in Sect.~\ref{sec:priors}, there is thus also a strong degeneracy between the AME dipole and the \K-band sidelobe. For polarization, the amplitude is mostly smaller than 1\muK, and therefore of minor importance for this channel. We note that only the intensity component of the \WMAP\ sidelobe model has been published, and polarized sidelobes are not accounted for in the current processing. However, as shown by \citet{barnes2003}'s Table 2, the amplitude of this contribution is relatively small, with a high-latitude mean of $0.8\,\mathrm{\mu K}$ for \K-band and $<0.1\,\mathrm{\mu K}$ for all other bands.

\subsection{Instrumental uncertainties in power spectrum domain }

We conclude this section by estimating the uncertainty of each instrumental effect in terms of angular power spectra, using the same methodology as \citet{bp10} for \Planck\ LFI. That is, we first compute the power spectra for each individual instrumental correction Markov chain sample, as illustrated in Fig.~\ref{fig:tod_corrections}, and then compute the standard deviation of all these samples. The results from this computation are summarized in Fig.~\ref{fig:corrmap_stddev} for four DAs, namely \K, \Ka, \Q1, and \W4. In each panel the black line shows the power spectrum of the full coadded frequency sky map (including both signal and noise), while the gray line shows the white noise level. The thick dark red line shows the sum of all variations, while the thin colored lines show the contribution from individual correction terms.

On large angular scales in the $TT$ spectrum, we see that the dominant uncertainty comes from the orbital dipole (blue lines), which trace gain uncertainties. This is reasonable, since these data are strongly signal-dominated; indeed, for \K-band even the sidelobe contribution (green lines) is higher than the correlated noise.

For large-scale $E$-mode polarization, the dominant term varies from channel to channel. Specifically, for \K-band the bandpass leakage (thin red lines) and gain fluctuations are significantly larger than the correlated noise, while for \Q1 and \W-band the correlated noise dominates for most multipoles, although the gain fluctuations are comparable for some $\ell$'s.

An important conclusion to be drawn from these measurements is that a simple uncertainty model that primarily accounts for correlated noise is likely to be suboptimal for detailed cosmological analysis of large-scale polarization. Both gain and bandpass uncertainties are at least as important for the lowest multipoles, and simultaneously accounting for all of these contributions is important in order to derive robust cosmological results.

\section{Frequency maps}
\label{sec:maps}

In this section, we discuss the reprocessed \WMAP\ frequency maps and their properties. In Sect.~\ref{ssec:means} we present the reprocessed \WMAP\ maps themselves, commenting on notable features in the maps themselves, as well as differences with the \WMAPnine\ results. In Sect.~\ref{sec:internal_consistency} we focus on the consistency of our new maps, both internally among the \WMAP\ channels, and with respect to \Planck. Section~\ref{subsec:imbalance_template} describes the efficiency of template-based transmission imbalance uncertainty propagation.

\subsection{Map survey}
\label{ssec:means}

\begin{figure*}
	\centering
	\includegraphics[width=0.75\textwidth]{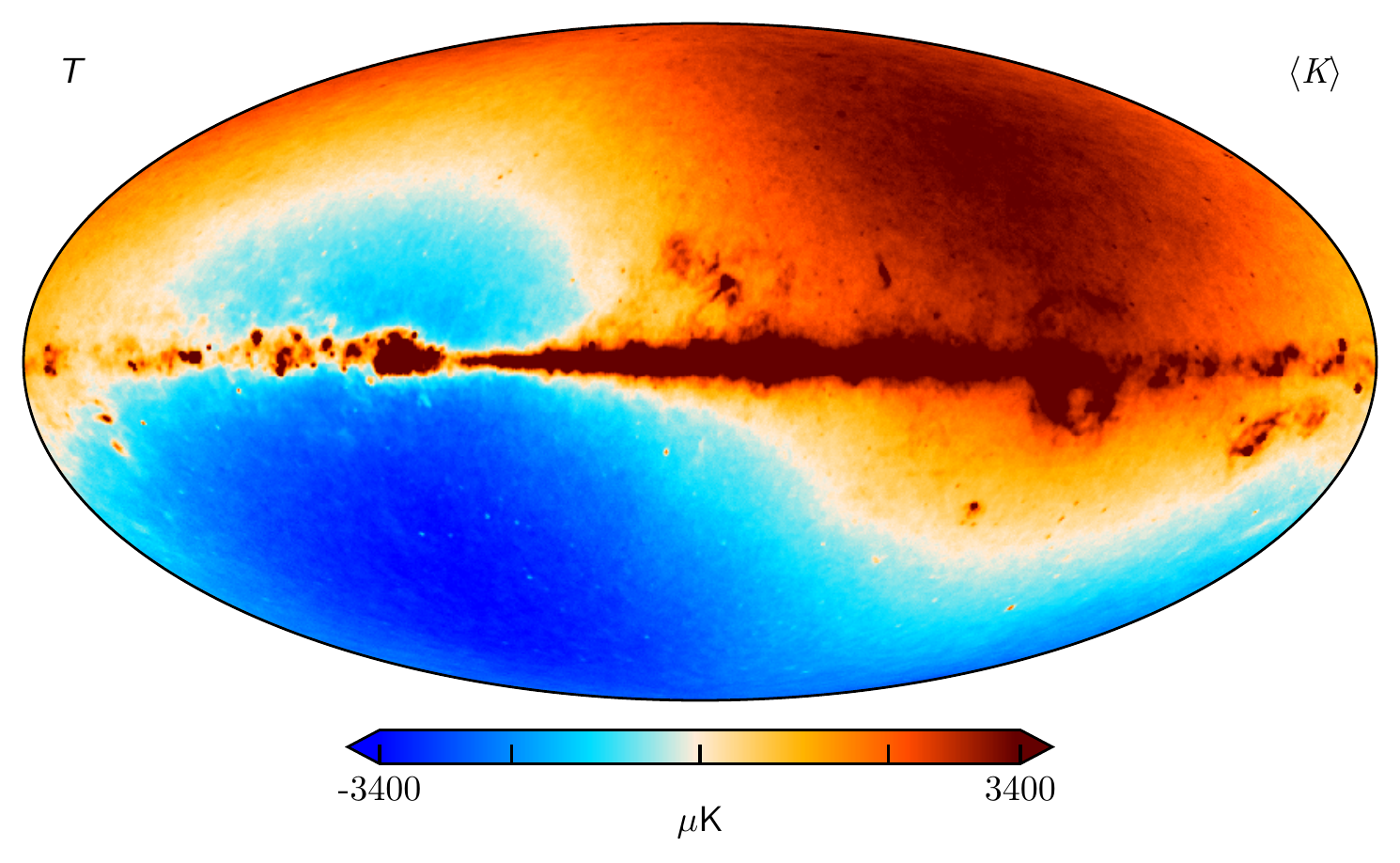}
	\includegraphics[width=0.75\textwidth]{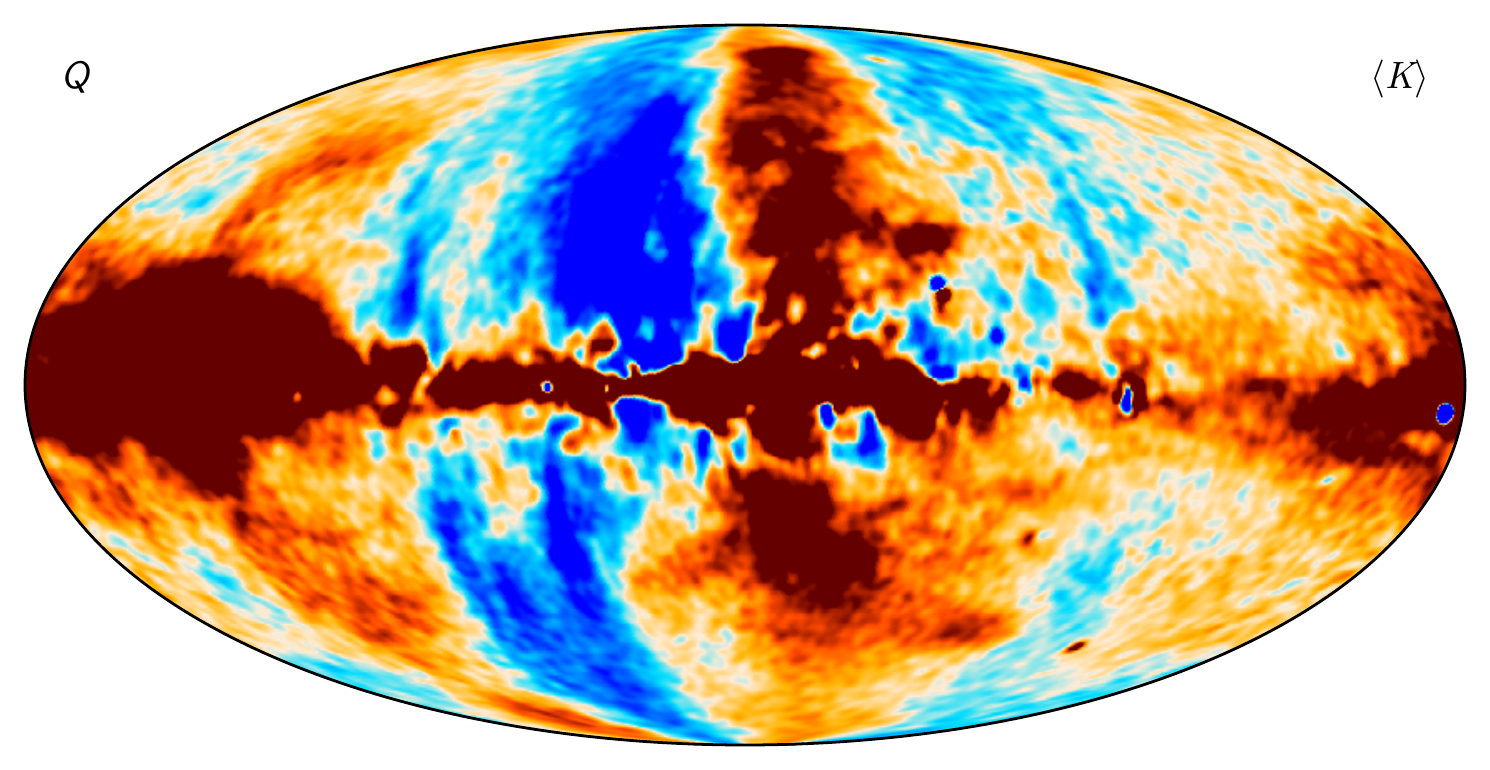}
	\includegraphics[width=0.75\textwidth]{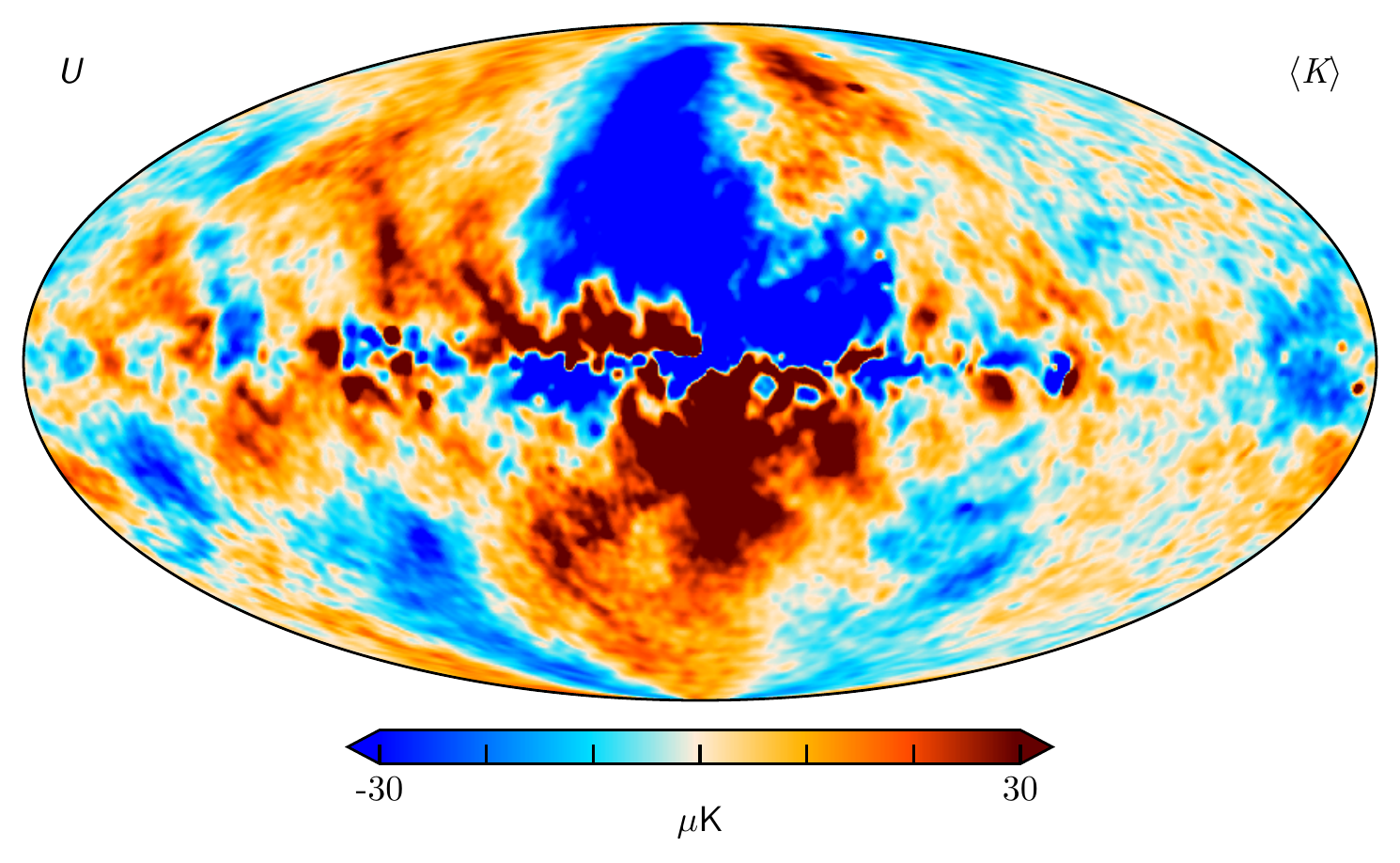}
	\caption{Posterior mean \K-band map produced with the \cosmoglobe\ pipeline. Rows show Stokes $T$, $Q$, and $U$, respectively. The temperature map is shown at full resolution, while the polarization maps are smoothed with a $2^{\circ}$ FWHM Gaussian beam to reduce small-scale noise.}
	\label{fig:kband}
\end{figure*}
\begin{figure*}
	\centering
	\includegraphics[width=0.75\textwidth]{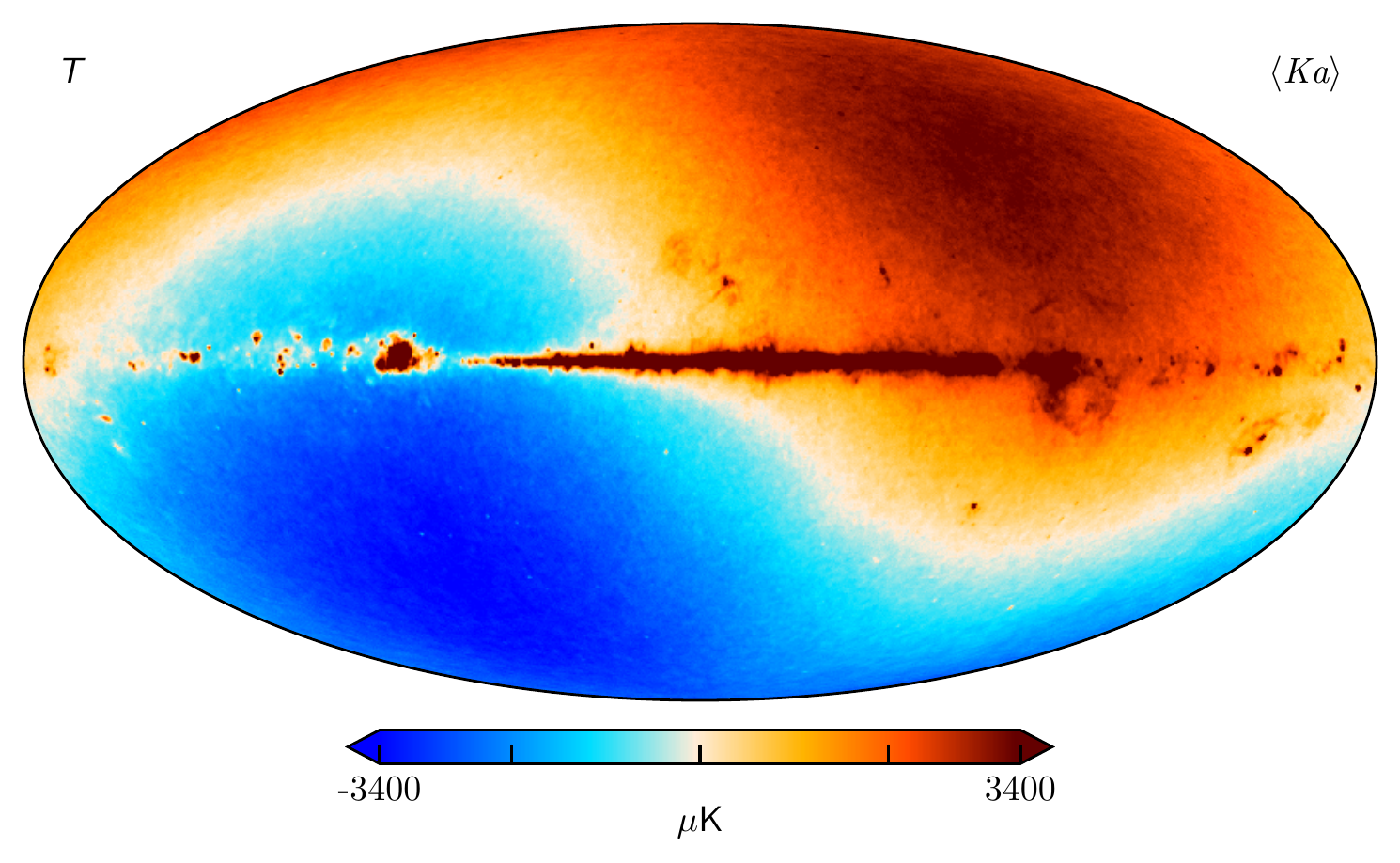}
	\includegraphics[width=0.75\textwidth]{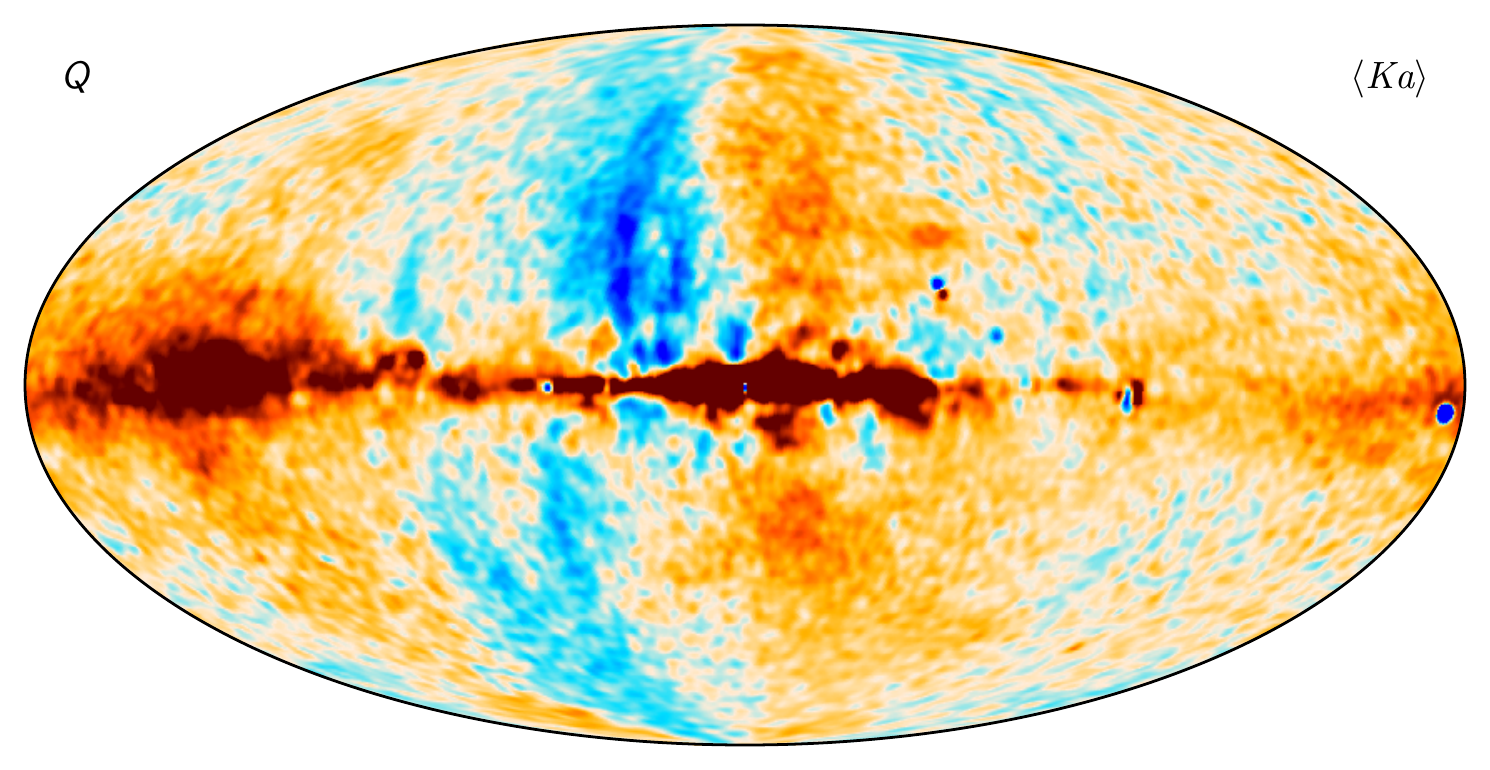}
	\includegraphics[width=0.75\textwidth]{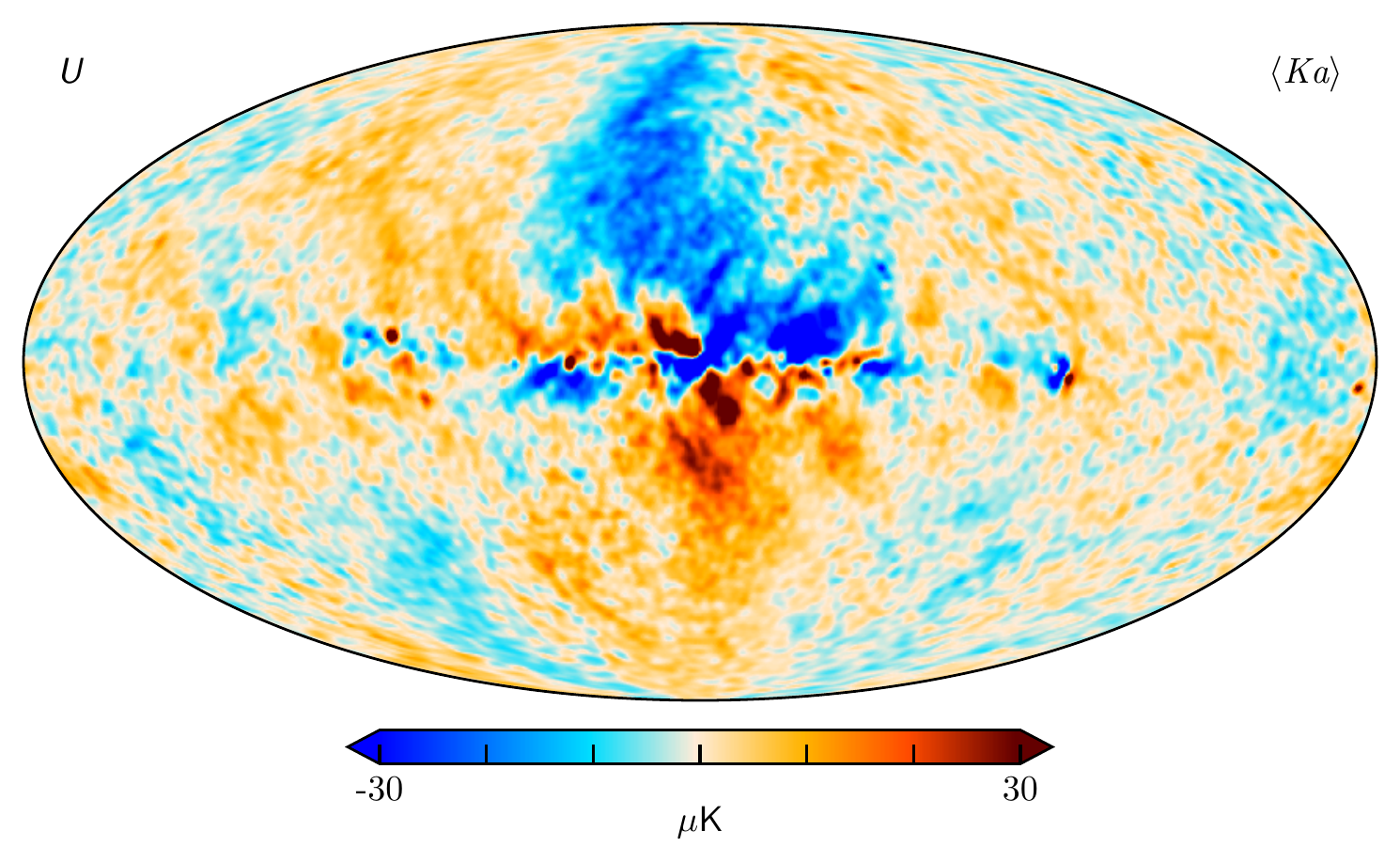}
	\caption{Same as Fig.~\ref{fig:kband}, but for \Ka-band.}
	\label{fig:kaband}
\end{figure*}
\begin{figure*}
	\centering
	\includegraphics[width=0.75\textwidth]{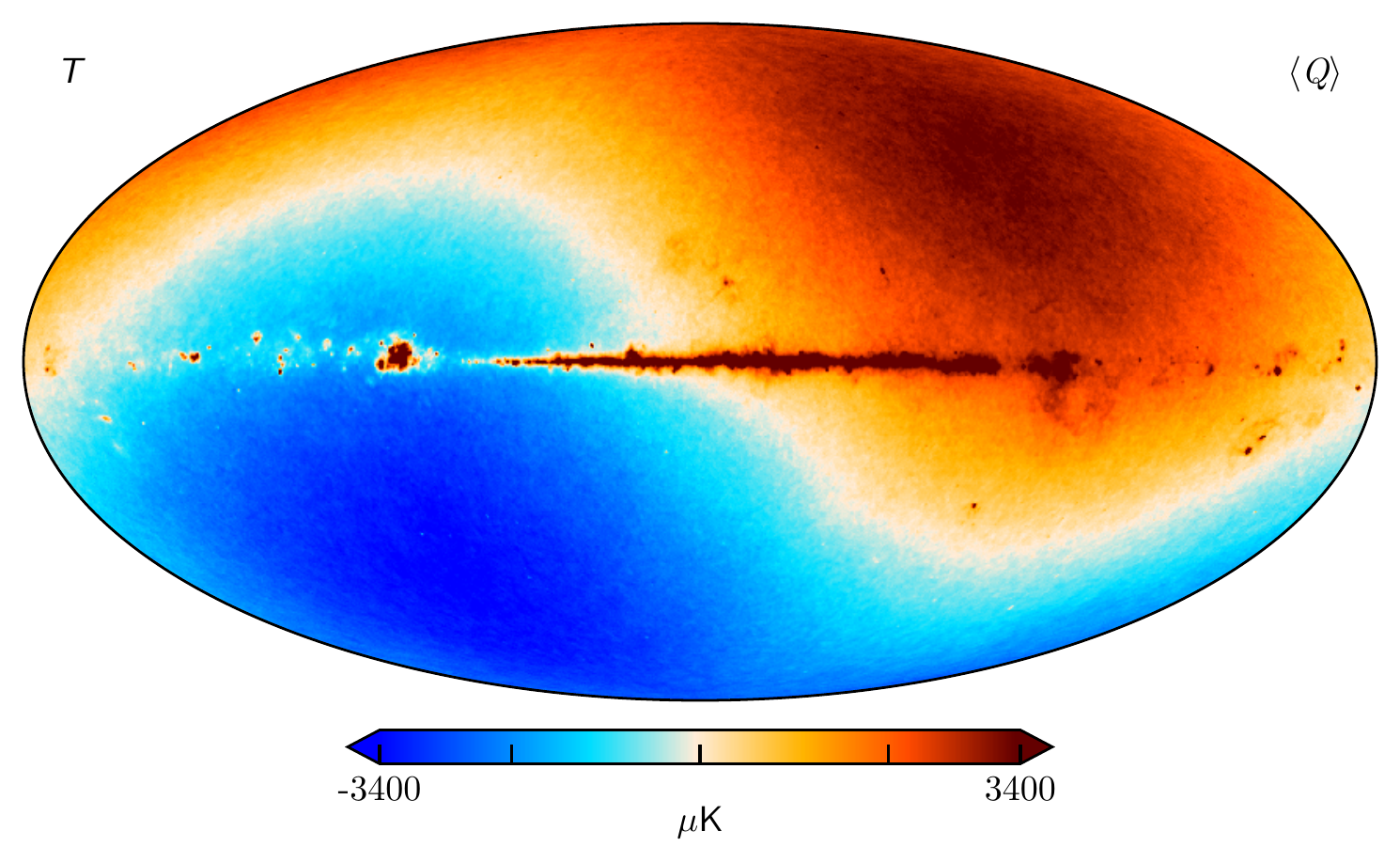}
	\includegraphics[width=0.75\textwidth]{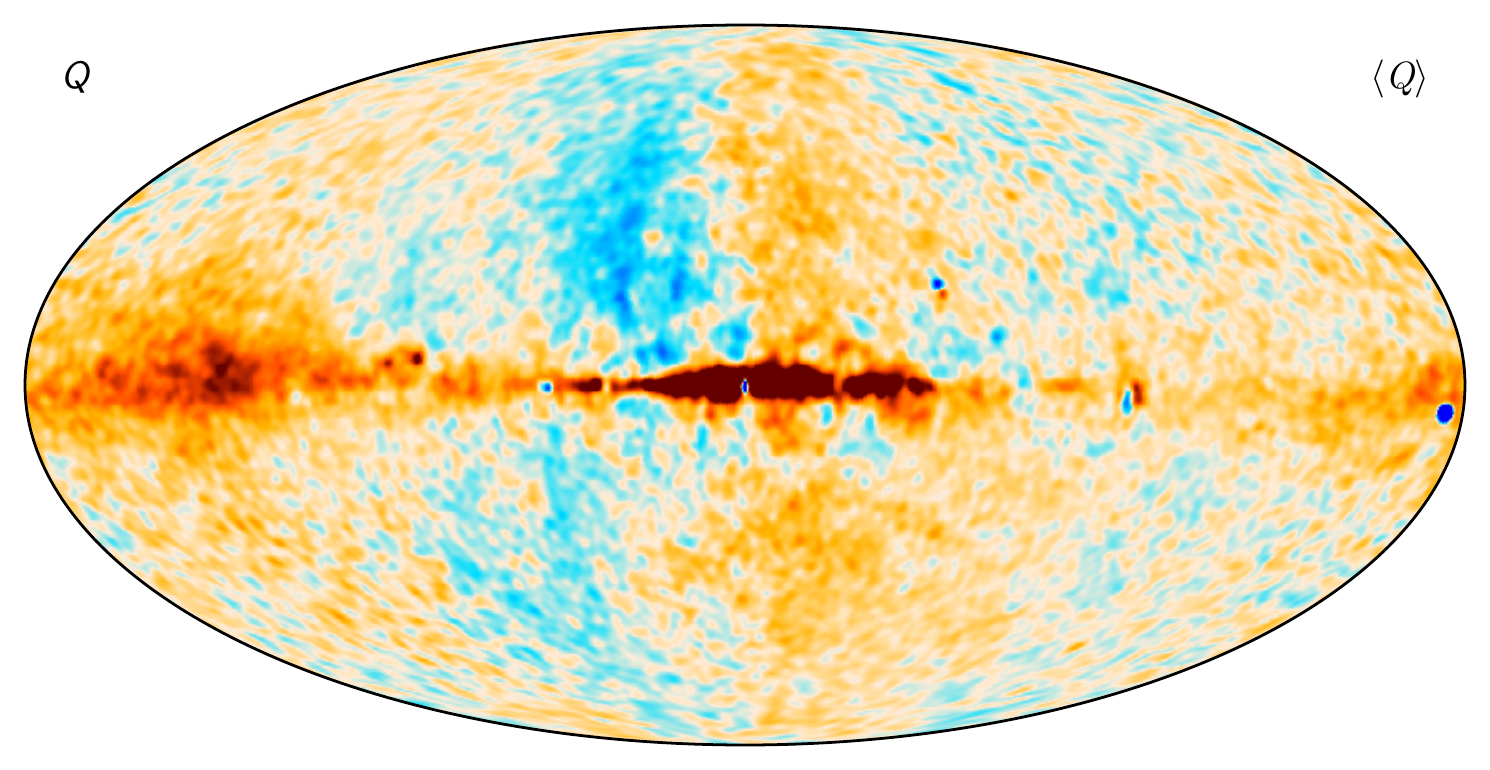}
	\includegraphics[width=0.75\textwidth]{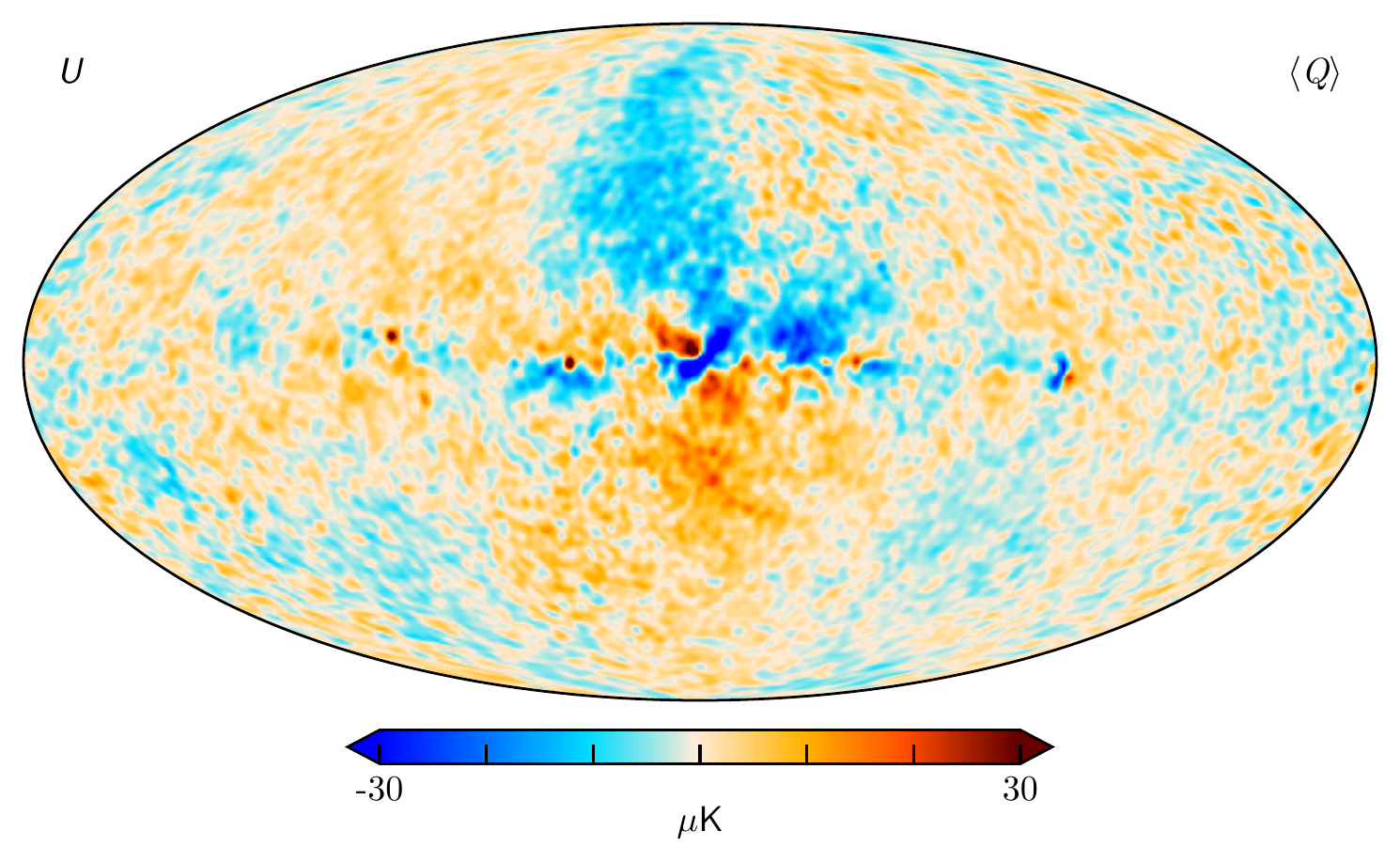}
        \caption{Same as Fig.~\ref{fig:kband}, but for \Q-band.}
	\label{fig:qband}
\end{figure*}
\begin{figure*}
	\centering
	\includegraphics[width=0.75\textwidth]{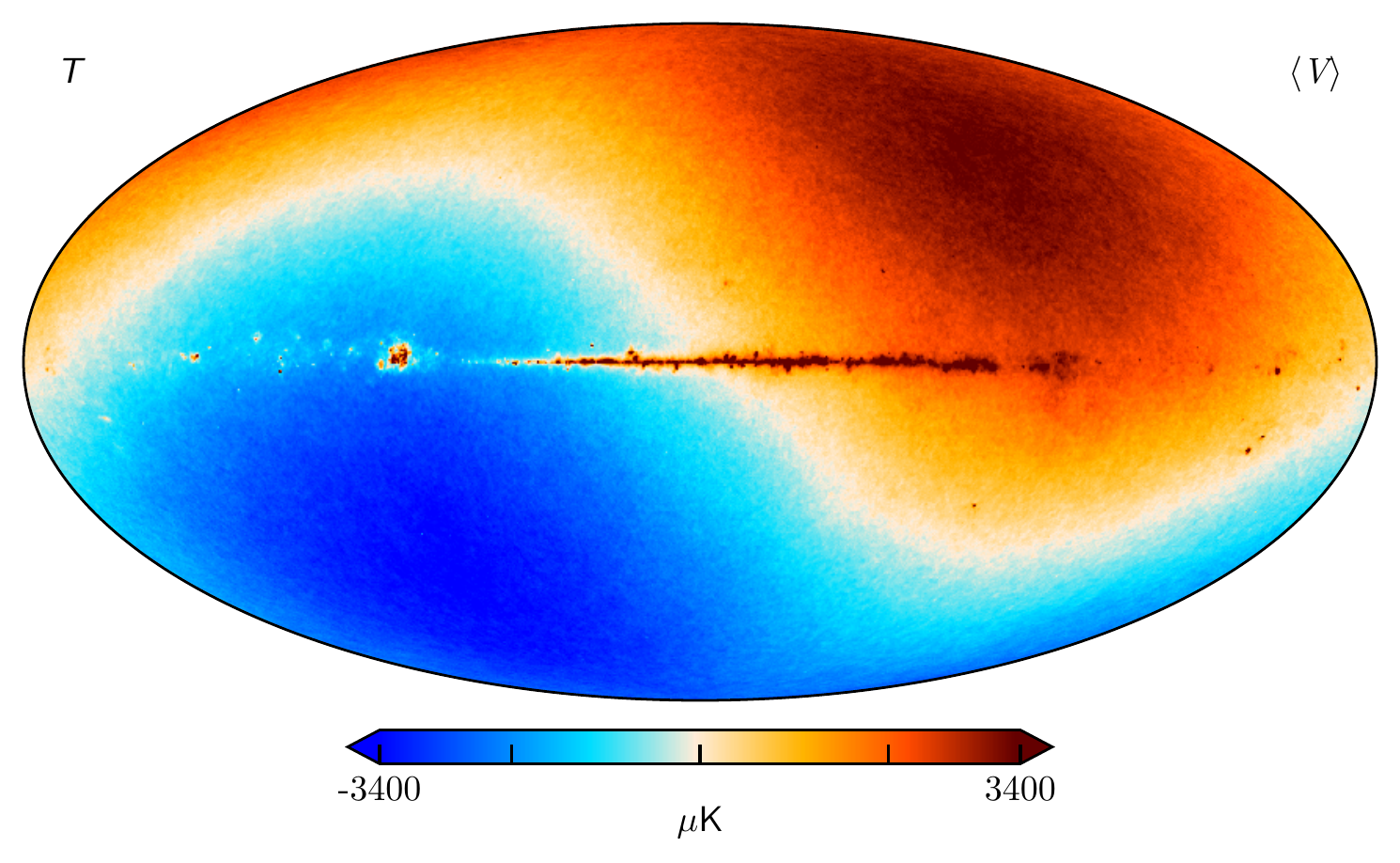}
	\includegraphics[width=0.75\textwidth]{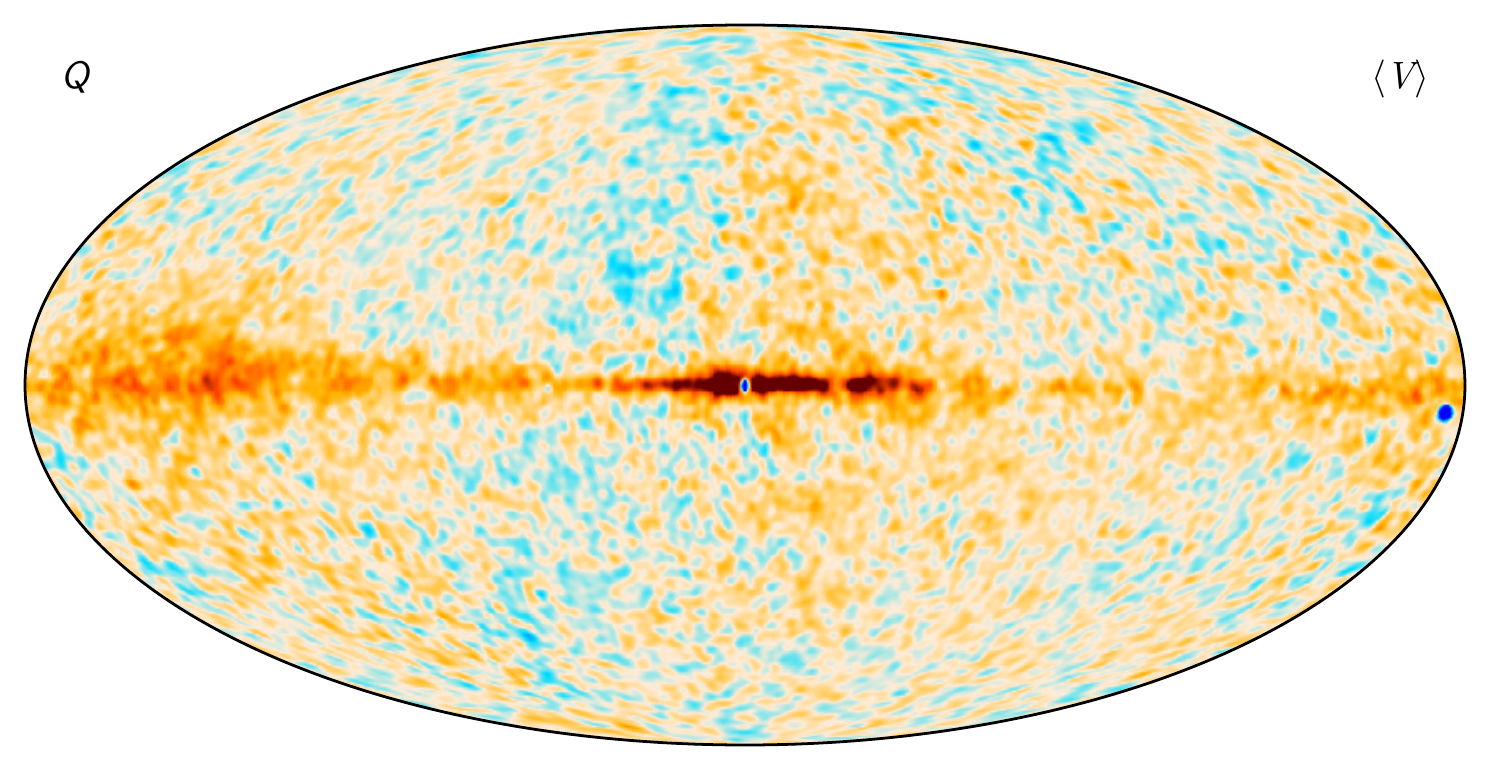}
	\includegraphics[width=0.75\textwidth]{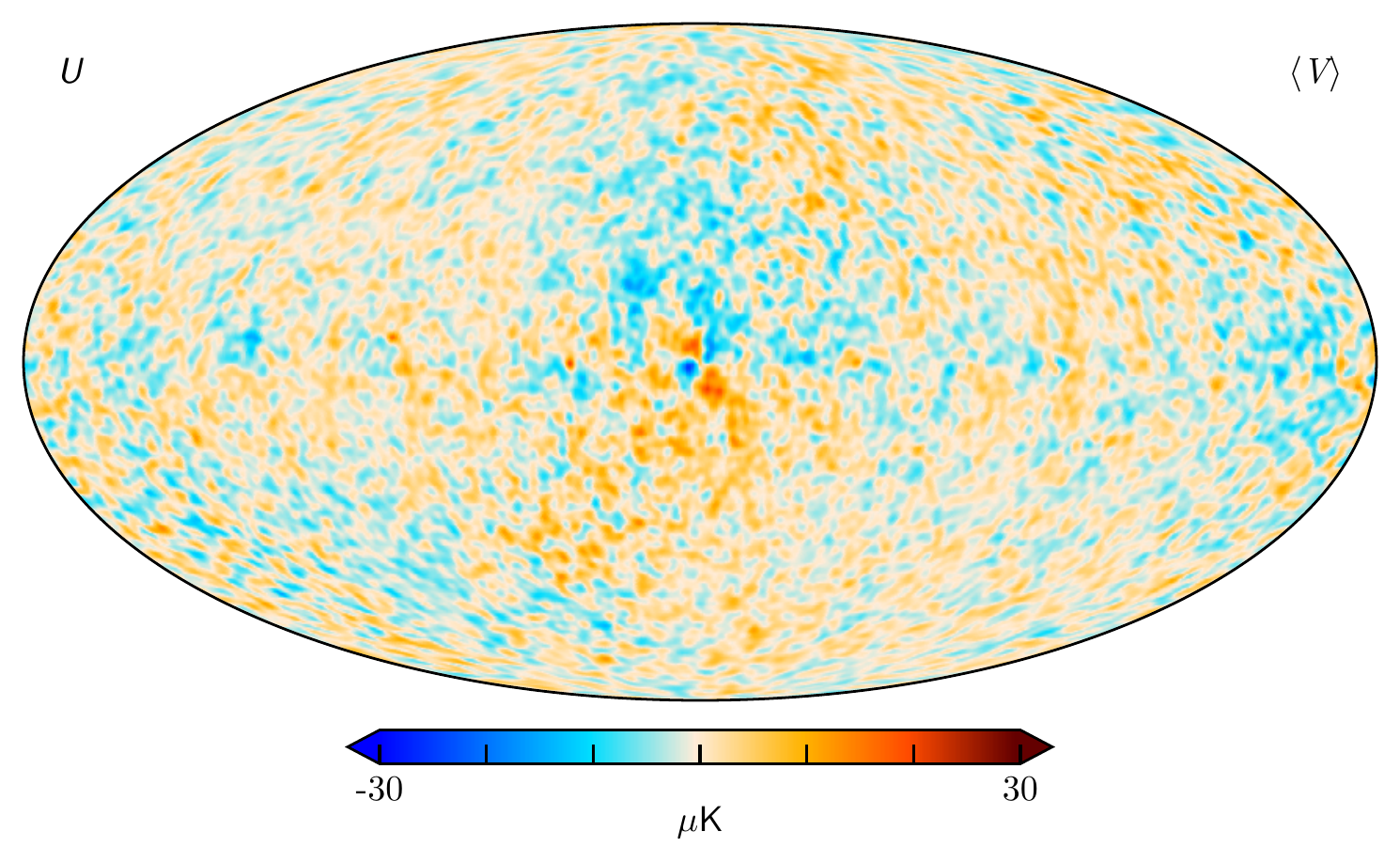}
        \caption{Same as Fig.~\ref{fig:kband}, but for \V-band.}        
	\label{fig:vband}
\end{figure*}
\begin{figure*}
	\centering
	\includegraphics[width=0.75\textwidth]{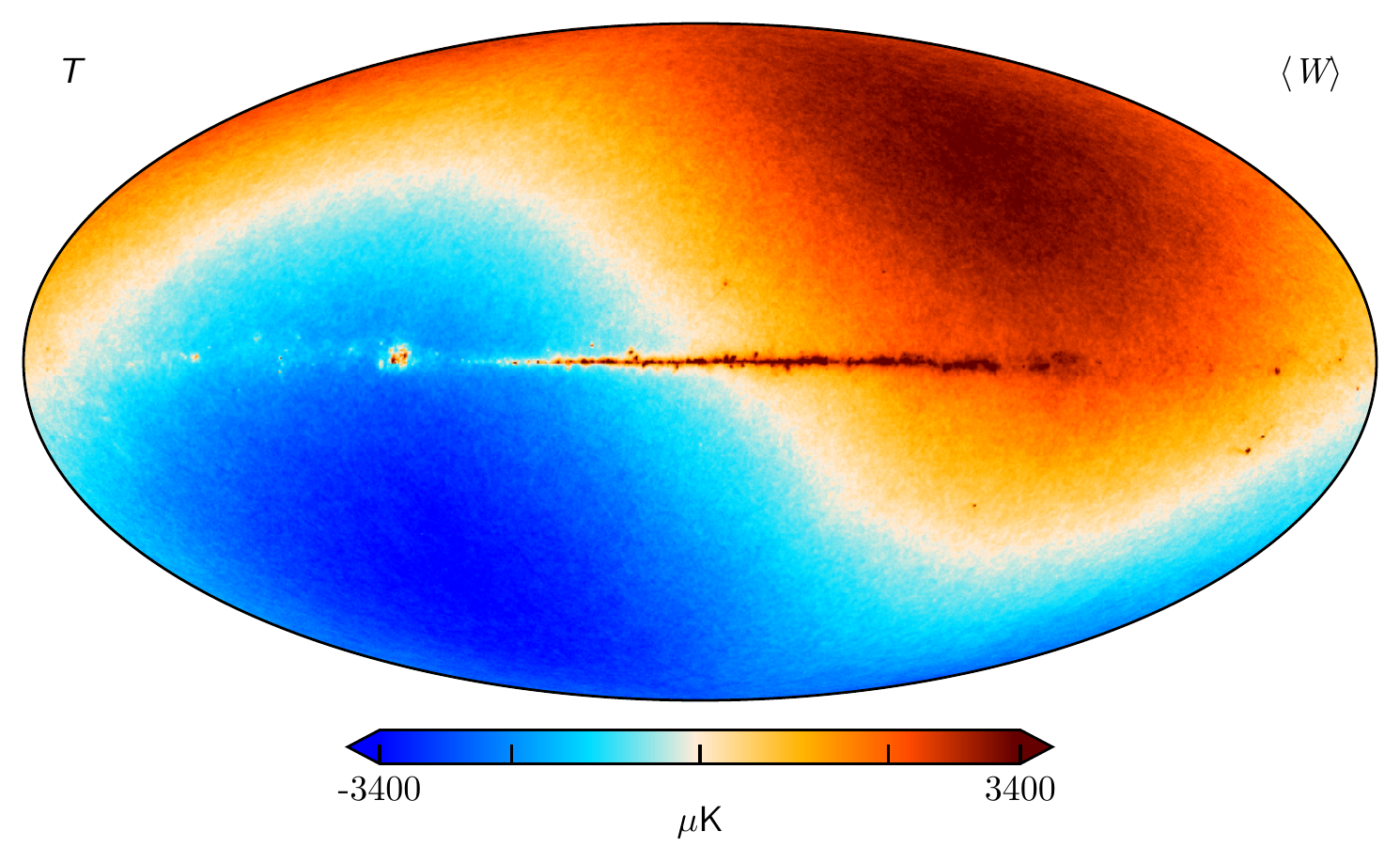}
	\includegraphics[width=0.75\textwidth]{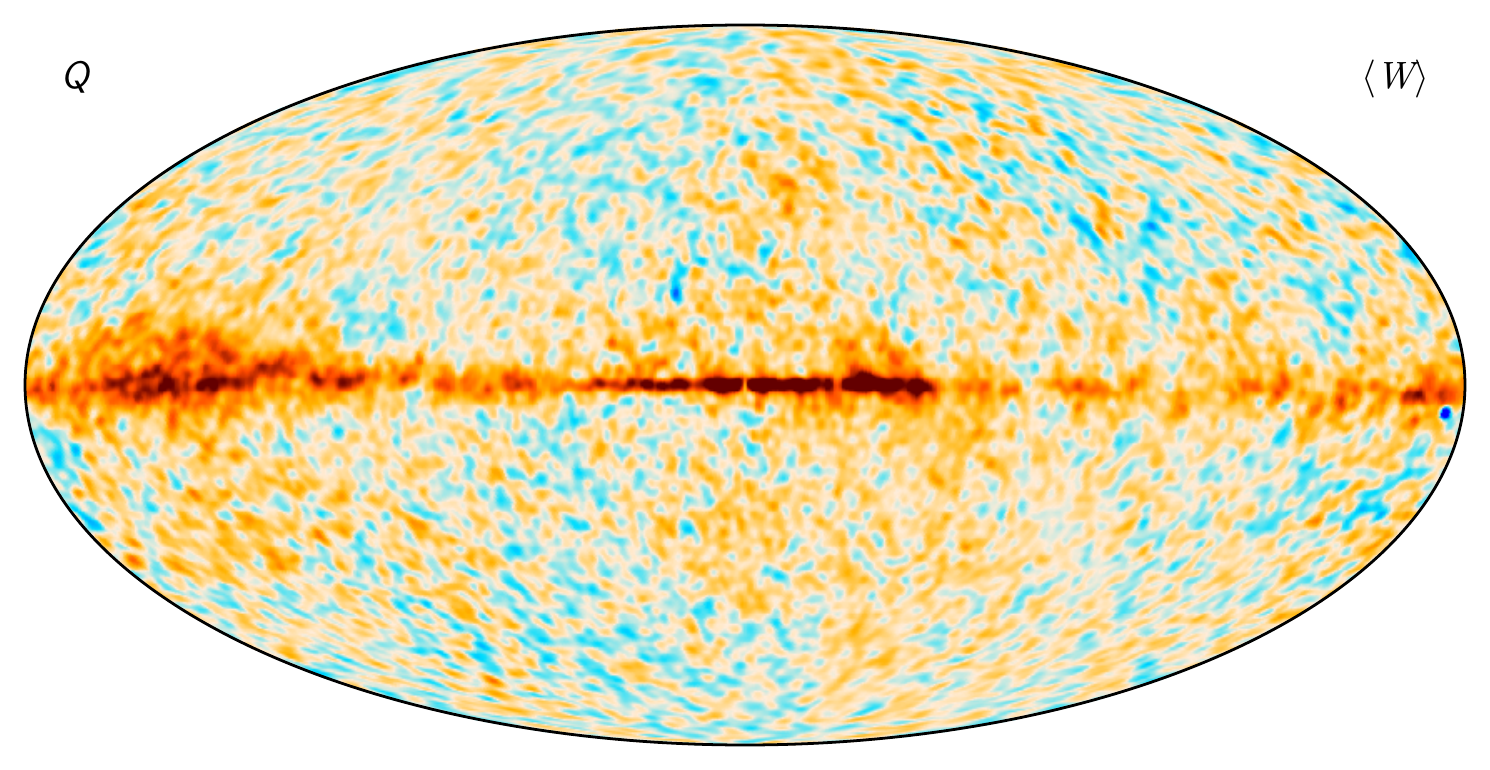}
	\includegraphics[width=0.75\textwidth]{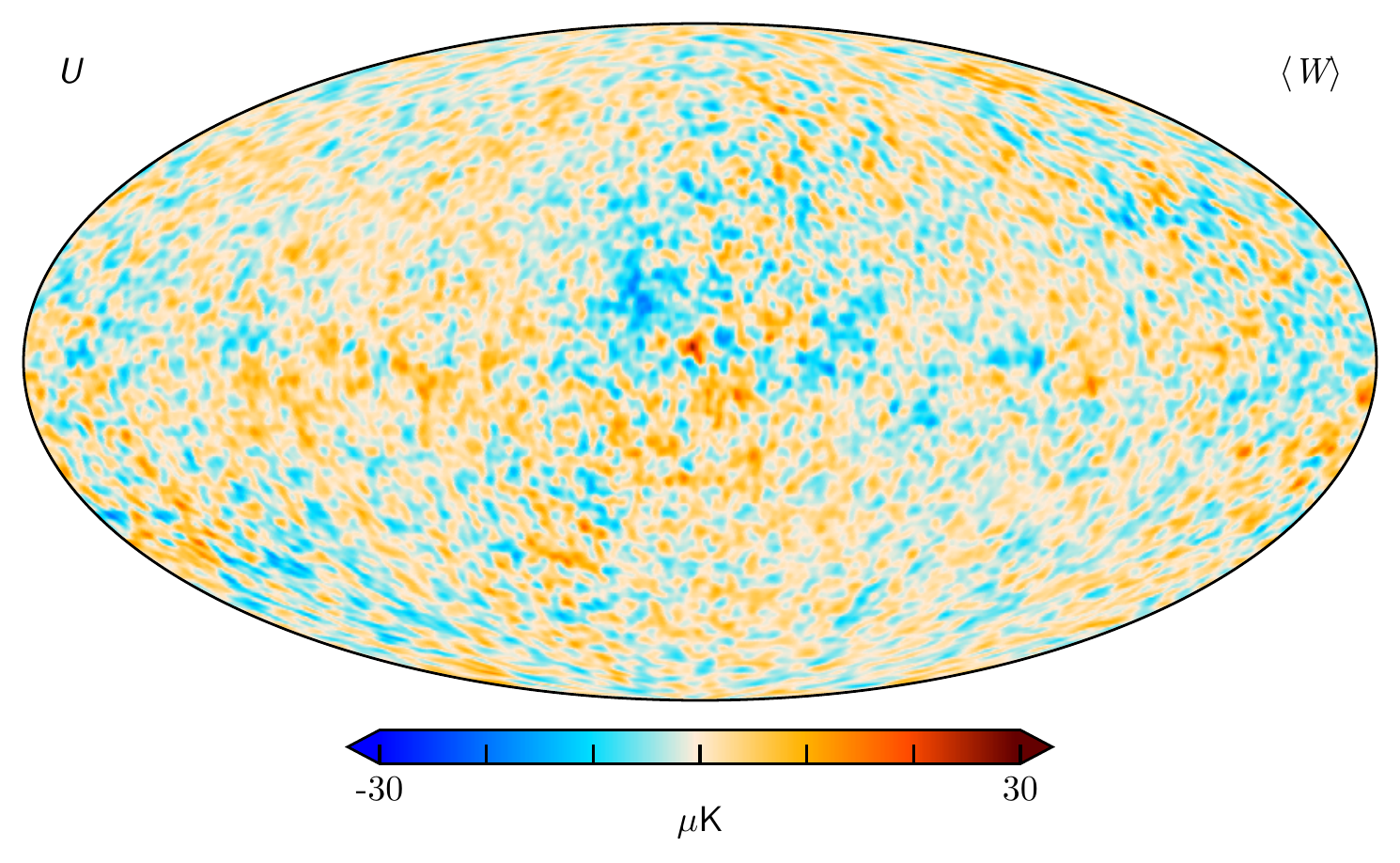}
        \caption{Same as Fig.~\ref{fig:kband}, but for \W-band.}                
	\label{fig:wband}
\end{figure*}

\begin{figure*}
	\centering
	\includegraphics[width=0.32\textwidth]{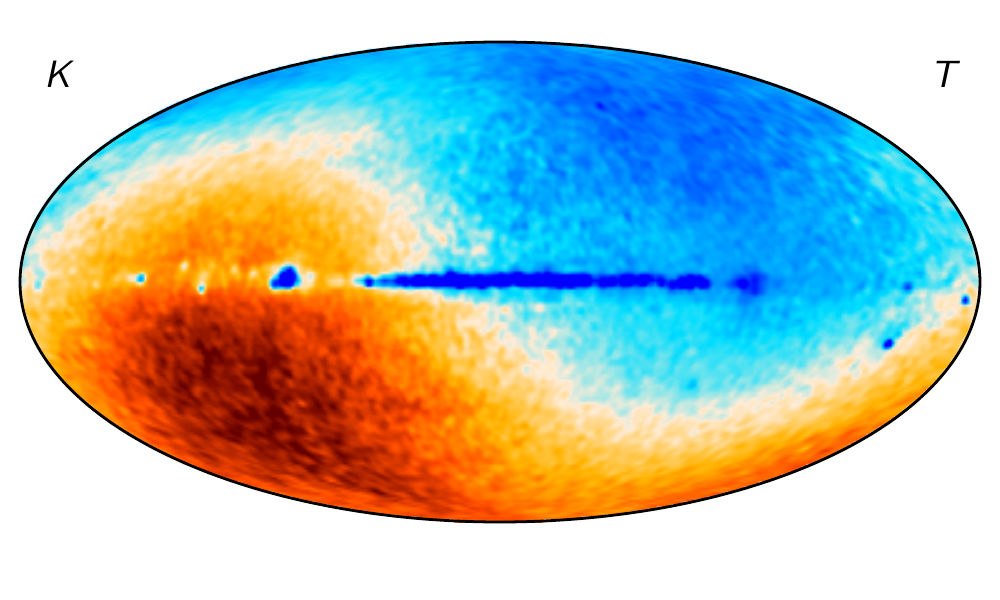}
        \includegraphics[width=0.32\textwidth]{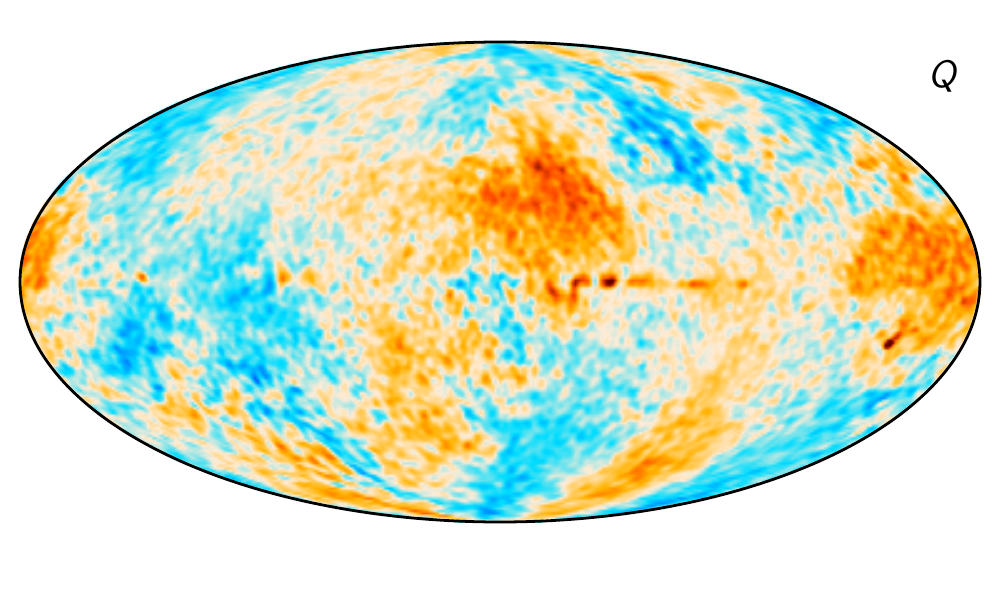}
        \includegraphics[width=0.32\textwidth]{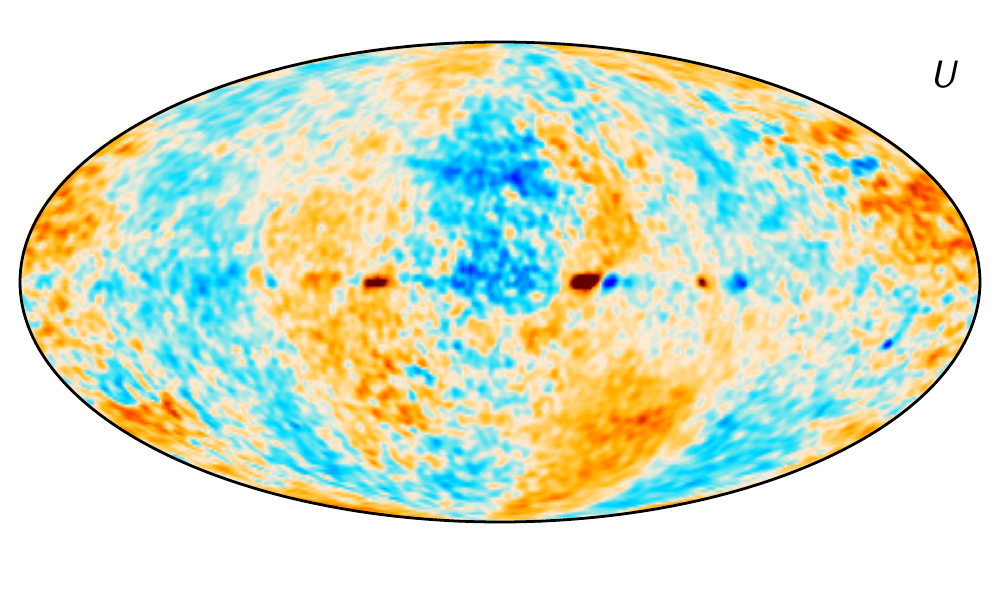}\\\vspace*{-4mm}
	\includegraphics[width=0.32\textwidth]{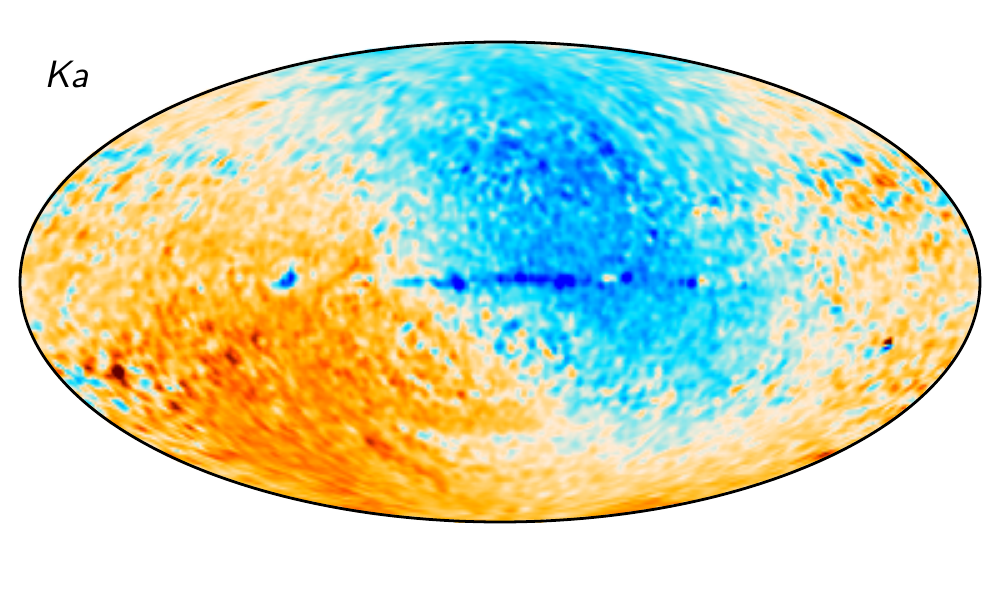}
        \includegraphics[width=0.32\textwidth]{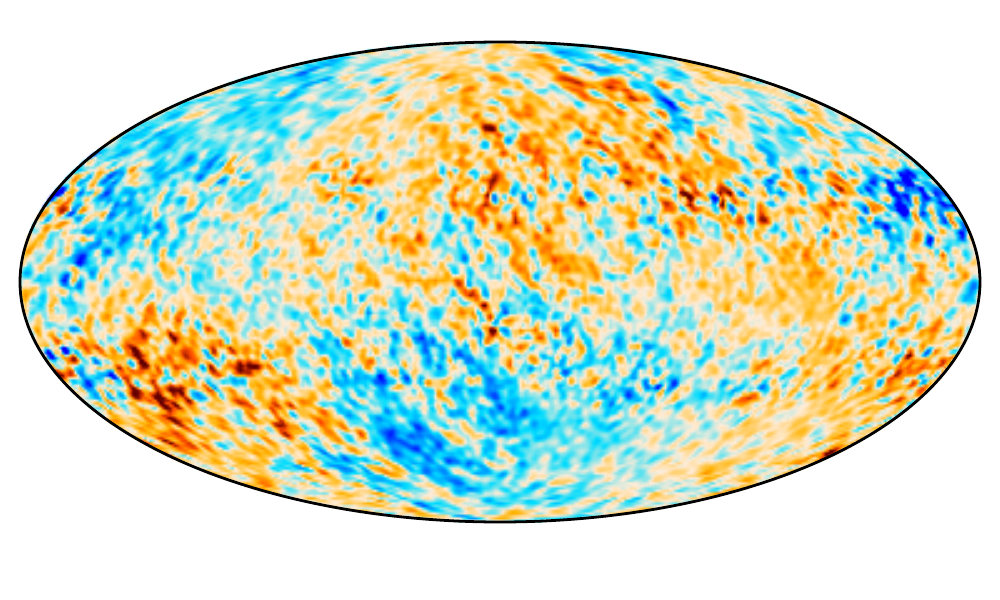}
        \includegraphics[width=0.32\textwidth]{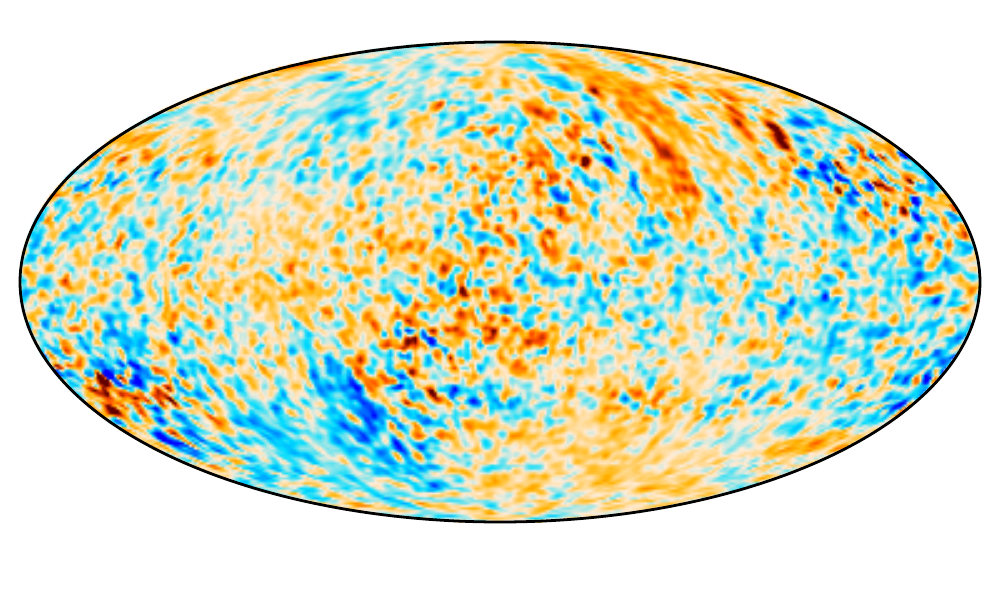}\\\vspace*{-4mm}
	\includegraphics[width=0.32\textwidth]{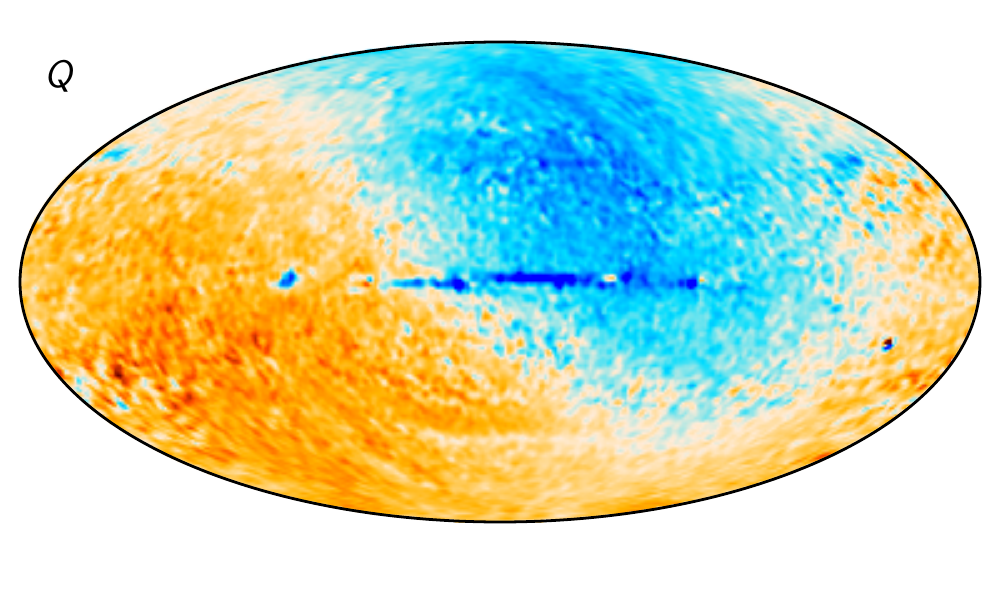}
        \includegraphics[width=0.32\textwidth]{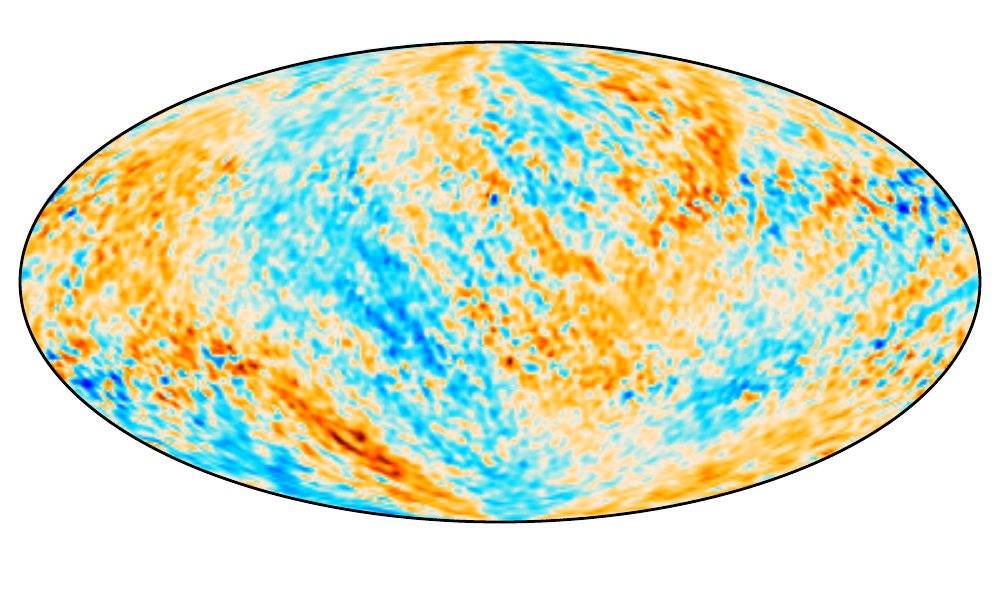}
        \includegraphics[width=0.32\textwidth]{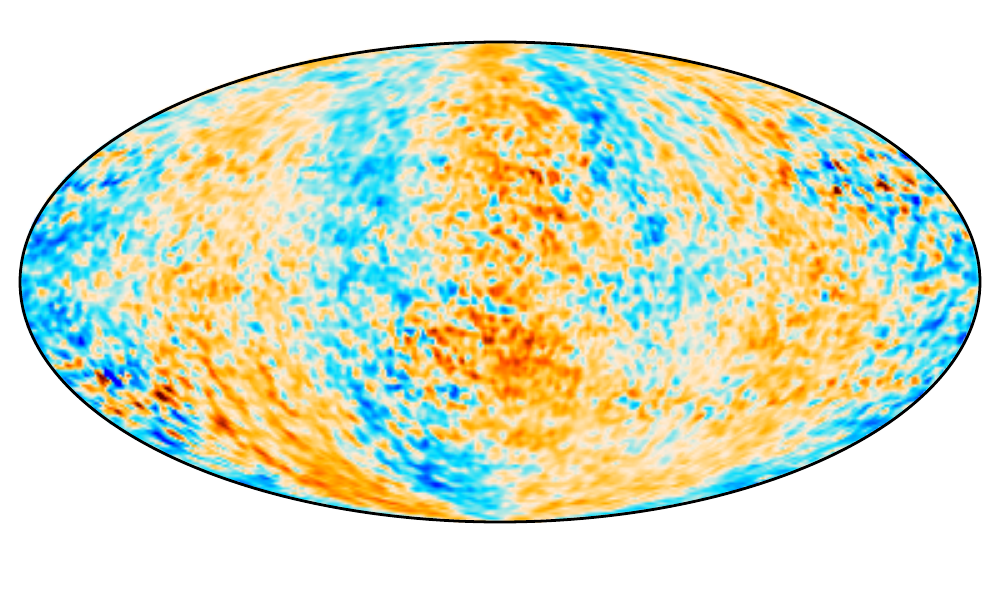}\\\vspace*{-4mm}
	\includegraphics[width=0.32\textwidth]{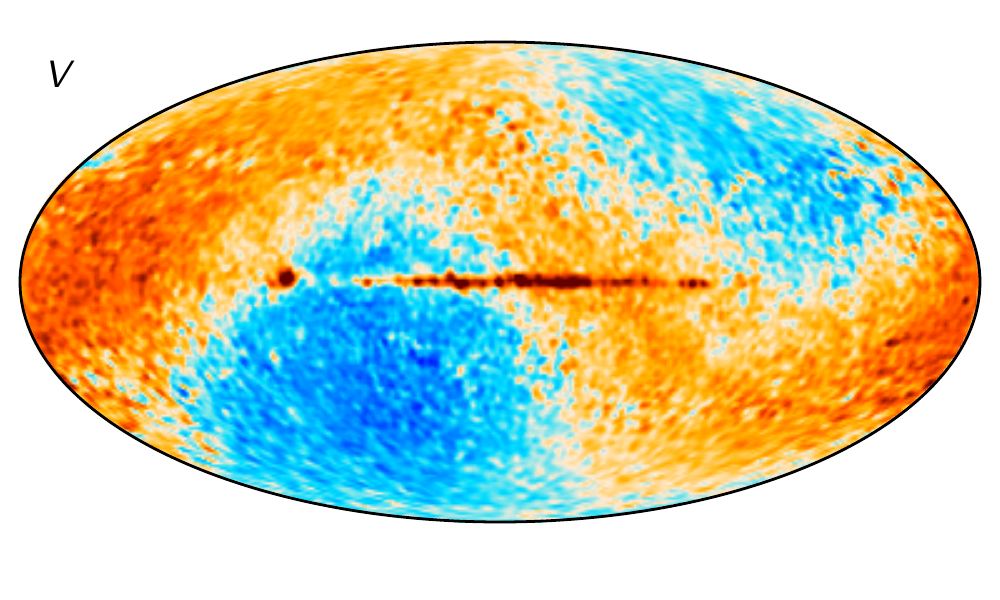}
        \includegraphics[width=0.32\textwidth]{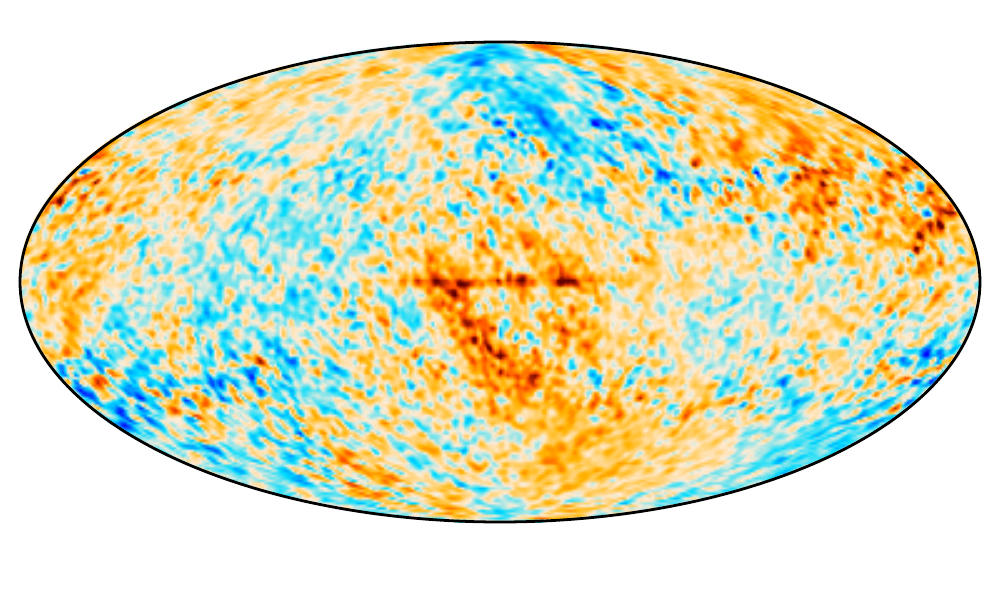}
        \includegraphics[width=0.32\textwidth]{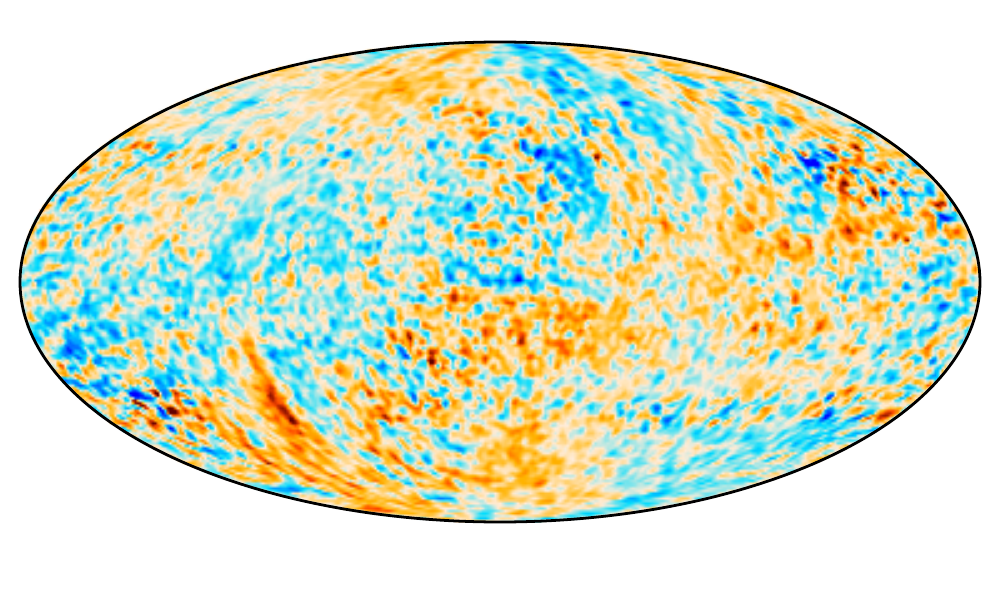}\\\vspace*{-4mm}
	\includegraphics[width=0.32\textwidth]{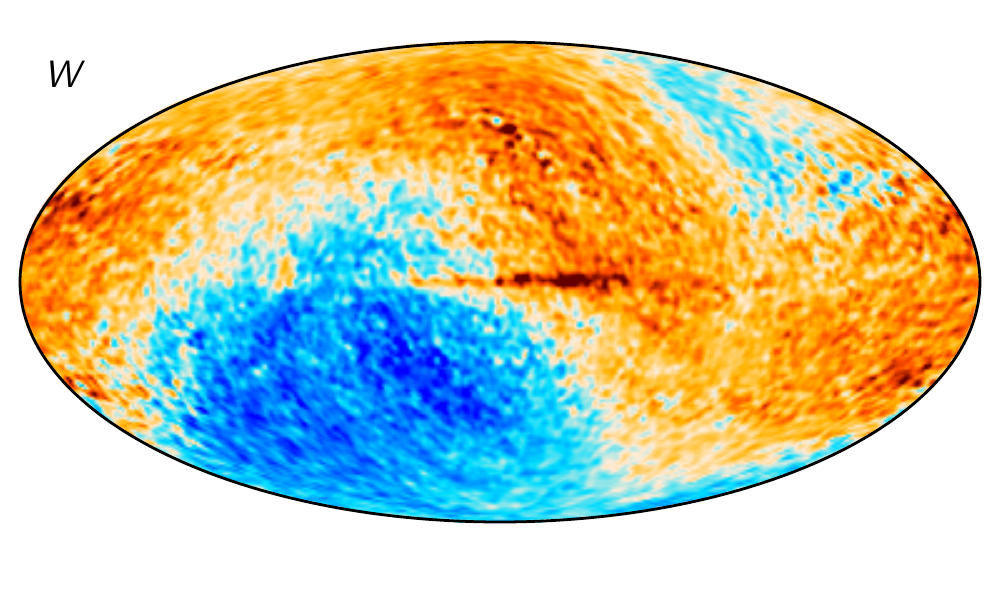}
        \includegraphics[width=0.32\textwidth]{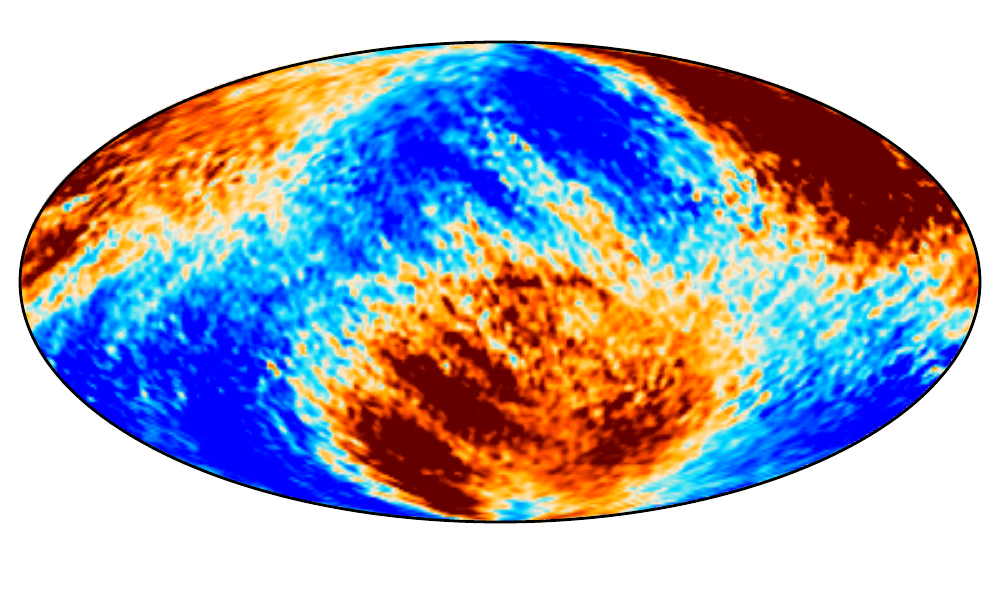}
        \includegraphics[width=0.32\textwidth]{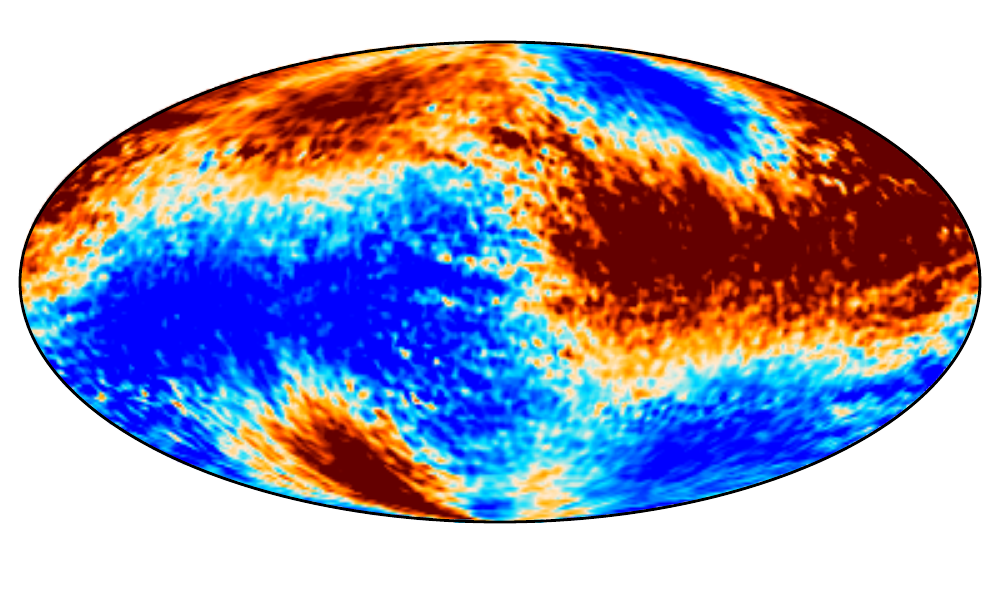}\\\vspace*{-4mm}
	\includegraphics[width=0.32\textwidth]{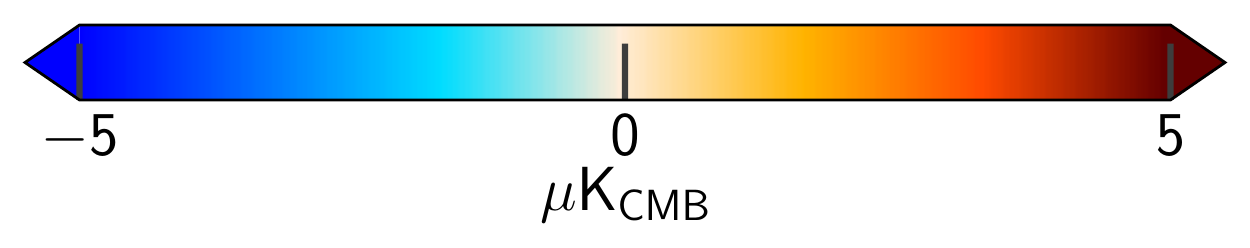}
	\caption{Difference maps between the \cosmoglobe\ and 9-year \WMAP\ frequency maps. Columns show Stokes $T$, $Q$, and $U$ parameter maps, while rows show \K-, \Ka-, \Q-, \V-, and \W-band maps. The maps are all smoothed with a $2^\circ$ FWHM Gaussian beam.}
        \label{fig:megadiff_wmap}
\end{figure*}

\begin{figure*}[t]
	\centering
	\includegraphics[width=\textwidth]{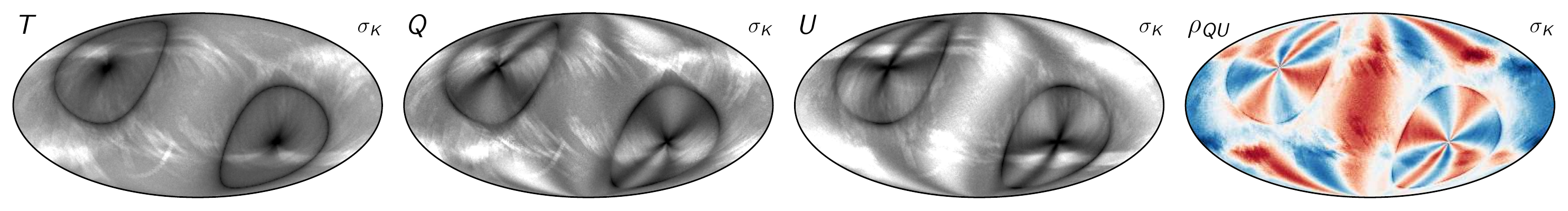}\\
	\includegraphics[width=0.245\textwidth]{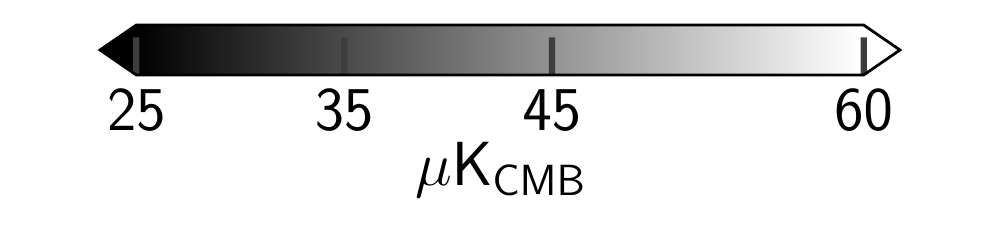}
	\includegraphics[width=0.245\textwidth]{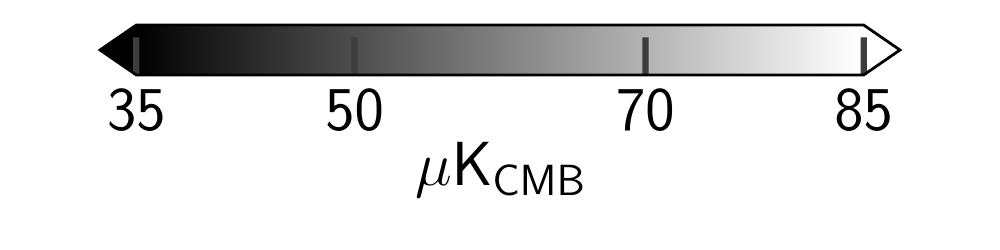}
	\includegraphics[width=0.245\textwidth]{figures/cbar_rms_P.pdf}
	\includegraphics[width=0.245\textwidth]{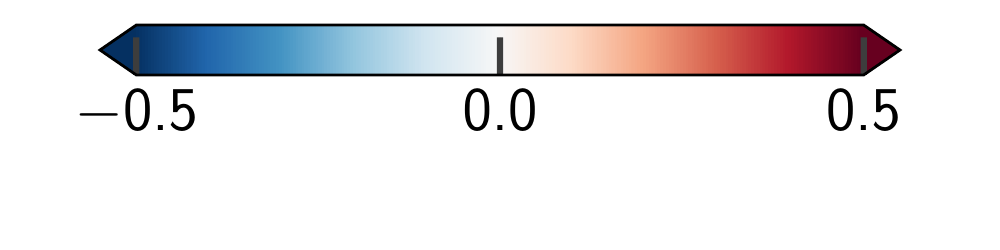}\\
	\includegraphics[width=\textwidth]{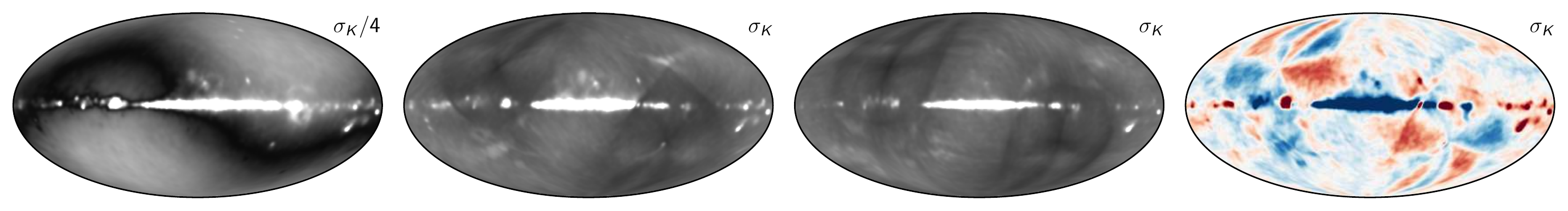}\\
	\includegraphics[width=0.245\textwidth]{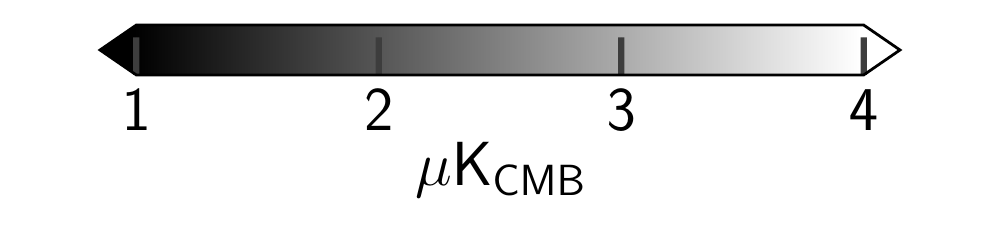}
	\includegraphics[width=0.245\textwidth]{figures/cbar_std.pdf}
	\includegraphics[width=0.245\textwidth]{figures/cbar_std.pdf}
	\includegraphics[width=0.245\textwidth]{figures/cbar_rho.pdf}\\
	\caption{Posterior variation maps for \K-band. Columns show the Stokes parameters and the correlation coefficient between $Q$ and $U$, while the rows show \textit{(top)} the white noise rms per pixel and \textit{(bottom)} the posterior standard deviation. The rms maps are unsmoothed, while the standard deviations have been smoothed to $7^\circ$.}
        \label{fig:K_rms_std}

	\centering
	\includegraphics[width=\textwidth]{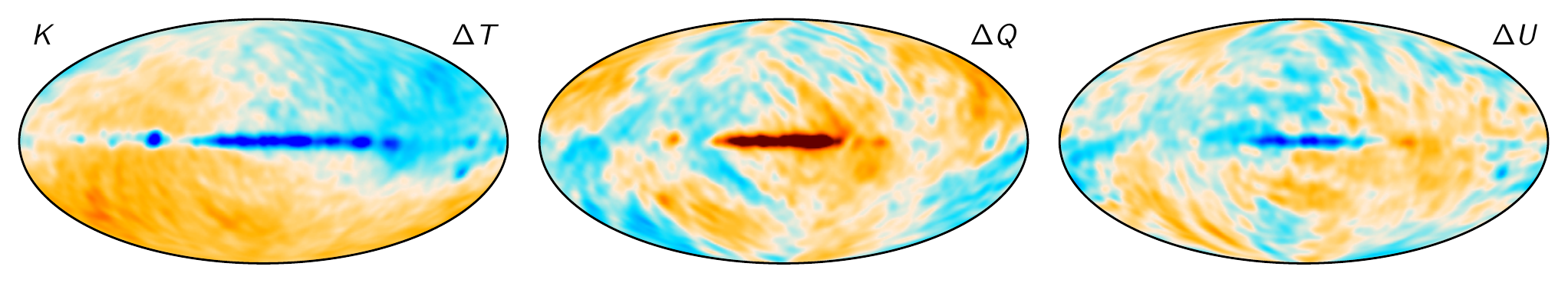}\\
	\includegraphics[width=0.30\textwidth]{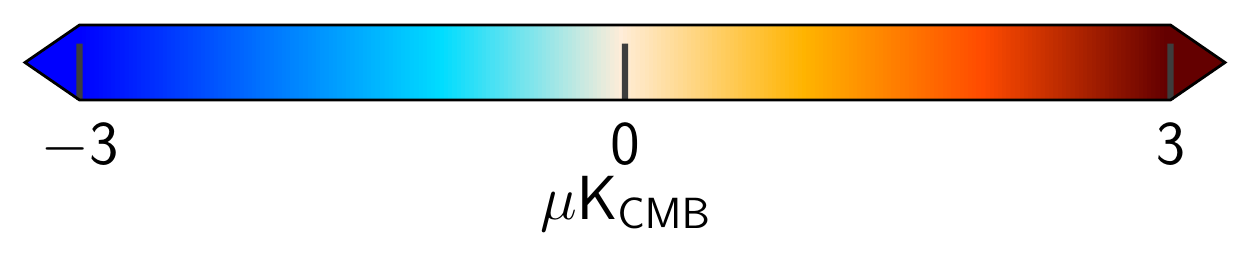}        
	\caption{Difference between two \K-band Gibbs samples, smoothed to $7^\circ$.}
        \label{fig:Ksampdiff}

	\centering
	\includegraphics[width=\textwidth]{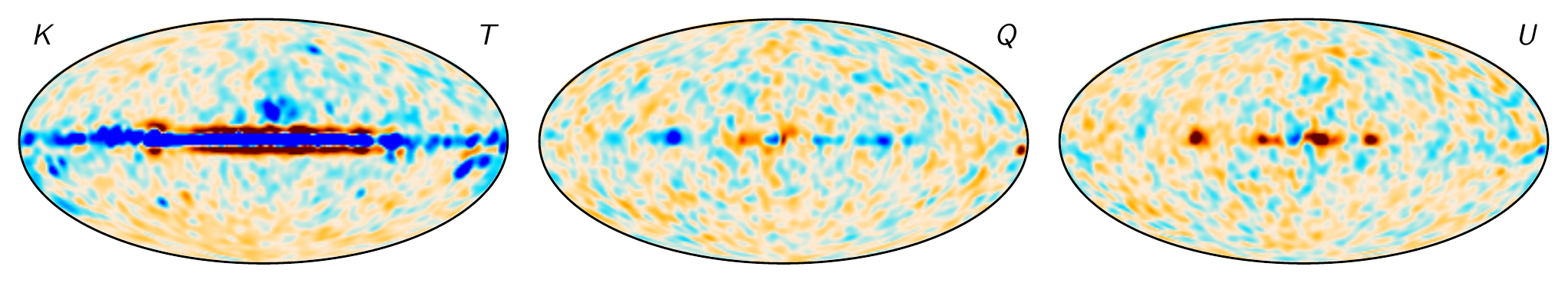}\\
	\includegraphics[width=0.30\textwidth]{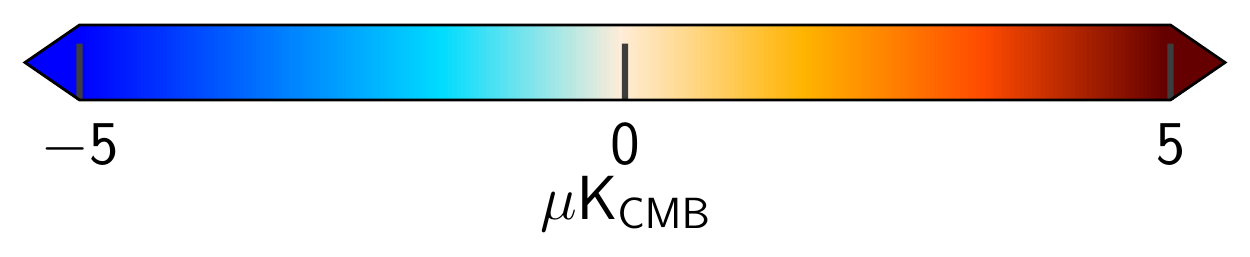}
	\caption{TOD-level residual map for \K-band, smoothed with a $5^\circ$ FWHM Gaussian beam.}
        \label{fig:todres_K}
\end{figure*}

We start by showing the co-added frequency \K-band, \Ka-band, \Q-band, \V-band, and \W-band posterior mean maps in Figs.~\ref{fig:kband}--\ref{fig:wband}, all defined in thermodynamic  $\mu\mathrm{K_{CMB}}$ units. All maps are produced at the DA level, and in these figures the \Q, \V, and \W-band maps are generated by inverse-variance weighting the individual DAs. The temperature maps are presented at full angular resolution, while the polarization maps have been smoothed with a $2^\circ$ Gaussian beam to reduce small scale structure and make the larger scale effects more apparent. Overall, the temperature maps behave as expected from the official \WMAP\ analysis, with falling foreground amplitudes with frequency. Furthermore, it is very difficult indeed, if not impossible, to see visual differences between the \cosmoglobe\ and \WMAP\ maps by eye when switching rapidly between them. However, it is important to note that the \cosmoglobe\ frequency maps retain the Solar CMB dipole, following \citet{npipe} and \citet{bp01}, while it is removed in the \WMAP\ official maps. Similarly, we see that the amplitudes of the polarized maps decrease as expected from \K--\V-band following the expected synchrotron behavior, with a slight increase at \W-band due to the contribution of thermal dust. %

Next, in Fig.~\ref{fig:megadiff_wmap}, we show corresponding difference maps between the official 9-year \WMAP\ maps and the maps produced in this work. The color scale in this plot is linear with range $\pm5\muK$, and we see that the differences are thus quite small, typically smaller than 1--2\muK\ for most channels. The main exception to this is \W-band polarization, for which the differences are generally larger than $5\muK$.

Going into greater detail and starting with total intensity, we see first that the \K-band difference is dominated by a dipole with a $\sim$\,2.5\muK\ amplitude that is anti-aligned with the CMB Solar dipole. In addition, the Galactic plane is slightly brighter in \WMAPnine\ than \cosmoglobe. Both of these suggest that our total absolute \K-band calibration is lower than the \WMAPnine\ value by about 0.1\,\%; given the major differences in methodology described in Sect.~\ref{sec:methods}, this degree of agreement is a major validation of the data and the two  pipelines. A similar small dipole difference is also seen in the \Q-band.

For the remaining channels, and in particular for the \V- and \W-bands, the main intensity difference takes the form of a quadrupole with an amplitude of 2--3\muK\ aligned with the CMB dipole. Naively, one could suspect this to be due to different treatments of the relativistic quadrupole. However, as noted by \citet{larson2014}, the \WMAPnine\ maps retain the kinematic quadrupole, as does \commanderthree; in our framework, this signal term is accounted for through the signal model defined in Eq.~\eqref{eq:cmb_astsky}. This is notably different from the \Planck\ maps, which do remove the relativistic quadrupole from the frequency maps \citep{planck2016-l02,planck2016-l03}. Additionally, even though the observed quadrupole has the expected shape, the frequency dependence is not consistent with the expected functional form $x\coth x$ where $x=h\nu/(2kT_\mathrm{CMB})$ \citep{Notari:2015}. For now, we speculate that these differences are rather due to second-order gain or baseline differences, possibly associated with the annual oscillatory structures seen in Fig.~\ref{fig:dgain}.

In polarization, we note large scale differences in both Stokes $Q$ and $U$. These differences do not match known Galactic component morphologies, but are more reminiscent of the poorly measured transmission imbalance modes discussed by \citet{jarosik2010}, although the map-space morphologies are not identical. In general, such large mode differences are due to at least three main effects: 1) incomplete polarization angle coverage for a few large-scale modes; 2) errors in transmission imbalance coupled with the Solar dipole; and 3) interplay between the transmission imbalance, the far sidelobe, and the Solar dipole, as briefly described in Sect.~\ref{sec:wmap_instmodel}. The scale of these effects is most pronounced in the \W-band polarization results, where we see the largest differences between the two processing pipelines.

From these differences alone, it is not possible to determine whether the excess structures are present in the \cosmoglobe\ or \WMAP\ maps, or both. However, Appendix~\ref{sec:map_survey} provides a complete survey of the \cosmoglobe\ frequency maps, and in particular Fig.~\ref{fig:skymaps} compares these with the \WMAPnine\ maps. In this case, one clearly sees that the large-scale modes are predominantly present in the \WMAP\ maps, rather than \cosmoglobe. 

Returning to the internal properties of the \cosmoglobe\ posterior distribution, we show in the top panel of Fig.~\ref{fig:K_rms_std} the \K-band white noise standard deviation per pixel in Stokes $T$, $Q$ and $U$; the fourth column shows the correlation coefficient between the $Q$ and $U$ coefficients. The bottom panel shows the corresponding posterior standard deviation per pixel. It is important to note that the white noise is not a free parameter in the data model, and there is no white noise component in the Gibbs sampler described by Eqs.~\eqref{eq:gain_samp_dist}--\eqref{eq:param_samp}. That also implies that there is no marginalization over white noise in the resulting frequency map ensemble. Rather, the full marginal uncertainty per frequency map pixel must be obtained by adding the two rows in Fig.~\ref{fig:K_rms_std} in quadrature. However, a preferable approach to performing error propagation for higher level scientific analyses is to analyze each sample separately, taking into account only white noise for each sample, and then use the full sample ensemble as the final result. This is the only robust way of fully accounting for all correlations between the various effects.  

The white noise pattern for $T$ follows the usual pattern with highest sensitivity at the North and South ecliptic poles, as well as circles around the poles corresponding to times when the partner horn is observing the opposite ecliptic pole. There are also regions of higher noise level corresponding to planets crossing the ecliptic, and regions of higher emission $\simeq140^\circ$ away from the Galactic center, which correspond to the times when the partner horn lies within the processing mask.

\begin{figure*}
	\centering
	\includegraphics{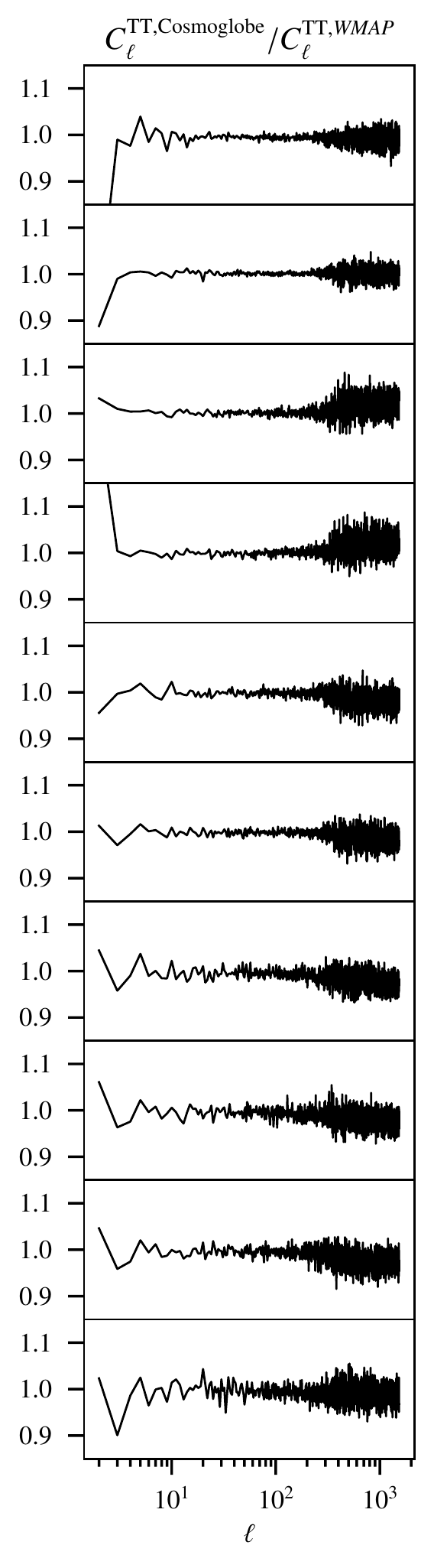}
	\includegraphics{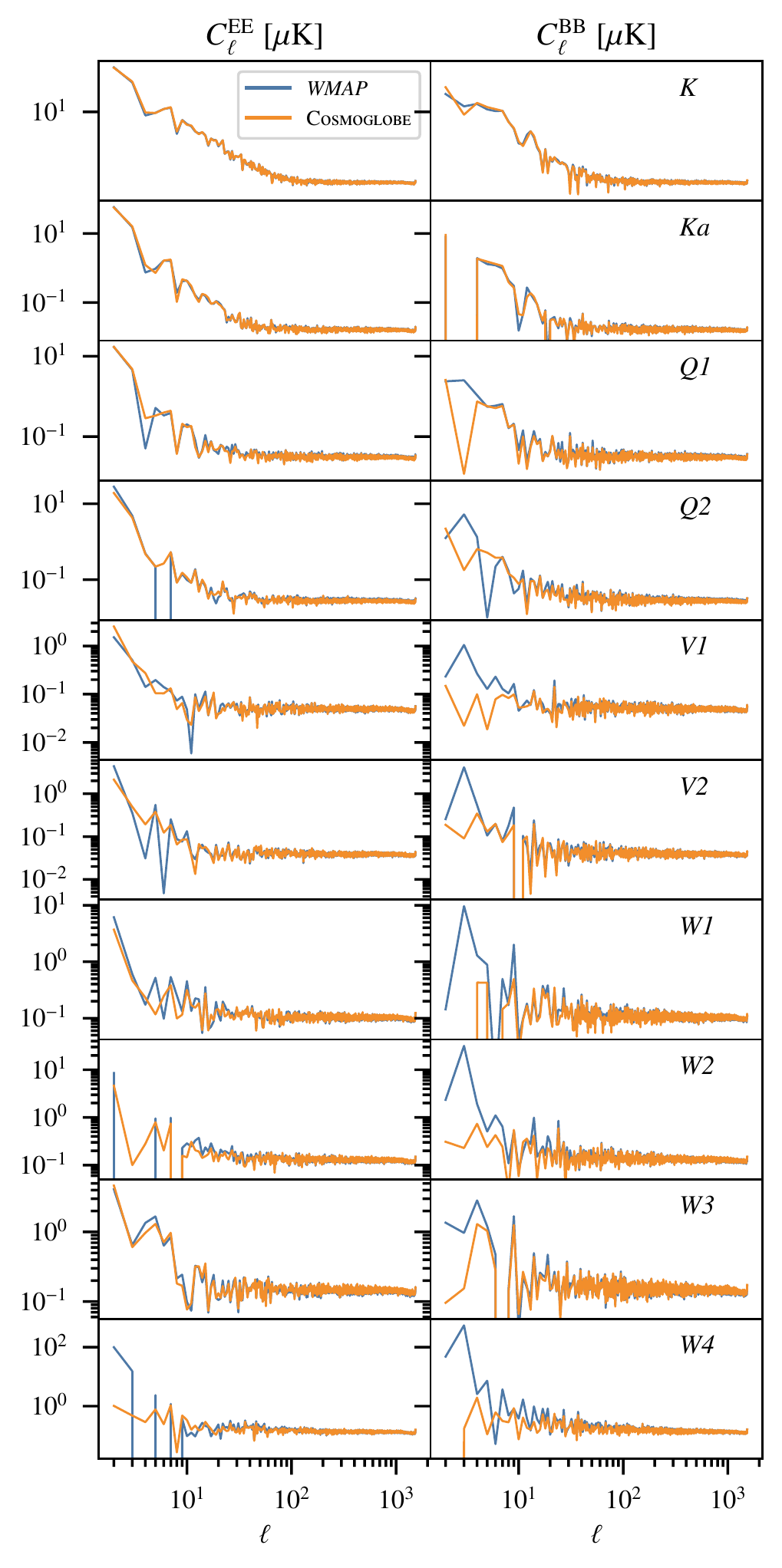}
	\caption{Comparison of the $C_\ell^\mathrm{TT}$, $C_\ell^\mathrm{EE}$, and $C_\ell^\mathrm{BB}$ from \WMAPnine\ and \cosmoglobe. Each row corresponds to a different DA, with frequency increasing from top to bottom. \textit{(left):} ratio of $C_\ell^\mathrm{TT}$ from \cosmoglobe\ compared to \WMAPnine. \textit{(middle/right):} $C_\ell^\mathrm{EE/BB}$ power spectra with \WMAPnine\ in blue and \cosmoglobe\ in orange.}
	\label{fig:map_spectra}
\end{figure*}

The polarized RMS maps share all of these characteristics, but with an overall amplitude shift due to polarization measurements having half the effective number of observations per pixel. In addition, the poles have a characteristic ``X''-like structure that is rotated $45^\circ$ degrees between $Q$ and $U$, corresponding to different polarization orientations. There are also characteristic large scale structures visible in Galactic coordinates, corresponding to polarization modes poorly constrained by the \WMAP\ scan strategy.

While the maps in the top row of Fig.~\ref{fig:K_rms_std} are directly comparable to the corresponding \WMAPnine\ products, the posterior standard deviation shown in the bottom row has no equivalent in the official \WMAP\ release. These maps can be considered the ``systematic'' error contributions, as their variation depends on the sampled instrumental parameters, i.e., gain, imbalance parameters, correlated noise, and sidelobe correction. The temperature map contains a clear quadrupole signature. This is due to the variation in the absolute calibration $g_0$, which changes the Solar dipole in the final map. In addition to the quadrupole, the Galactic plane also varies due to the gain solution's fluctuations. As expected, the white noise patterns associated with the scan strategy also appear in the polarization maps, which have much lower signal-to-noise ratio than the temperature map. %

Another useful quantity is the difference between two arbitrary samples, which we show in Fig.~\ref{fig:Ksampdiff}. In temperature, the most striking term is a dipole, corresponding to the absolute gain difference, and the Galactic plane. There are also additional weaker lines associated with the scanning strategy, corresponding to different correlated noise and time variable gain realizations. In polarization, gain variations,  bandpass uncertainties, and correlated noise dominate the differences between two samples, as quantified in Fig.~\ref{fig:corrmap_stddev}. The polarization differences are aligned with \WMAP's scans, modulated by the polarization angle.

Finally, the quality of the model in map space can be evaluated quite well by looking at the calibrated residual map, i.e., mapping the time-ordered residual $\boldsymbol r\equiv(\data-\ncorr)/\boldsymbol g-\boldsymbol s^\mathrm{tot}$. We display this TOD residual for the \K-band in the bottom panel of Fig.~\ref{fig:tod_corrections} ($2^\circ$~FWHM) and Fig.~\ref{fig:todres_K} ($5^\circ$~FWHM). In temperature there is a large residual along the Galactic plane, which is to be expected for both temperature and polarization, due to the complexity of the Galactic center. The temperature residual also contains a $\sim3\,\mathrm{\mu K}$ dipole due to the prior sampling of $g_0$ -- drawing a sample based on the sky model would track the dipole much more closely, whereas the prior sampling by definition does not use the sky model to draw $g_0$. Other than the Solar dipole, Galactic plane, and point sources, both the temperature and polarization maps are visually consistent with white noise across the entire sky.

Again, we have for brevity primarily focused on the \K-band in this discussion. For completeness, however, similar plots for all DAs are shown in Appendix~\ref{sec:map_survey}. In particular, Fig.~\ref{fig:skymaps} compares the \cosmoglobe\ and \WMAP\ DA polarization maps, Figs.~\ref{fig:rms} and \ref{fig:std} shows the white noise and posterior rms's, Fig.~\ref{fig:sampdiff} shows sample differences, and Fig.~\ref{fig:todres} shows TOD residual maps.

In Fig.~\ref{fig:map_spectra} we compare angular power spectra computed from both  \cosmoglobe\ and \WMAPnine\ frequency maps. These spectra are derived using the \texttt{NaMaster} \citep{namaster}\footnote{\url{https://github.com/LSSTDESC/NaMaster}} \texttt{compute\_full\_master} routine, while applying the extended \WMAP\ temperature analysis mask which allows a sky fraction of 68.8\,\%. As the $TT$ power spectrum is strongly signal-dominated for $\ell\lesssim200$ for all DAs, it is particularly informative to consider ratios, which we show in the left column of Fig.~\ref{fig:map_spectra}. Here we see that the $TT$ spectra derived from the two pipelines are consistent to sub-percent level at all but the very largest and smallest scales for all DAs. We speculate that the large scale differences are due to different CMB Solar dipoles -- as noted above, the \cosmoglobe\ maps retain the Solar CMB dipole, and an estimate of this must be subtracted before evaluating these spectra. In contrast, the \WMAP\ maps have this contribution removed at the TOD level; small differences due to these different treatments are not unexpected. The small scale differences above $\ell\sim200$ can be attributed to different data selections and low-level processing. For instance, the \cosmoglobe\ maps exploit about 1\,\% less data than \WMAPnine; \cosmoglobe\ fits one $\sigma_0$ parameter per scan, while \WMAPnine\ assumes it to be constant for each year; the \WMAP\ gain model varies every 23\,sec, while the \cosmoglobe\ model assumes constant gain per scan etc.

\begin{figure*}

	\centering
	\includegraphics[width=0.24\textwidth]{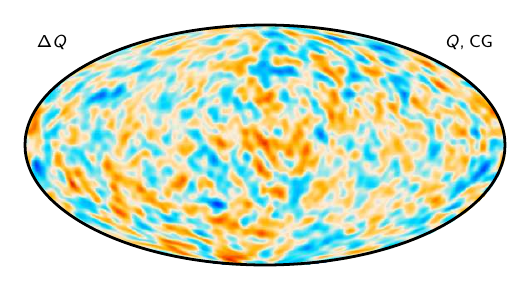}
	\includegraphics[width=0.24\textwidth]{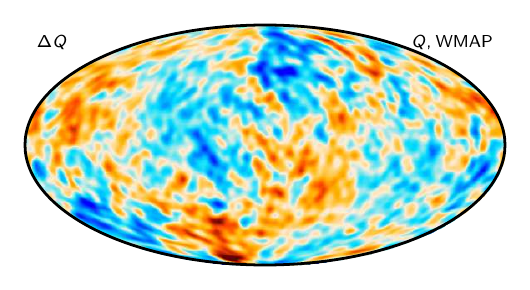}
	\includegraphics[width=0.24\textwidth]{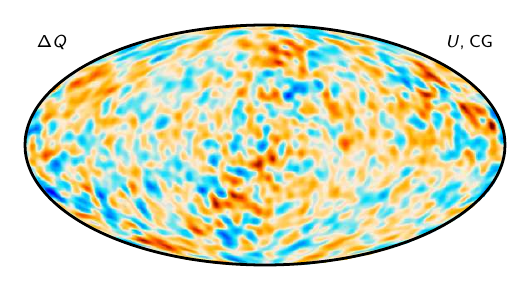}
	\includegraphics[width=0.24\textwidth]{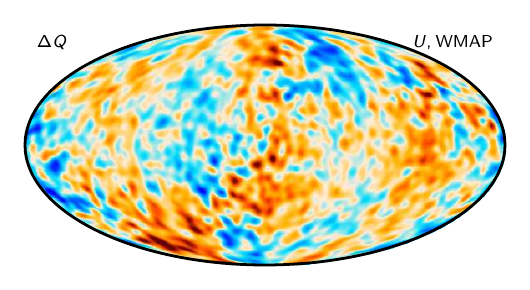}\\
	\includegraphics[width=0.24\textwidth]{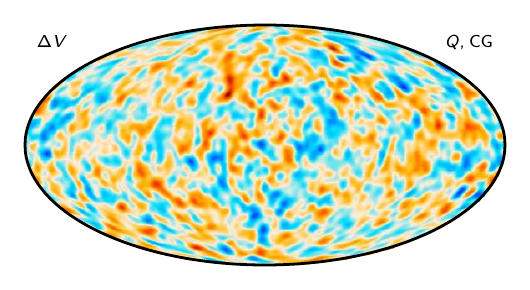}
	\includegraphics[width=0.24\textwidth]{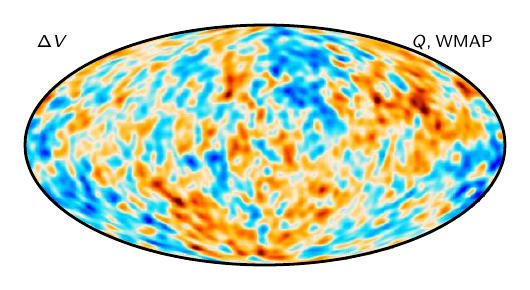}
	\includegraphics[width=0.24\textwidth]{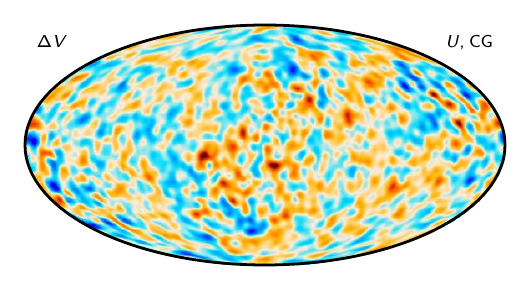}
	\includegraphics[width=0.24\textwidth]{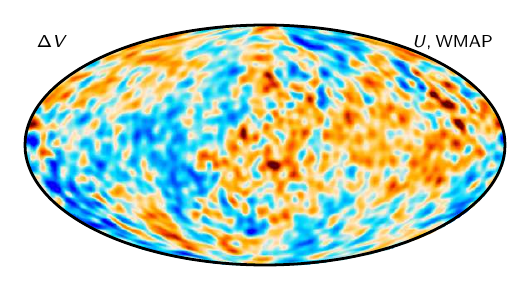}\\
	\includegraphics[width=0.24\textwidth]{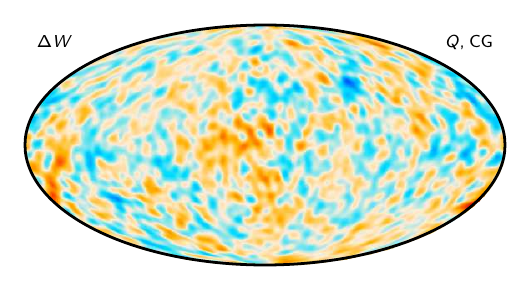}
	\includegraphics[width=0.24\textwidth]{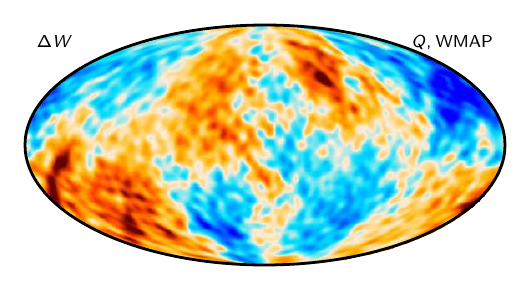}
	\includegraphics[width=0.24\textwidth]{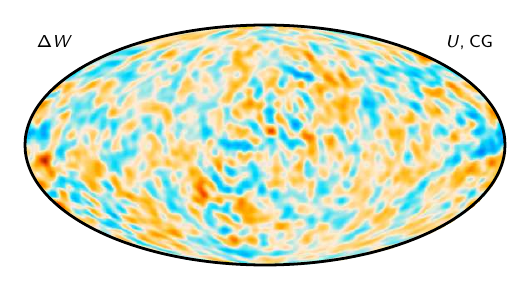}
	\includegraphics[width=0.24\textwidth]{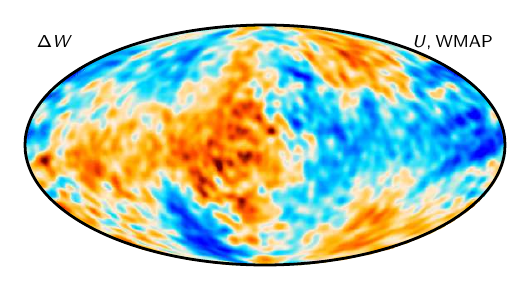}
        \includegraphics[width=0.25\textwidth]{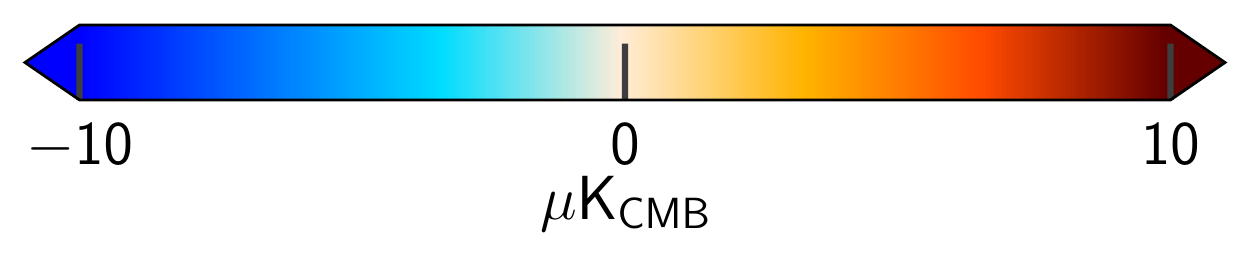}
	\caption{Internal \WMAP\ difference maps, smoothed by $10^\circ$. The two left columns are Stokes $Q$, and the two right columns are Stokes $U$, with the \cosmoglobe\ and \WMAPnine\ maps alternating between columns. The top to bottom rows are difference maps in increasing frequency.}
	\label{fig:internal_diff}

	\centering
	\includegraphics[width=0.24\textwidth]{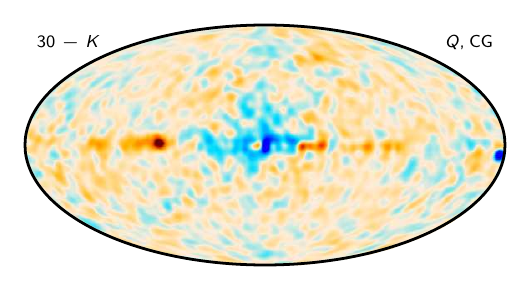}
	\includegraphics[width=0.24\textwidth]{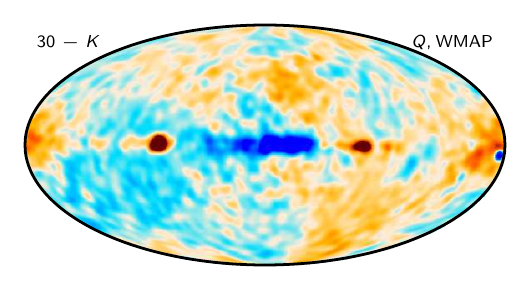}
	\includegraphics[width=0.24\textwidth]{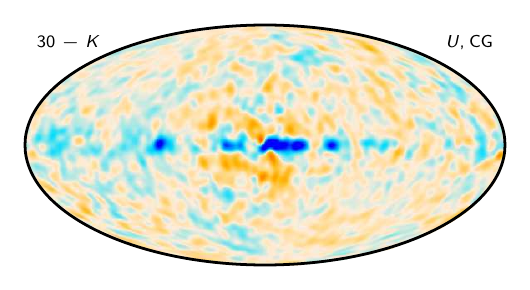}
	\includegraphics[width=0.24\textwidth]{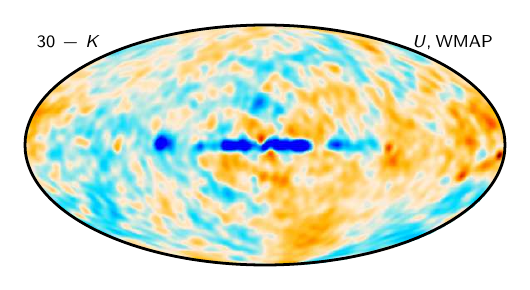}\\
	\includegraphics[width=0.24\textwidth]{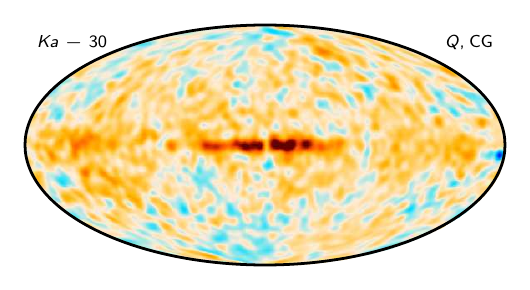}
	\includegraphics[width=0.24\textwidth]{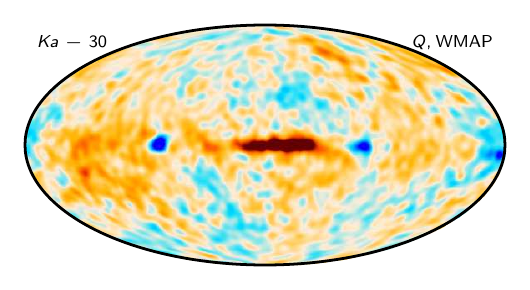}
	\includegraphics[width=0.24\textwidth]{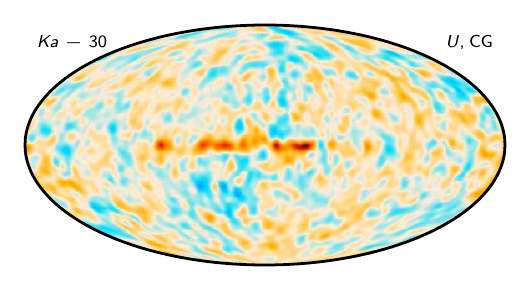}
	\includegraphics[width=0.24\textwidth]{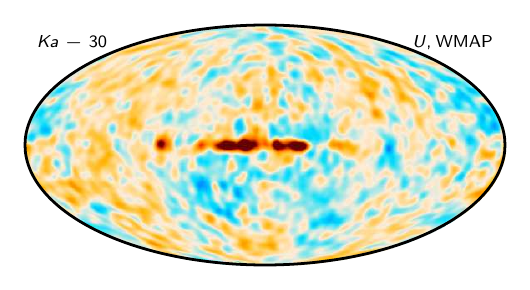}\\
	\includegraphics[width=0.24\textwidth]{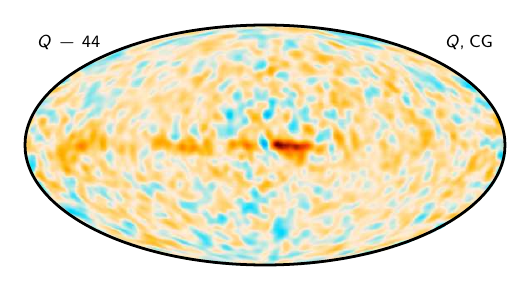}
	\includegraphics[width=0.24\textwidth]{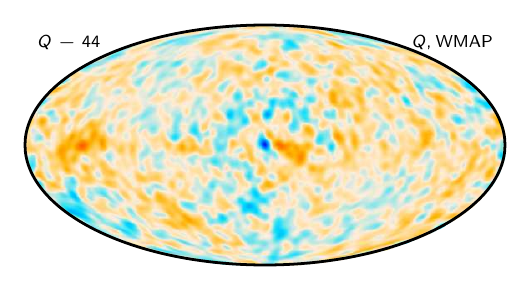}
	\includegraphics[width=0.24\textwidth]{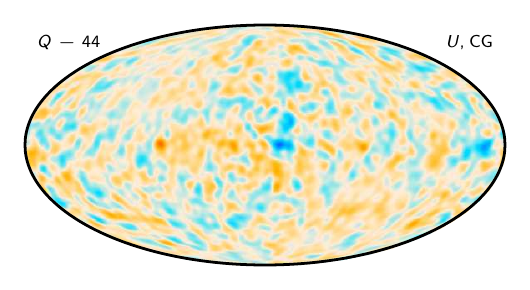}
	\includegraphics[width=0.24\textwidth]{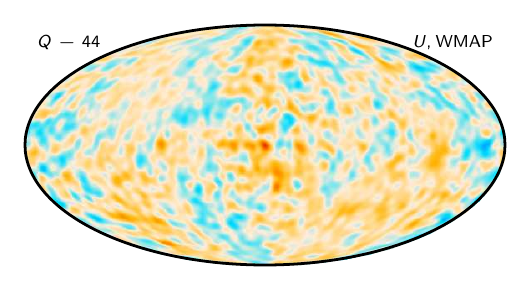}\\
	\includegraphics[width=0.24\textwidth]{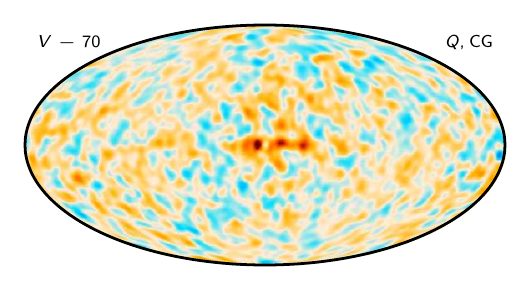}
	\includegraphics[width=0.24\textwidth]{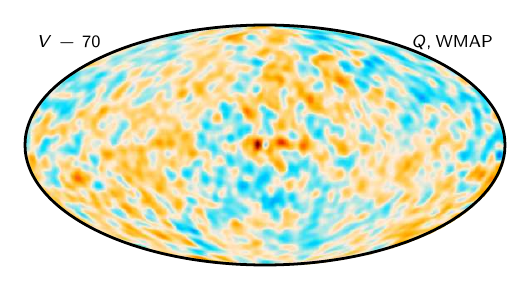}
	\includegraphics[width=0.24\textwidth]{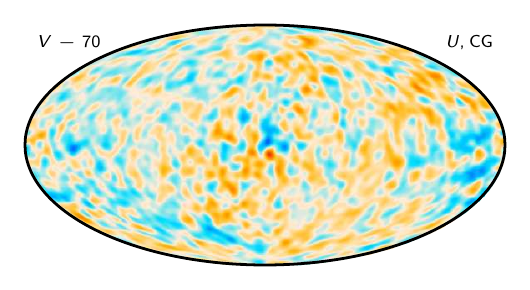}
	\includegraphics[width=0.24\textwidth]{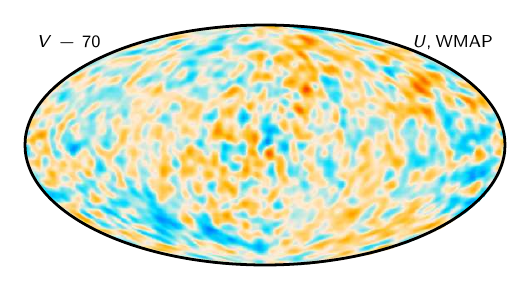}\\
	\includegraphics[width=0.24\textwidth]{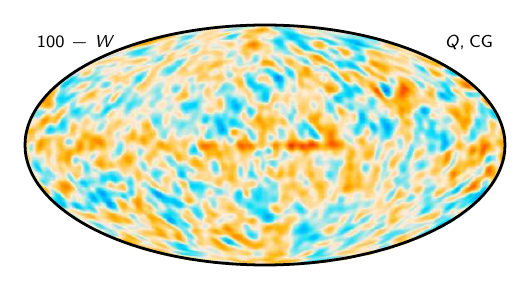}
	\includegraphics[width=0.24\textwidth]{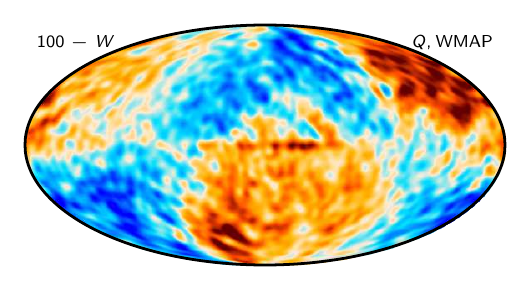}
	\includegraphics[width=0.24\textwidth]{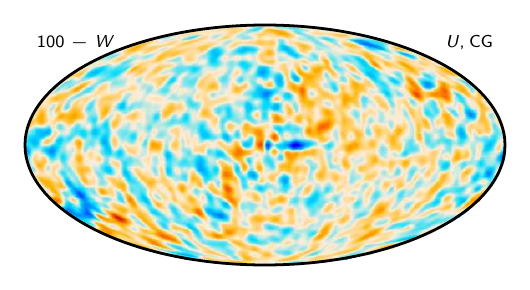}
	\includegraphics[width=0.24\textwidth]{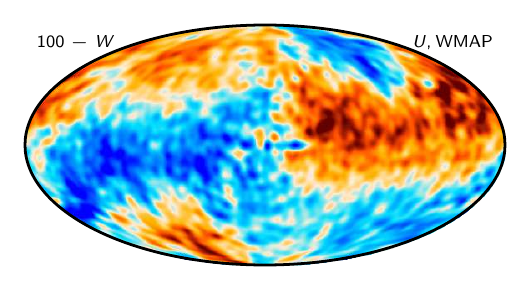}
	
        \includegraphics[width=0.25\textwidth]{figures/cbar_10uK.pdf}
	\caption{Difference maps between similar \WMAP\ and \Planck\ frequency maps. The comparison plots go, by column: Stokes $Q$ for the \cosmoglobe-produced \WMAP\ and \Planck\ sky maps, Stokes $Q$ for official \WMAP\ and \BP\ data products, Stokes $U$ for the \cosmoglobe\ sky maps, and Stokes $U$ for the official data products. \textit{(Top row):} \WMAP\ LFI 30 GHz minus \K-band, scaled by the synchrotron power-law. \textit{(Top middle row):} \WMAP\ \Ka-band minus LFI 30 GHz, also scaled by the synchrotron power-law. \textit{(Middle row):} \WMAP\ \Q-band compared to the LFI 44 GHz sky maps, scaled by the synchrotron power-law. \textit{(Bottom middle row):} \WMAP\ \V-band minus LFI 70 GHz, with unit scalings for each band. \textit{(Bottom row):} The \Planck\ DR4 100 GHz map minus the \WMAP\ \W-band\, also with unit scalings for each band.}
	\label{fig:wmap_lfi_compare}
\end{figure*}

The $E$-mode power spectra, displayed in the second column of Fig.~\ref{fig:map_spectra}, are mainly dominated by noise and polarized synchrotron emission. As expected, the large scale foreground-dominated multipoles decrease in amplitude according to the relative amplitude of the synchrotron spectrum. Overall, the \cosmoglobe\ and \WMAPnine\ power spectra appear fairly consistent for the \K--\Q\ channels, while at \V- and \W-band there is noticeably more scatter at low multipoles in the \WMAPnine\ spectra than in the \cosmoglobe\ spectra.%

The $B$-mode power spectra, displayed in the third column of Fig.~\ref{fig:map_spectra}, are expected to follow a similar pattern, but since foregrounds are generally reduced by a factor of $\simeq2$--4 \citep{bennett2012,planck2016-l04}, this spectrum is less signal-dominated, and therefore more susceptible to instrumental systematics. For instance, the $C_{\ell=3}^\mathrm{BB}$ mode has been identified as being particularly poorly constrained due to its symmetry aligning with $\gtrsim10\,\mathrm{min}$ signals in the TOD induced by the \WMAP\ scan strategy \citep[e.g.,][]{jarosik2010}. In this figure, it appears that these low-$\ell$ modes appear significantly better constrained in the \cosmoglobe\ maps than in \WMAPnine\ for \V- and \W-bands, and the overall large scale noise level is lower by one or two orders of magnitude. 

\begin{figure}
	\centering
	\includegraphics[width=\linewidth]{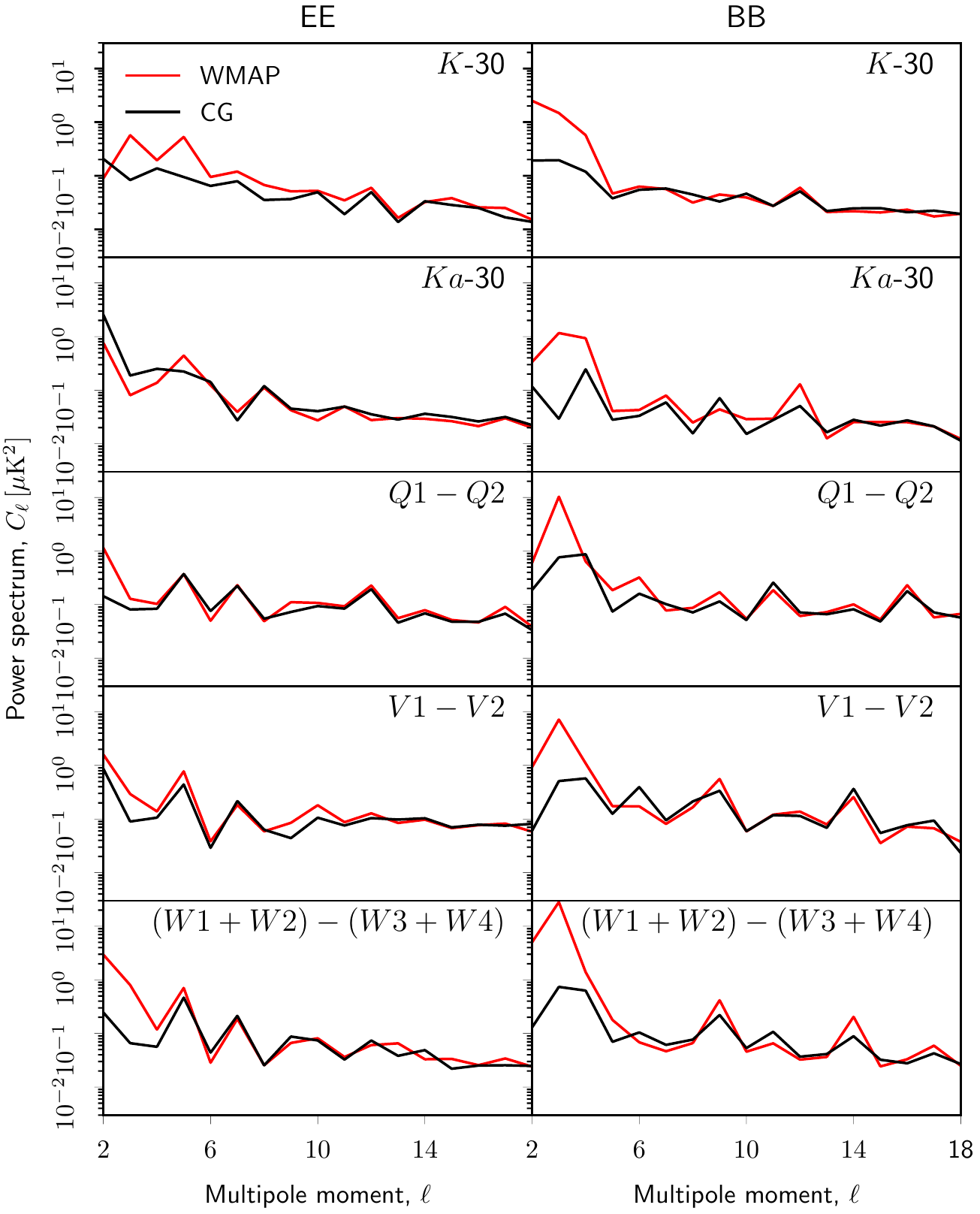}
	\caption{Full sky half-difference spectra. The red lines are the power spectra of the \WMAPnine\ difference maps, while the black lines are the same for the reprocessed \cosmoglobe\ maps.}
        \label{fig:cl_halfdiff}
\end{figure}

\subsection{Consistency tests through inter-channel difference maps}
\label{sec:internal_consistency}

As described in Sect.~\ref{sec:wmap_instmodel}, the \Q- and \V-bands each had two DAs, while the \W-band had four DAs, and computing differences between the corresponding DA maps can highlight mismodeled systematics. While the \K-band and \Ka-band have different central frequencies, they are close enough that we can compare them by scaling \K-band assuming a polarized synchrotron power law SED of $\beta_\mathrm s=-3.1$. Similarly, internal differences between scaled \K, \Ka, and LFI 30\,GHz maps provide an important null test. In particular, the \K--30 difference has received significant attention ever since the \Planck\ 2015 data release \citep{planck2014-a01}, showing clear signatures of instrumental systematics. These were gradually reduced through improved \Planck\ processing in the \Planck\ 2018 \citep{planck2016-l02}, PR4 \citep{npipe}, and \bp\ \citep{bp01} data releases. Still, even after all these developments, large-scale residuals remained that were difficult to resolve through further LFI improvements \citep{bp07}. In this section, we revisit this question for the \cosmoglobe\ products.

We start by inspecting internal \WMAP\ half-difference maps of the form $(Q1-Q2)/2$ etc. These are plotted in Fig.~\ref{fig:internal_diff}. Here we see that the \Q-band and \V-band half-difference maps from \cosmoglobe\ have virtually no trace of poorly measured modes, and the differences appear to be well-traced by the rms maps. In contrast, the \WMAP\ half-difference maps show clear evidence of large-scale residuals. The largest visual improvement is in the \W-band, where the \cosmoglobe\ case is almost entirely consistent with instrumental noise, as opposed to the \WMAPnine\ difference that is dominated by large-scale residuals.  

Next, Fig.~\ref{fig:wmap_lfi_compare} shows comparisons between the \WMAP\ $K$-
and $Ka$-bands and the LFI 30 GHz channel, between the \WMAP\ $Q$-band and LFI
44 GHz, and finally between \WMAP\ $V$-band and LFI 70 GHz. When comparing
\WMAPnine\ maps with \Planck\ LFI, we use \BP\ products, which represent the
cleanest version of \Planck\ LFI published to date. For the \cosmoglobe\ map
comparison, both \WMAP\ and \Planck\ maps were produced by this joint analysis.
Additionally, we compare the mean \W-band maps with the \Planck\ HFI DR4
100\,GHz channel. It is worth noting that this 100\,GHz map has had no input
from \commanderthree\ so this difference map is an independent comparison
between two datasets and processing methods.

Starting with the \cosmoglobe\ maps, we see in the first and third columns of
Fig.~\ref{fig:wmap_lfi_compare} that the magnitude of the differences are small
in both Stokes $Q$ and $U$. Overall, across all five frequency map comparisons
we see small levels of variation, with structure contained to the Galactic
plane. Notably, however, there is a larger sky signal within the
$\mathit{Ka}-30$ Stokes $Q$ comparison.
This large-scale difference also exists in the $\mathit Q-44$ Stokes
$Q$, but it did not appear in the internal \Q\ half-difference map.

Columns two and four of Fig.~\ref{fig:wmap_lfi_compare} show corresponding differences
between the official \WMAPnine\ and \BP\ LFI frequency maps. Similar to the
\cosmoglobe\ sky map comparisons, we see differences in the Galactic center,
and to a lesser degree along the Galactic plane due to the slight differences
in the frequency coverage. When comparing the official \WMAP\ maps,
particularly for \K-band, we see structures sweeping across large angular
scales across the sky, likely due to the poorly measured modes in \K-band.

Of particular note is the $100-\mathit W$ difference map. The \cosmoglobe\
difference maps here have a similar level of white noise and Galactic
contamination as the $\mathit V-70$ maps,  whereas the \WMAPnine\ differences
are driven by obvious transmission imbalance modes, each with an opposite sign and
magnitude. The difference between 100\,GHz and \W\ demonstrates that the good
agreement between the \WMAP\ and \Planck\ LFI is not simply due to fitting
low-level parameters in a joint analysis framework -- by obtaining \W-band maps
that are consistent with an independent 100\,GHz polarization map, we have
shown that the \WMAP-LFI agreement is the result of a genuine improvement in data processing.

\begin{table}
\newdimen\tblskip \tblskip=5pt
\caption{Transmission imbalance template amplitudes for each \WMAP\ radiometer as estimated by fitting the official templates to low-resolution difference maps between \cosmoglobe\ and \WMAP. The templates are provided in mK, and the template amplitudes are therefore dimensionless. The fourth column lists the fractional decrease in standard deviation, $\Delta\sigma/\sigma\equiv\sqrt{\sigma_{\mathrm{raw}}^2 - \sigma_{\mathrm{corr}}^2}/\sigma_{\mathrm{raw}}$,  after subtracting the best-fit templates. }
\label{tab:transmission}
\vskip -4mm
\footnotesize
\setbox\tablebox=\vbox{
 \newdimen\digitwidth
 \setbox0=\hbox{\rm 0}
 \digitwidth=\wd0
 \catcode`*=\active
 \def*{\kern\digitwidth}
  \newdimen\dpwidth
  \setbox0=\hbox{.}
  \dpwidth=\wd0
  \catcode`!=\active
  \def!{\kern\dpwidth}
  \halign{\hbox to 1.8cm{#\leaderfil}\tabskip 2em&
    \hfil$#$\hfil \tabskip 2em&
    \hfil$#$\hfil \tabskip 2em&    
    \hfil$#$\hfil \tabskip 0em\cr
\noalign{\doubleline}
\omit\hfil\sc DA \hfil& a_1 & a_2 & \Delta \sigma/\sigma \cr
\noalign{\vskip 3pt\hrule\vskip 5pt}
\K1 &   -27.5*  &  -50.6* & 0.30 \cr
\Ka1 &   -1.4  &   -1.9 & 0.25 \cr
\Q1 &   -30.0* &  -71.6* & 0.11 \cr
\Q2 &   -7.1  &  -1.5 & 0.20 \cr
\V1 &   -32.8*  &  -53.4* & 0.06 \cr
\V2 &   *8.8  &  -4.1 & 0.16 \cr
\W1 &   -2.8  &  *4.6 & 0.08 \cr
\W2 &   -6.9  & -3.5 & 0.11 \cr
\W3 &   29.1  & 53.4 & 0.12 \cr
\W4 &   15.5  & -6.8 & 0.52 \cr
\noalign{\vskip 5pt\hrule\vskip 5pt}}}
\endPlancktablewide
\end{table}

\begin{figure*}[t]
  \centering
        \includegraphics[width=0.16\linewidth]{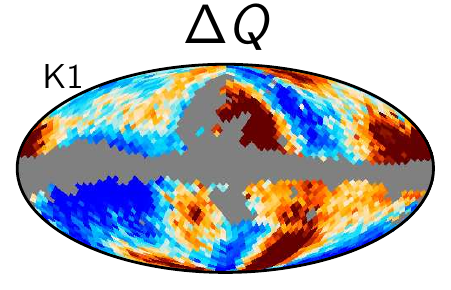}
        \includegraphics[width=0.16\linewidth]{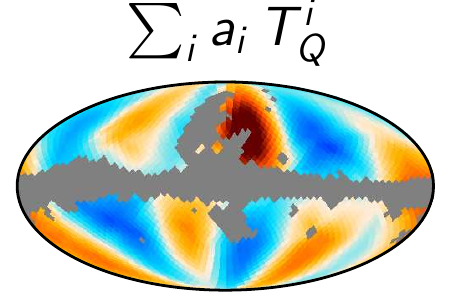}
        \includegraphics[width=0.16\linewidth]{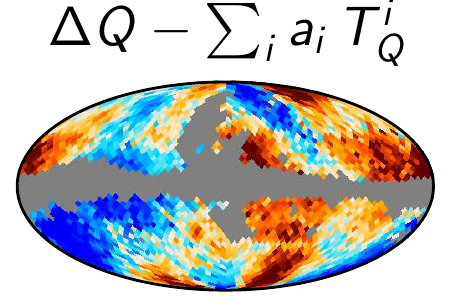}\hspace*{2mm}
        \includegraphics[width=0.16\linewidth]{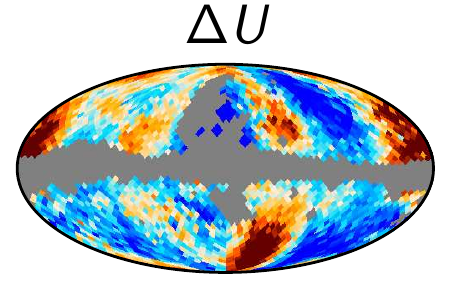}
        \includegraphics[width=0.16\linewidth]{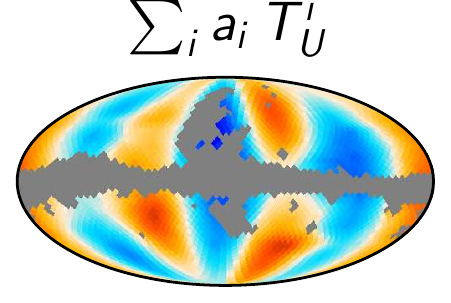}
        \includegraphics[width=0.16\linewidth]{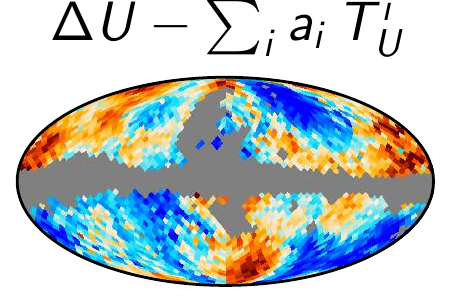}\\
        \includegraphics[width=0.16\linewidth]{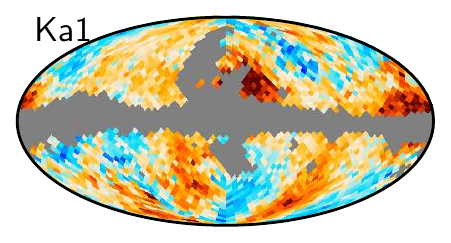}
        \includegraphics[width=0.16\linewidth]{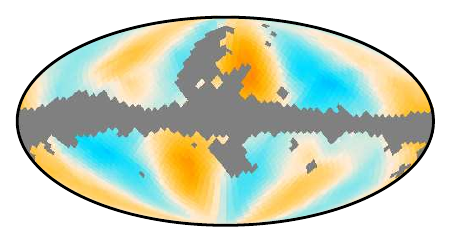}
        \includegraphics[width=0.16\linewidth]{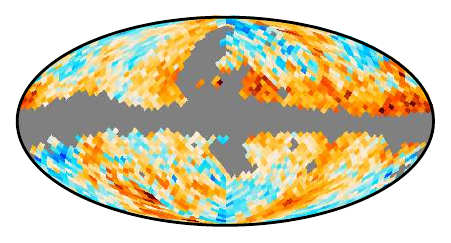}\hspace*{2mm}
        \includegraphics[width=0.16\linewidth]{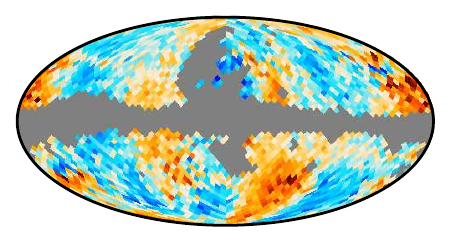}
        \includegraphics[width=0.16\linewidth]{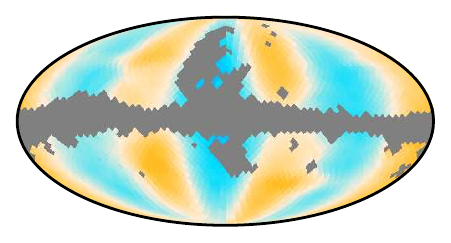}
        \includegraphics[width=0.16\linewidth]{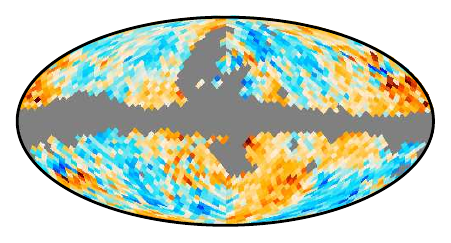}\\
        \includegraphics[width=0.16\linewidth]{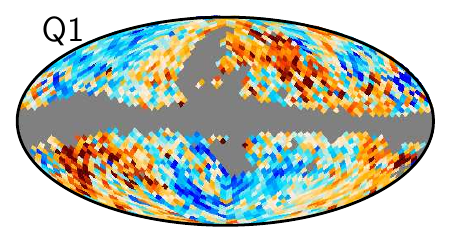}
        \includegraphics[width=0.16\linewidth]{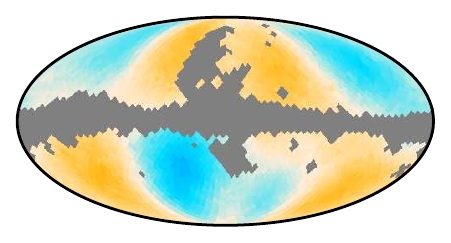}
        \includegraphics[width=0.16\linewidth]{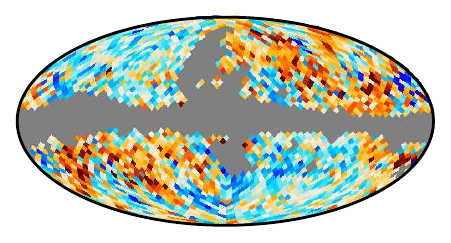}\hspace*{2mm}
        \includegraphics[width=0.16\linewidth]{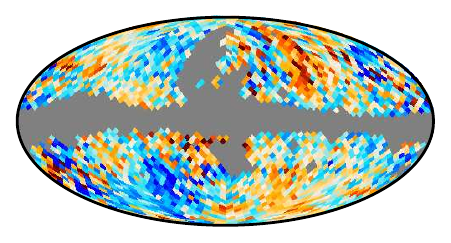}
        \includegraphics[width=0.16\linewidth]{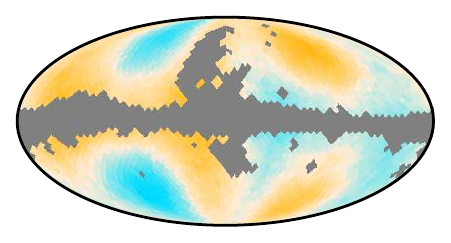}
        \includegraphics[width=0.16\linewidth]{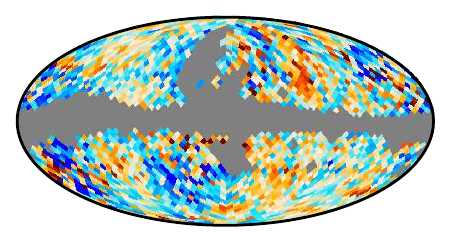}\\
        \includegraphics[width=0.16\linewidth]{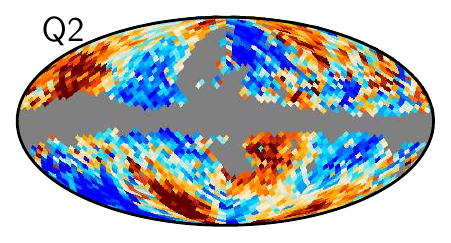}
        \includegraphics[width=0.16\linewidth]{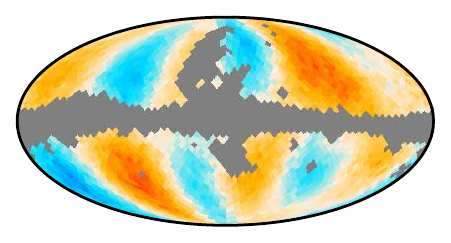}
        \includegraphics[width=0.16\linewidth]{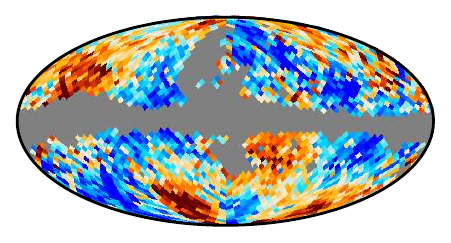}\hspace*{2mm}
        \includegraphics[width=0.16\linewidth]{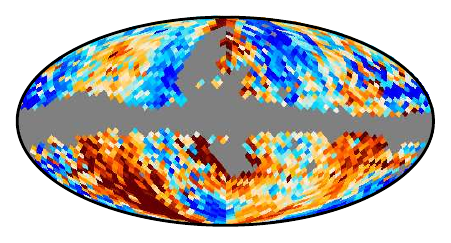}
        \includegraphics[width=0.16\linewidth]{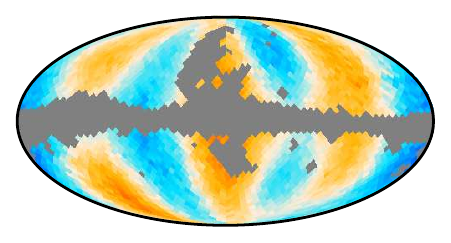}
        \includegraphics[width=0.16\linewidth]{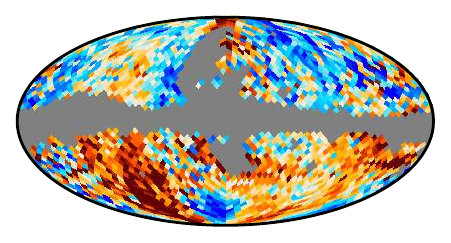}\\
        \includegraphics[width=0.16\linewidth]{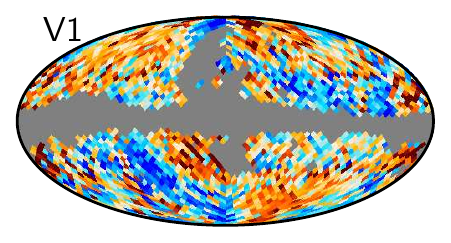}
        \includegraphics[width=0.16\linewidth]{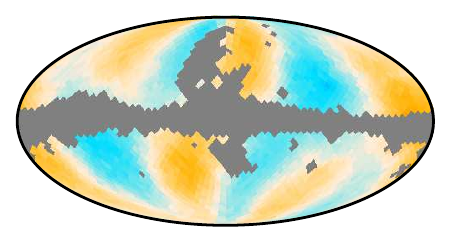}
        \includegraphics[width=0.16\linewidth]{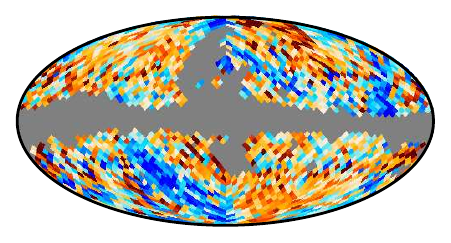}\hspace*{2mm}
        \includegraphics[width=0.16\linewidth]{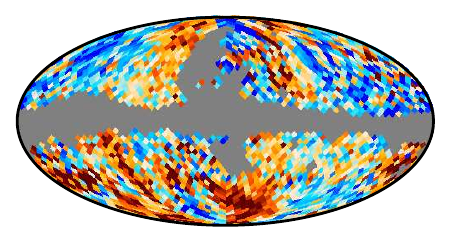}
        \includegraphics[width=0.16\linewidth]{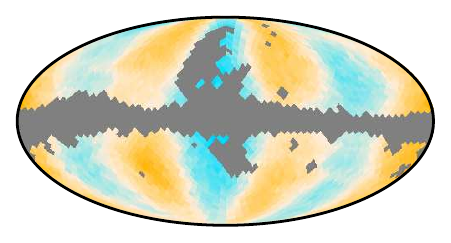}
        \includegraphics[width=0.16\linewidth]{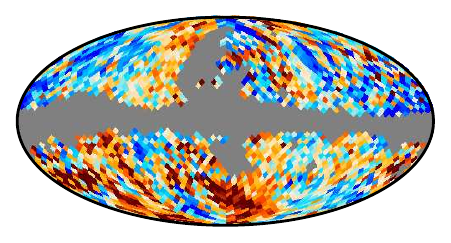}\\
        \includegraphics[width=0.16\linewidth]{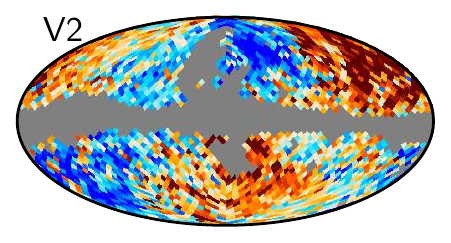}
        \includegraphics[width=0.16\linewidth]{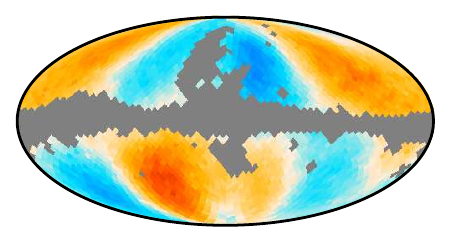}
        \includegraphics[width=0.16\linewidth]{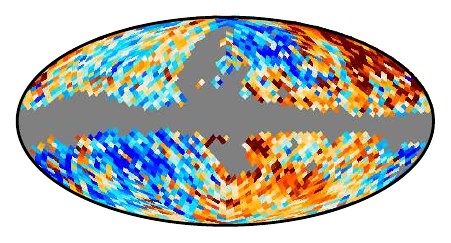}\hspace*{2mm}
        \includegraphics[width=0.16\linewidth]{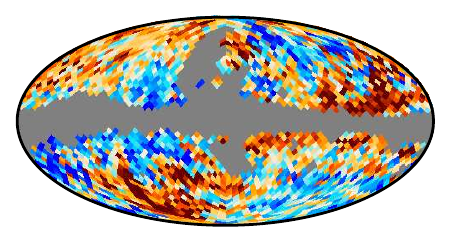}
        \includegraphics[width=0.16\linewidth]{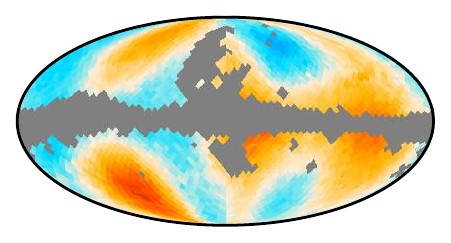}
        \includegraphics[width=0.16\linewidth]{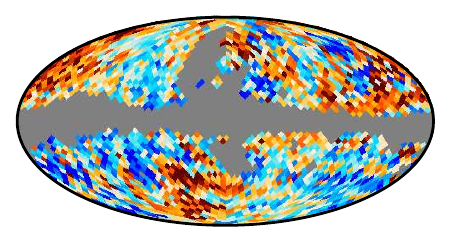}\\
        \includegraphics[width=0.16\linewidth]{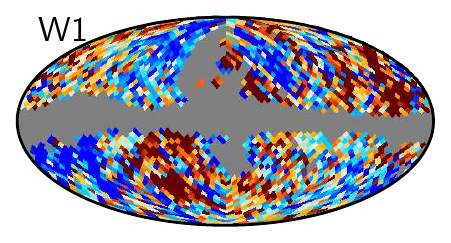}
        \includegraphics[width=0.16\linewidth]{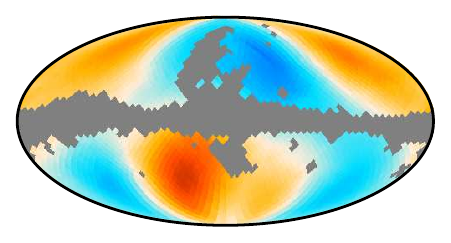}
        \includegraphics[width=0.16\linewidth]{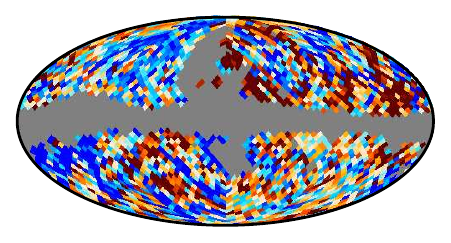}\hspace*{2mm}
        \includegraphics[width=0.16\linewidth]{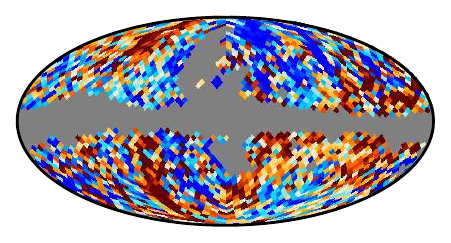}
        \includegraphics[width=0.16\linewidth]{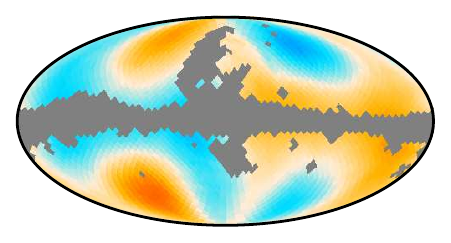}
        \includegraphics[width=0.16\linewidth]{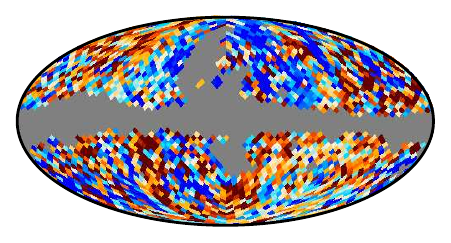}\\
        \includegraphics[width=0.16\linewidth]{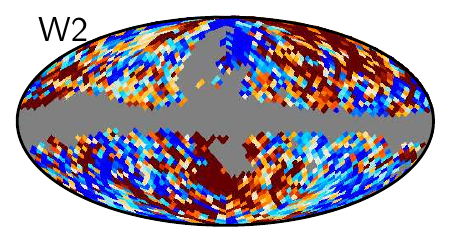}
        \includegraphics[width=0.16\linewidth]{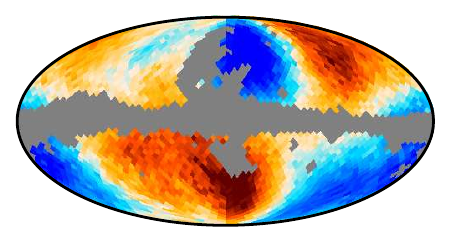}
        \includegraphics[width=0.16\linewidth]{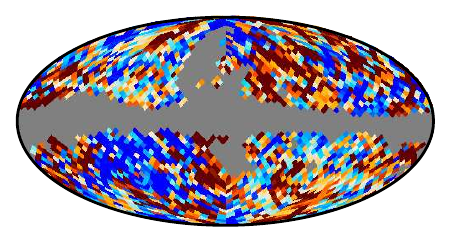}\hspace*{2mm}
        \includegraphics[width=0.16\linewidth]{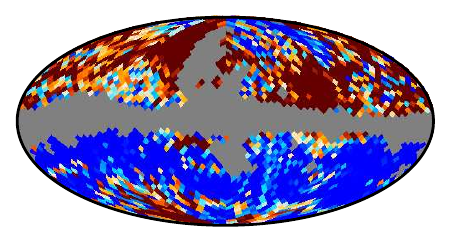}
        \includegraphics[width=0.16\linewidth]{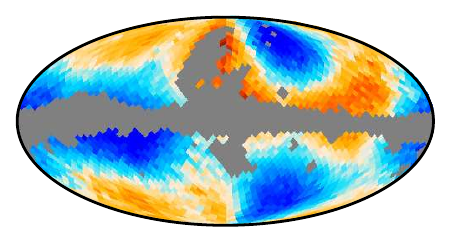}
        \includegraphics[width=0.16\linewidth]{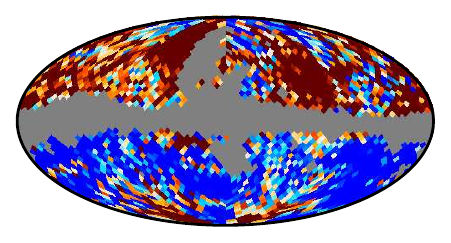}\\
        \includegraphics[width=0.16\linewidth]{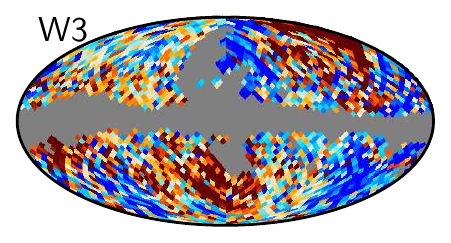}
        \includegraphics[width=0.16\linewidth]{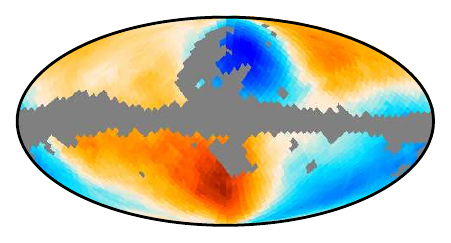}
        \includegraphics[width=0.16\linewidth]{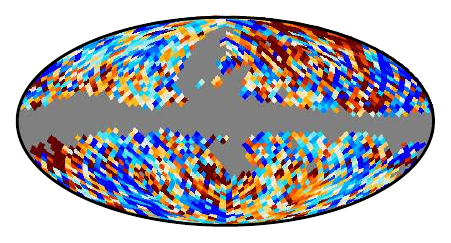}\hspace*{2mm}
        \includegraphics[width=0.16\linewidth]{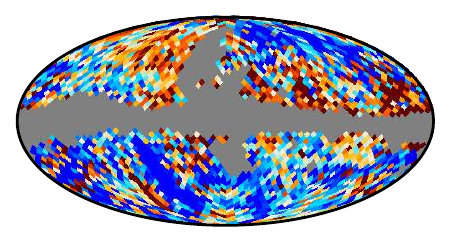}
        \includegraphics[width=0.16\linewidth]{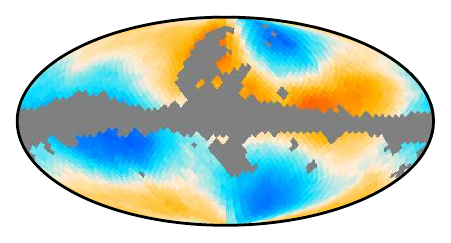}
        \includegraphics[width=0.16\linewidth]{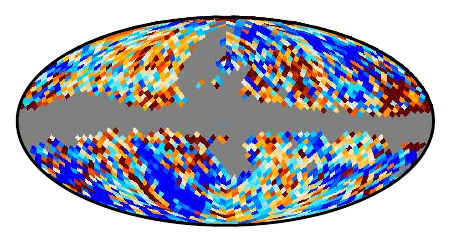}\\
        \includegraphics[width=0.16\linewidth]{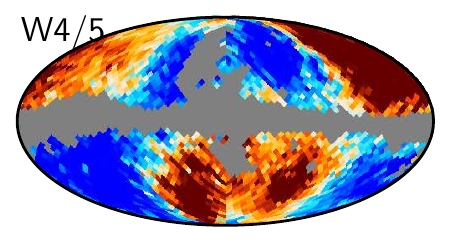}
        \includegraphics[width=0.16\linewidth]{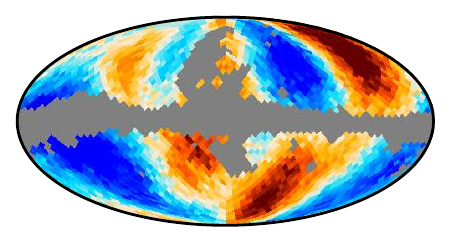}
        \includegraphics[width=0.16\linewidth]{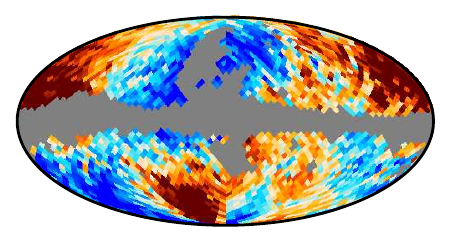}\hspace*{2mm}
        \includegraphics[width=0.16\linewidth]{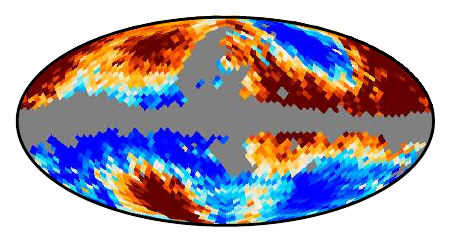}
        \includegraphics[width=0.16\linewidth]{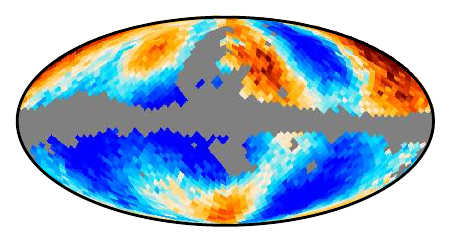}
        \includegraphics[width=0.16\linewidth]{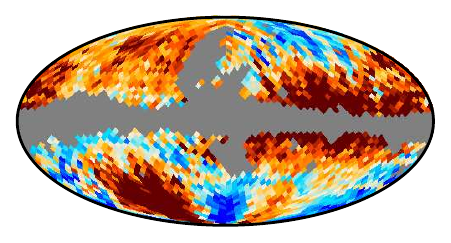}\\
        \includegraphics[width=0.30\linewidth]{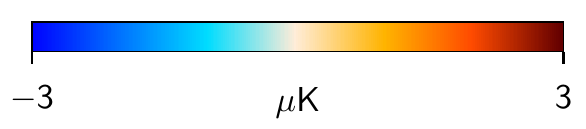}\\         
	\caption{Efficiency assessment of the \WMAP\ approach for transmission imbalance error propagation. The left and right sections of the figure correspond to Stokes $Q$ and $U$ parameters, respectively, while rows show different DAs. Within each section, the left panel shows the raw difference between the \WMAPnine\ and \cosmoglobe\ DA maps, while the middle panel shows the best-fit \WMAP\ transmission imbalance template combination; the right panel shows the difference between the two. Only one template amplitude is fit for both $Q$ and $U$.}
	\label{fig:imbal}
\end{figure*}

\begin{figure*}[t]
  \centering
  \includegraphics[width=0.33\linewidth]{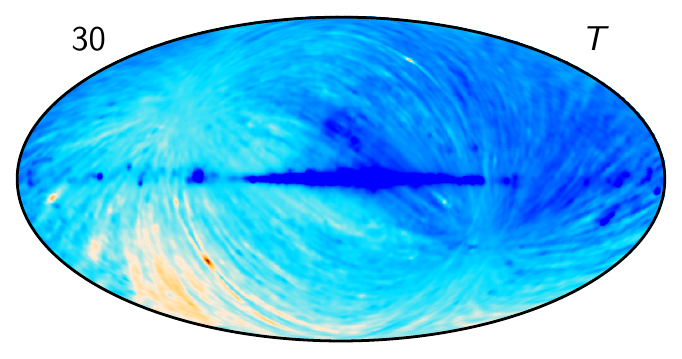}
  \includegraphics[width=0.33\linewidth]{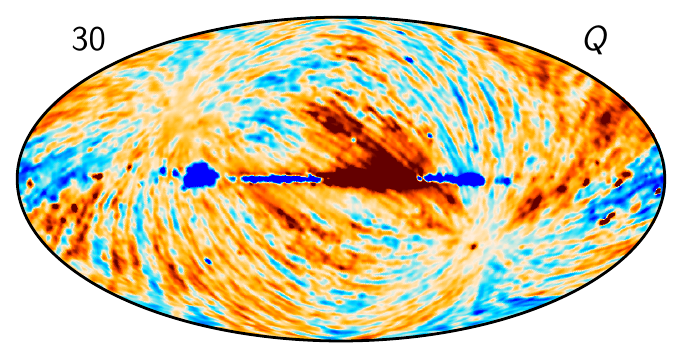}
  \includegraphics[width=0.33\linewidth]{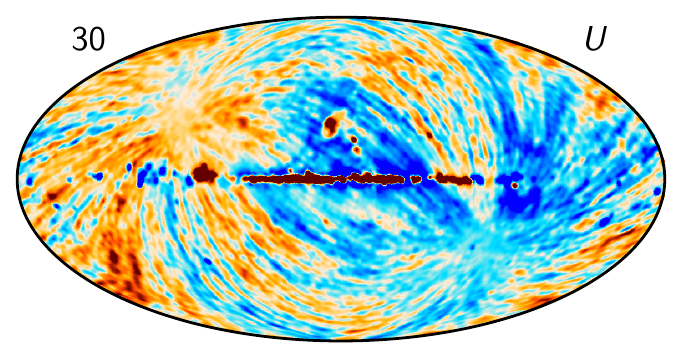}\\
  \includegraphics[width=0.33\linewidth]{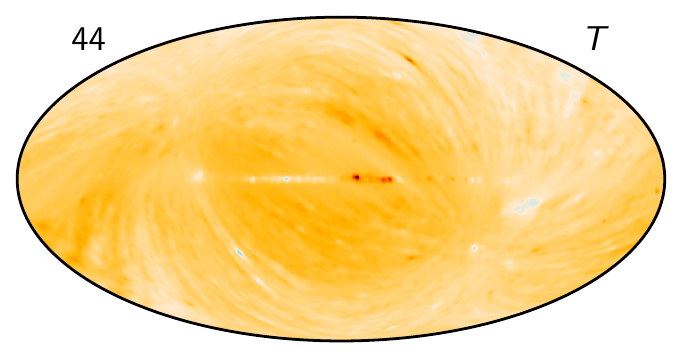}
  \includegraphics[width=0.33\linewidth]{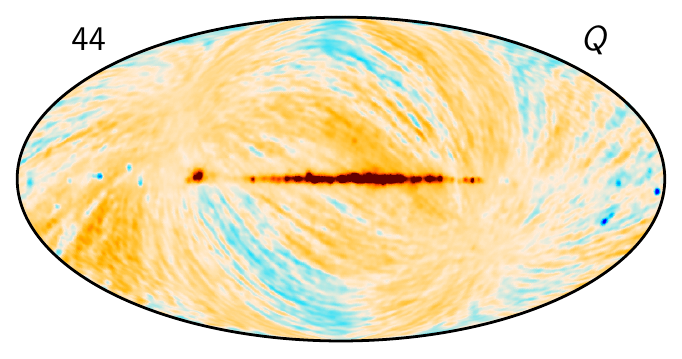}
  \includegraphics[width=0.33\linewidth]{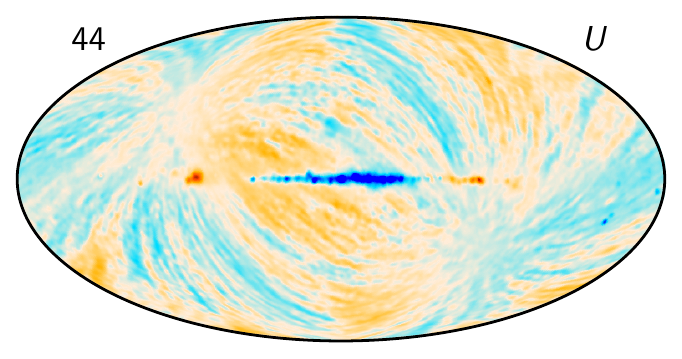}\\
  \includegraphics[width=0.33\linewidth]{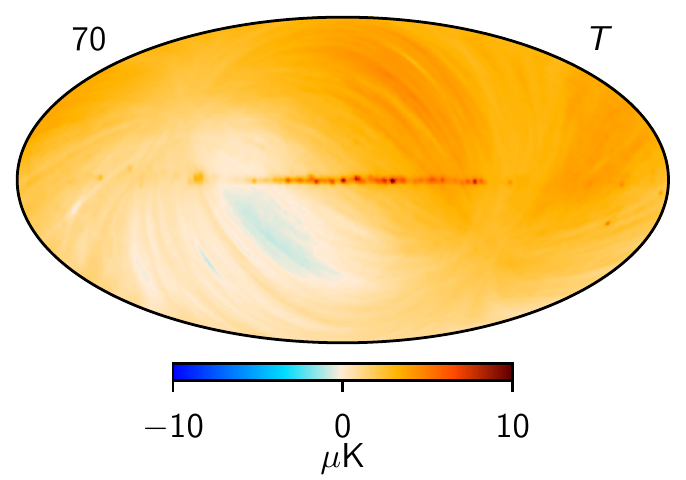}
  \includegraphics[width=0.33\linewidth]{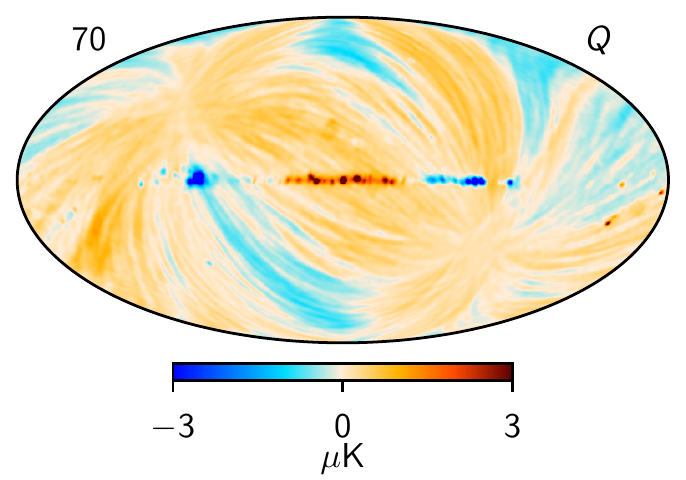}
  \includegraphics[width=0.33\linewidth]{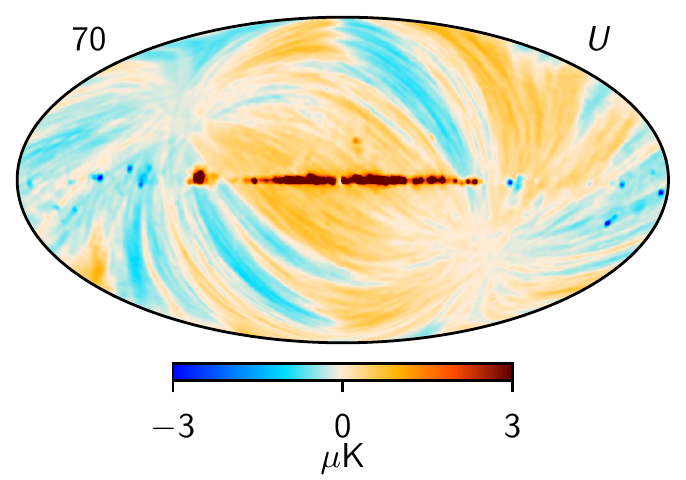}  
  \caption{Difference between \cosmoglobe\ and \bp\ posterior mean frequency maps for 30 (top row), 44 (middle row), and 70\,GHz (bottom row). Columns show Stokes $T$, $Q$, and $U$ parameters. }
  \label{fig:lfi_comparison}
\end{figure*}

Finally, as noted by \citet{jarosik2010} the low-$\ell$ \W-band polarization data were excluded entirely from the cosmological analysis due to excess variance in the $\ell\leq7$ multipoles. To test the \cosmoglobe\ maps' performance at these scales, we take the power spectrum of the full-sky difference maps using the standard \texttt{anafast} routine in Fig.~\ref{fig:cl_halfdiff}. With very few exceptions, the \WMAPnine\ power spectra have much more power at $\ell\leq7$ than the \cosmoglobe\ maps in both the $E$-modes and $B$-modes. Of particular note is the $\ell=3$ $B$-mode, which has consistently been identified as poorly measured in the \WMAP\ scan strategy, and has been reduced in every difference spectrum. Based on these power spectra alone, there does not appear to be a strong justification for excluding the reprocessed \W-band polarization data in future cosmological analyses.

\begin{figure*}
	\centering
	\includegraphics[width=0.45\textwidth]{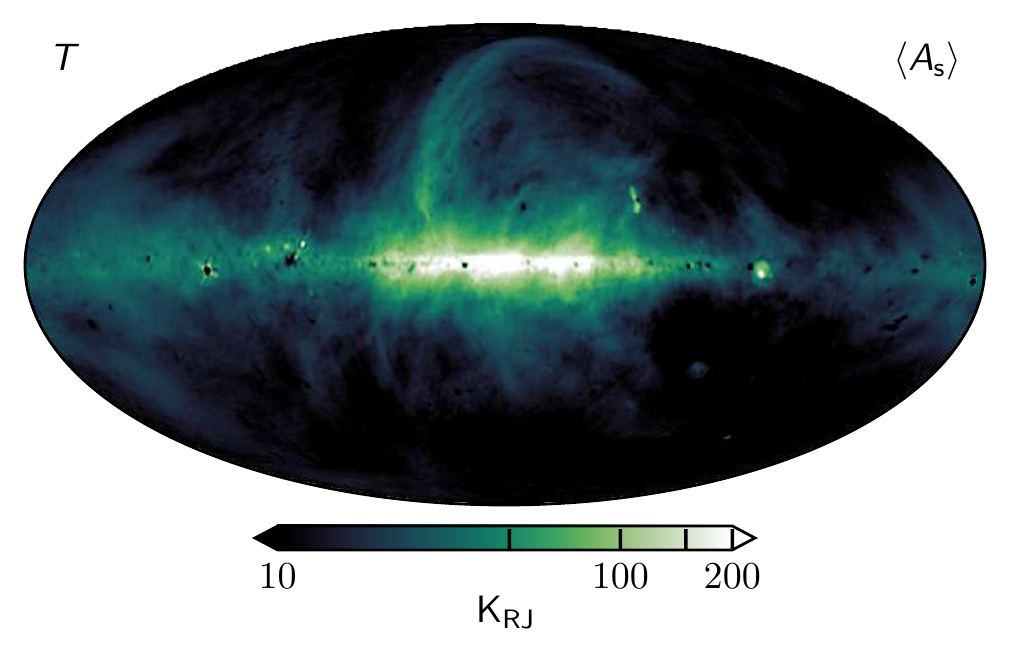}
	\includegraphics[width=0.45\textwidth]{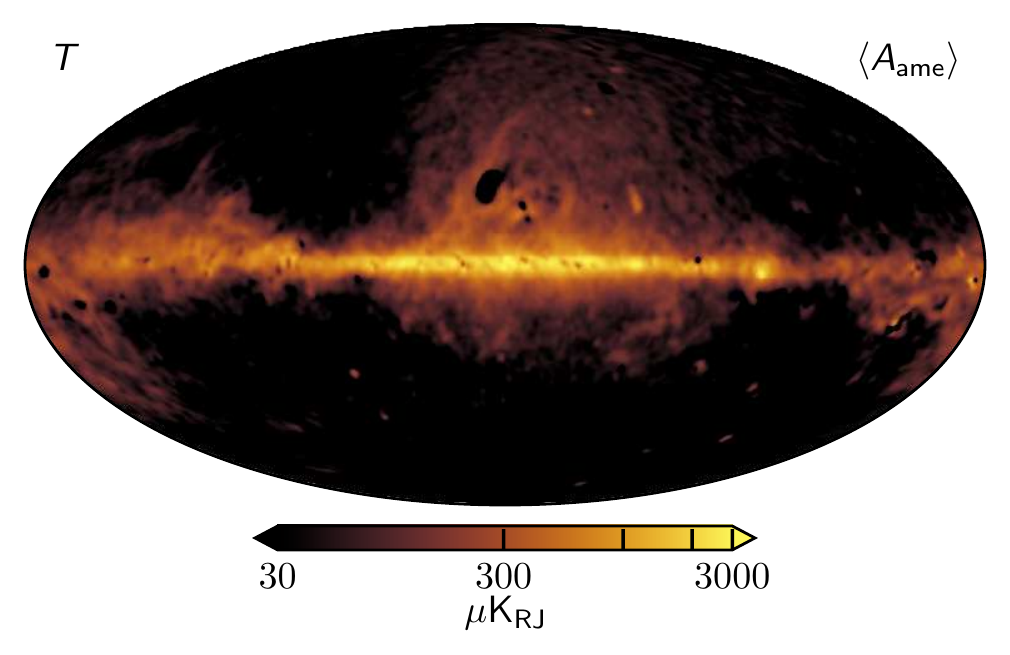}\\
	\includegraphics[width=0.45\textwidth]{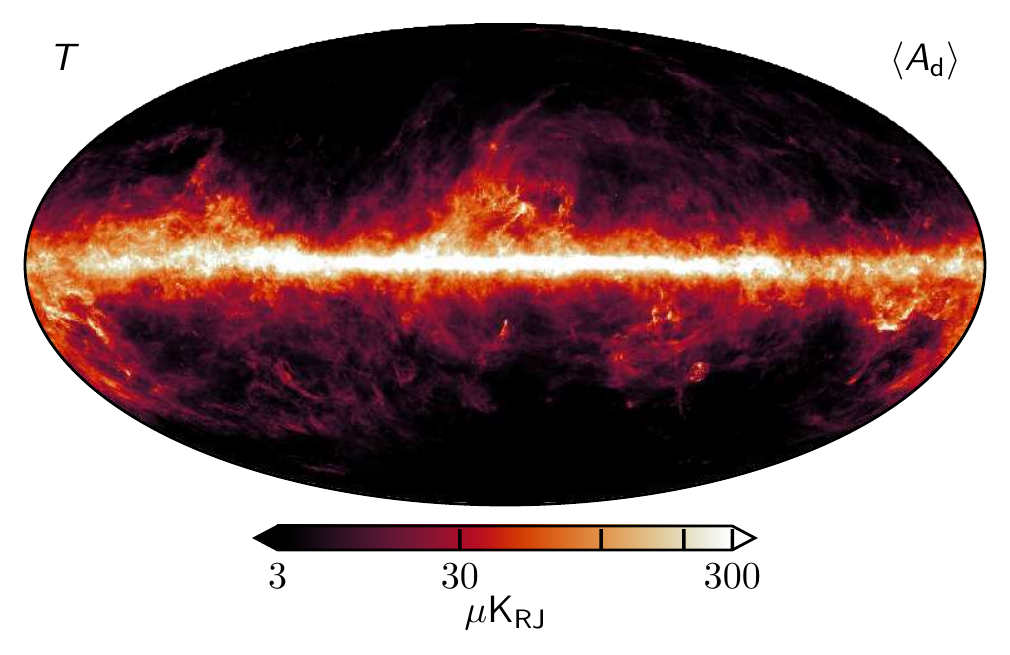}
	\includegraphics[width=0.45\textwidth]{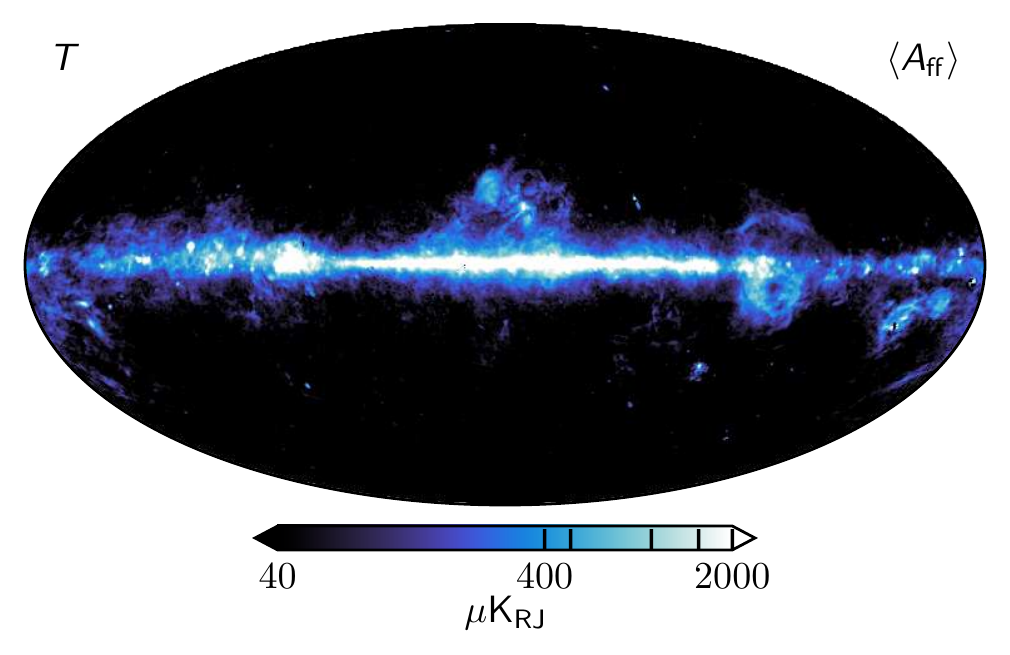}\\
	\caption{Foreground intensity maps, evaluated at their respective reference frequencies. 
	\textit{(Top left:)} Synchrotron emission evaluated at 408\,MHz.
	\textit{(Top right:)} Anomalous microwave emission evaluated at 22\,GHz. 
	\textit{(Bottom left:)} Free-free emission at 40\,GHz. 
	\textit{(Bottom right:)} Thermal dust emission at 70\,GHz. 
	}\label{fig:intensity_foregrounds}
\end{figure*}

\begin{figure}
	\centering
	\includegraphics[width=\columnwidth]{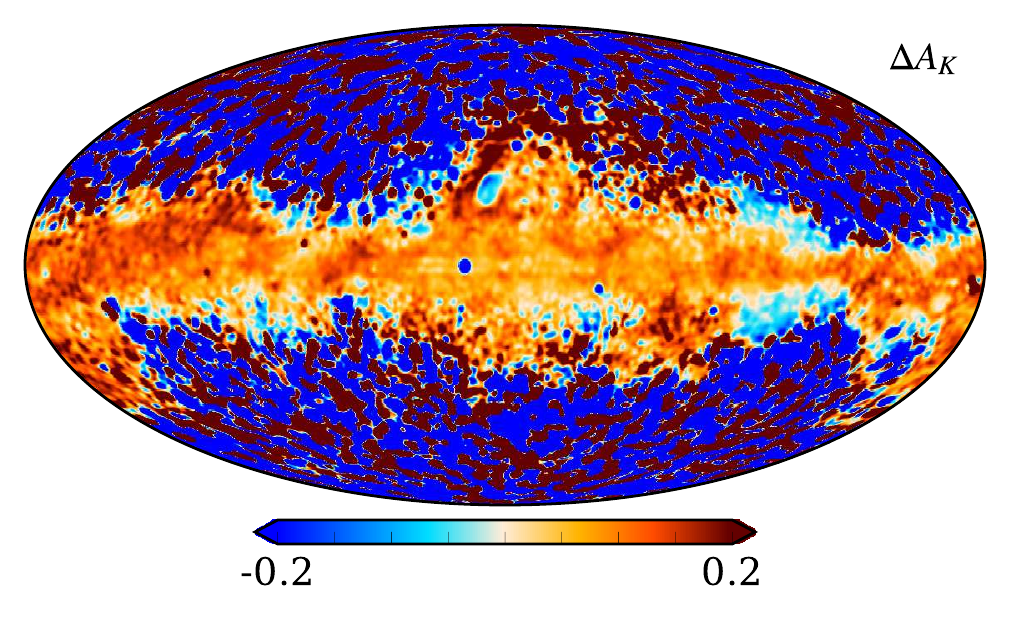}
	\caption{Fractional difference map between the sum of AME and free-free emission at 22\,GHz between the \cosmoglobe\ and \bp\ foreground modes, i.e., $\Delta A_{\mathrm{K}}=s^{\mathrm{CG}}_{\mathrm{AME}+\mathrm{ff}}-s^{\mathrm{BP}}_{\mathrm{AME}+\mathrm{ff}})/s^{\mathrm{BP}}_{\mathrm{AME}+\mathrm{ff}}$.}
	\label{fig:intensity_foregrounds_diff}
\end{figure}

Based on these calculations, we conclude that the modes that are nearly degenerate by the \WMAP\ scanning strategy, and have represented a major challenge for the official \WMAP\ processing for more than a decade, appear to have been properly regularized by the global \cosmoglobe\ processing. The frequency maps do not show any evidence of either poorly constrained transmission imbalance modes or other large-scale artifacts, and they are more self-consistent than the \WMAPnine\ frequency maps.

\subsection{Efficiency of template-based transmission imbalance uncertainty propagation}
\label{subsec:imbalance_template}

Next, we revisit the issue of transmission imbalance error propagation in the official \WMAP\ approach. The \WMAPnine\ likelihood deweights estimates of the transmission imbalance modes in the low resolution polarization likelihood covariance matrix, so that accurate power spectra could be obtained from the maps \citep{jarosik2010}. As discussed by \citet{jarosik2007} and as  seen in Fig.~\ref{fig:x_im} in this paper, $x_{\mathrm{im}}$ is associated with significant measurement uncertainties, and these uncertainties translate directly into correlated large-scale polarization residuals. To account for these uncertainties, the \WMAP\ pipeline produced a spatial template by calculating a frequency map for which both $x_{\textrm{im}}$ values were increased by 10\,\% from their nominal values in a given DA, and subtracting this from the baseline map. A second template was generated by increasing one value by 10\,\% and decreasing the other by 10\,\%. The modes corresponding to the resulting spatial templates were then accordingly downweighted through the low-resolution noise covariance matrix using the Woodbury formula.

This approach is effectively equivalent to assuming that the major impact of transmission imbalance may be described in terms of a bilinear vector space, and that this vector space is statistically independent from other error contributions, such as the correlated noise and gain. This is obviously an approximation, and given the new maps presented in this paper it is possible to derive at least a rough estimate of how well this approximation works.

Specifically, considering the large improvement in half-difference maps seen in Fig.~\ref{fig:internal_diff}, and the morphology of the full frequency difference maps seen in Fig.~\ref{fig:skymaps}, for the purposes of this section it is reasonable to assume that most of the large-scale polarization differences in Fig.~\ref{fig:megadiff_wmap} are due to \WMAPnine\ rather than \cosmoglobe. We therefore fit the pair of \WMAP\ transmission imbalance templates for each DA to the difference map between \WMAPnine\ and \cosmoglobe, and we subtract this from the difference map. The best-fit template amplitudes are listed in the second and third column of Table~\ref{tab:transmission} for each DA. In this table, we see that the coefficients within each DA tend to be quite similar, both in magnitude and sign. This is due to the fact that the two individual templates also tend to be highly correlated in terms of spatial structure but with opposite sign. As such, there are strong degeneracies between the two resulting coefficients, and only the sum over both templates carries physical meaning. 

\begin{figure*}
	\centering
	\includegraphics[width=0.45\textwidth]{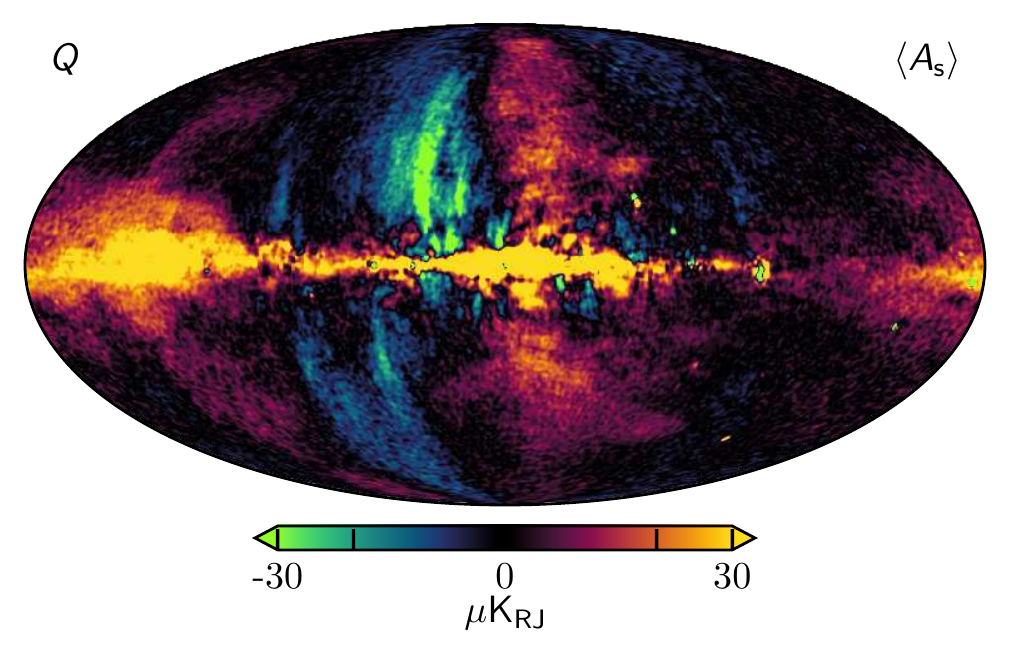}
	\includegraphics[width=0.45\textwidth]{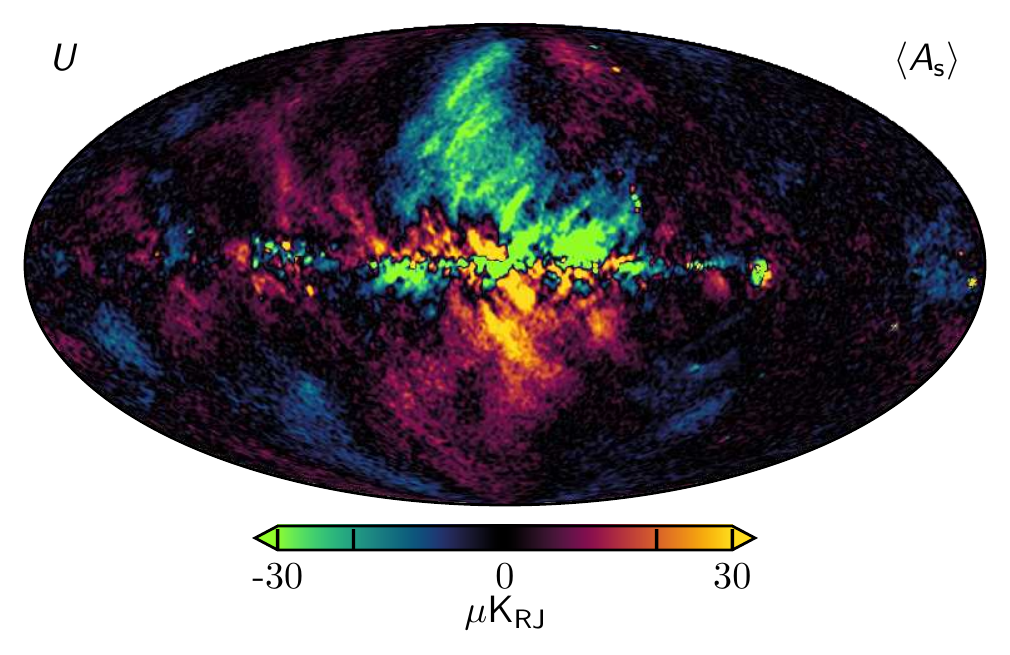}\\
	\includegraphics[width=0.45\textwidth]{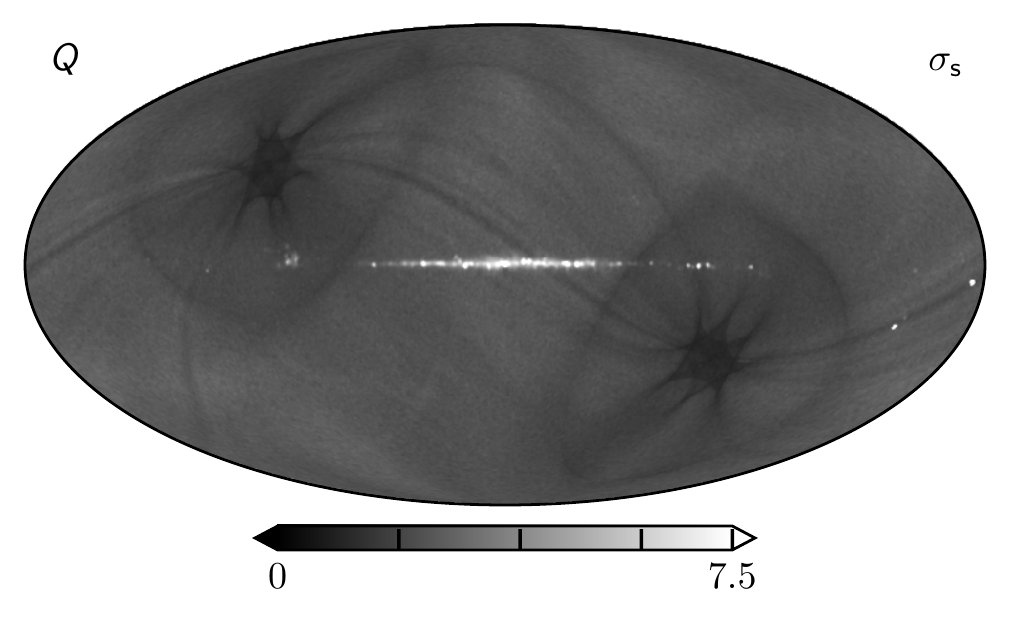}
	\includegraphics[width=0.45\textwidth]{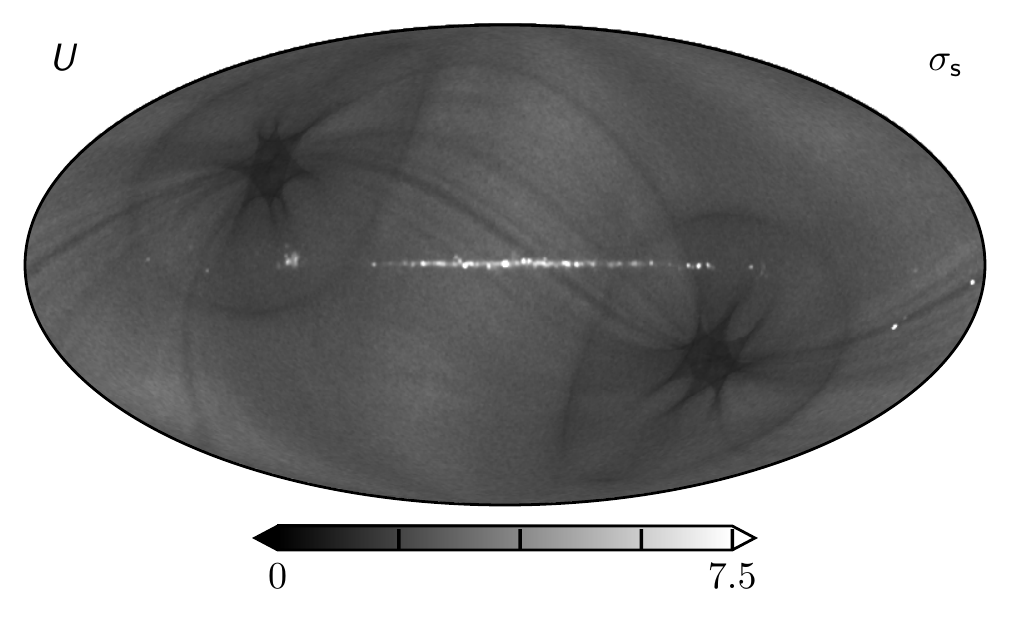}\\
	\caption{Polarized synchrotron maps and their standard deviations evaluated at 30\,\GHz.}\label{fig:polarized_foregrounds}
\end{figure*}

Figure~\ref{fig:imbal} summarizes the results from these calculations in map domain. Each row corresponds to one DA, while the left and right half-sections of the figure correspond to Stokes $Q$ and $U$, respectively. (One common parameter is fitted per template for both Stokes parameters, but they are for convenience visualized separately.) Within each section, the left column shows the raw difference, the middle column shows the sum of the best-fit templates, and the right column shows the residual obtained after subtracting the best-fit templates from the raw difference.

Comparing the left and middle columns in each section, we see that -- at least at a visual level -- the \WMAP\ templates do indeed trace the residuals to a high degree for many channels, e.g., \K, \Q1, \V2 etc. For some channels, such as \Ka1 and \Q2, the agreement is less convincing. However, as seen in the right-most column, even for the well-fitting channels the templates are unable to explain all of the difference.

To roughly estimate how much of the full difference may be described by the \WMAP\ templates, we compute the relative decrease in standard deviation between the full difference map (left columns) and the template-corrected map (right columns) of the form $\sqrt{\sigma_{\mathrm{raw}}^2 - \sigma_{\mathrm{corr}}^2}/\sigma_{\mathrm{raw}}$; in the case that the template correction happened to account for the full difference, this quantity would be unity, while in the case in which it did not account for anything, it would be 0.

The results from this calculation are summarized in the fourth column in Table~\ref{tab:transmission}. Here we see that the two linear templates are able to account for between 6 and 52\,\% of the full large-scale difference between \cosmoglobe\ and \WMAP; the remaining power must either be due to implicit but unmodeled nonlinear couplings between $x_{\mathrm{im}}$ and other parameters, such as the gain, in the \WMAP\ pipeline that are accounted for through the Markov chain sampling in \cosmoglobe; other large-scale systematic effects in \WMAP\ that are unrelated to transmission imbalance; or not-yet-identified systematic uncertainties in \cosmoglobe. Overall, it seems likely that the template-based correction of the low-resolution covariance matrices used in the cosmological \WMAP\ likelihood does not provide a complete description of the full large-scale uncertainties. This could potentially be relevant for the differing estimates of the optical depth of reionization between \WMAPnine\ \citep{hinshaw2012} and \Planck\ \citep{planck2016-l05}, and resolving this point is clearly a high priority for a future second \cosmoglobe\ data release.

\subsection{Comparison with \BP\ LFI frequency maps}
\label{subsec:lfi_comparison}

Before concluding this section, we briefly compare the updated \cosmoglobe\ LFI maps with the previous \bp\ products. It is important to keep in mind that whenever new data sets or astrophysical components are added to a global analysis framework such as \cosmoglobe, all frequency channels and astrophysical components are affected by the recent additions. Figure~\ref{fig:lfi_comparison} illustrates this in terms of differences between the \cosmoglobe\ and \bp\ posterior mean maps, all of which are smoothed to a common resolution of $2^{\circ}$ FWHM. No additional postprocessing is performed.

Starting with the temperature case, we see that the color scale spans
$\pm10\muK$. However, most of this range is spent on capturing the monopole
component across channels. the monopoles are in general determined by the
foreground model, and this is obviously strongly affected by the introduction of
the \WMAP\ \K-band. The second most important change is a dipole. Specifically,
the 30\,GHz dipole has changed by 2--3\muK, and this change is also a direct consequence of the addition of \K-band, which now dominates the free-free and AME components. Furthermore, we see that the direction of this dipole is in the opposite direction compared to the CMB Solar dipole, and that the Galactic plane is negative; overall, the absolute calibration of the 30\,GHz channel has decreased by about 0.1\,\%. Also for the 70\,GHz we see a dipole difference of about 3--4\muK, which we will return to the astrophysical implications of in Sect.~\ref{sec:dipole}.

In polarization, we find differences of about 2--3\muK\ in the 30\,GHz channel, while they are generally below 1\muK\ in the 44 and 70\,GHz channels. The general morphology of these difference correspond to gain differences, with obvious striping along the \Planck\ scanning path. Thus, this plot serves as yet another powerful demonstration of the tight coupling between gain calibration, temperature component separation, and large-scale polarization systematics.

\section{Preliminary astrophysical results}
\label{sec:astrophysics}

The main scientific goal of the current paper is to derive and publish new low systematic state-of-the-art \WMAP\ frequency sky maps through end-to-end Bayesian analysis. Ideally, these maps should be accompanied with a fully converged posterior distribution that allows derivation of all relevant scientific applications, including low-$\ell$ polarization. However, as discussed in Sect.~\ref{sec:resources}, producing a sufficient number of samples for estimating the optical depth of reionization will require about nine months of continuous runtime. At the same time, any scientific applications that do not require such a large number of samples will benefit greatly already from the currently available data.

In this section, we present a number of typical applications for which this is the case. In particular, in Sect.~\ref{subsec:foregrounds} we present Galactic foreground maps derived in the current analysis, and in Sect.~\ref{sec:cmb} we present \WMAP+LFI CMB results, including an updated dipole estimate, a temperature power spectrum, and a reassessment of various low-$\ell$ anomalies. In Sect.~\ref{subsub:compsep_chisq} we quantify the goodness-of-fit of the current \cosmoglobe\ sky model in terms of frequency residual maps and $\chi^2$ statistics, before concluding with a comparison of the relative signal-to-noise ratio of \WMAP\ and LFI to each physical component in Sect.~\ref{sec:s2n}.

\subsection{Galactic foregrounds}\label{subsec:foregrounds}

As described in Sect.~\ref{subsec:sky_model}, and defined by Eqs.~\eqref{eq:cmb_astsky}--\eqref{eq:sum_ptsrc}, the Galactic sky model we adopt in this analysis is very similar to that of \citet{bp01}. Explicitly, it includes synchrotron, free-free, AME, and thermal dust emission in intensity, and synchrotron and thermal dust emission in polarization, and we fit the individual amplitude of each component per pixel. However, there are two notable changes. First, we adopt an exponential SED model for AME rather than an \texttt{SpDust}-based SED; this is motivated by the observation that the \texttt{SpDust} models appears to underestimate the AME amplitude at frequencies between 40 and 60\,GHz. Second, as discussed in Sect.~\ref{sec:priors}, we impose stronger (data-informed) priors on the SED parameters. The motivation for this is to reduce degeneracies between the foreground model and overall gains, which otherwise can lead to very long Markov chain correlation lengths. In the future, these priors should be removed after adding additional data that breaks these degeneracies directly, in particular from \Planck\ HFI, QUIJOTE, and C-BASS. As shown in Sect.~\ref{subsub:compsep_chisq}, no significant foreground correlated artifacts arise from these priors (as would be the case if the priors were poorly chosen), so these priors have a small impact on the \WMAP\ frequency maps themselves, which are the main scientific targets in this paper. On the other hand, this does imply that the SED parameters that are sampled as part of the Gibbs chain are non-informative. Rather, spectral parameters must be estimated through external analyses from the frequency maps, and this is for instance done for polarized synchrotron emission in a companion paper by \citet{fuskeland:2023}.

\begin{figure*}
	\centering
	\includegraphics[width=0.8\textwidth]{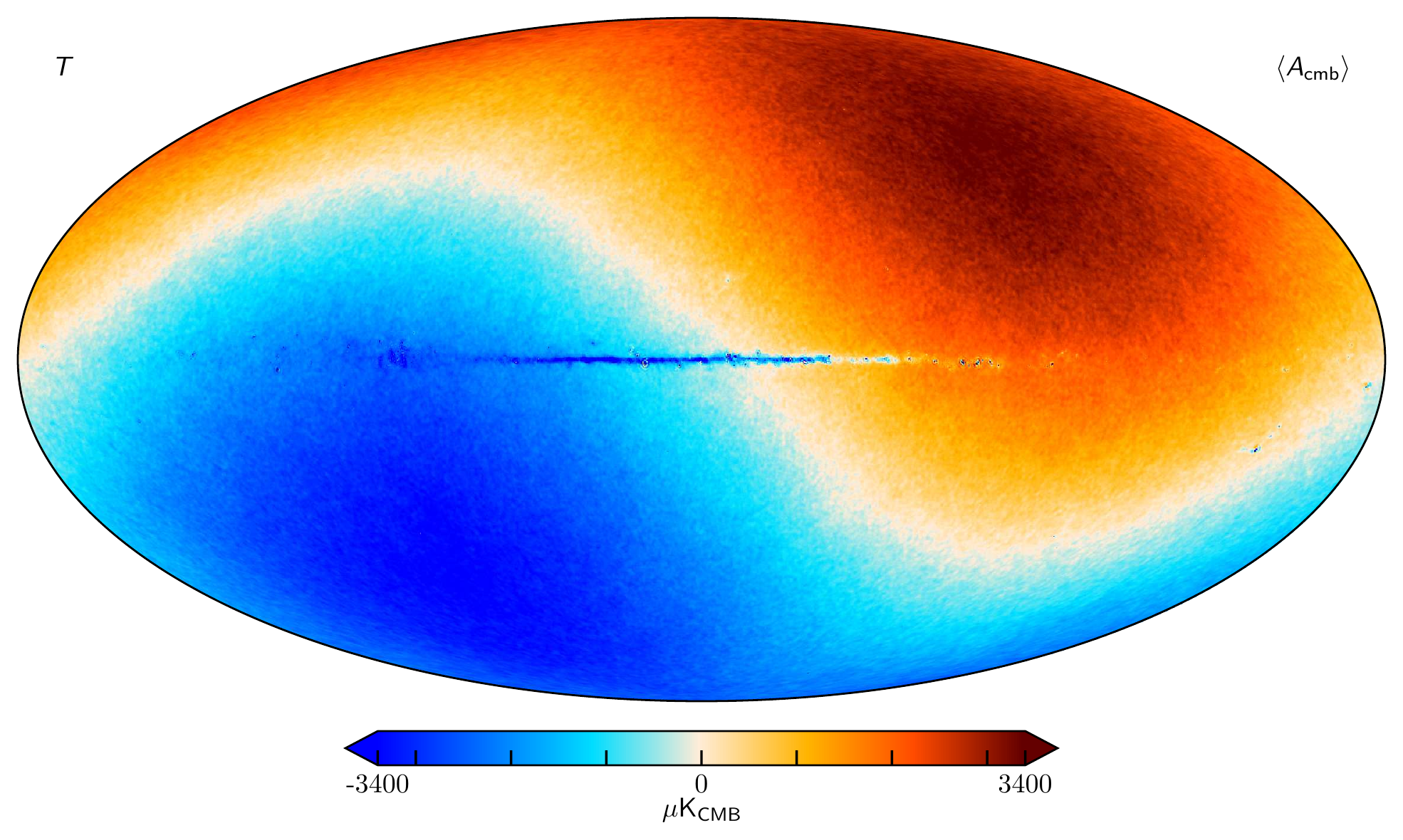}
	\caption{Posterior mean CMB \cosmoglobe\ temperature map, smoothed to an angular resolution of $14'$ FWHM.}
        \label{fig:cmb_mean}
\end{figure*}

\begin{figure*}
	\includegraphics[width=0.33\textwidth]{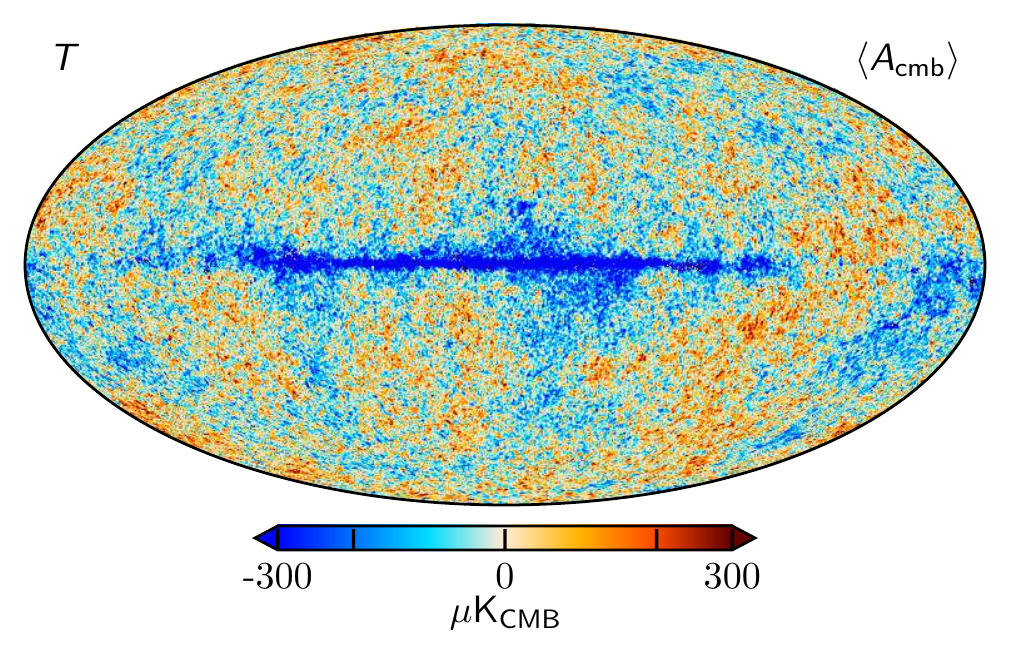}
	\includegraphics[width=0.33\textwidth]{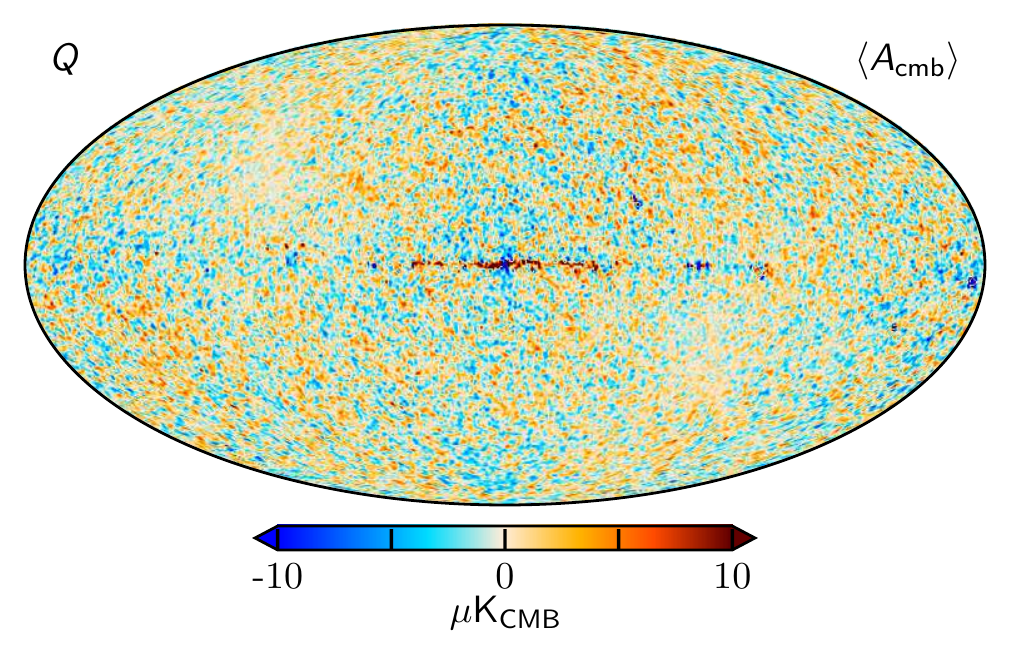}
	\includegraphics[width=0.33\textwidth]{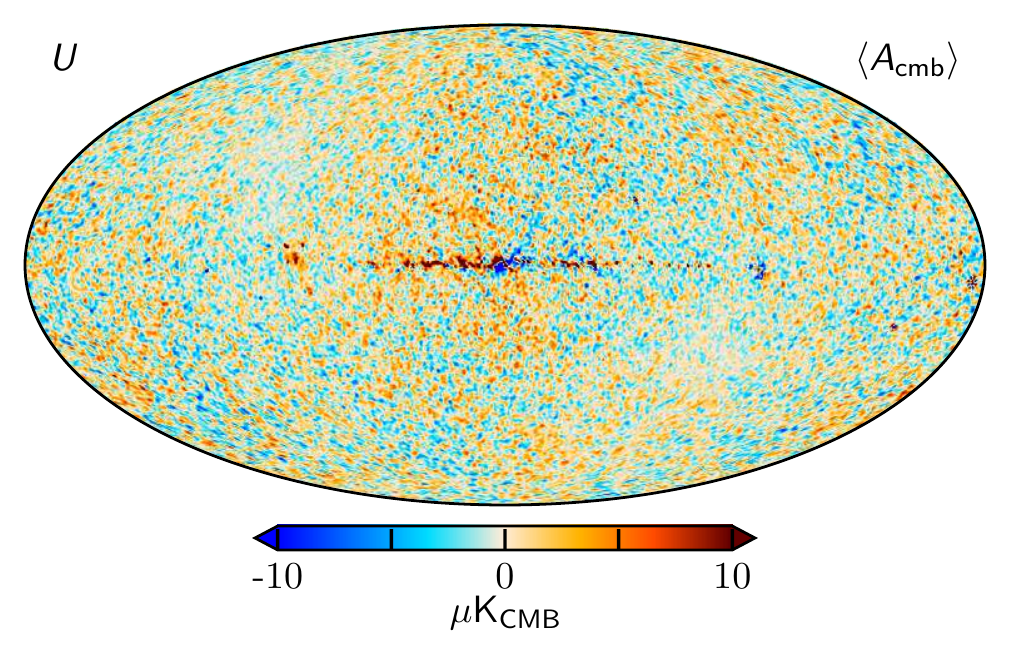}\\
	\includegraphics[width=0.33\textwidth]{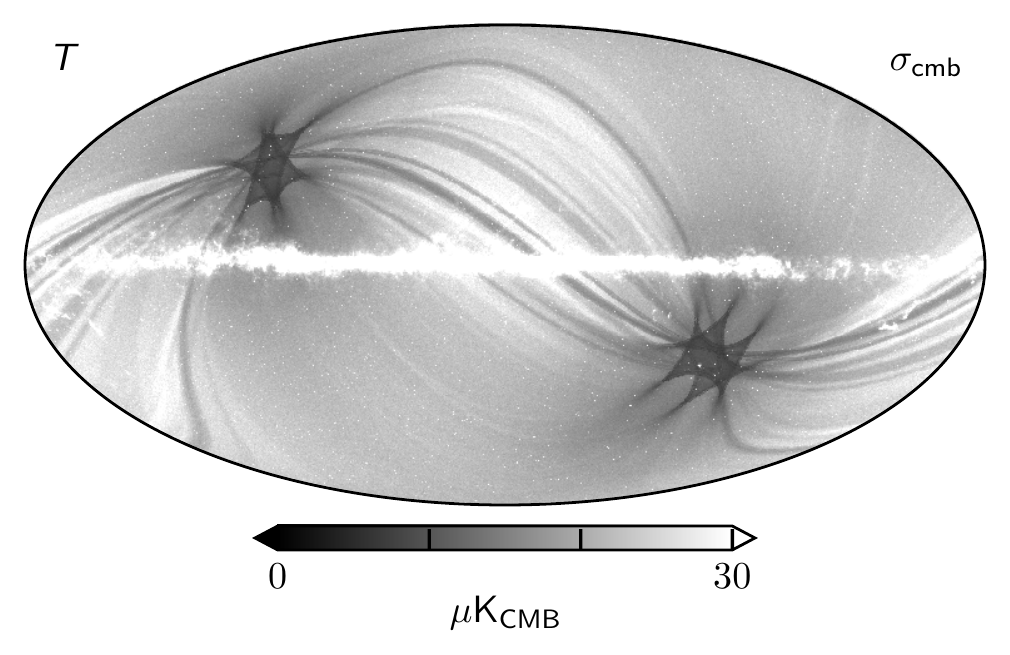}
	\includegraphics[width=0.33\textwidth]{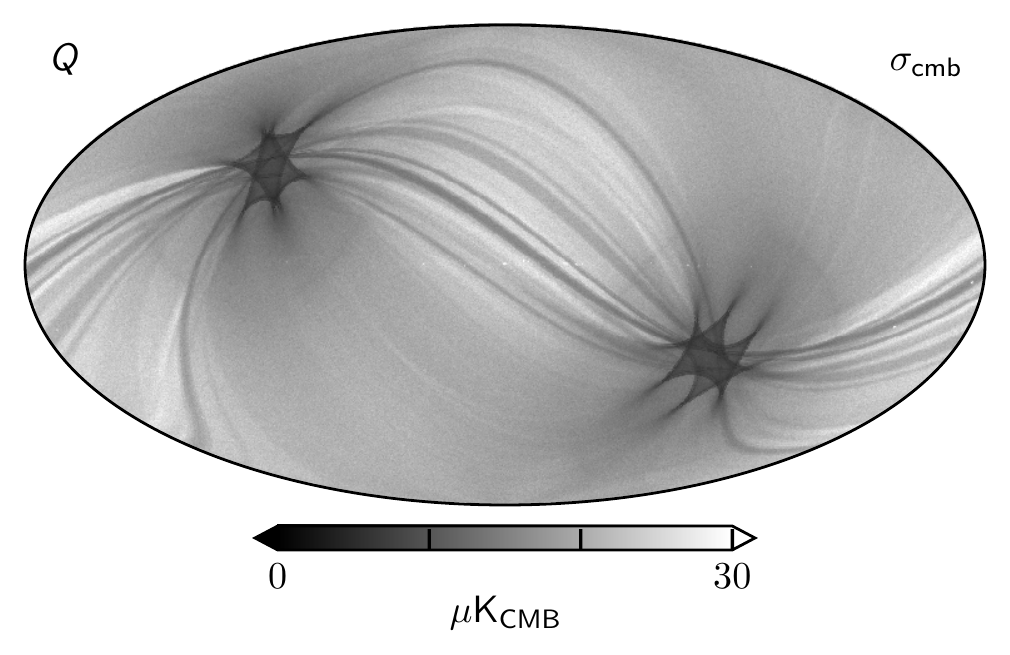}
	\includegraphics[width=0.33\textwidth]{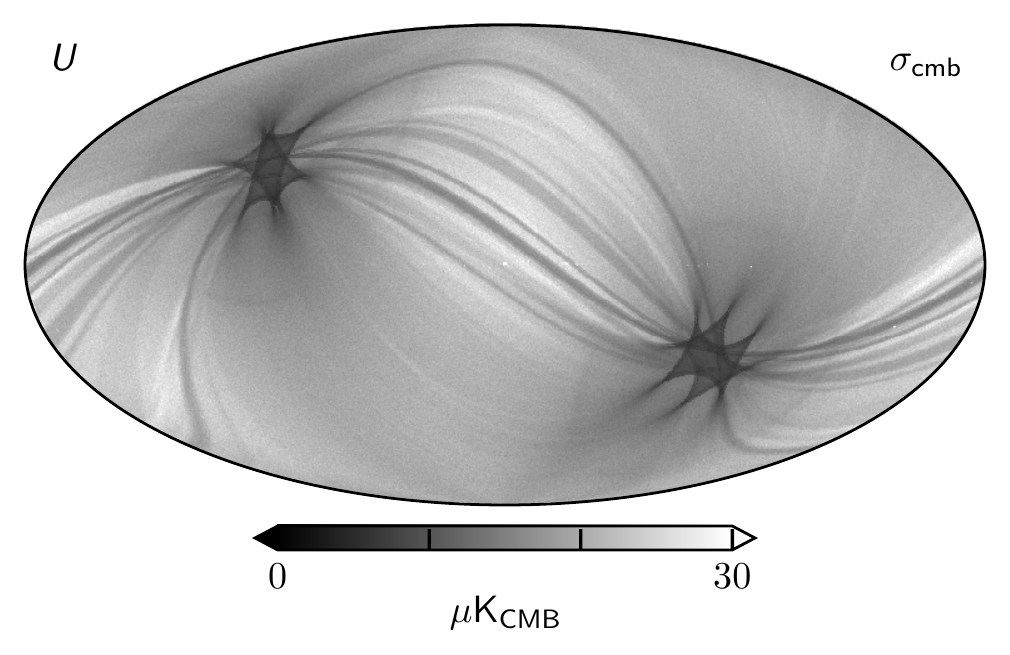}\\
	\caption{Posterior mean CMB \cosmoglobe\ maps for Stokes $T$, $Q$, and $U$, and their corresponding standard deviation. The polarization maps have been smoothed to an angular resolution of $2^{\circ}$ FWHM.}
        \label{fig:cmb_posterior}
\end{figure*}

\begin{table*}
\newdimen\tblskip \tblskip=5pt
\caption{Comparison of Solar dipole measurements from \COBE, \WMAP, and \Planck. }
\label{tab:dipole}
\vskip -4mm
\footnotesize
\setbox\tablebox=\vbox{
 \newdimen\digitwidth
 \setbox0=\hbox{\rm 0}
 \digitwidth=\wd0
 \catcode`*=\active
 \def*{\kern\digitwidth}
  \newdimen\dpwidth
  \setbox0=\hbox{.}
  \dpwidth=\wd0
  \catcode`!=\active
  \def!{\kern\dpwidth}
  \halign{\hbox to 2.5cm{#\leaderfil}\tabskip 2em&
    \hfil$#$\hfil \tabskip 2em&
    \hfil$#$\hfil \tabskip 2em&
    \hfil$#$\hfil \tabskip 2em&
    #\hfil \tabskip 0em\cr
\noalign{\doubleline}
\omit&&\multispan2\hfil\sc Galactic coordinates\hfil\cr
\noalign{\vskip -3pt}
\omit&\omit&\multispan2\hrulefill\cr
\noalign{\vskip 3pt} 
\omit&\omit\hfil\sc Amplitude\hfil&l&b\cr
\omit\hfil\sc Experiment\hfil&[\muK_{\rm
CMB}]&\omit\hfil[deg]\hfil&\omit\hfil[deg]\hfil&\hfil\sc Reference\hfil\cr
\noalign{\vskip 3pt\hrule\vskip 5pt}
\COBE \rlap{$^{\rm a,b}$}&                  3358!**\pm23!**&     264.31*\pm0.16*&
     48.05*\pm0.09*&\citet{lineweaver1996}\cr
\WMAP\ \rlap{$^{\rm c}$}&                  3355!**\pm*8!**&     263.99*\pm0.14*&
     48.26*\pm0.03*&\citet{hinshaw2009}\cr
\noalign{\vskip 3pt}
LFI 2015 \rlap{$^{\rm b}$}&              3365.5*\pm*3.0*&     264.01*\pm0.05*&
     48.26*\pm0.02*&\citet{planck2014-a03}\cr
HFI 2015 \rlap{$^{\rm d}$}&              3364.29\pm*1.1*&     263.914\pm0.013&
     48.265\pm0.002&\citet{planck2014-a09}\cr
\noalign{\vskip 3pt}
LFI 2018 \rlap{$^{\rm b}$}&              3364.4*\pm*3.1*&     263.998\pm0.051&
     48.265\pm0.015&\citet{planck2016-l02}\cr
HFI 2018 \rlap{$^{\rm d}$}&              3362.08\pm*0.99&     264.021\pm0.011&
     48.253\pm0.005&\citet{planck2016-l03}\cr
\noalign{\vskip 3pt}
Bware & 3361.90\pm*0.40 & 263.959\pm0.019 & 48.260\pm0.008  & \citet{delouis:2021}  \cr
\Planck\ PR4\ \rlap{$^{\rm a,c}$}& 3366.6*\pm*2.6*& 263.986\pm0.035&
48.247\pm0.023&\citet{planck2020-LVII}\cr
\noalign{\vskip 3pt}
\BP\ \rlap{$^{\rm e}$} & 3362.7*\pm*1.4*& 264.11*\pm0.07*&
 48.279\pm0.026&\citet{bp11}\cr
\noalign{\vskip 3pt}
\bf\cosmoglobe\ \rlap{$^{\rm e}$} & \bf3366.2*\pm*1.4*& \bf264.08*\pm0.07*&
 \bf48.273\pm0.024&This work\cr
\noalign{\vskip 5pt\hrule\vskip 5pt}}}
\endPlancktablewide
\tablenote {{\rm a}} Statistical and systematic uncertainty estimates are added in quadrature.\par
\tablenote {{\rm b}} Computed with a naive dipole estimator that does not account for higher-order CMB fluctuations.\par
\tablenote {{\rm c}} Computed with a Wiener-filter estimator that estimates, and marginalizes over, higher-order CMB fluctuations jointly with the dipole.\par
\tablenote {{\rm d}} Higher-order fluctuations as estimated by subtracting a dipole-adjusted CMB-fluctuation map from frequency maps prior to dipole evaluation. \par
\tablenote {{\rm e}} Estimated with a sky fraction of 68\,\%. Error bars include only statistical uncertainties, as defined by the global \cosmoglobe\ posterior framework, and they thus account for instrumental noise, gain fluctuations, parametric foreground variations etc. 
\par
\end{table*}
\begin{figure}
	\includegraphics[width=\columnwidth]{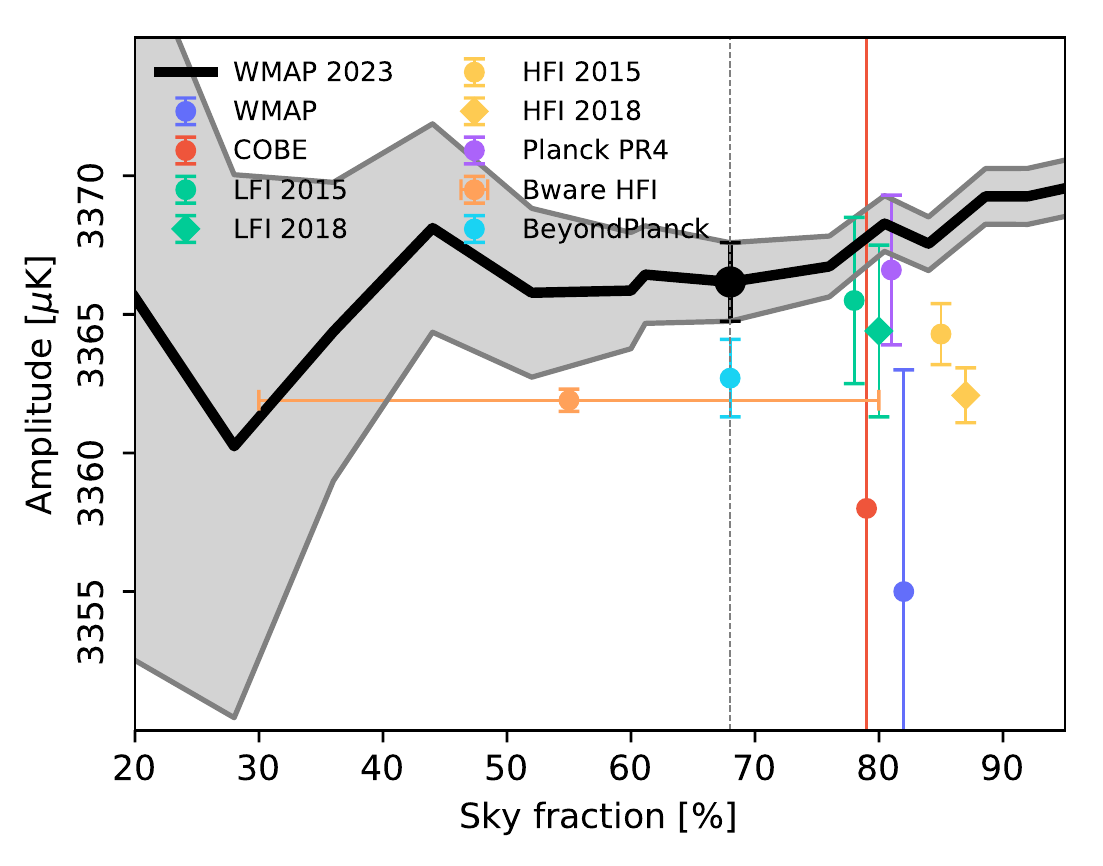}
	\caption{CMB dipole amplitude as a function of sky fraction. The gray band indicates the 68\,\% posterior confidence region.}
	\label{fig:dip_amp}
\end{figure}

\begin{figure*}
	\includegraphics[width=\textwidth]{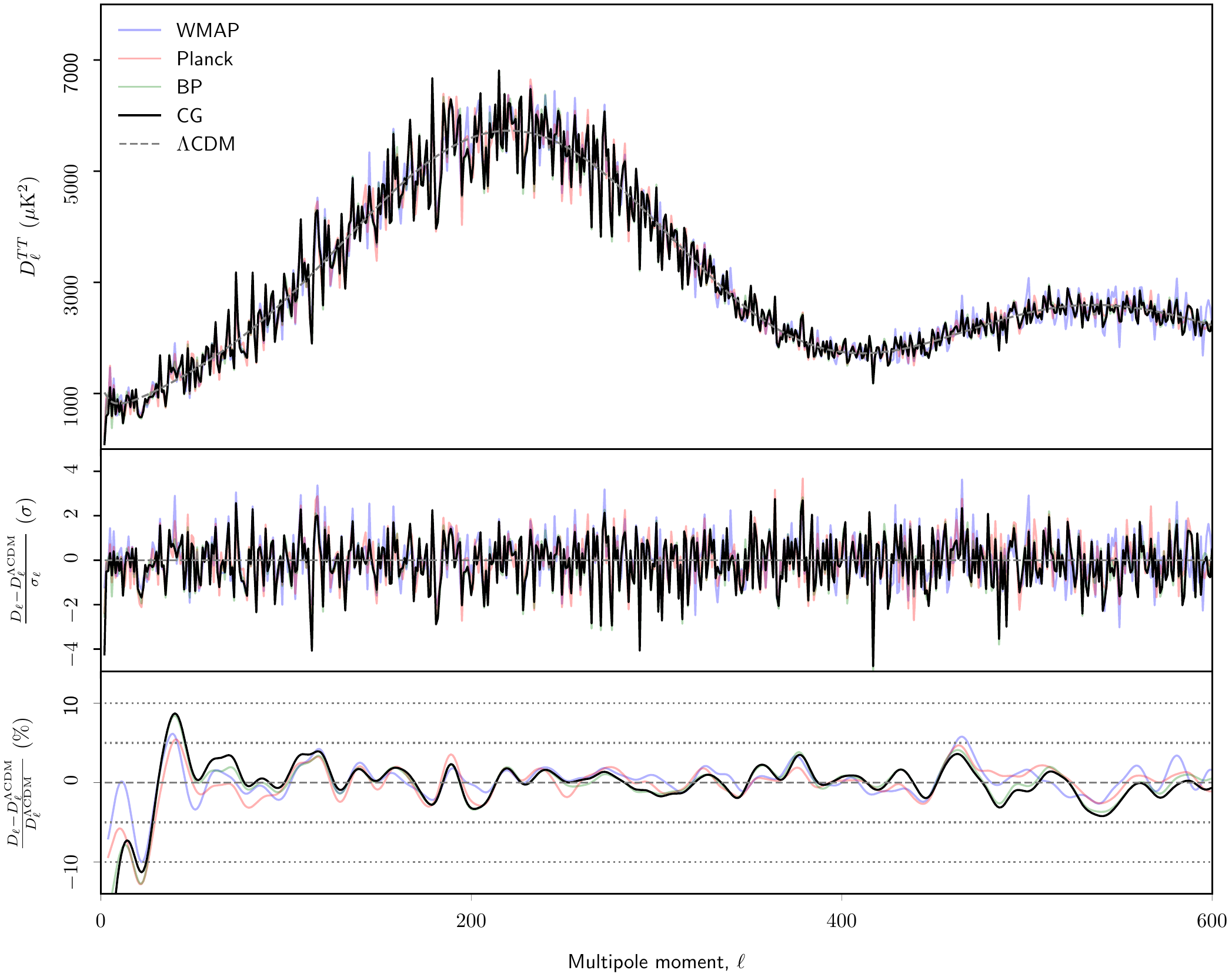}
	\caption{\textit{(Top:)} Angular CMB temperature power spectrum, $D_\ell^\mathrm{TT}$, as derived by \cosmoglobe\ (black), \bp\ (green), \planck\ (red), and \WMAPnine\ (blue). The best-fit \planck\ 2018 \LCDM\ spectrum is showed in dashed gray. \textit{(Middle:)} Residual power spectrum relative to \LCDM, measured relative to the quoted error bars, $(D_\ell-D_\ell^\mathrm{\Lambda CDM})/\sigma_\ell$. For pipelines that report asymmetric error bars, $\sigma_\ell$ is taken to be the average of the upper and lower error bar. \textit{(Bottom:)} Fractional difference with respect to the \planck\ \LCDM\ spectrum. In this panel, each curve has been boxcar averaged with a window of $\Delta\ell=100$ to suppress random fluctuations.}
	\label{fig:cl_tt}
\end{figure*}

In this section, we therefore focus only on the foreground amplitude parameters. Specifically, Fig.~\ref{fig:intensity_foregrounds} shows posterior mean intensity maps for all four components, while Fig.~\ref{fig:polarized_foregrounds} shows the posterior mean and standard deviation for the polarized synchrotron component.

Starting with the free-free intensity component shown in the top left panel of Fig.~\ref{fig:intensity_foregrounds}, we observe good agreement with previous full sky component separation studies \citep{planck2014-a12,bp13}. We note that compared to the \cite{planck2014-a12} analysis, there is less diffuse structure in the free-free component, which is driven by the imposition of a prior on the component amplitude at high Galactic latitudes \citep{bp13}. However, in high emission regions, such as the Galactic plane and the Gum Nebula, we see good agreement. %

For the AME, shown in the top right panel, we see a slightly differing morphology compared to both \cite{planck2014-a12} and \cite{bp13}. The most notable difference is the lack of extended diffuse structure in this work, with a marginal shift in the overall direction of the component's dipole. These differences are due to the different SED model, as well as the degeneracy between the \K-band gain and the AME dipole, as described at length in Sect.~\ref{sec:ame_Kband}. Both this analysis and \cite{bp13} differ from the \cite{planck2014-a12} AME solution by showing less extended diffuse structure, and most visibly notable is the $\rho$-Ophiuchi complex, which appears as a hole in the AME component in this work. For further details regarding the AME SED, we refer the interested reader to \citet{watts2023_ame}.

Next, regarding synchrotron emission in total intensity, the reprocessed full sky Haslam map \citep{remazeilles2014} at 408\,MHz is used as an anchor for the full sky synchrotron emission in both \Planck\ 2015, \bp, and \cosmoglobe. As such, the estimate shown in the bottom right panel of Fig.~\ref{fig:intensity_foregrounds} shares very similar morphology to both these previous analyses, although there are some slight deviations around point sources. Similar observations apply to the thermal dust model, which is strongly dominated by the \Planck\ 857\,GHz, which is common to all these mentioned analyses. 

Though the \BP\ analysis did not utilize the \WMAP\ \K-band, the foreground model derived in \cite{bp13} still predicted the sky model at \K-band. As such, it is therefore interesting to check how well the \BP\ model was able to predict the current \K-band signal. To this end, we compare the AME-plus-free-free contribution at 22~\GHz\ between this work and the sky model derived in \BP. (Synchrotron and thermal dust emission is omitted in this particular calculation since these are strongly dominated by the same data sets in the two analyses, namely Haslam 408\,MHz and \Planck\ 857\,GHz.) The combined sky model, smoothed to $2^\circ$, is shown as a fractional difference $\Delta A_{\mathrm{K}}=s^{\mathrm{CG}}_{\mathrm{AME}+\mathrm{ff}}-s^{\mathrm{BP}}_{\mathrm{AME}+\mathrm{ff}})/s^{\mathrm{BP}}_{\mathrm{AME}+\mathrm{ff}}$ in Fig.~\ref{fig:intensity_foregrounds_diff}. Here we see that the addition of \K-band\ data has altered the sky model at 22\,GHz in the high-signal-to-noise region by 5--10\,\%, and the new model exhibits a stronger foreground amplitude at 22\,GHz than the \bp\ model. Detailed inspection of the individual free-free and AME components indicate that they have typically changed by about 20\,\%, which is also partially due to the exponential SED model used for AME in the present analysis.

Finally, for the polarized synchrotron amplitude, shown in Fig.~\ref{fig:polarized_foregrounds}, we also find good agreement with previous \bp\ results \citep{bp14}. However, the morphology of the standard deviation map shows a stronger imprint of the \WMAP\ scanning strategy than in \bp, because the \K-band data were omitted from that analysis. In this updated work, the two experiments have more comparable signal-to-noise ratios to polarized synchrotron emission, and which experiment is stronger depends now on position on the sky. As a result of finally combining all \WMAP\ and LFI data, this updated map represents the most sensitive full-sky polarized synchrotron map published to date.

\subsection{CMB results}
\label{sec:cmb}

Next, we consider various CMB results, the most important scientific products from both \WMAP\ and \Planck. In this paper, we focus primarily on intensity results, as far fewer Markov chain samples are required to produce robust results for these than large-scale polarization. In addition, cosmological parameter estimation is also left for future work, simply because we find that the angular temperature power spectrum derived in this work is fully consistent with that derived in \BP, and no significant changes are therefore expected. Once a robust low-$\ell$ polarization likelihood has been established, this issue will of course be revisited.

Figure~\ref{fig:cmb_mean} shows the posterior mean CMB intensity map including the dipole, while Fig.~\ref{fig:cmb_posterior} shows the posterior mean and standard deviation for all three Stokes parameters; in this figure, the best-fit CMB Solar dipole has been subtracted from the temperature map. Overall, these maps look visually very similar to those presented by \bp\ \citep{bp13}, and we therefore adopt the same confidence masks and analysis configuration as described there.

\subsubsection{Solar dipole}
\label{sec:dipole}

We start our discussion with the largest angular scales, namely the CMB dipole. As discussed by \citet{thommesen:2019}, estimating the Solar dipole is arguably one of the most difficult parameters to constrain accurately. This is due to the strong degeneracy with the gain model, as well as the effect of mode coupling when masking the Galactic plane. Calibration misestimation propagates directly into an incorrect CMB dipole, and vice-versa.

Our Solar dipole estimates are summarized in Table~\ref{tab:dipole} and Fig.~\ref{fig:dip_amp}. First, we find that the dipole direction is very consistent with \bp\ \citep{bp11}, and also statistically consistent with most previous analyses within the quoted uncertainties. Strong agreement with \bp\ is of course expected, since the data selection is very similar (the only difference is that \K-band has been added in \cosmoglobe), and the processing pipelines are very similar.

It is therefore interesting to note that the dipole amplitude is in fact 3.5\muK\ (or $2.5\,\sigma$) higher than in \bp. This is clearly a larger change than one would expect simply by adding one more data set. It is also interesting to note that this new mean value of 3366.2\muK\ is $11\,\mathrm{\mu K}$ higher than the \WMAPnine\ result \citep{hinshaw2009}, and also 4.1\muK\ higher than the \Planck\ HFI 2018 result \citep{planck2016-l02}. On the other hand, it is now consistent with the latest \Planck\ PR4 result, with a difference of only 0.4\muK.

One plausible explanation for this behaviour is the following: \bp\ used the official \WMAPnine\ frequency maps directly in the analysis, and these data are known to have a lower CMB dipole than \Planck. It is therefore natural to assume that the \BP\ dipole was pulled toward low values by these maps. In the new \cosmoglobe\ analysis, however, the \WMAP\ and \Planck\ data are forced to agree on a common dipole prior during the global calibration process. In particular, this has increased the \WMAP\ dipole, and the previous tension has been released. The net result is that the \WMAP+LFI-dominated \cosmoglobe\ estimate now finally agree with the highly independent \Planck\ HFI-dominated PR4 result.

\subsubsection{Angular temperature power spectrum}
\label{sec:cls}

Next, in Fig.~\ref{fig:cl_tt} we show the angular temperature power spectrum derived from the CMB samples from the main Gibbs chain, obtained using a Gaussianized Blackwell-Rao estimator \citep{chu2005,rudjord:2009} with an identical analysis setup and mask as in \bp\ \citep{bp11}. We compare with the official \WMAP\ \citep{hinshaw2012} and \planck\ \citep{planck2016-l05} power spectra, as well as the \bp\ \citep{bp11} spectrum. For reference, the best-fit \planck\ 2018 \LCDM\ spectrum is also plotted along side them. The middle panel shows the deviation from the \planck\ \LCDM\ solution, in units of $\sigma_\ell$ from each individual pipeline, while the bottom panel shows the fractional difference with respect to the \planck\ \LCDM\ spectrum. 

\begin{figure}[t]
	\includegraphics[width=\columnwidth]{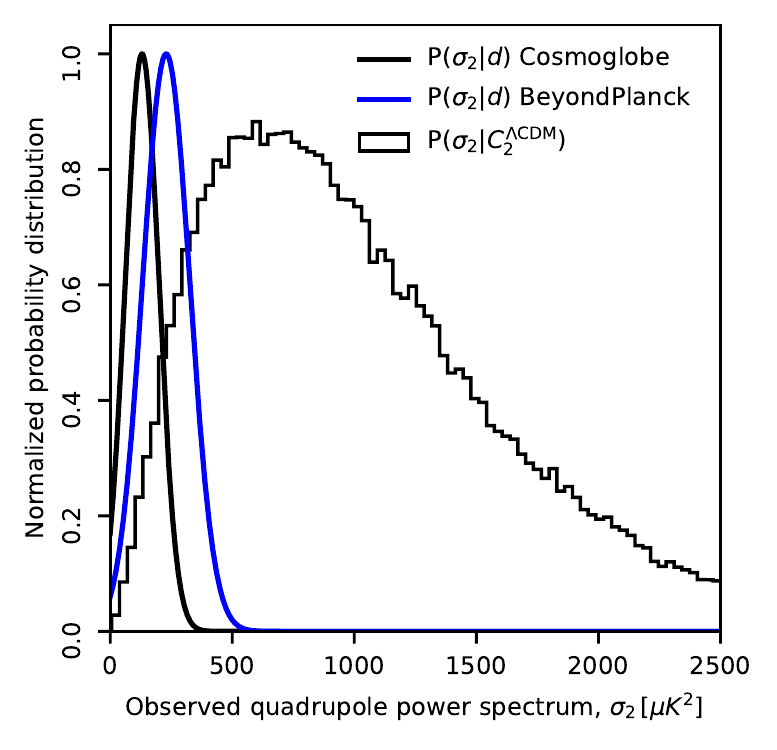}
	\caption{Temperature quadrupole power amplitude posterior distribution as computed by \cosmoglobe\ (solid black line) and \bp\ (solid blue line). For comparison, the histogram shows 100 000 realizations of $\sigma_2$ given the best-fit \Planck\ 2018 ensemble-averaged prediction of ${C_2^{\Lambda \mathrm{CDM}} = 1064.7\,\mathrm{\mu K^2}}$.}
	\label{fig:sigma_2_hist}
\end{figure}

\begin{figure}[t]
	\includegraphics[width=\columnwidth]{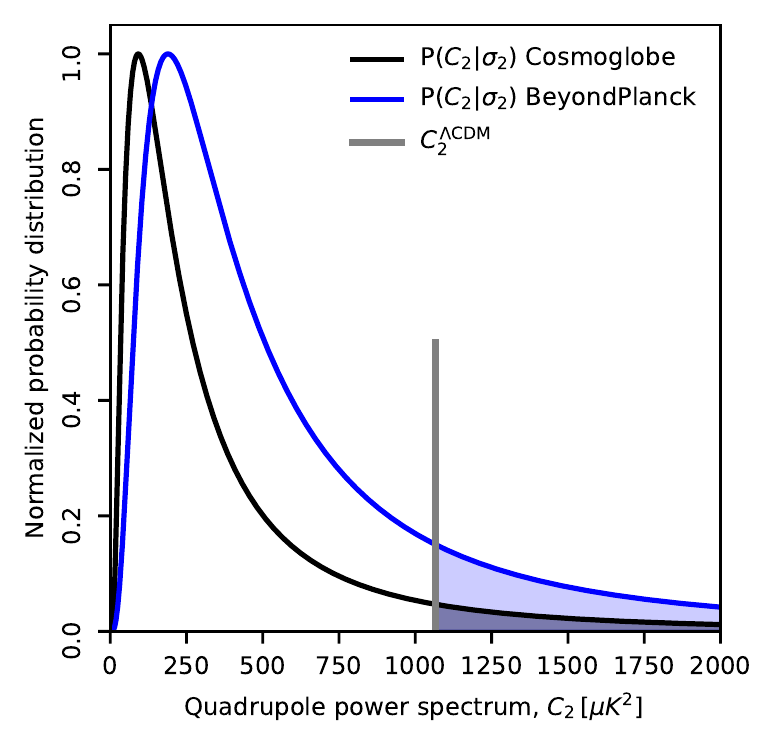}
	\caption{Marginal probability distribution of the ensemble-averaged $C_2$ given the data, $P(C_2\mid\data)$, as measured by \Cosmoglobe\ (black) and \BP\ (blue).}
	\label{fig:blackwell_rao_sigma_2}
\end{figure}

At $\ell\lesssim500$, each of these datasets are signal-dominated, and all spectra agree very well. At higher multipoles, more samples are needed in order to obtain a robust Blackwell-Rao estimator. %
Given this good agreement, we do not anticipate any significant difference in terms of $\Lambda$CDM parameters, and we therefore postpone a full cosmological parameter reanalysis to future work.

\subsubsection{Low-$\ell$ anomalies}
\label{sec:anomalies}

Although the CMB power spectrum agrees exceedingly well with a \LCDM\ model \citep[e.g.,][]{hinshaw2012,planck2016-l06,bp12}, several anomalies have been reported that appear to be in tension with this model, in particular on large angular scales \citep[e.g.,][and references therein]{planck2016-l07}. Generally, the presence of these anomalies in the CMB map is not debated as such; however, their statistical significances are highly debated. In particular, some authors argue that the correct interpretation of these anomalies are likely to be described by the so called look-elsewhere effect \citep[e.g.,][]{bennett2010}. %

The traditional approach to studying CMB anomalies is to compute a single maximum-likelihood CMB map and a corresponding ensemble of \LCDM\ simulations processed with similar instrumental properties. Then one derives a single value for a given anomaly statistic of interest, and compares the true value with the histogram of simulated values. By counting how many simulations exceed the true value, one obtains a probability-to-exceed (PTE) value that quantifies the level of agreement. 

In our case, however, we do not have only a single best-fit CMB likelihood map, but rather a full posterior distribution of such maps produced through the Gibbs chain. These can then be used to assess the significance of any given anomaly using exactly the same approach as in a traditional analysis, except that one now obtains a histogram for the real data as well. The main advantage of this approach is that systematic uncertainties are propagated with much greater fidelity than with a single maximum-likelihood map. This is particularly important for very low-$\ell$ anomalies, which tend to be sensitive to the instrument calibration, and propagating these uncertainties properly is nontrivial using traditional approaches \citep{bp04}. %

\begin{figure}[t]
	\includegraphics[width=\columnwidth]{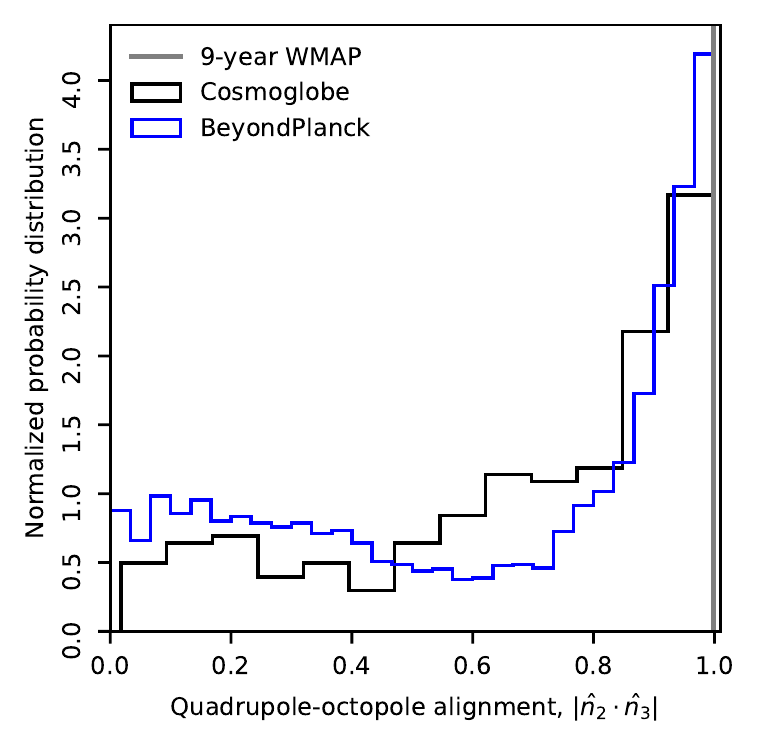}
	\caption{The quadrupole-octopole alignment of \Cosmoglobe\ compared with \BP\ and 9-year \WMAP.}
        \label{fig:alignment}
\end{figure}

\begin{figure}[t]
	\includegraphics[width=\columnwidth]{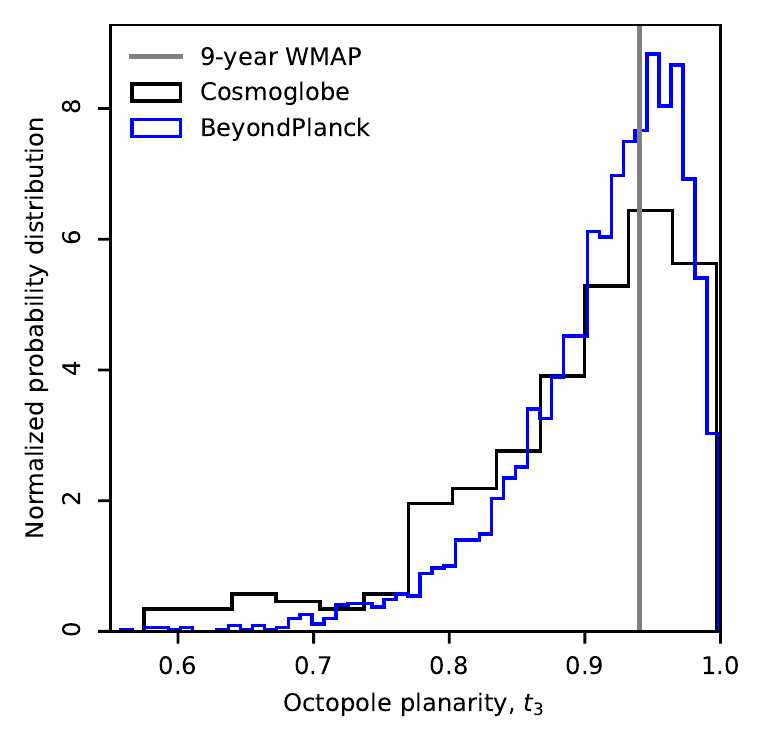}
	\caption{The octopole planarity statistics $t_3$ compared with the \BP\ analysis (blue).}
        \label{fig:planarity}        
\end{figure}

In this section, we revisit a few well known low-$\ell$ anomalies regarding the two lowest multipoles, $\ell=2$ and 3, and compare our findings with similar results reported by \bp\ \citep{bp11}. The main difference between these two analyses is thus that the \WMAP\ data are now analyzed in time domain, rather than in the form of preprocessed maps. It is reasonable to assume that this recalibration can modify the amplitude and morphology of these lowest multipoles, in particular given the notable differences in the CMB dipole amplitude reported in Sect.~\ref{sec:dipole}.

First, we start with the absolute amplitude of the temperature quadrupole,\footnote{Note that $\sigma_2$ is the realization-specific quadrupole amplitude of our universe, i.e., $\sigma_2=\tfrac1{2\cdot2+1}\sum_m|a_{2,m}|^2$, while $C_2\equiv\langle |a_{2,m}|^2\rangle$ is the mean prediction of \LCDM\ over all potential realizations of our universe.} which has been noted to be low compared to the theoretical prediction ever since \COBE-DMR \citep{bennett:1992}. This was later confirmed by both \WMAP\ \citep{hinshaw2003a} and \planck\ \citep{planck2013-XV}, but with large discrepancies in mean value and error bars, both within and between experiments \citep{bp11}. For instance, the \WMAP\ team reported in their 7-year analysis a best-fit value of $201\,\mu\mathrm{K}^2$ \citep{larson2010}, which decreased to $151\,\mu\mathrm{K}^2$ in the 9-year analysis \citep{hinshaw2012}. The naive Fisher uncertainty on $\sigma_2$ was reported by \citet{hinshaw2012} to be $9\,\mu \mathrm{K}^2$ which only accounted for a noise-only estimate. As such, this relative change between the two algorithmically very similar 7- and 9-year analyses corresponded to a roughly $5~\sigma$ discrepancy in terms of Fisher uncertainties. Similarly, \Planck\ later found in 2013 and 2018 $\sigma_2$ to be 299 and 226 $\mu \mathrm{K}^2$, respectively, which corresponds to an internal $8\,\sigma$ discrepancies in terms of Fisher uncertainties \citep{planck2016-l05}.

These large variations indicate that instrumental noise is not the dominant source of uncertainty regarding $\sigma_2$. Indeed, this observation was demonstrated in practice through the end-to-end \bp\ analysis, which found an amplitude of $229\pm97\muK^2$. The important point about this estimate is that the uncertainty is almost an order of magnitude larger than the Fisher uncertainty, and this is likely driven by the additional marginalization over calibration uncertainties. 

With the new set of \cosmoglobe\ CMB maps derived in this paper, we are in a position that allows us to improve further on the \bp\ result, by additionally marginalizing over \WMAP\ instrumental effects. This is quantified in terms of the marginal posterior distribution, $P(\sigma_2\,|\,\data)$, which is shown in Fig.~\ref{fig:sigma_2_hist}. The \cosmoglobe\ estimate may be summarized in terms of a Gaussian distribution with $\sigma_2 = 131 \pm 69\ \mu\mathrm{K}^2$. For comparison, the corresponding \bp\ result is plotted as a blue curve in the same figure, while the histogram shows $10^5$ realizations of $\sigma_2$ given the Planck 2018 best-fit $C^{\Lambda \mathrm{CDM}}_2 = 1064.6\, \mu\mathrm{K}^2$ \citep{planck2016-l06}. It is interesting to note that this updated central value is almost a factor of two lower than the previous \bp\ results, which suggests that the largest scales have indeed changed sufficiently in the updated \WMAP\ to affect the low-$\ell$ anomalies. Furthermore, the low quadrupole amplitude anomaly has become more anomalous through these modifications, and is now almost as low as the 1-year \WMAP\ result.

To quantify the statistical significance of the low $\sigma_2$ value, we first compute the probability of obtaining an ensemble-averaged power coefficient, $C_2$, equal to or larger than the \LCDM\ prediction given the observed realization-specific power coefficient, $\sigma_2$. This can be done by evaluating full marginal posterior distribution $P(C_2 \mid\sigma_2)$ as a function of $C_2$ through the Blackwell-Rao estimator \citep{chu2005}. This is shown as a solid black line in Fig.~\ref{fig:blackwell_rao_sigma_2}, while the solid blue line shows the corresponding \bp\ result; the vertical gray line shows the \Planck\ best-fit value of $C^{\Lambda \mathrm{CDM}}_2 = 1064.6\, \mu\mathrm{K}^2$. Computing the integrals above this value, we find that the probability for $C_2$ to exceed $C^{\Lambda \mathrm{CDM}}_2$ is $11.0\,\%$ for \cosmoglobe\ and $21.7\,\%$ for \bp, both indicated as shaded areas.

Next, in Fig.~\ref{fig:alignment} we revisit the so-called quadrupole-octopole alignment statistic, $|\hat{n}_2\cdot \hat{n}_3|$, introduced by \citet{deOliveira-Costa2004}. This statistic quantifies the angular distance between the vectors that maximize the angular momentum of each multipole, and is thus a measure of the relative alignment between the plane of these modes on the sky. Again, the black line shows the posterior distribution derived from the \cosmoglobe\ samples, while the blue histogram shows the same for \bp; the vertical gray line shows the best-fit value derived from \WMAPnine\ data. Our updated results are fully consistent with those reported by \citet{bp11} for \bp; while the new results that implement full end-to-end error propagation are statistically consistent with the classical pipelines in terms of a single best-fit value, the total posterior uncertainty is now much larger, both because of marginalization over a more complete instrumental model and a more conservative confidence mask \citep{bp11}, to the point that the evidence for this effect is no longer compelling. 

Figure~\ref{fig:planarity} shows similar result for the octopole planarity statistic, also introduced by \citet{deOliveira-Costa2004}. As for \bp, we also in this case observe a broad distribution of allowed values, and the \cosmoglobe\ distribution is even a little broader than the \bp\ distribution; this is of course expected, since we now marginalize over a larger set of instrumental parameters. At the same time, it is intriguing to note that the maximum posterior value is actually even closer to one in \cosmoglobe\ than in \WMAPnine, which indicates that it is in fact possible to attribute all the octopole power into one single azimuthal mode, $a_{33}$. In order to shed more light on this effect, the overall error budget must be decreased significantly by adding more data, in particular \Planck\ HFI observations. 

\begin{figure}
	\includegraphics[width=\columnwidth]{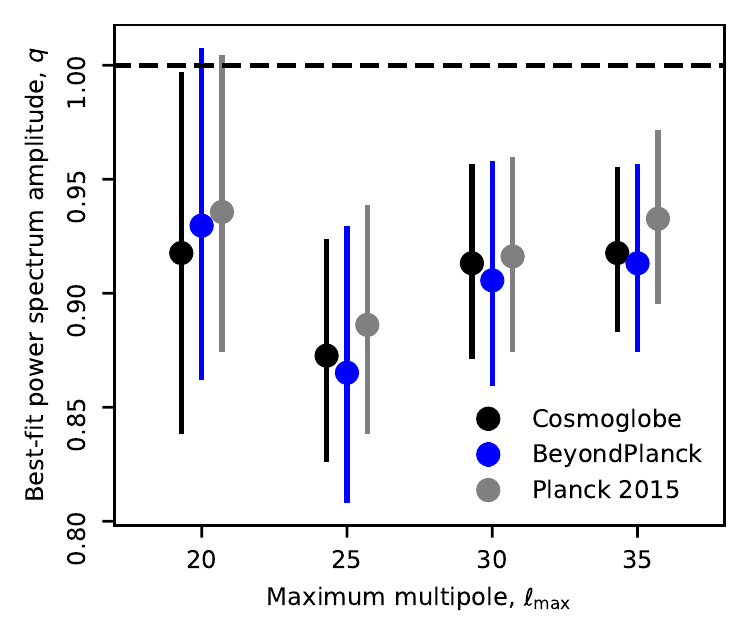}
	\caption{Best-fit amplitude, $q$, of the low multipole power spectrum $C_{\ell} = q C^{\Lambda \mathrm{CDM}}_{\ell}$, $2 \leq \ell \leq \ell_{\mathrm{max}}$ compared to \planck\ 2015 (grey) and \bp\ (blue).}
        \label{fig:lowl}
\end{figure}

Finally, Fig.~\ref{fig:lowl} provides an update of the so called low multipole power anomaly first presented by \citet{planck2014-a13}. In this case, we fit a scaling factor, $q$, relative to the \LCDM\ spectrum to multipoles between $2\le\ell\le\ell_{\mathrm{max}}$, and vary $\ell_{\mathrm{max}}$ between 20 and 35. In this figure, we see that the low-$\ell$ power increases very slightly from \bp\ to \cosmoglobe, and it is now even closer to \Planck\ 2015. Overall, the significance of this effect is similar to previously reported results.

\subsection{Goodness-of-fit: Map-level residuals and $\chi^2$ statistics}\label{subsub:compsep_chisq}

\begin{figure}
	\centering
	\includegraphics[width=\linewidth]{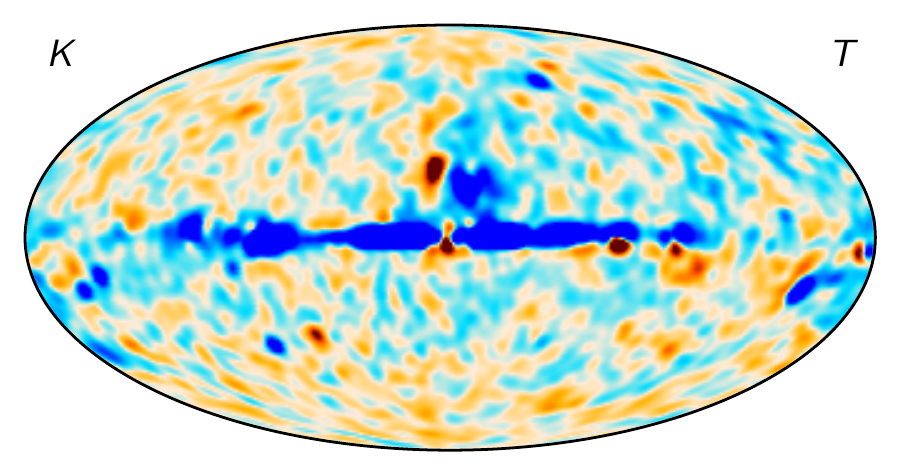}\\
        \includegraphics[width=\linewidth]{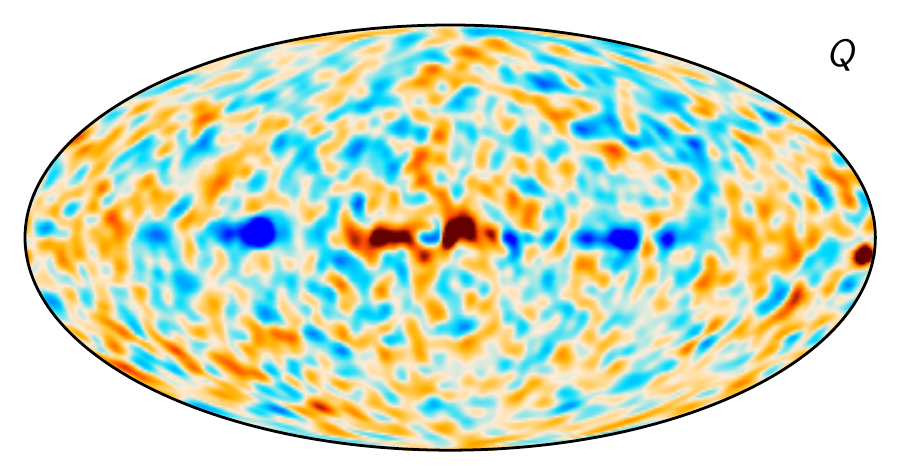}\\
	\includegraphics[width=\linewidth]{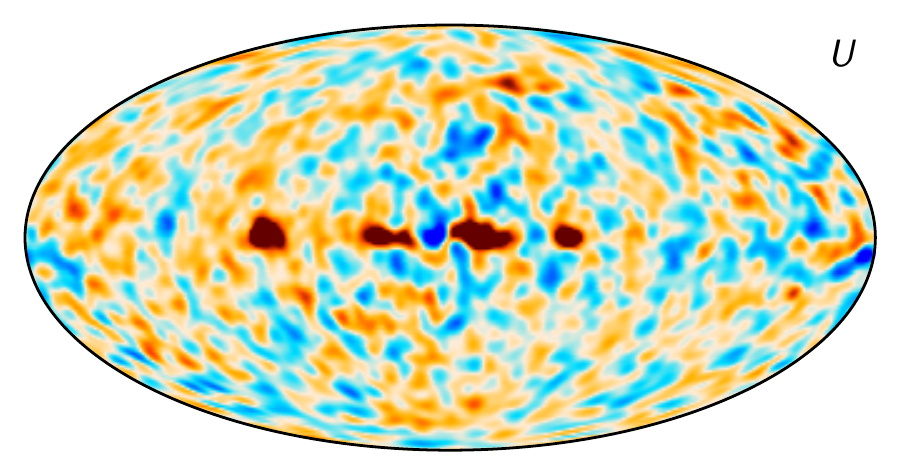}\\        
	\includegraphics[width=0.7\linewidth]{figures/cbar_5uK.pdf}
	\caption{Frequency map residual (DA map minus sky model) for \K-band. Panels show, from top to bottom, Stokes $T$, $Q$, and $U$, and all maps are smoothed by $5^\circ$ FWHM.}
	\label{fig:K_compsep_residual}
\end{figure}

\begin{figure}
	\centering
	\includegraphics[width=\linewidth]{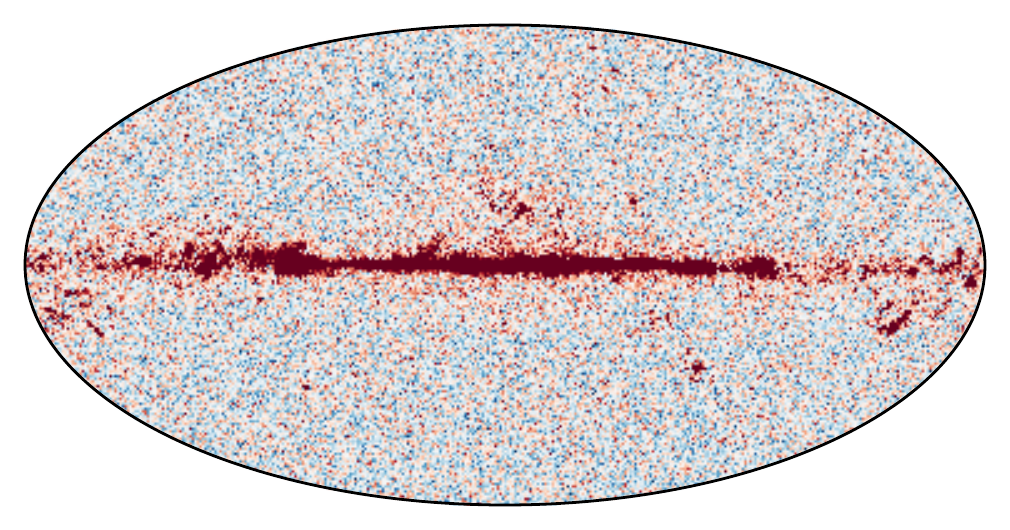}\\
	\includegraphics[width=0.3\textwidth]{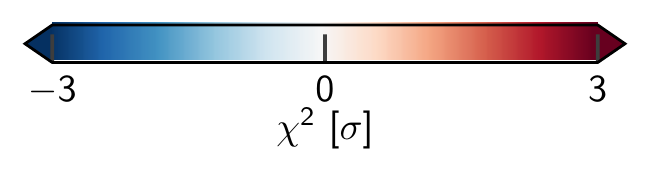}
	\caption{Pixel-space reduced normalized $\chi^2$. In this figure, $n_\mathrm{dof}=300$, which is obtained from fitting to the regions outside of the \K-band processing mask; for further information regarding this statistic, see \citet{bp13}.}\label{fig:reduced_chisq}
\end{figure}

The quality of the component separation procedure is evaluated through map space residuals (i.e., frequency map minus sky model) and a spatial map of the reduced $\chi^2$. Residual maps for all three \K-band Stokes parameters are shown in Fig.~\ref{fig:K_compsep_residual}, while Fig.~\ref{fig:compsep_residual} in Appendix~\ref{sec:map_survey} shows the same for all DAs. Figure~\ref{fig:reduced_chisq} shows the total reduced normalized $\chi^2$, summed over all frequency channels. Overall, we see that the magnitude of the map level residuals are generally well below 5\muK, and the morphology at high Galactic latitudes are generally consistent with instrumental noise. However, in intensity, the Galactic plane stands out with statistically significant residuals, in particular at the Galactic center and bright regions such as the Orion region, $\rho$ Ophiuchus, and the Large Magellanic Cloud. Implementing support for a more complex and realistic foreground model, and therefore necessarily also adding support for additional datasets, is an important goal of the general \cosmoglobe\ framework, and this will hopefully reduce these residuals in the future. In particular, integrating the \Planck\ HFI data, and thereby being able to fit thermal dust emission pixel-by-pixel represents a key milestone in this program. 

Inspecting the residual maps for all channels in Fig.~\ref{fig:compsep_residual}, we first of all see that the polarization maps are generally consistent with instrumental noise at all channels. This is even true for the \W-band, although in this case the impact of correlated noise (due to its higher $f_{\mathrm{knee}}$ parameter) is clearly evident. In temperature, the situation is less clear, as there are weak large-scale residuals at the $\sim$\,2\muK\ level present in most channels, but with different morphologies in each case. Clearly, these residuals indicate the presence of some very low-level residual systematics that have not been perfectly modeled in the current processing. Considering the spatial structure of these residuals, it is natural to suspect the small non-idealities in the gain or baseline model; this seems particularly plausible given the pronounced annual temperature variations seen in Figs.~\ref{fig:baseline} and \ref{fig:gain}. Of course, considering that the amplitude of these residuals is indeed very low compared to the overall sky signal, they are not likely to have any significant impact on any cosmological or astrophysical residuals, but they should nevertheless be understood through future work in order to reach the full white-noise potential of the experiment.

\subsection{\WMAP\ and LFI signal-to-noise ratio comparison}
\label{sec:s2n}

\begin{figure}
  \center	
  \includegraphics[width=\linewidth]{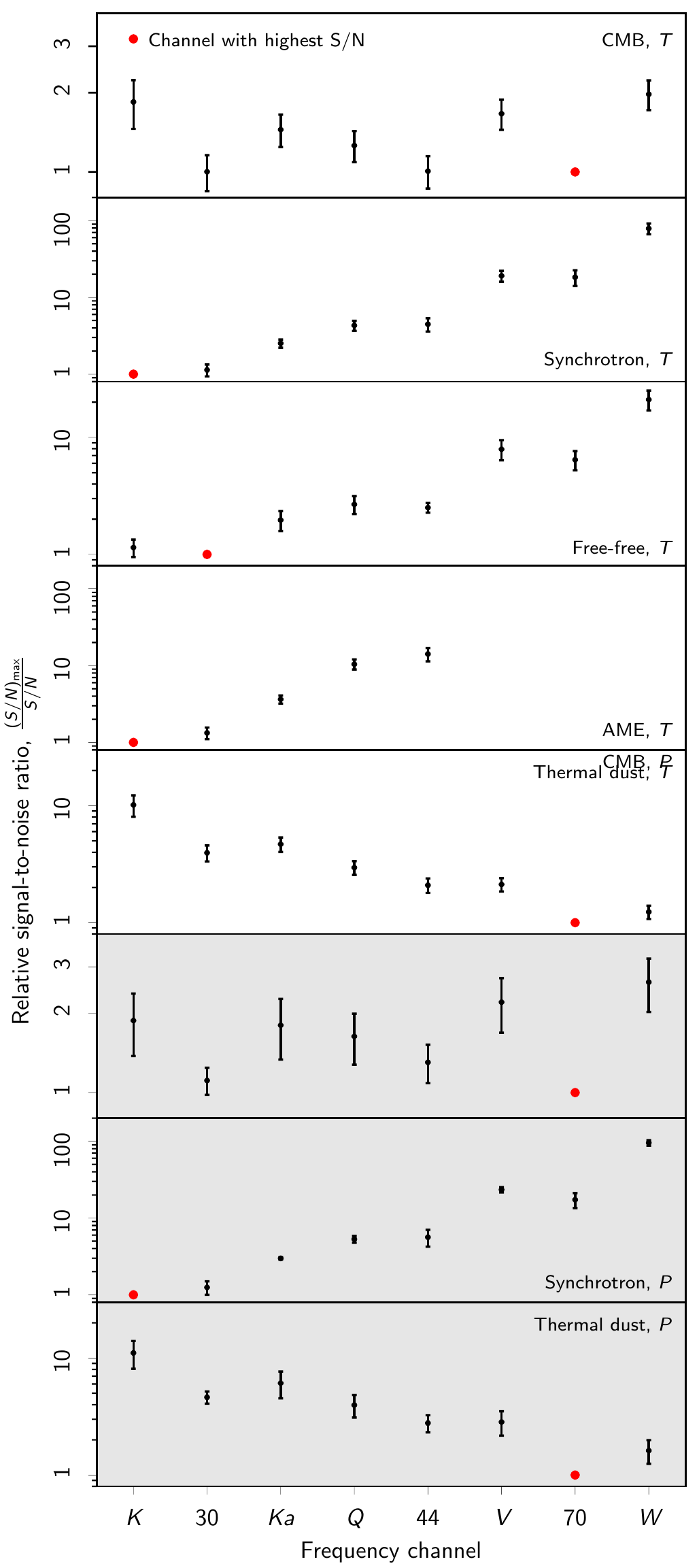}
  \caption{Relative signal-to-noise ratios for \WMAP\ and LFI channels and various components, as defined in terms of the ratio between the mixing matrix and the total instrumental uncertainty, $\psi = \M/\sigma$. The total instrumental uncertainty is derived by adding the white noise and instrumental uncertainty maps (as given by Figs.~\ref{fig:rms} and \ref{fig:std}) in quadrature. Values are reported as the ratio between the most sensitive channel (marked by a red dot) and the given channel; points with error bars correspond to mean and standard deviations evaluated over the full sky. Panels with white background indicate intensity results, while panels with gray background indicates polarization results. All quantities are evaluated at a common angular resolution of $2^{\circ}$ FWHM.}
  \label{fig:fg_s2n}
\end{figure}

We conclude this section with a comparison of the relative signal-to-noise ratios to each astrophysical component between \WMAP\ and \Planck\ LFI. To quantify this, we first define the normalized signal-to-noise ratio for a given channel $\nu$ and pixel $p$ to be the ratio between the mixing matrix and the total noise rms,  $\psi \equiv M_{c}(\nu, p)/\sigma(\nu, p)$, all evaluated at a common angular resolution of $2^{\circ}$ FWHM. The total noise level is estimated by adding in quadrature the white noise and posterior rms levels, similar to those shown in Figs.~\ref{fig:rms} and \ref{fig:std} but both evaluated at $2^{\circ}$ FWHM. 
We then identify the channel with the highest mean value of $\psi$ for a given component, and adopt this as a reference channel. Finally, we compute $r=\psi_{\mathrm{max}}(p)/\psi(p)$ for all pixels,and report the mean and standard deviation over the full sky.

The results from this calculation are summarized in Fig.~\ref{fig:fg_s2n} for all \WMAP\ and LFI channels. The top five panels (with white background) show temperature results, while the bottom three panels (with gray background) show polarization results. In each case, the red dot indicates the reference channel with highest signal-to-noise ratio, which by definition has value equal to one. For a given other channel, $X$, the values in Fig.~\ref{fig:fg_s2n} should be interpreted as ``The reference channel has $r$ times higher signal-to-noise ratio with respect to component $c$ than channel $X$.''

Starting from the CMB intensity case shown in the top panel, we first note that both LFI and \WMAP\ internally have fairly similar signal-to-noise ratio, in agreement with their design specifications. Furthermore, we see that each of the LFI channels generally is about 1.5--2 times more sensitive that each of the \WMAP\ channels, and taking into account the fact that \WMAP\ has five channels, while LFI only has three, the total raw sensitivity of the two experiments is therefore quite comparable. This, in combination with the inter-leaved center frequencies, make the two experiments highly complementary.

Next, the second panel shows the results for synchrotron emission in intensity. Since the synchrotron SED scales roughly as $\mathcal{O}(\nu^{-3})$, \K-band is the overall strongest synchrotron tracer, although the LFI 30\,GHz is lower only by about 13\,\%. A similar observation is true also for AME, while for free-free emission LFI 30\,GHz is about 15,\% more sensitive than \K-band. In contrast, the \Planck\ 70\,GHz channel is strongest with respect to thermal dust emission, and it is about 23\,\% more sensitive than \W-band and two times more sensitive than \V-band.

Finally, we see very similar behaviour in polarization. The LFI 70\,GHz channel is strongest for both CMB and thermal dust emission, while for polarized synchrotron emission, \K-band is stronger than LFI 30\,GHz by about 25\,\%. Again, we note that all these calculations refer to an angular resolution of $2^{\circ}$ FWHM, and the results will vary with angular scale due to the different angular resolutions of the two experiments.

\section{Outstanding issues}
\label{sec:issues}

As shown in the previous sections, there are very few residuals, artifacts, or systematics within this jointly processed dataset, hereafter referred to as \cosmoglobe\ data release 1 (CG1). However, the global nature of this analysis allows us to identify issues in the data processing that will otherwise have gone unnoticed. In this section, we enumerate the issues we have encountered in CG1, and which we plan to improve upon in future data releases.

\subsection{Noise modeling}
\label{sec:noisemodel}

As discussed in Sect.~\ref{sec:noise}, and shown explicitly in Fig.~\ref{fig:chisq}, the TOD-level $\chi^2$ is discrepant up to the $10\sigma$ level for several \WMAP\ diodes. The main driver of this model failure lies in the fact that our current noise estimation algorithm fits the white noise level by computing the standard deviation of pairwise signal-subtracted sample differences. The motivation for this is to prevent slowly varying model failures from contaminating the white level, which worked very well for \Planck\ LFI \citep{bp06}. However, the \WMAP\ time-ordered data exhibits a low level of colored noise at very high temporal frequencies, and is not supported by a rigid $1/f$ model. This is illustrated in Fig.~\ref{fig:W413_psd_zoom}, which shows that the high-frequency noise is fixed to the noise PSD at the sampling frequency. In contrast, if $\sigma_0$ were a free parameter in this particular parametric fit, it would be driven by the intermediate frequencies 2--6\,Hz at the expense of a good fit at the highest frequencies.

The particular case of \W413's PSD is a noise spectrum that could easily be modeled as a spectrum that is continuing to drop beyond the sampling rate, not dissimilar to the two-pole Bessel filter implemented in \WMAP's electronics \citep{jarosik2003:MAP}. In practice, the white noise can be identified with the flat portion of the spectrum well above $f_\mathrm{knee}$, but in the case of these noise spectra, there is no such flat portion, challenging the very existence of ``white noise'' for this particular diode. Additionally, a Bessel filter tail could affect the signal band as well, requiring more detailed modeling of the noise.

In practice, the decomposition of instrumental noise into a ``white'' component and a correlated component is very useful, and provides a stringent test for the final data products. Indeed, the particular model failure was so subtle that such a description of noise being split into scale-dependent and scale-independent would have made it nearly impossible to detect such an issue.
For the case of \WMAP\ data, there is a natural need to improve the noise PSD modeling, especially when a successful parametrization was found by the \WMAP\ team in time space. In practice, this will likely be useful for the analysis of other CMB experiments, and will be of broad use in the future.

\begin{figure}
	\includegraphics[width=\columnwidth]{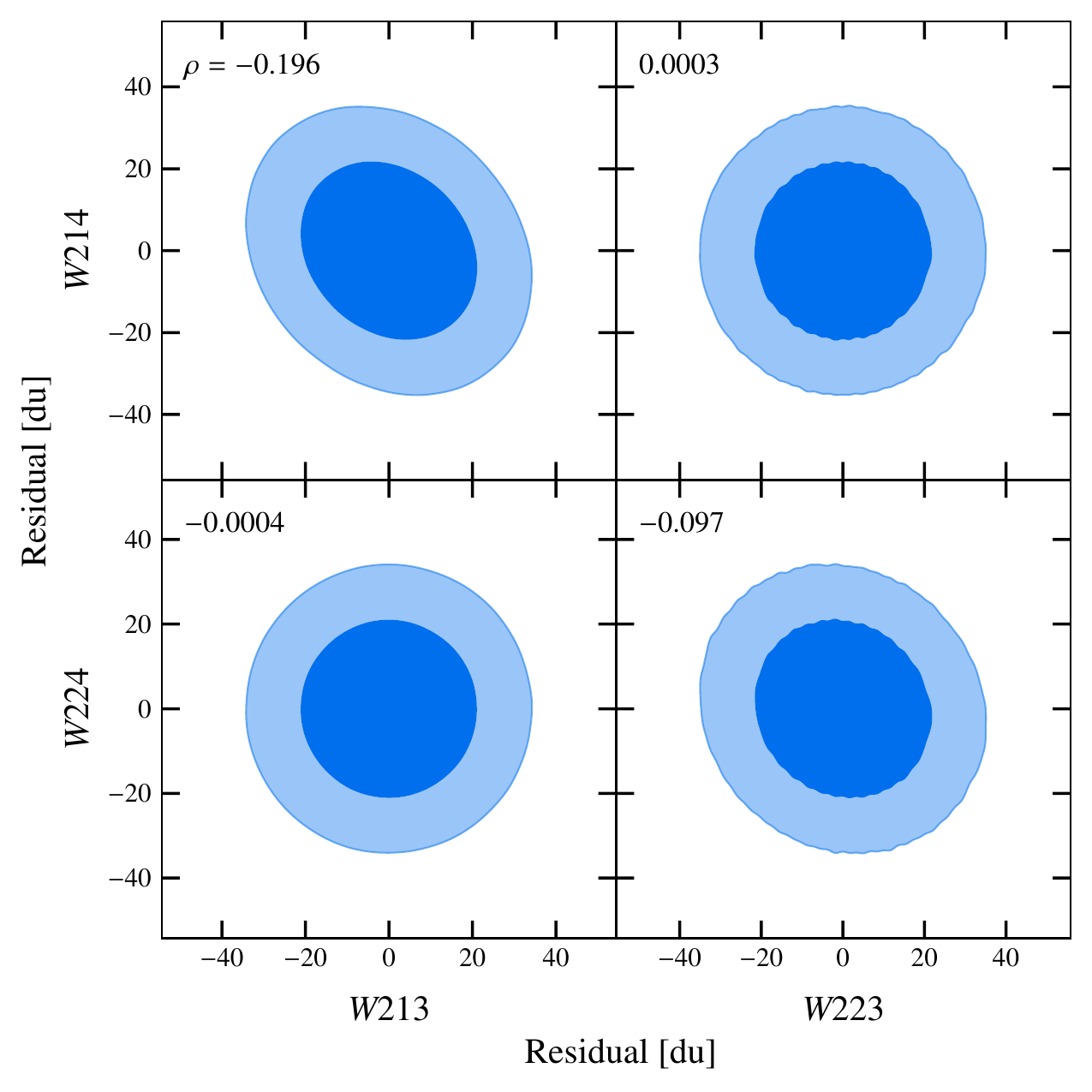}
	\caption{Pearson correlation coefficients between diode residuals for data spanning MJDs 55355.75--55358.24. This estimates the amount of correlated noise shared between detectors.}
	\label{fig:example_corr}
\end{figure}

An additional issue is that of correlation between pairs of diodes. In general, the correlation between a diode pair is close to 5\,\%, but for some radiometers the correlation is up to 25\,\%. In Fig.~\ref{fig:example_corr}, we plot an example of this, showing the correlation between the TOD residuals in a single scan for \W2. In this case, the correlation between diodes within \W21 is ${\sim20\,\%}$, compared to $\sim0.01\,\%$ for diodes corresponding to different radiometers, i.e., \W213 and \W224. Explicitly modeling this effect will be an important step for future data releases, both as a goal in itself, but also for developing the tools for future high-precision $B$-mode searches from, e.g., \textit{LiteBIRD}.

\subsection{Large-scale intensity residuals}

\label{sec:quadres}

As shown in the residual maps in Fig.~\ref{fig:compsep_residual}, we identify large-scale residuals at the level of 1--2\muK\ in intensity in all DA maps. The detailed morphology of these residuals vary from channel to channel, and it is difficult to pinpoint their origin to a given physical effect. However, due to their long-range coherence, it is natural to speculate that they are associated with either the gain or baseline model. In this respect, it is worth recalling that while the \WMAP\ gain model includes gain fluctuations on all time-scales down to 23\,sec, as constrained by housekeeping data, the \cosmoglobe\ gain model is constant within each scan, and the durations of these are several days. While the overall improvement in large-scale consistency in \cosmoglobe\ strongly suggests that gain fluctuations on time-scales of minutes or hours cannot be large, they could potentially be relevant for residuals at the 1--2\muK\ level. It might therefore be useful in future work to integrate housekeeping data also in \cosmoglobe; if not with a resolution of 23\,sec, then perhaps smoothed to minutes or hours.

A similar issue was discovered by \citet{jarosik2007} in the form of an $8\,\mathrm{\mu K}$ dipole, and this was determined to be due to an inadequacy in the gain model. As mentioned earlier, we assume a linear baseline trend throughout a given scan, and allow correlated noise residuals to pick up longer scale fluctuations. Compared to the \WMAP\ team's approach of fitting cubic polynomials every hour, there is much more room for unmodeled temporal variation in zero-level. As the gain, correlated noise, and baseline are all deeply correlated, a subtle error in the baseline determination could easily induce a small quadrupolar signal.

It is also worth noting that in an early stage of this analysis, a large quadrupolar signal was induced due to an error in the orbital dipole calculation. Essentially, a single satellite velocity was assumed for an entire scan, which proved to be a poor approximation over $\sim$$3$-day period. A linear interpolation between scans improved this issue, and a cubic interpolation provided a negligible improvement. This observation points generally to a strong correlation between long-timescale effects and quadrupolar residuals. In particular, if one assumes that a given unmodelled systematic error can generate any type of large-scale pattern, then the dipole component of that pattern can typically be mostly accommodated for in the model as a gain or sky signal dipole variation. Spurious quadrupoles, on the other hand, are much harder to account for in our parametric model, and the leading modes in both residual and channel difference maps therefore often take a quadrupolar shape; this is indeed seen in Figs.~\ref{fig:megadiff_wmap} and \ref{fig:compsep_residual}.

\subsection{Degeneracy between \K-band calibration and AME dipole}
\label{sec:ame_Kband}

\begin{figure}
	\includegraphics[width=\columnwidth]{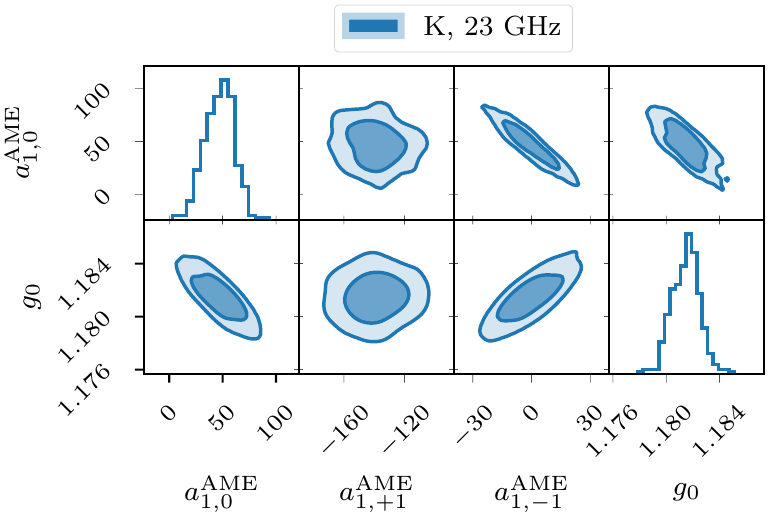}
	\caption{Correlation between \K-band's absolute calibration $g_0$ [$\mathrm{du\,mK^{-1}}$] and AME dipole's spherical harmonic coefficients [$\mathrm{\mu K_{RJ}}$].}
	\label{fig:ame_g0}
\end{figure}

As discussed in Sect.~\ref{sec:priors}, we have identified a strong degeneracy between \K-band's absolute calibration and the AME dipole that requires external information to break. In this work, we implemented a prior on the absolute calibration based on its effect on the AME dipole. To illustrate this degeneracy, we can compare the absolute \K-band calibration with the AME dipole values in terms of the posterior distribution, a slice of which is shown in Fig.~\ref{fig:ame_g0}. Here the degeneracy between $g_0$, $a_{1,0}^\mathrm{AME}$, and $g_{1,-1}^\mathrm{AME}$ is quite apparent. Because there is no causal connection between $g_0$ and the AME dipole, we apply a prior the analysis in this work that results in a physically plausible AME dipole.

In the official \WMAP\ pipeline, the degeneracy was effectively broken by using a preliminary \K-band sky map and removing it from the timestream.  In practice, both solutions are the result of scientific intuition solving an algorithmic issue. The \cosmoglobe\ approach of using a prior on $g_0$ comes from the strong prior that Galactic emission should not have a dipole aligned with the CMB's Solar dipole. The \WMAP\ team's approach of using a previous iteration's map as a sky model comes from the strong prior that errors in the first iteration of the sky map are uncorrelated with the orbital dipole in the timestream.

Even though the \cosmoglobe\ \K-band absolute calibration is informed by the requirement of obtaining a physically reasonable AME dipole, the resulting instrumental solution still generates maps that are consistent with the sky model at the $1\,\mathrm{\mu K}$ level at high Galactic latitudes. Conversely, the \WMAPnine\ solution does not rely on any knowledge of the sky, but as a result induces poorly measured modes with a $2.5\,\mathrm{\mu K}$ amplitude.

Regardless of the details, an accurate model of the sky as observed by the \K-band is a necessary condition for obtaining an accurate measurement of the gain. The difficulty of obtaining an accurate AME model is of course compounded by the fact that the AME is brighter in \K-band than any of the other \WMAP\ or \Planck\ bands. This may be mitigated in the near future, following a joint \WMAP+LFI+QUIJOTE analysis, but this of course depends on the signal-to-noise of AME in QUIJOTE's frequencies, and is further hindered by QUIJOTE's partial sky coverage. Relatedly, also the introduction of \Planck\ HFI data may help breaking this degeneracy by providing much stronger constraints on CMB and free-free emission, both of which are significantly degenerate with AME for the current data selection \citep{bp13}.

A future analysis involving the most robust parts of the \WMAPnine\ and \cosmoglobe\ analysis also has the potential to solve the $g_0$-AME degeneracy. In particular, as noted above, the \cosmoglobe\ analysis did not directly use the housekeeping data to estimate the gain model. There is no a priori reason that the parameters in Eq.~\eqref{eq:wmap_gain} cannot be included in the Gibbs chain. This would of course require detailed knowledge of the \WMAP\ satellite's hardware, and we hope that a joint effort between the \WMAP\ team and \cosmoglobe\ will help to solve this outstanding issue.

\subsection{Other minor effects}
\label{sec:minor}

The issues listed above are known problems in the analysis that will be fixed in the future. Below, we discuss parts of the analyses that we know exist, but have not yet made an attempt to correct because they have not posed direct problems yet.

\subsubsection{Time-variable bandpass modeling}

The \WMAP\ team discovered year-to-year variations in the Galactic plane of the \K, \Ka, \Q, and \V\ maps \citep[Appendix A]{bennett2012}. They determined that the central frequency drifted by 0.13\,\%, 0.12\,\%, 0.11\,\%, and 0.06\,\%, respectively, with a maximum jump of $\sim0.01\,\%$. This was not incorporated in the \WMAPnine\ mapmaking, as each year of data were processed separately, so that each map could be considered to have a single effective frequency.

The \cosmoglobe\ mapmaking procedure has incorporated no correction for this effect. In principle, this could be problematic, as the relative gain solution is obtained by comparing to a bandpass-integrated map of the sky for each DA. However, we have not noticed a sign of this in our analyses, in large part because so much of the sky signal is dominated by the Solar dipole, whose amplitude is not affected by bandpass shifts.

This effect could potentially be modeled using the housekeeping data, as \citet{bennett2012} posit that the instrument's physical temperature changes may have induced changes in the onboard electronics causing the bandpass shift. In this way we could model the bandpass shift and modify the sky model as a function of scan. Ideally, a parametric model for the bandpass shift would be implemented and then sampled for as part of the Gibbs chain. Practically, this effect is subdominant to all other effects we have described in this work, and will not be a priority for the foreseeable future. That said, time variation in the effective bandpass could induce spurious polarization signals in future experiments attempting to measure the tensor-to-scalar ratio $r$. In this context, a full understanding of the temporal dependence on \WMAP's bandpass would be invaluable as preparation for the data analysis of future experiments.

\subsubsection{Polarized sidelobe modeling}

As shown by \citet{barnes2003} and \citet{bp17}, unpolarized sky signals can generate spurious polarized signals, through radiometer mismatch and transmission imbalance, respectively. \citet{barnes2003} also reported the results from lab-based measurements, in which the differential polarized pickup from horns A and B were quantified. Polarized sidelobes could in principle channel a polarized sky signal into the final maps, but \citet{barnes2003} reported that the radiometer mismatch signal dominated the sky across all regions except the Galactic center.

To our knowledge, the polarized sidelobe response has  never been made publicly available in a digital format, thus making the relevant calculation impossible to carry out without the relevant laboratory measurements and results. Again, we hope that a joint effort between the \WMAP\ and \cosmoglobe\ teams may resolve this issue.

\section{Conclusions}
\label{sec:conclusions}

\begin{figure*}
	\includegraphics[height=0.15\textheight]{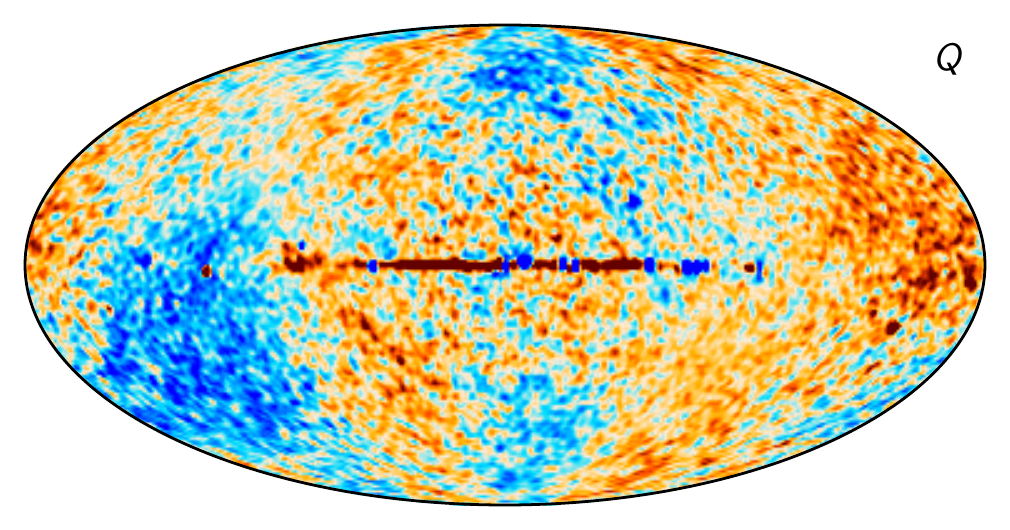}
	\includegraphics[height=0.15\textheight]{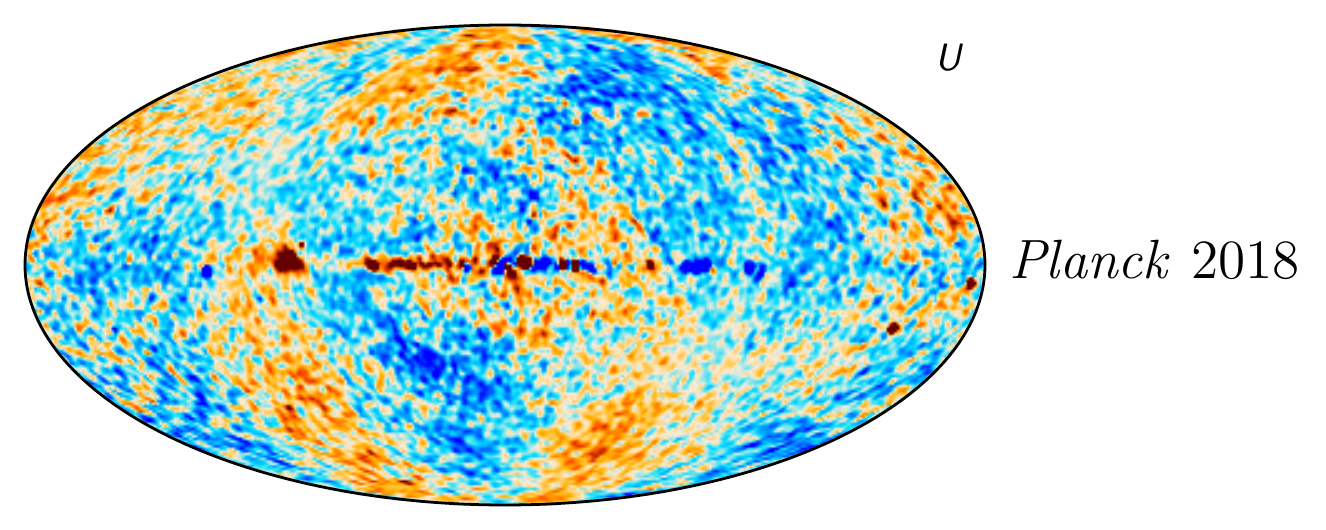}
	\newline
	\includegraphics[height=0.15\textheight]{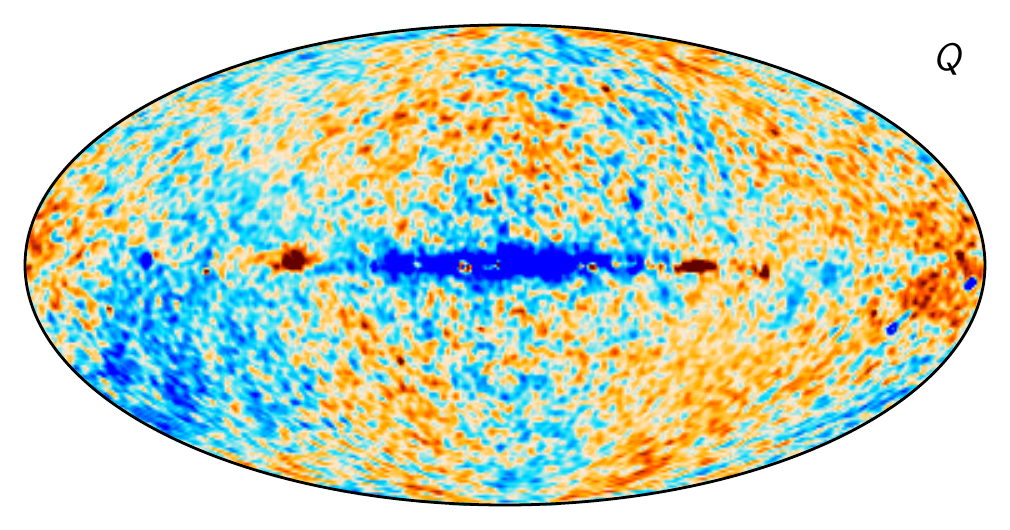}
	\includegraphics[height=0.15\textheight]{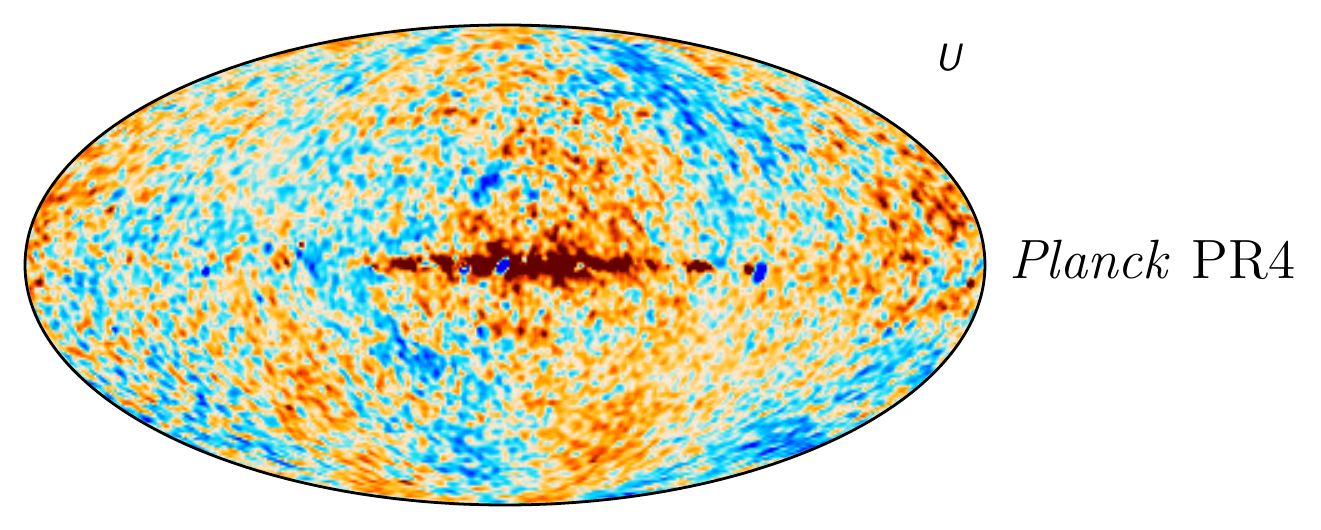}
	\newline
	\includegraphics[height=0.15\textheight]{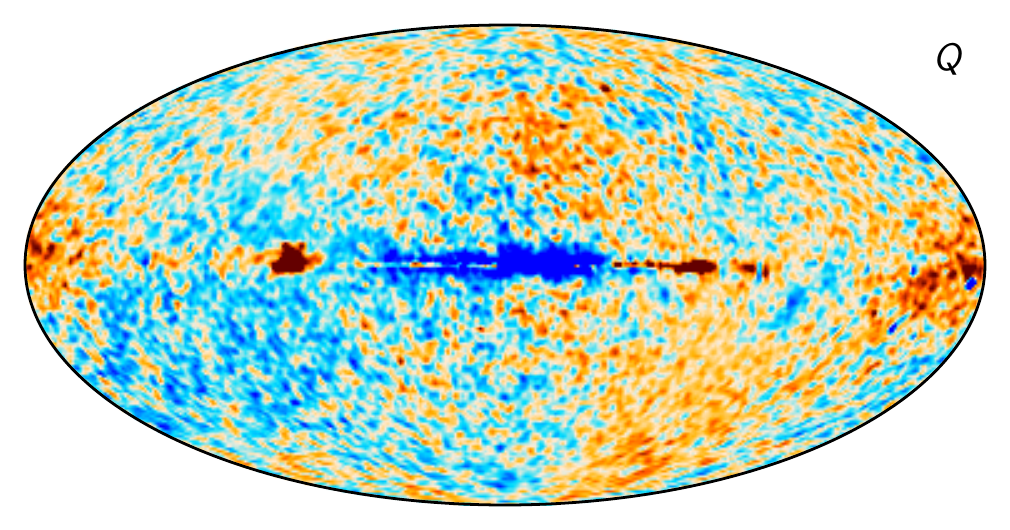}
	\includegraphics[height=0.15\textheight]{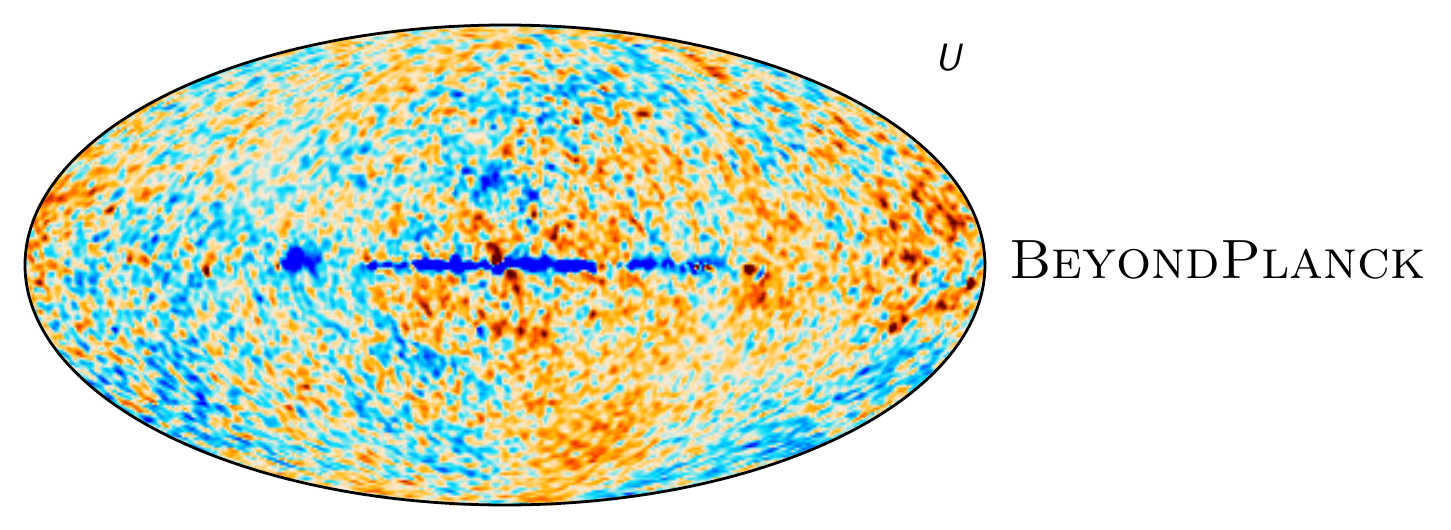}
	\newline
	\includegraphics[height=0.15\textheight]{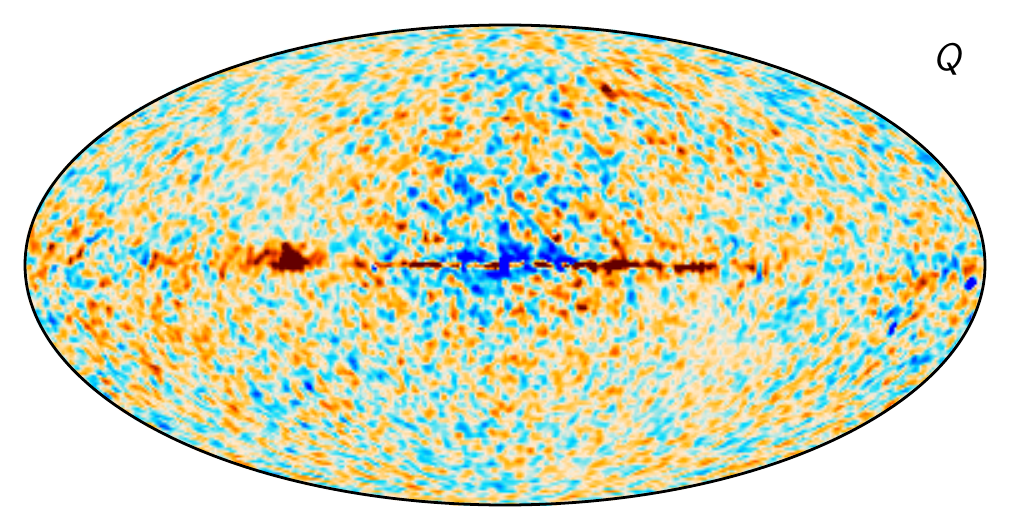}
	\includegraphics[height=0.15\textheight]{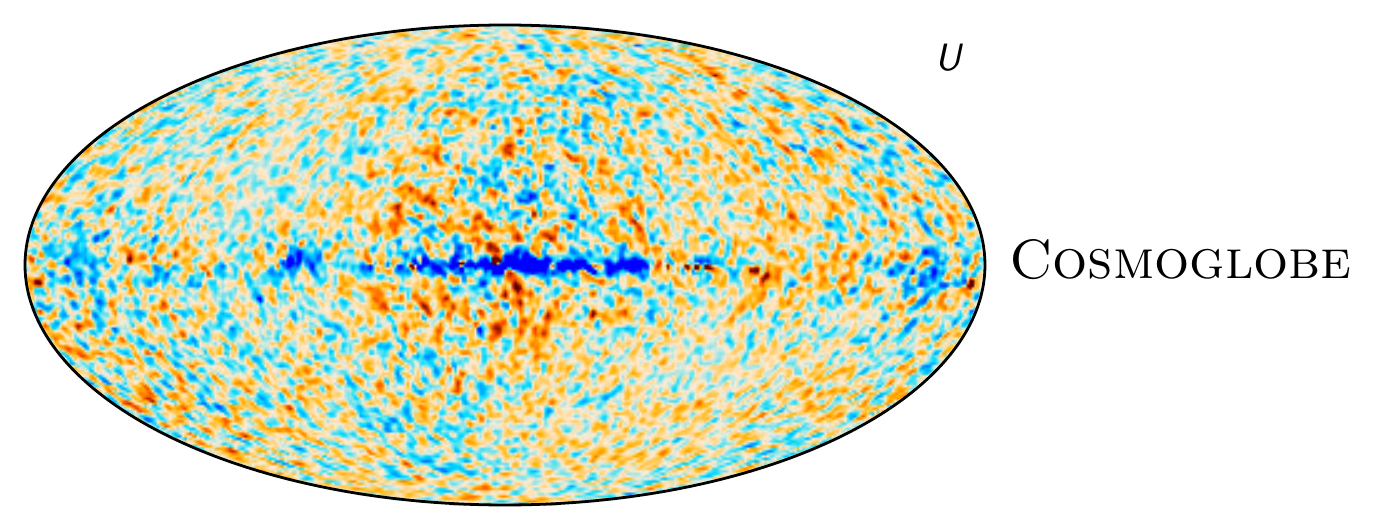}\\
        \hspace*{40mm}\includegraphics[width=0.25\textheight]{figures/cbar_10uK.pdf}
	\caption{Difference maps between the \Planck\ 30\,GHz and \WMAP\ \K-band maps. The columns are \textit{(1)} \Planck\ 2018 v.~\WMAPnine, \textit{(2)} \Planck\ PR4 v.~\WMAPnine, \textit{(3)} \BP\ v.~\WMAPnine, and \textit{(4)} \cosmoglobe\ \Planck\ 30\,GHz and \WMAP\ \K-band both produced in this paper. All maps have been smoothed to a common resolution of $2^\circ$ FWHM, and the \K-band map has been scaled by $0.495$ to account for different central frequencies, assuming a synchrotron spectral index $\beta_\mathrm s=-3.1$.}
	\label{fig:diff_history}
\end{figure*}

Over the last half century, a long series of technological breakthroughs have revolutionized our understanding of both cosmology and astrophysics through increasingly detailed measurements of the radio, microwave and submillimeter sky. Typically, these developments have taken place within individual experiments, each constructing novel equipment and techniques to better measure and constrain new physics while mitigating the relevant systematic effects. Each new generation of experiments has moved the frontier in terms of sensitivity, angular resolution, frequency coverage, and systematics control. This steady improvement in technological capability comes at a non-negligible cost. A typical budget for an astrophysics satellite mission ranges between 100 and 1000 million euros or dollars, and a few next-generation ground-based experiments have budgets in the hundreds of million dollar range. To make new breakthroughs in the future, we cannot afford to perform science without fully utilizing the data from experiments that have come before.  Given that all experiments fundamentally observe the same sky, it is therefore essential for the field as a whole to optimally reuse all already existing state-of-the-art measurements whenever designing, fielding, and analyzing a new experiment.

This insight is the defining goal of the \cosmoglobe\ project; namely to integrate the world's best data from radio to submillimeter wavelengths through global analysis. In this sentence, the word ``global'' carries three different meanings. First, as demonstrated by the \Planck\ experience, it is of critical importance to analyze all aspects of a given experiment globally -- instrument calibration, component separation, and cosmological interpretation -- in order to understand and mitigate all relevant systematic effects. Otherwise, degeneracies between the calibration and sky model will invariably dominate the final error budget. Secondly, as demonstrated by both \Planck\ and \WMAP, virtually all experiments have some blind spots to which they are uniquely insensitive, whether it is due to frequency range, angular resolution, scanning strategy, or raw sensitivity. It is therefore always, at least in principle, advantageous to analyze multiple experiments jointly, in order to use the strengths of one experiment to break the degeneracy of another; multi-experiment analysis is thus the second aspect of global analysis. Thirdly, in order for this ambitious program to succeed, it is essential that large fractions of the community work together, and expertise from different collaborations, groups, and countries are optimally combined.  International collaboration is thus the third aspect of global analysis.

This massive program would not be organizationally feasible without decades of investments in instrumentation and algorithm development by the entire field. The current \cosmoglobe\ implementation relies heavily on work done within the \Planck\ collaboration, both in terms of developing the general understanding of both integrated and Bayesian parametric analysis \citep[e.g.,][]{planck2014-a12,planck2016-l02,npipe}, as well as the specific code implementation (\texttt{Commander}; \citealt{eriksen:2004,eriksen:2008}). After the conclusion of the official \Planck\ collaboration, this work was continued within the \bp\ collaboration \citep{bp01}, which culminated in the \commanderthree\ code \citep{bp03}, which for the first time allowed true integrated end-to-end Bayesian analysis of a major CMB experiment, namely \Planck\ LFI. This framework provides a mature computational foundation for the \cosmoglobe\ analysis, which aims to apply the same process to all available state-of-the-art datasets.

The current paper represents the first step in this long process. Specifically, we have analyzed the \WMAP\ measurements from raw time-ordered data to final CMB power spectra within one single computer code, which is a notable milestone by itself. However, this analysis accounts in fact for both \WMAP\ and \Planck\ LFI time-ordered data at the same time, marking an even larger milestone. It demonstrates that joint multi-experiment analysis is indeed both computationally and practically feasible. Furthermore, while the new (signal-dominated) temperature sky maps appear to be consistent with the previous state-of-the-art \WMAPnine\ products, the new \cosmoglobe\ polarization maps appear to be of higher quality than the \WMAPnine\ maps. This provides a strong testament to the original \cosmoglobe\ idea; better results are obtained when exploiting synergies between complementary experiments. Even moreso, we believe that this analysis demonstrates that it is in fact easier to analyze multiple experiments together because of fewer degeneracies.

Going into greater detail, we find that our gain model agrees with the \WMAPnine\ estimates to within 1\,\%, despite the fact that the two approaches use very different calibration methods. This is a testament to the performance of both. In addition, our parametric noise model results in knee frequencies that are mostly consistent with the first-year \WMAP\ values, while still tracking temporal variations on shorter timescales than those considered by the \WMAP\ team. The transmission imbalance parameters are statistically consistent within the \WMAPnine\ error budget, although the two methods disagree on the magnitude of the uncertainty for individual radiometers. Finally, we are able to track the goodness of fit per scan for each individual radiometer, and find that the raw $\chi^2$ is within 0.3\,\% of $n_\mathrm{TOD}$ throughout the entire mission.

Turning to the map domain, we note that our temperature maps are consistent with \WMAPnine\ to about $2.5\,\mathrm{\mu K}$ on large angular scales, and the remaining differences are most likely due to the different baseline and gain modeling. In polarization, we see differences of up to $\sim10\,\mathrm{\mu K}$. A large fraction of this is due to the poorly measured transmission imbalance modes identified by the \WMAP\ team; however, we have also shown that the bilinear template marginalization method used in \WMAPnine\ is unable to account for the full observed differences. This is of course not unexpected, given that the impact of transmission imbalance is a nonlinear effect that couples to a wide range of instrumental parameters, and two linearly dependent templates cannot account for the full posterior volume of these parameters. In contrast, by virtue of tracing all these nonlinear couplings through an explicit data model, an MCMC sampler is able to model these correlated modes naturally. As a result, we find that the large-scale residuals are visually present in internal \WMAPnine\ half-difference DA maps, but not in the corresponding \cosmoglobe\ maps. Correspondingly, we find that the \WMAPnine\ and \cosmoglobe\ temperature power spectra agree very well, while the $E$-mode and $B$-mode power spectra from \cosmoglobe\ are better behaved for nearly every multipole.

This work provides a natural resolution to the long-standing discrepancy between the polarized \WMAP\ \K-band and \Planck\ LFI 30\,GHz observations first reported by \citet{planck2014-a12}. These two channels are sufficiently close in frequency that any uncertainty in the synchrotron SED should be subdominant to instrumental noise. However, when differencing these two maps (after scaling one by the synchrotron SED), large-scale residuals correlated with the scanning strategy appeared. The \Planck\ LFI team worked hard to improve their data processing through the \Planck\ 2018 and PR4 release, gradually reducing systematic errors in their products. This process continued into the \bp\ epoch, but even after that obvious residuals remained \citep{bp07}. It is only now, when both LFI and \WMAP\ are processed from scratch, that the two datasets agree to a level compatible with instrumental noise. This progress is illustrated in Fig.~\ref{fig:diff_history}, which shows the \K--30 difference maps for \Planck\ 2018, \Planck\ PR4, \bp, and \cosmoglobe. In all but the last case, the \Planck\ maps are differenced with the \WMAPnine\ \K-band map.

In turn, the successful resolution of these long-standing problems have important and direct implications for a wide range of polarization-based applications. One example of this is estimation of the synchrotron spectral index, which is addressed in a separate paper by \citet{fuskeland:2023}; it is only with these new data products that it is possible to derive physically meaningful estimates of $\beta_{\mathrm{s}}$ with combined LFI and \WMAP\ measurements on all angular scales. Correspondingly, this is the first time LFI and \WMAP\ constraining power have been successfully combined into a single polarized synchrotron amplitude map, and the \cosmoglobe\ version of this map therefore represents both the most sensitive and systematically cleanest full-sky tracer of polarized emission available today. We recommend that this map should be the preferred synchrotron template in the foreseeable future, for instance when simulating the radio sky with \texttt{PySM} \citep{pysm} or forecasting the performance of next-generation experiments \citep[e.g.,][]{ptep,aurlien:2022}. In general, we believe that these maps redefine the state-of-the-art in terms of \WMAP\ data quality.

In terms of CMB intensity science, there are also a few interesting observations worth pointing out. First, we find a CMB Solar dipole amplitude of $3366.2\pm1.4\,\mathrm{\mu K}$, which is $2.5\,\sigma$ higher than the closely related \bp\ analysis \citep{bp11}. This is very close to the final \Planck\ PR4 value of $3366.6\pm2.6\muK$, and the combination of \WMAP\ and LFI now agree very well with the independent HFI measurements. Second, we find a lower quadrupole amplitude than latest \Planck\ results of $\sigma_2 = 131 \pm 69\,\mathrm{\mu K^2}$. Although statistically consistent with the \LCDM\ prediction at the 11\,\% level, it is still interesting to note that this is lower than all previous estimates, except for the first-year \WMAP\ release. Third, it is also intriguing to note that the peak of the octopole alignment statistic posterior peaks at unity, which, if true, could potentially hint towards physics beyond the \LCDM\ model, for instance in the form of nontrivial topology. An important goal for future work is to decrease this -- and all other -- uncertainty by adding additional data and improved data models. In particular the integration of \Planck\ HFI measurements is a high priority for future work.

We believe that this first \cosmoglobe\ data release successfully demonstrates the advantages of global data analysis, and we hope and anticipate that it is only the first among many, each adding support for one or more new experiments. We invite all interested parties to join this effort, whether it is by providing access to experimental data, novel algorithmic ideas, or just sheer work force; for further details on how to contribute, we refer the interested reader to the \cosmoglobe\ webpage.\footnote{\url{https://cosmoglobe.uio.no}}

\begin{acknowledgements}
  We thank Prof.~Charles Bennett, Dr.~Janet Weiland, Prof.~Lyman Page, and
  Prof.~Eiichiro Komatsu for useful suggestions, comments, and discussions.
  We thank the entire \Planck\ and \WMAP\ teams for
  invaluable support and discussions, and for their dedicated efforts
  through several decades without which this work would not be
  possible. The current work has received funding from the European
  Union’s Horizon 2020 research and innovation programme under grant
  agreement numbers 819478 (ERC; \textsc{Cosmoglobe}) and 772253 (ERC;
  \textsc{bits2cosmology}).
  In
  addition, the collaboration acknowledges support from
  RCN (Norway; grant no.\ 274990).
  We acknowledge the use of the Legacy Archive for Microwave Background Data
  Analysis (LAMBDA), part of the High Energy Astrophysics Science Archive Center
  (HEASARC). HEASARC/LAMBDA is a service of the Astrophysics Science Division at
  the NASA Goddard Space Flight Center.
  Some of the results in this paper have been derived using the \texttt{healpy}
  and \texttt{HEALPix}\footnote{\url{http://healpix.sf.net}} packages
  \citep{gorski2005, Zonca2019}.  This work made use of
  Astropy:\footnote{\url{http://www.astropy.org}} a community-developed
  core Python package and an ecosystem of tools and resources for
  astronomy \citep{astropy:2013, astropy:2018, astropy:2022}.
\end{acknowledgements}

\bibliographystyle{aa}

\bibliography{Planck_bib,CG_bibliography}

\appendix

\section{Survey of instrumental parameters}
\label{sec:survey}

\begin{figure*}[p]
	\centering
	\includegraphics[width=\textwidth]{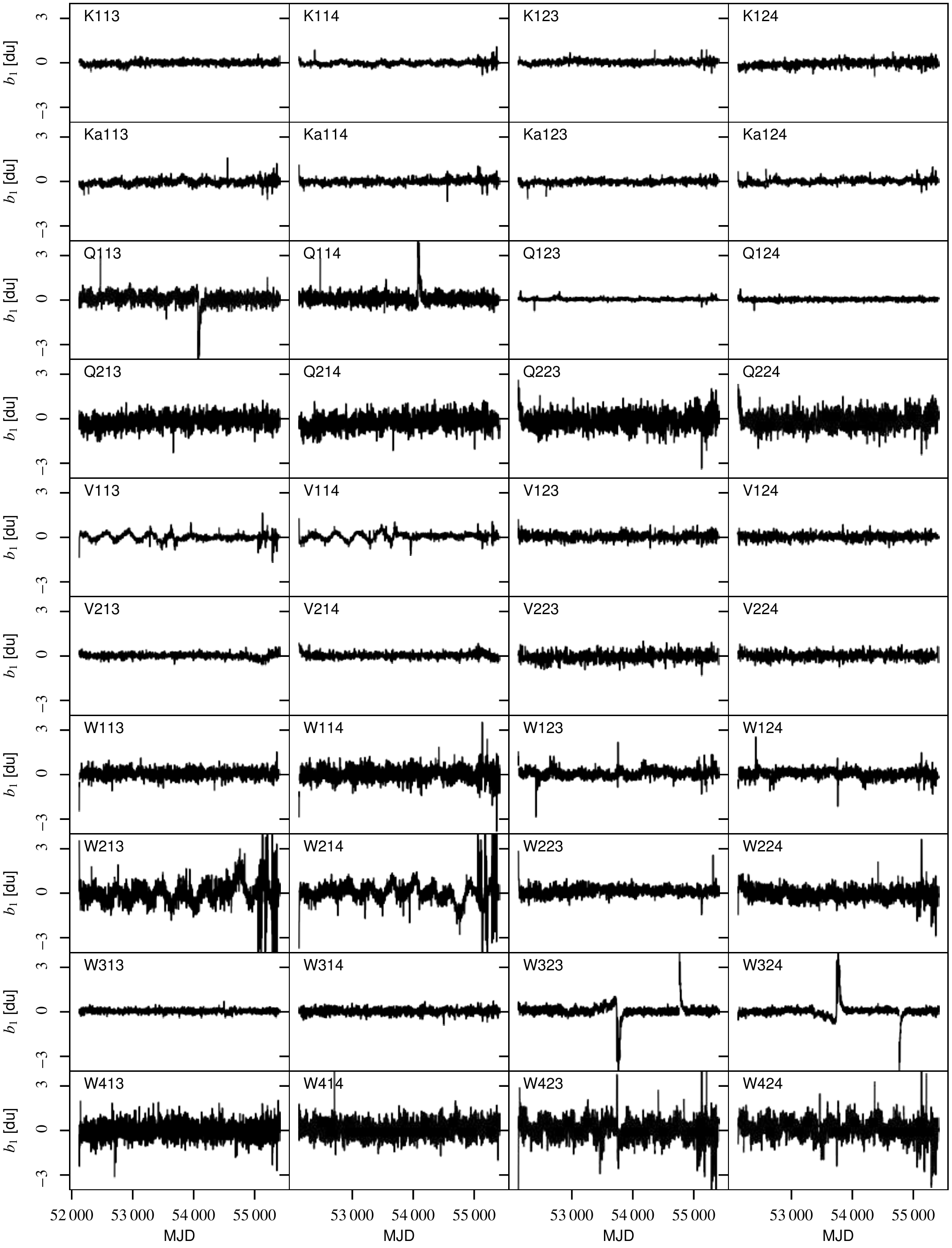}
	\caption{\cosmoglobe\ first-order baseline correction (i.e., slope) for each diode.}
	\label{fig:baseslope}
\end{figure*}

\begin{figure*}[p]
	\centering
	\includegraphics[width=\textwidth]{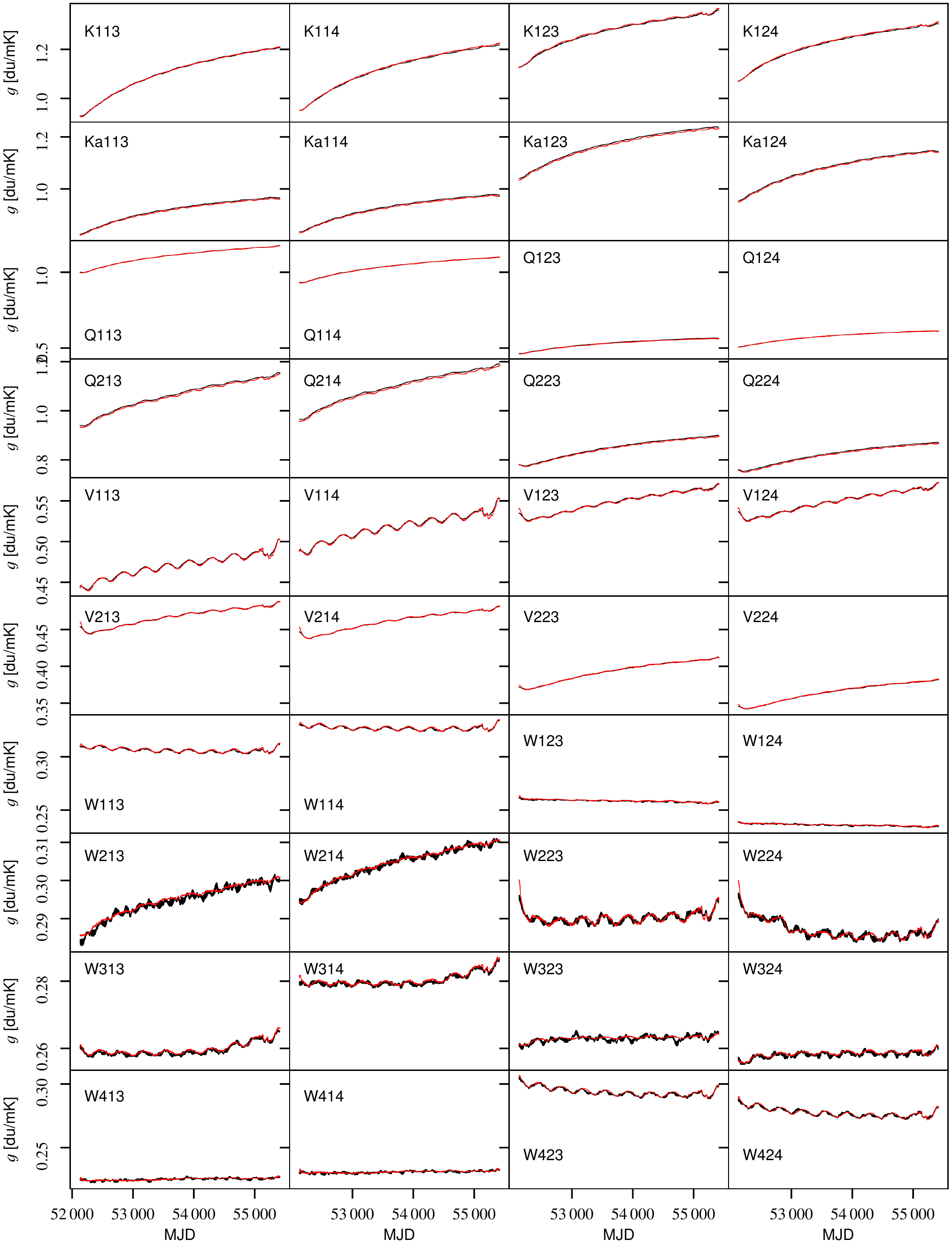}
	\caption{Time-variable gain model for each diode for \cosmoglobe\ (black) and \WMAPnine\ (red).}
	\label{fig:gain}
\end{figure*}

\begin{figure*}[p]
	\centering
	\includegraphics[width=\textwidth]{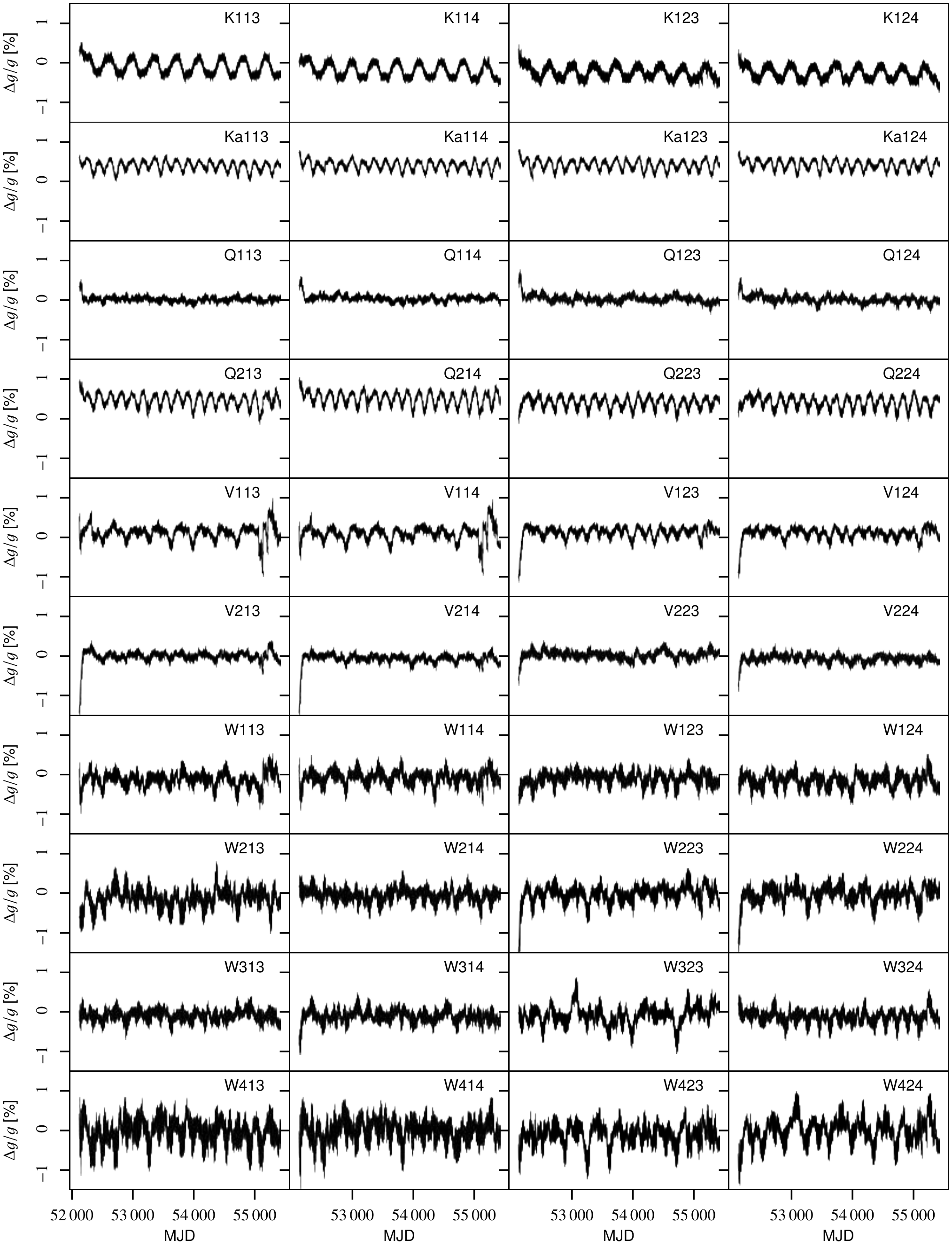}
	\caption{Relative gain difference between \cosmoglobe\ and \WMAPnine, $(g^\mathrm{CG}-g^\mathit{WMAP})/g^\mathit{WMAP}$.}
	\label{fig:dgain}
\end{figure*}

\begin{figure*}[p]
	\centering
	\includegraphics[width=\textwidth]{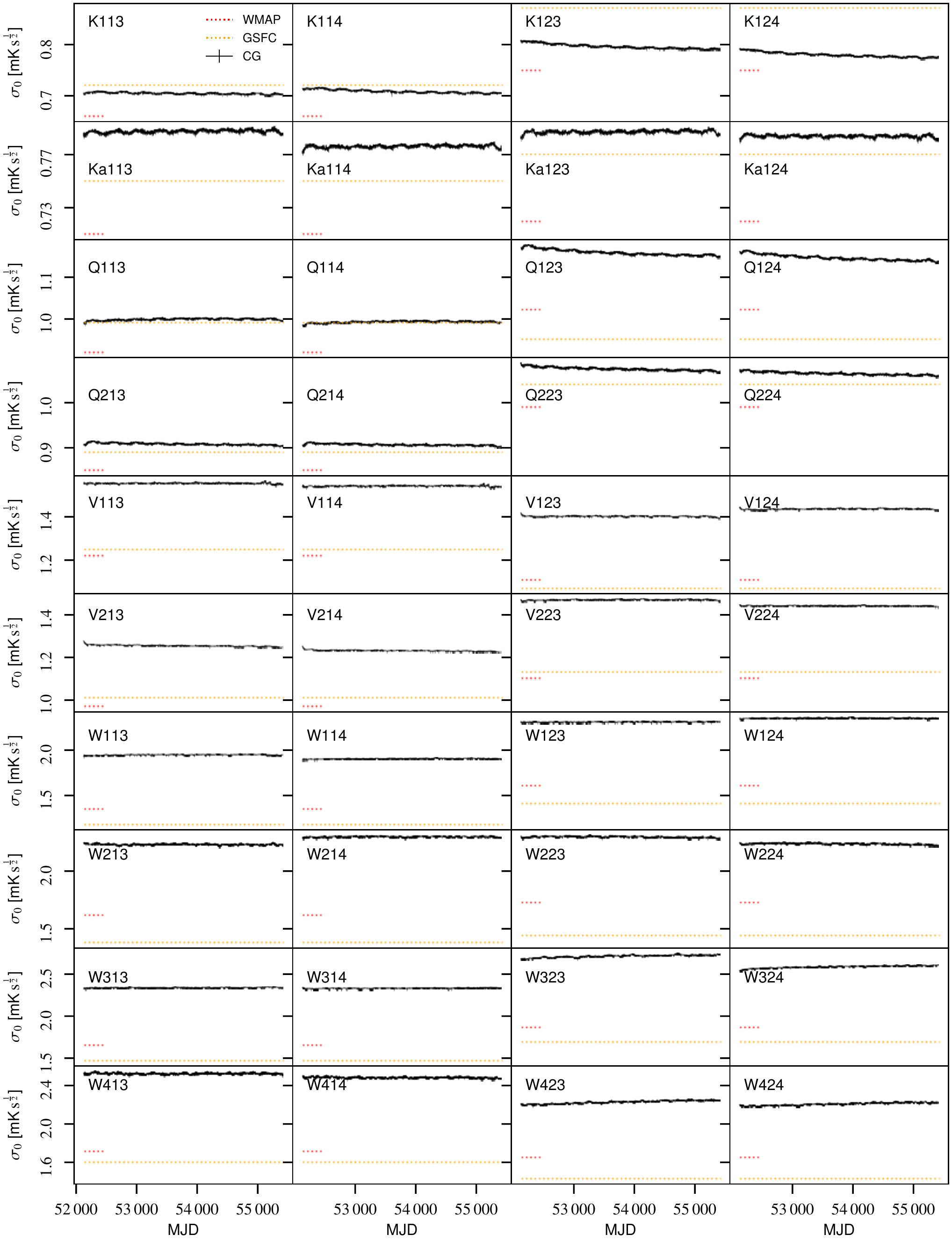}
	\caption{White noise rms per TOD sample, $\sigma_0$. Black lines show \cosmoglobe\ estimates, while dotted red and orange lines show \WMAP\ first-year in-flight and GSFC laboratory measurements.}

	\label{fig:sigma0}
\end{figure*}

\begin{figure*}[p]
	\centering
	\includegraphics[width=\textwidth]{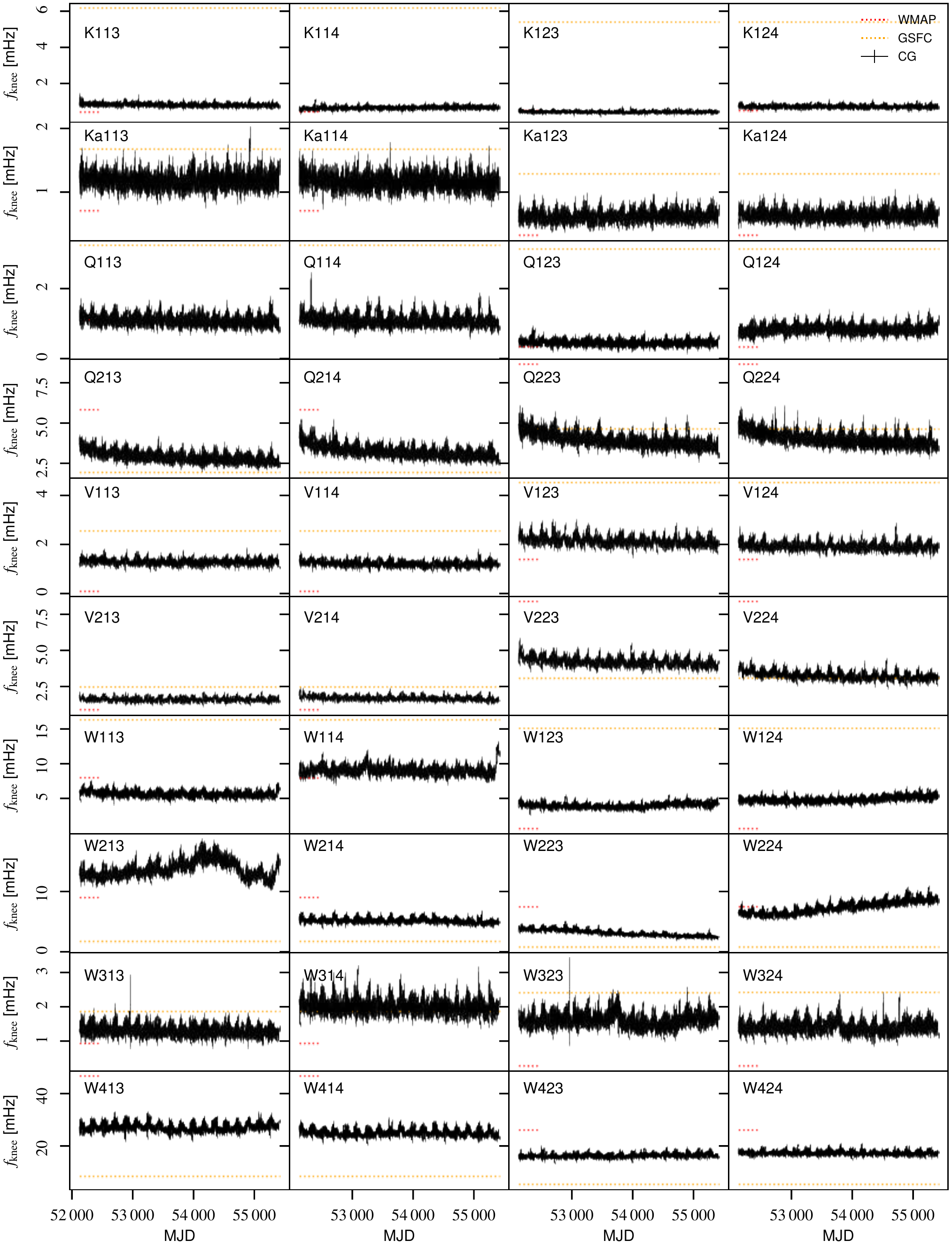}
	\caption{Correlated noise knee frequency, $f_{\mathrm{knee}}$, for each diode. Black lines show \cosmoglobe\ estimates while dotted red and orange lines show \WMAP\ first-year in-flight and GSFC laboratory measurements.}

	\label{fig:fknee}
\end{figure*}

\begin{figure*}[p]
	\centering
	\includegraphics[width=\textwidth]{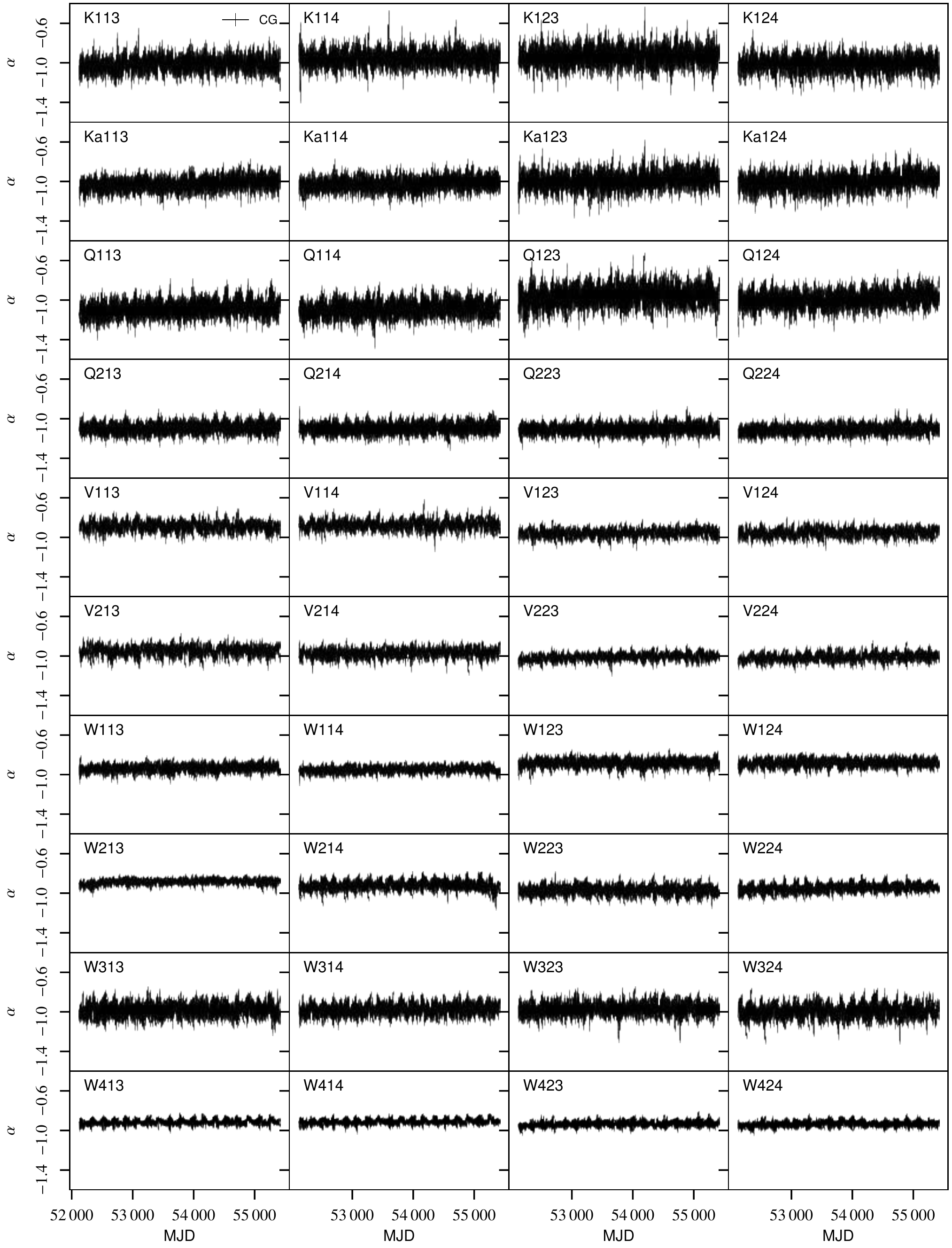}
	\caption{Correlated noise slope, $\alpha$, for each diode. Black lines show \cosmoglobe\ estimates, while dotted orange and red lines show \WMAP\ first-year in-flight and GSFC laboratory measurements.}        
	\label{fig:alpha}
\end{figure*}

\begin{figure*}[p]
	\centering
	\includegraphics[width=\textwidth]{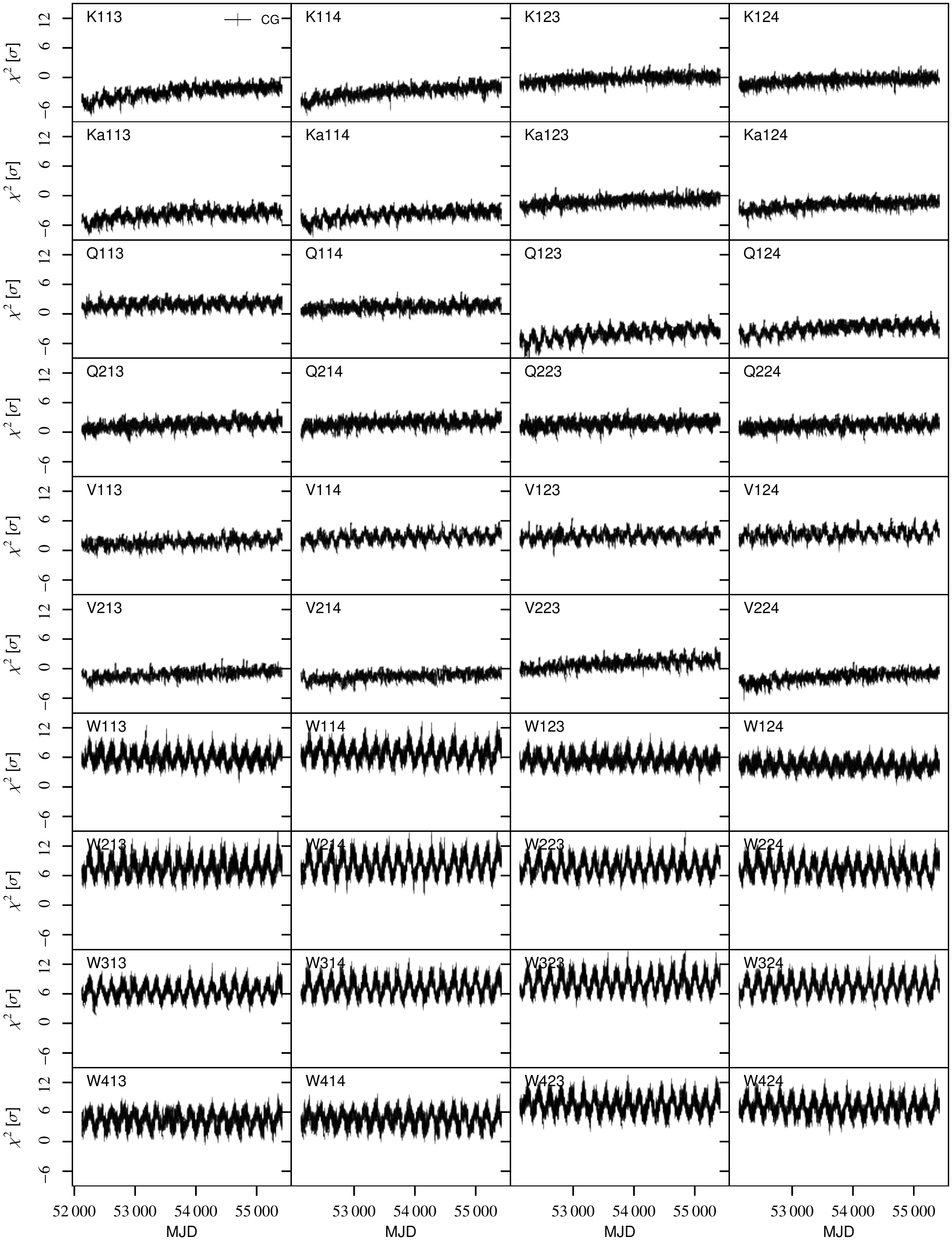}
	\caption{TOD-level reduced normalized $\chi^2$ as defined by Eq.~\eqref{eq:chisq}.}
	\label{fig:chisq}
\end{figure*}

\noindent\begin{minipage}{\textwidth}
In this Appendix, we provide for reference purposes a complete survey of the posterior mean estimates for all time-dependent instrumental parameters. In each figure, columns correspond to individual diodes, while rows correspond to DA. Figure~\ref{fig:baseline} shows the zeroth order baseline (plotted as the difference between the full time-variable baseline and its own time average), while Fig.~\ref{fig:baseslope} shows the corresponding baseline slopes. Figure~\ref{fig:gain} shows the full time-dependent gain, and Fig.~\ref{fig:dgain} shows the fractional gain difference between \cosmoglobe\ and \WMAP\ in units of percent. Figures~\ref{fig:sigma0}--\ref{fig:alpha} shows the noise model parameters, $\sigma_0$, $\fknee$, and $\alpha$, respectively. Finally, Fig.~\ref{fig:chisq} shows the TOD-level reduced normalized $\chi^2$ in units of standard deviation. Black lines show \cosmoglobe\ results, while solid red lines show (where available) official \WMAP\ results derived by linear regression between the raw and calibrated \WMAP\ TODs. Orange and red dotted lines show \WMAP\ first-year in-flight and GSFC laboratory measurements, respectively. For further discussion regarding these plots, see Sect.~\ref{sec:instrument}.
\end{minipage}

\noindent\begin{minipage}{0.83\textwidth}
\vspace*{1mm}
\centering
\hspace*{1.5cm}\includegraphics[width=\textwidth]{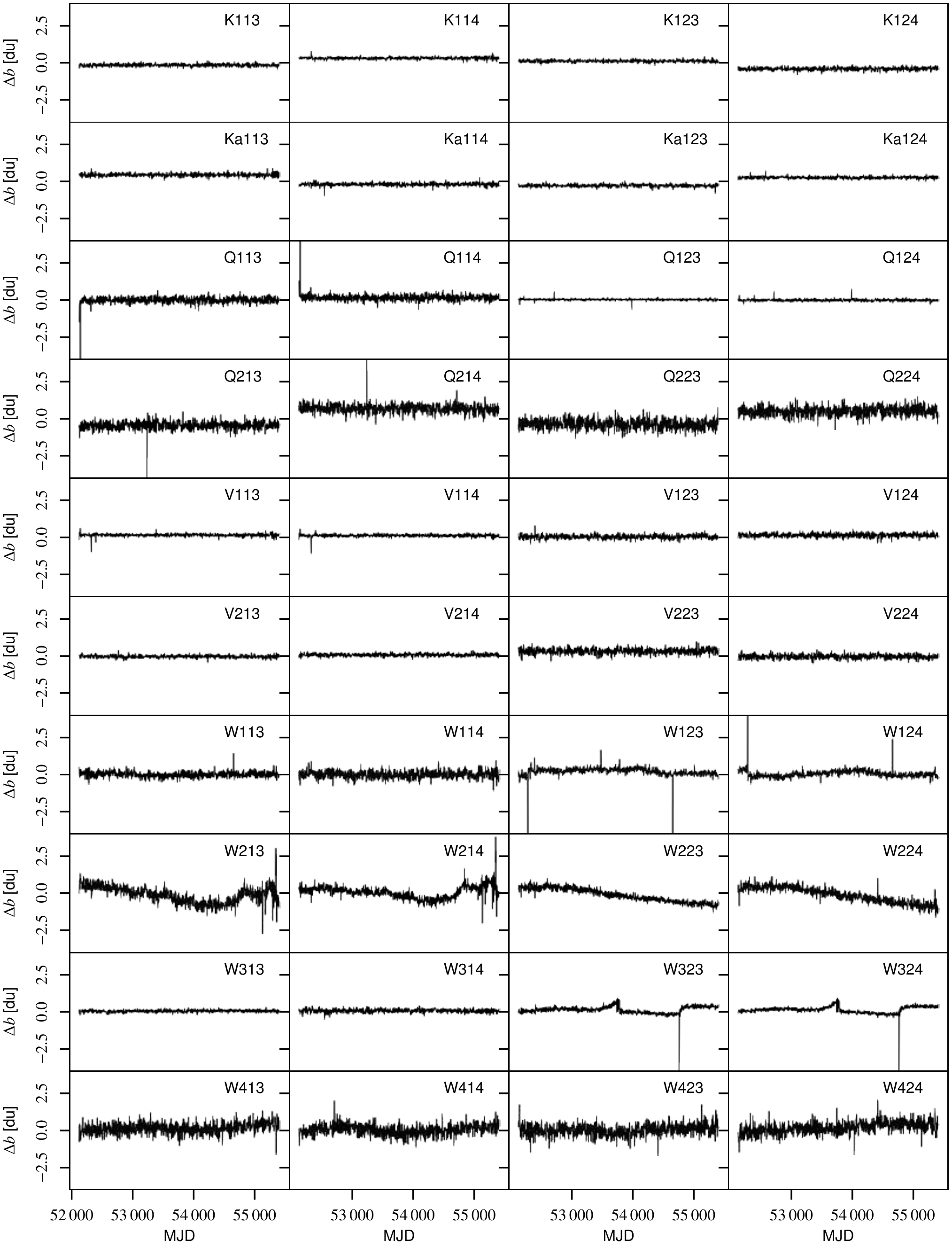}
\captionof{figure}{Difference in zero-level baseline between \cosmoglobe\ and \WMAPnine, $b_0^\mathrm{CG}-b_0^\mathit{WMAP}$.}
\label{fig:baseline}
\end{minipage}

\clearpage
\section{\WMAP\ frequency map survey}
\label{sec:map_survey}

\begin{figure*}[p]
	\centering
	\includegraphics[width=0.9\textwidth]{figures/023-WMAP_K_rms.pdf}
	\includegraphics[width=0.9\textwidth]{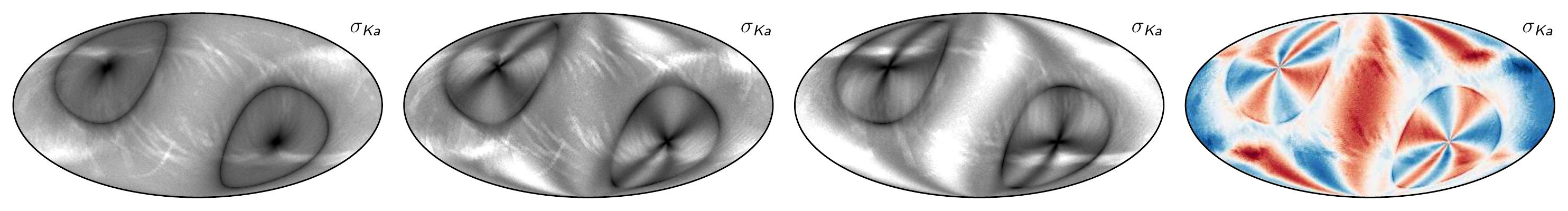}
	\includegraphics[width=0.9\textwidth]{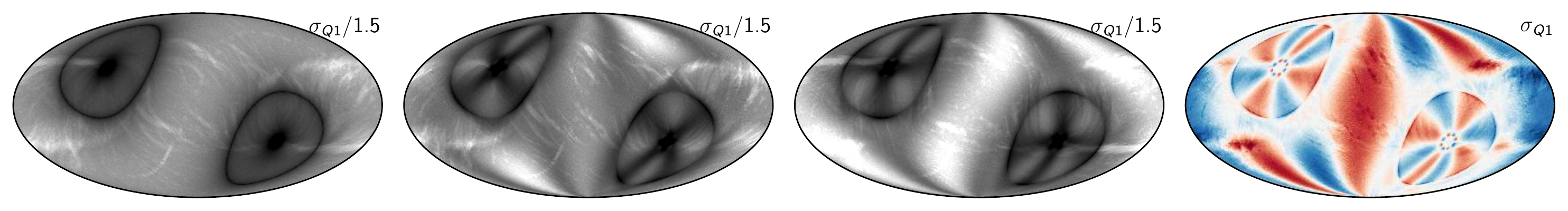}
	\includegraphics[width=0.9\textwidth]{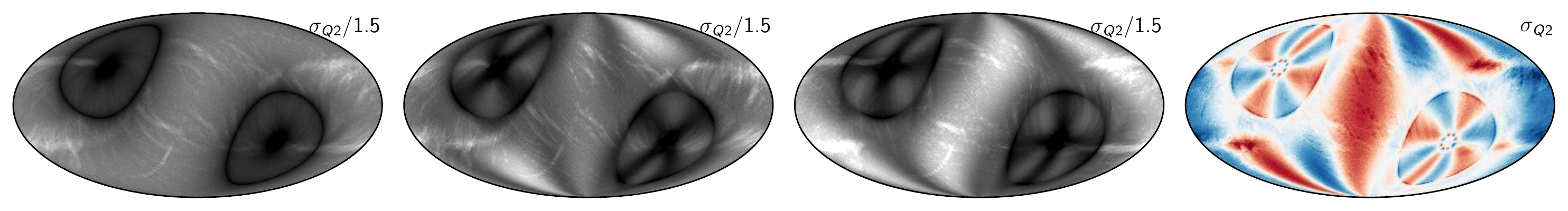}
	\includegraphics[width=0.9\textwidth]{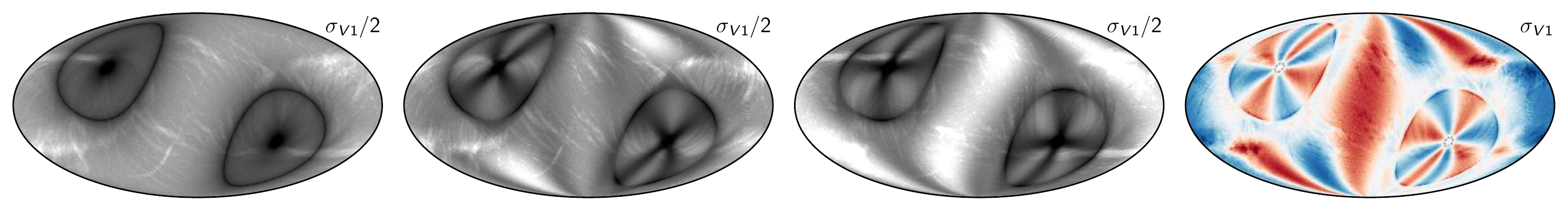}
	\includegraphics[width=0.9\textwidth]{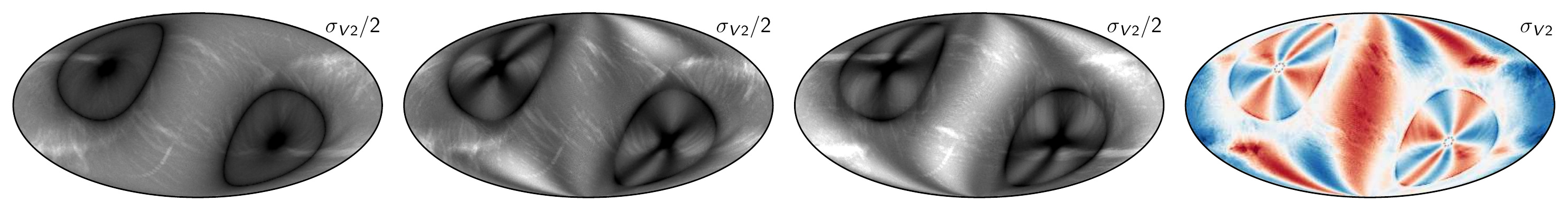}
	\includegraphics[width=0.9\textwidth]{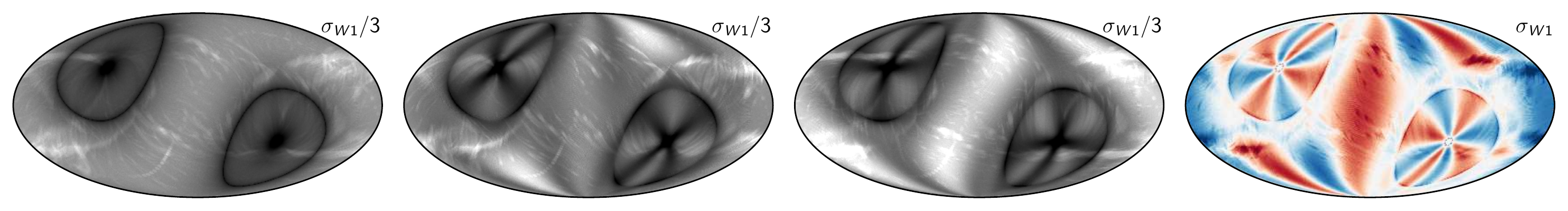}
	\includegraphics[width=0.9\textwidth]{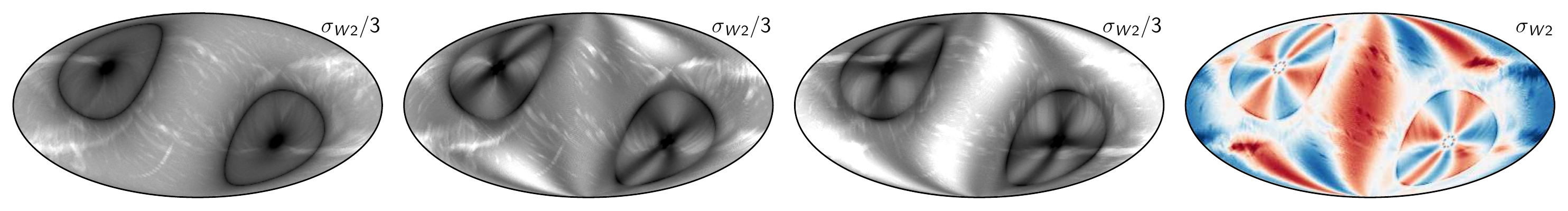}
	\includegraphics[width=0.9\textwidth]{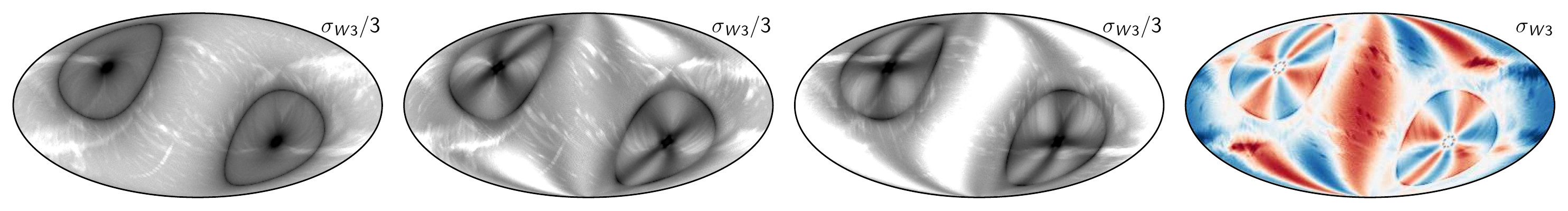}
	\includegraphics[width=0.9\textwidth]{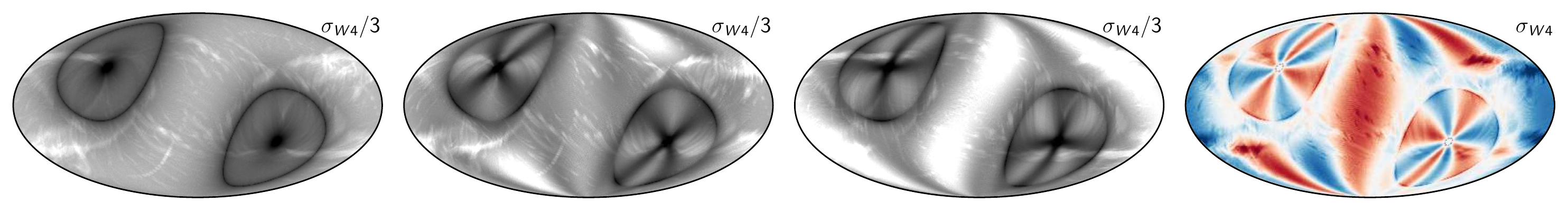}
	\includegraphics[width=0.22\textwidth]{figures/cbar_rms_I.pdf}
	\includegraphics[width=0.22\textwidth]{figures/cbar_rms_P.pdf}
	\includegraphics[width=0.22\textwidth]{figures/cbar_rms_P.pdf}
	\includegraphics[width=0.22\textwidth]{figures/cbar_rho.pdf}
	\caption{\cosmoglobe\ white noise rms per pixel, $\sigma_p$, for each DA and Stokes parameters. The rightmost column shows the cross-correlation between Stokes $Q$ and $U$ due to the \WMAP\ scanning strategy. }
        \label{fig:rms}
\end{figure*}

\begin{figure*}[p]
	\centering
	\includegraphics[width=0.9\textwidth]{figures/023-WMAP_K_std.pdf}
	\includegraphics[width=0.9\textwidth]{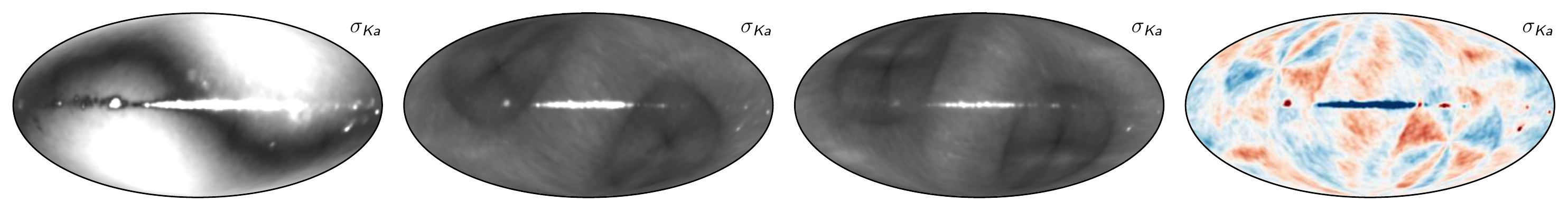}
	\includegraphics[width=0.9\textwidth]{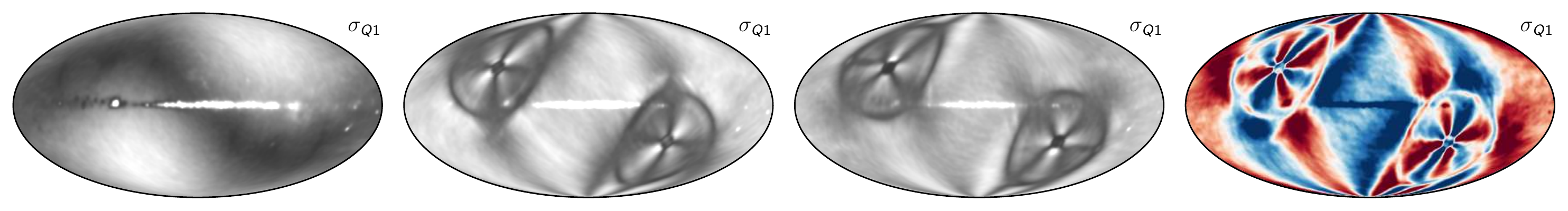}
	\includegraphics[width=0.9\textwidth]{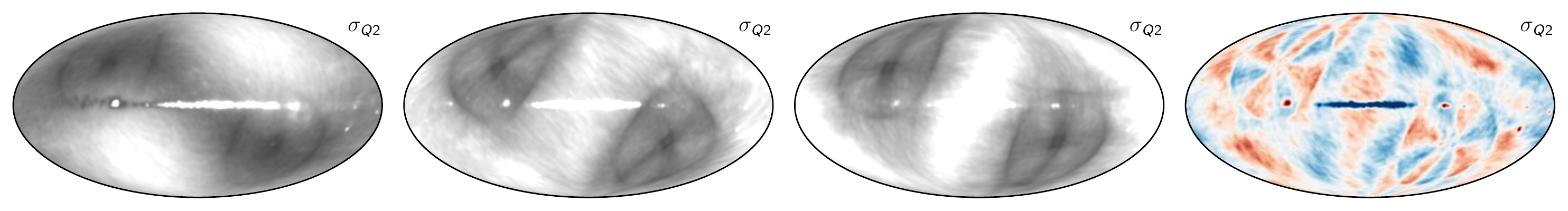}
	\includegraphics[width=0.9\textwidth]{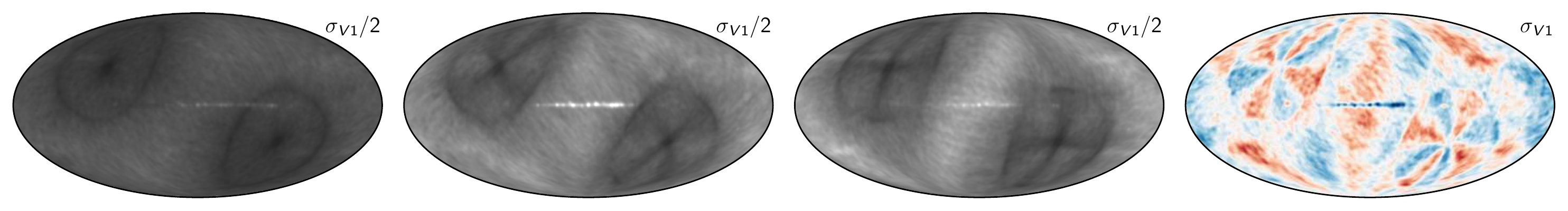}
	\includegraphics[width=0.9\textwidth]{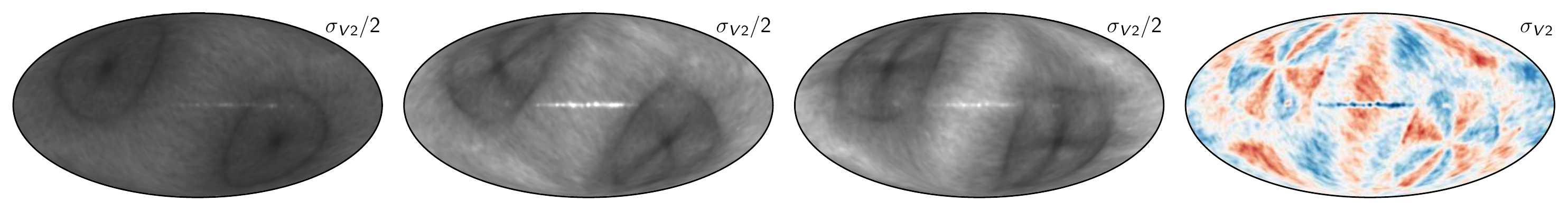}
	\includegraphics[width=0.9\textwidth]{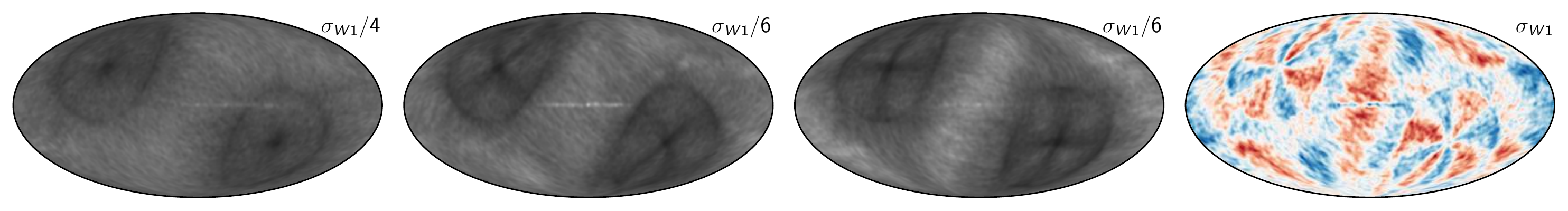}
	\includegraphics[width=0.9\textwidth]{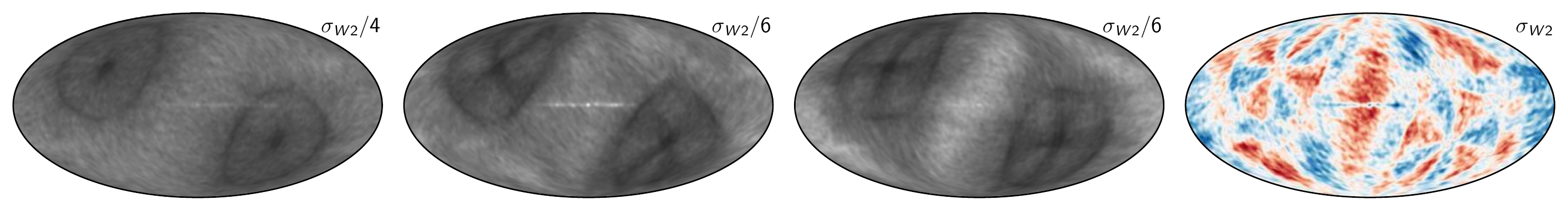}
	\includegraphics[width=0.9\textwidth]{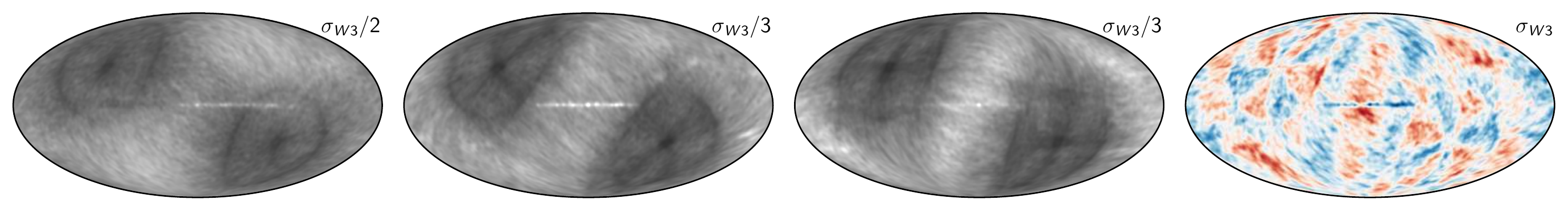}
	\includegraphics[width=0.9\textwidth]{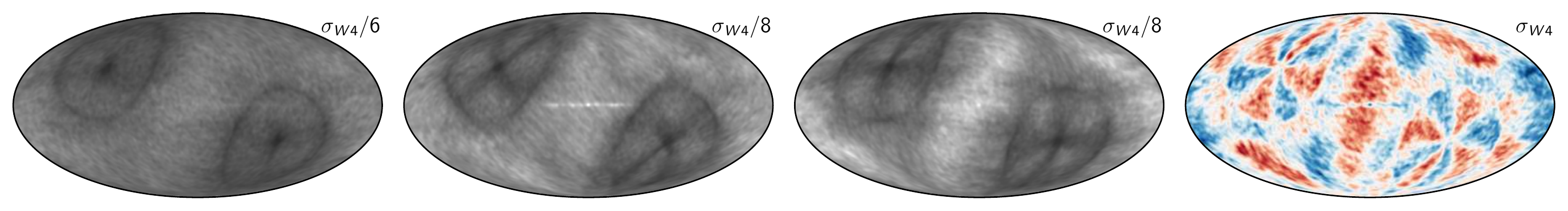}
	\includegraphics[width=0.22\textwidth]{figures/cbar_std.pdf}
	\includegraphics[width=0.22\textwidth]{figures/cbar_std.pdf}
	\includegraphics[width=0.22\textwidth]{figures/cbar_std.pdf}
	\includegraphics[width=0.22\textwidth]{figures/cbar_rho.pdf}
	\caption{\cosmoglobe\ posterior standard deviation for each DA and Stokes parameter, evaluated at an angular resolution of $2^{\circ}$ FWHM. The rightmost column shows the correlation coefficient between Stokes $Q$ and $U$.}
        \label{fig:std}
\end{figure*}

\begin{figure*}
	\centering
	\includegraphics[width=0.7\textwidth]{figures/023-WMAP_K_sampdiff.pdf}
	\includegraphics[width=0.7\textwidth]{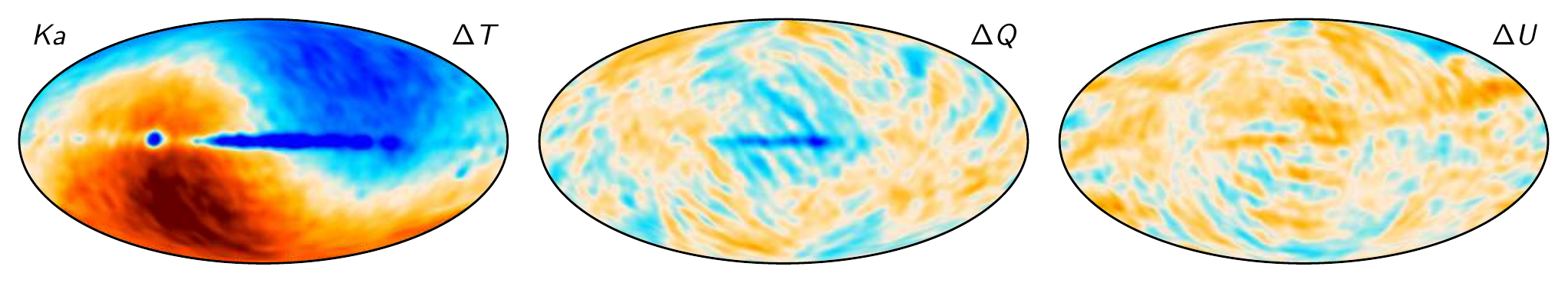}
	\includegraphics[width=0.7\textwidth]{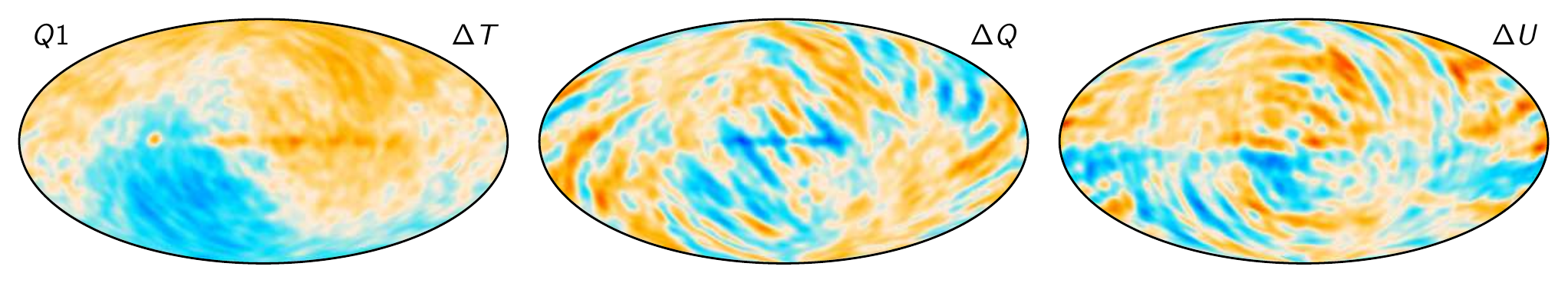}
	\includegraphics[width=0.7\textwidth]{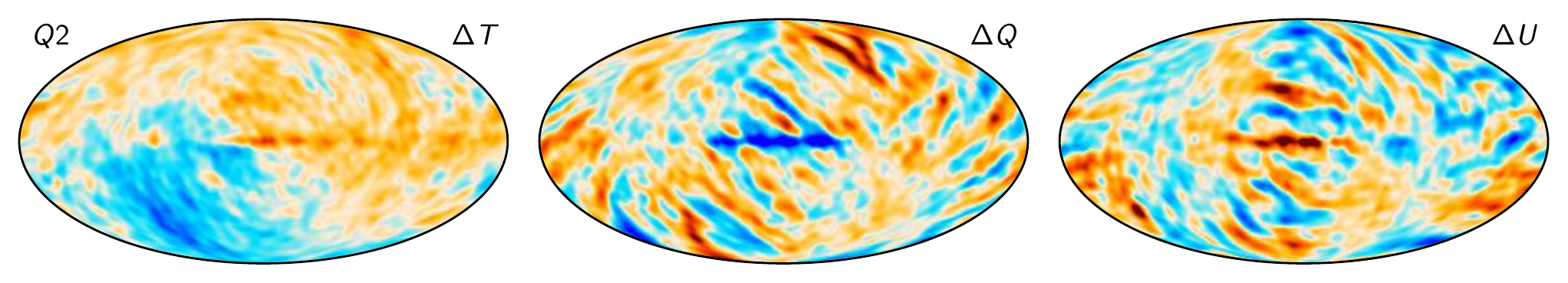}
	\includegraphics[width=0.7\textwidth]{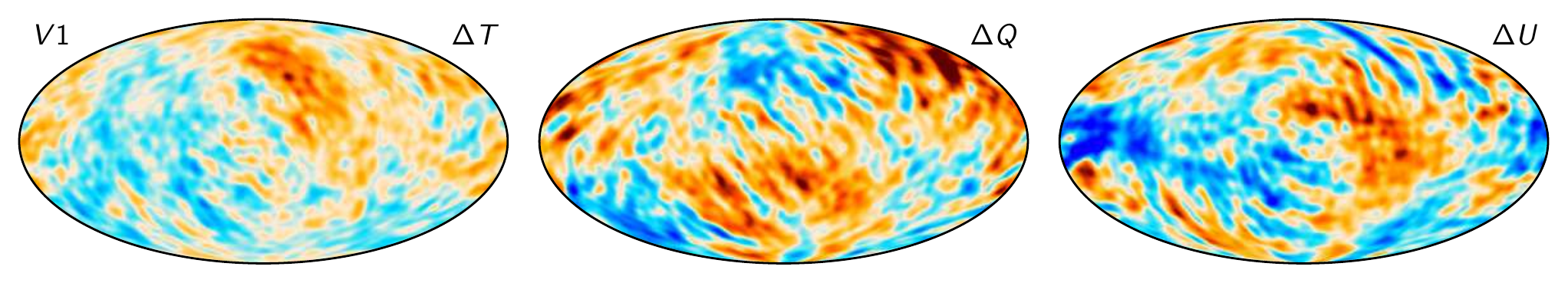}
	\includegraphics[width=0.7\textwidth]{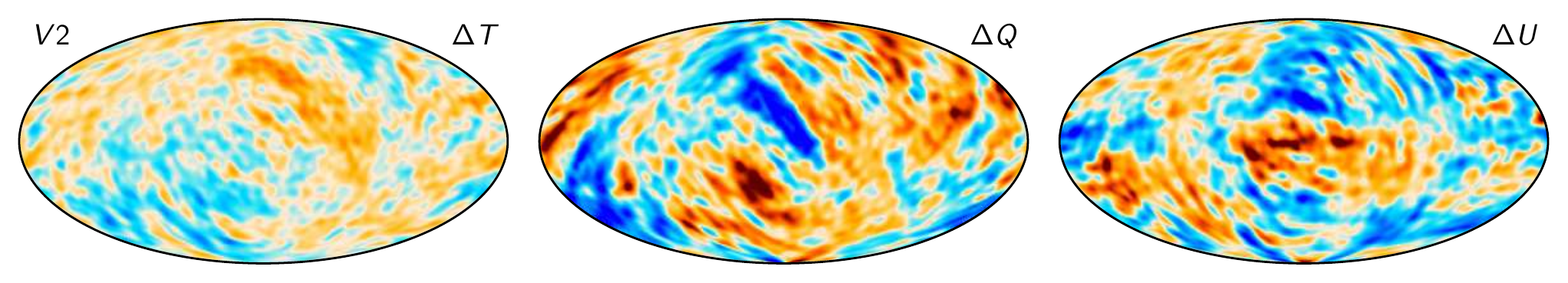}
	\includegraphics[width=0.7\textwidth]{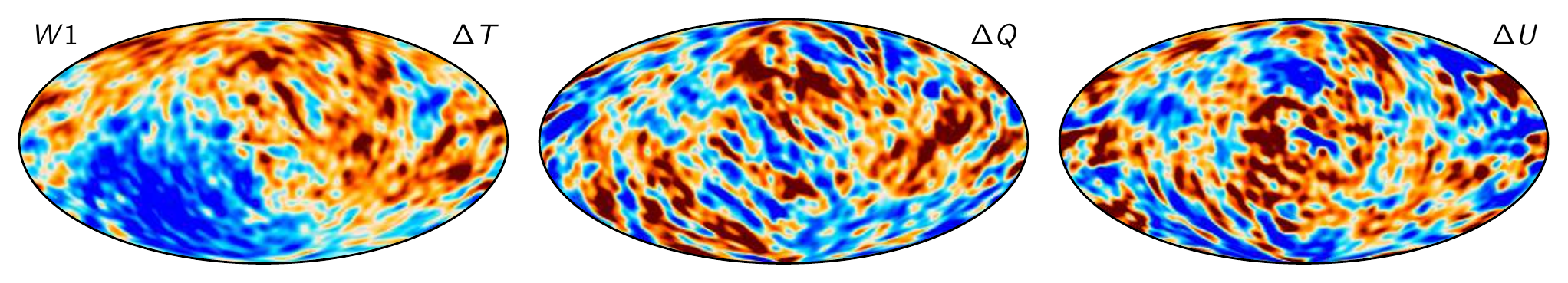}
	\includegraphics[width=0.7\textwidth]{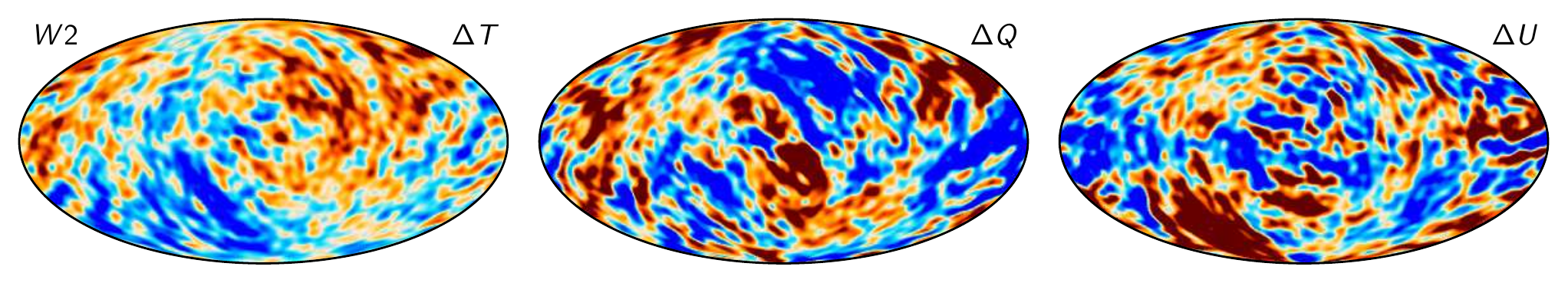}
	\includegraphics[width=0.7\textwidth]{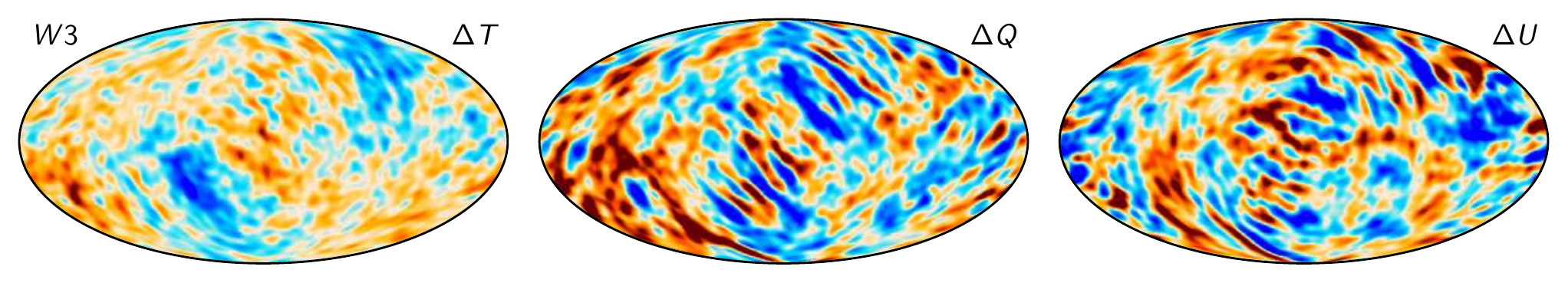}
	\includegraphics[width=0.7\textwidth]{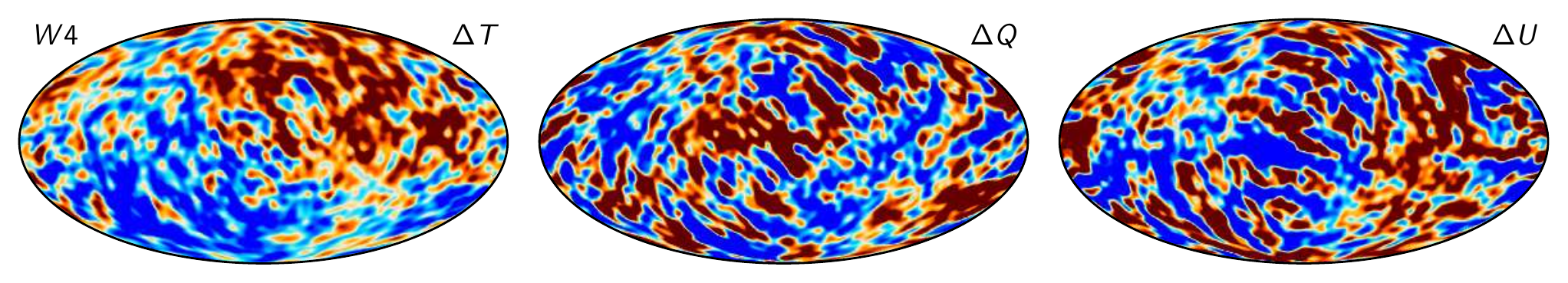}\\
	\includegraphics[width=0.30\textwidth]{figures/cbar_3uK.pdf}
	\caption{Differences between two \cosmoglobe\ frequency map samples, smoothed with a $5^\circ$ FWHM Gaussian beam.}
        \label{fig:sampdiff}
\end{figure*}

\begin{figure*}
	\centering
	\includegraphics[width=0.7\textwidth]{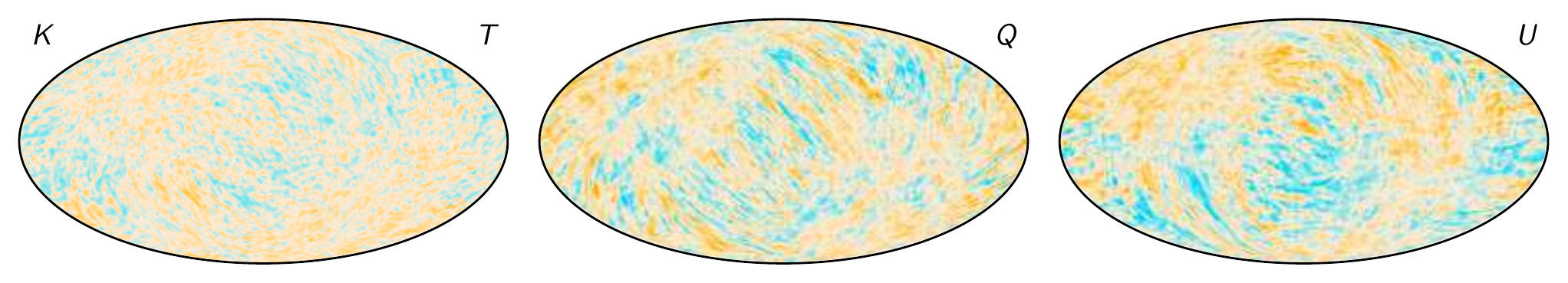}\\
	\includegraphics[width=0.7\textwidth]{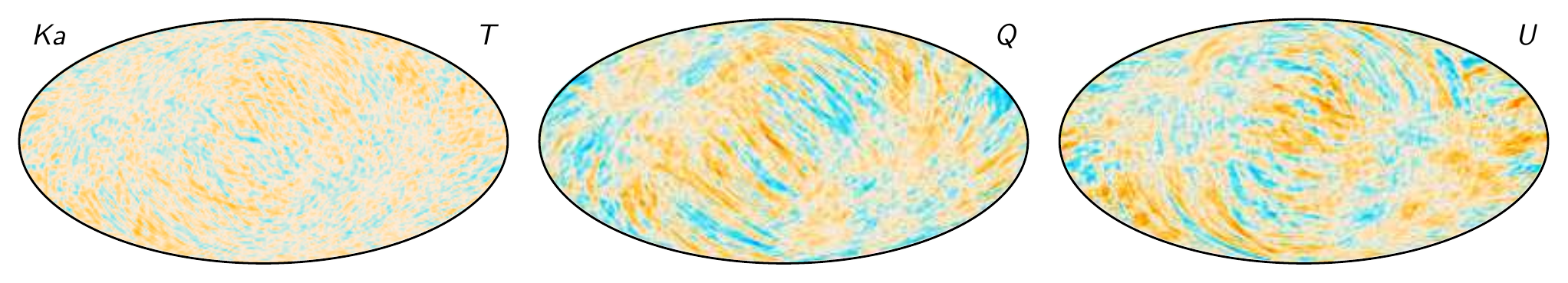}\\
	\includegraphics[width=0.7\textwidth]{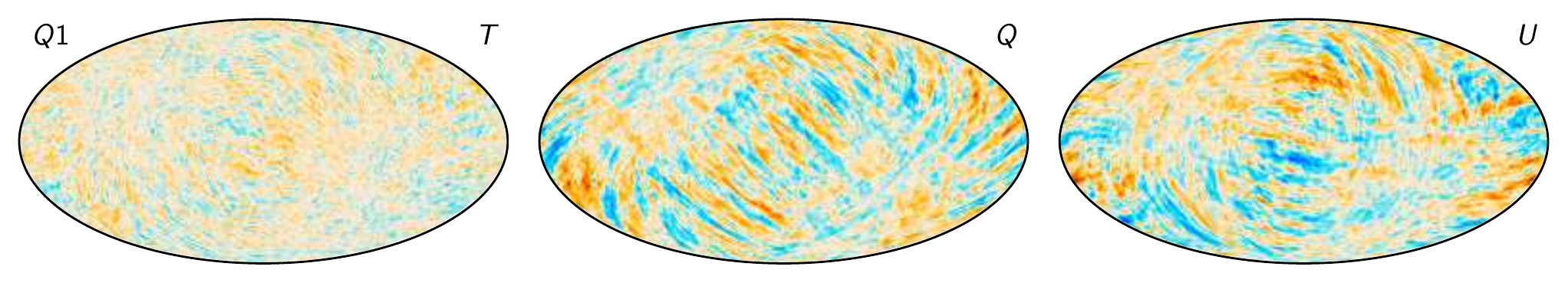}\\
	\includegraphics[width=0.7\textwidth]{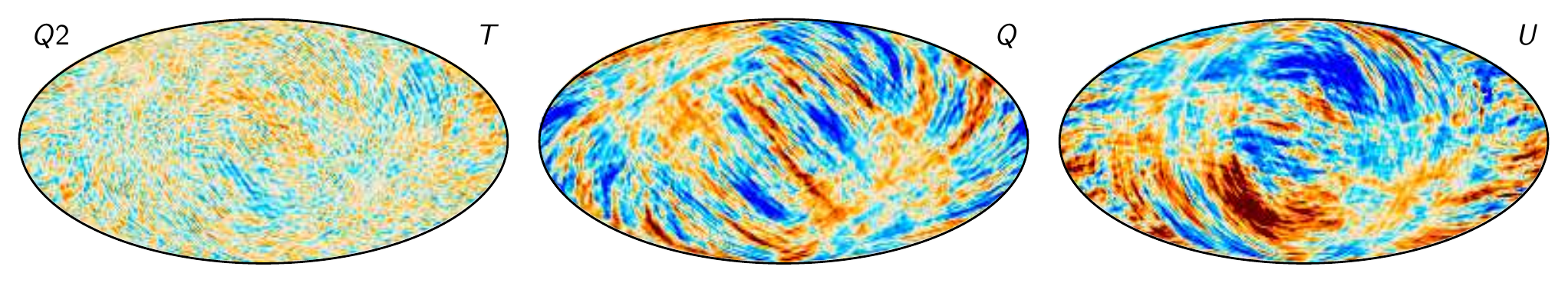}\\
	\includegraphics[width=0.7\textwidth]{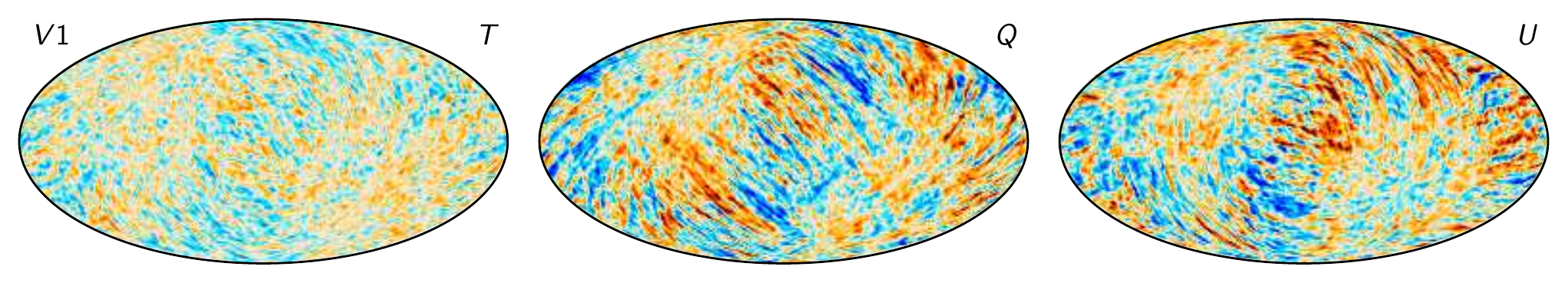}\\
	\includegraphics[width=0.7\textwidth]{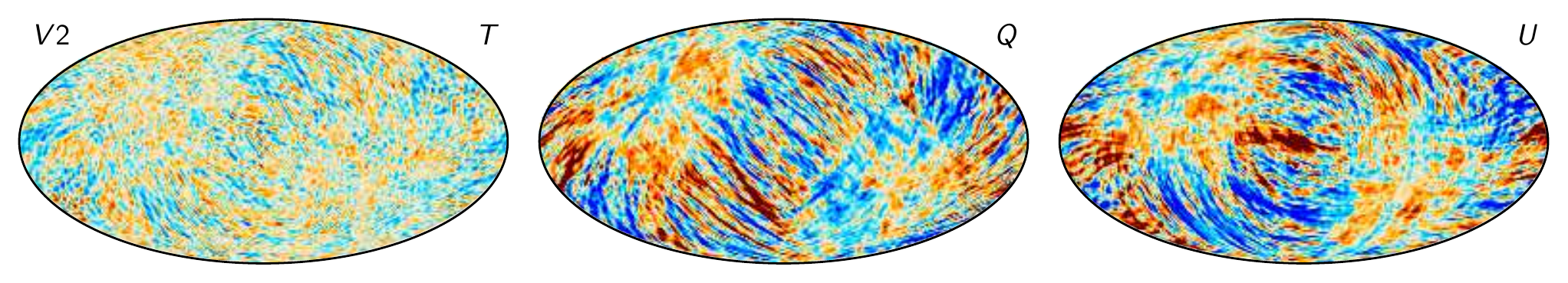}\\
	\includegraphics[width=0.7\textwidth]{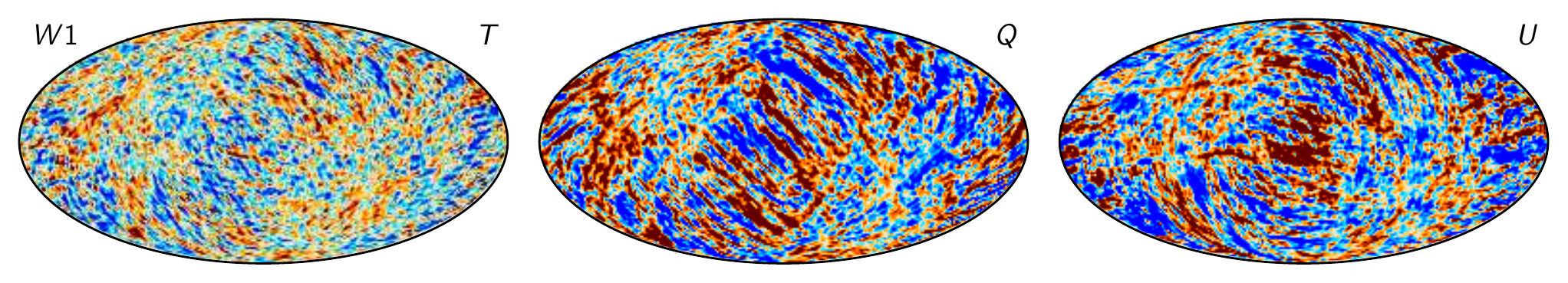}\\
	\includegraphics[width=0.7\textwidth]{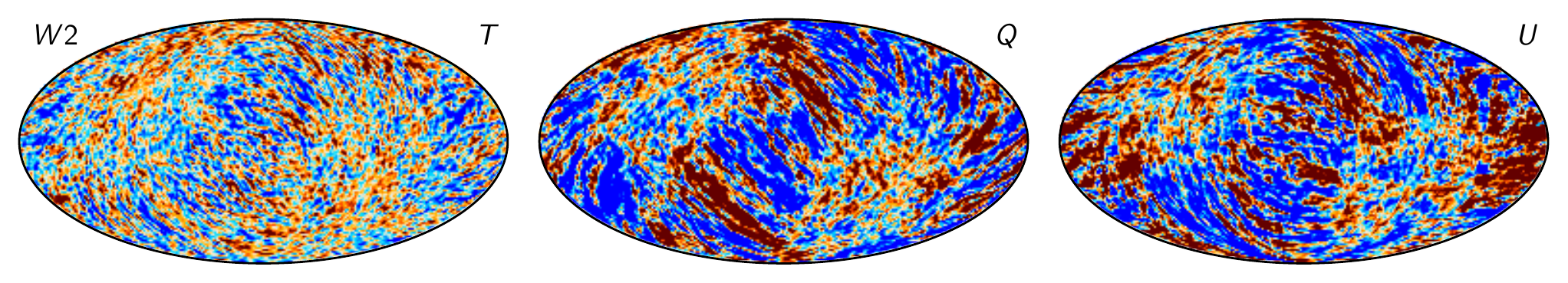}\\
	\includegraphics[width=0.7\textwidth]{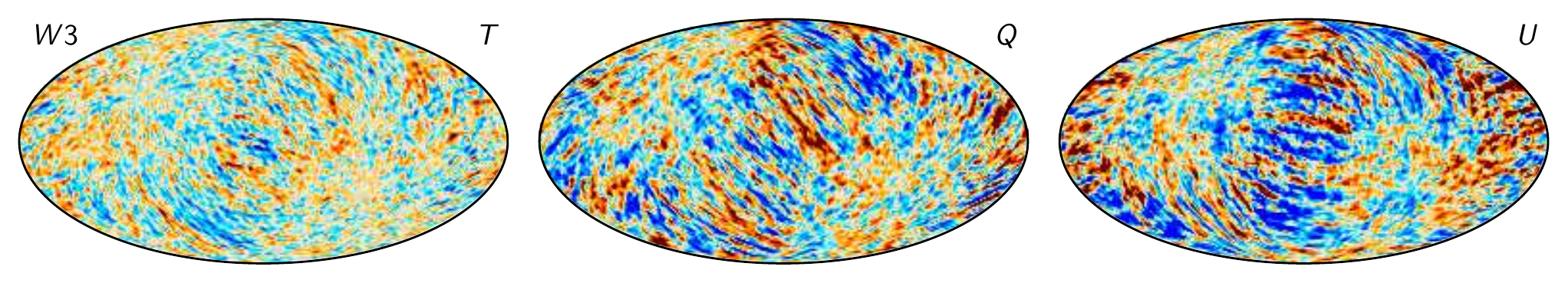}\\
	\includegraphics[width=0.7\textwidth]{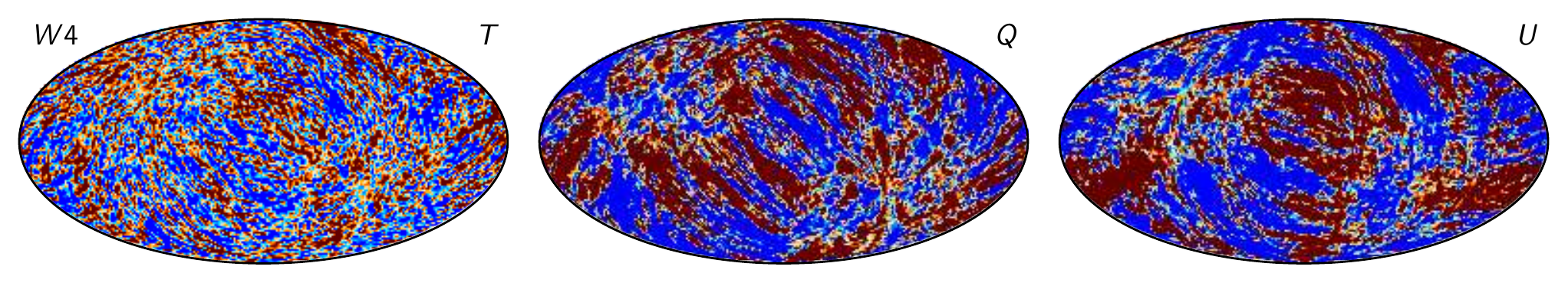}\\
	\includegraphics[width=0.30\textwidth]{figures/cbar_3uK.pdf}
	\caption{Binned correlated noise timestreams for each DA, smoothed with a $2^\circ$ FWHM Gaussian beam.}
        \label{fig:ncorr}
\end{figure*}

\begin{figure*}
	\centering
	\includegraphics[width=0.7\textwidth]{figures/tod_res_K_IQU.pdf}\\
	\includegraphics[width=0.7\textwidth]{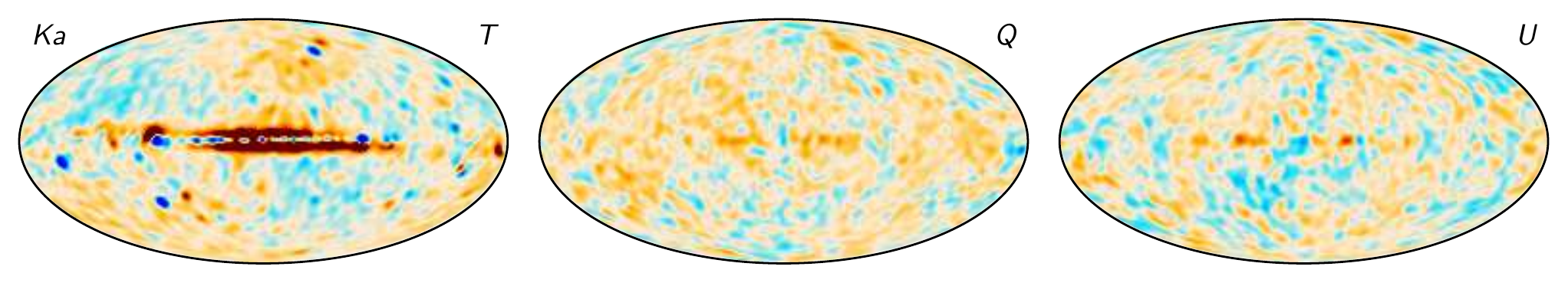}\\
	\includegraphics[width=0.7\textwidth]{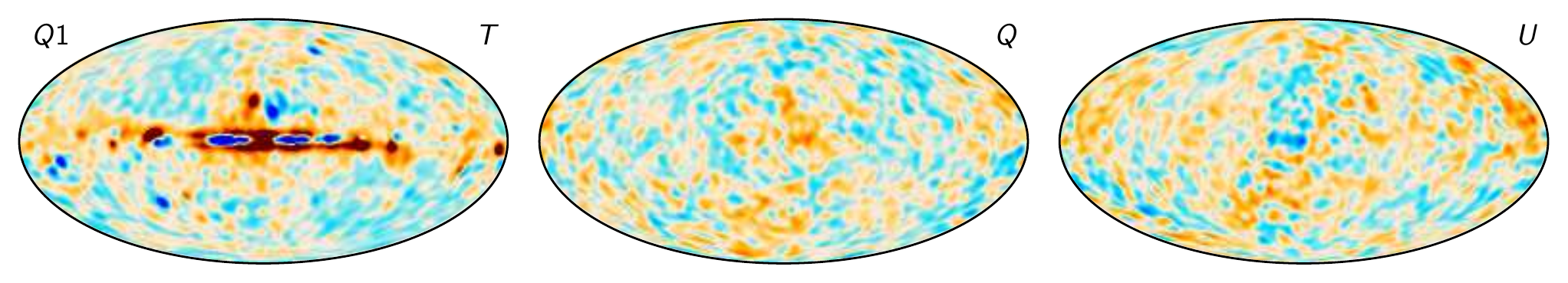}\\
	\includegraphics[width=0.7\textwidth]{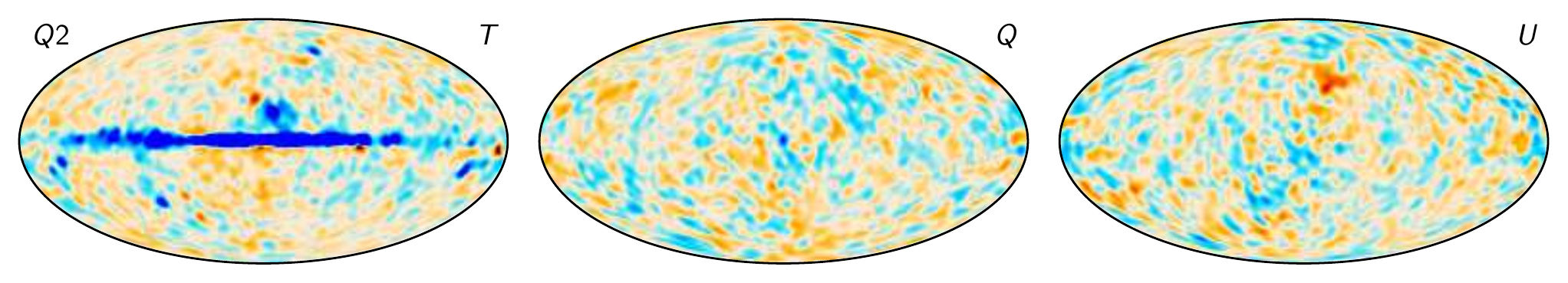}\\
	\includegraphics[width=0.7\textwidth]{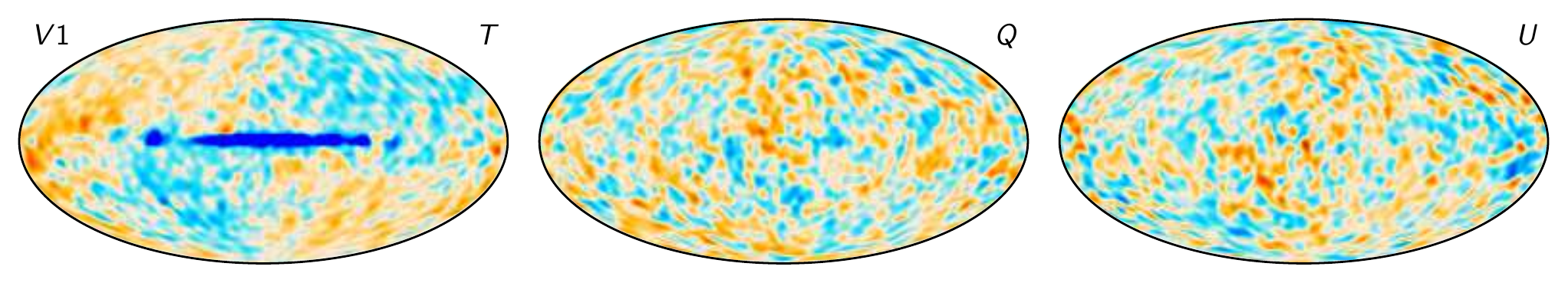}\\
	\includegraphics[width=0.7\textwidth]{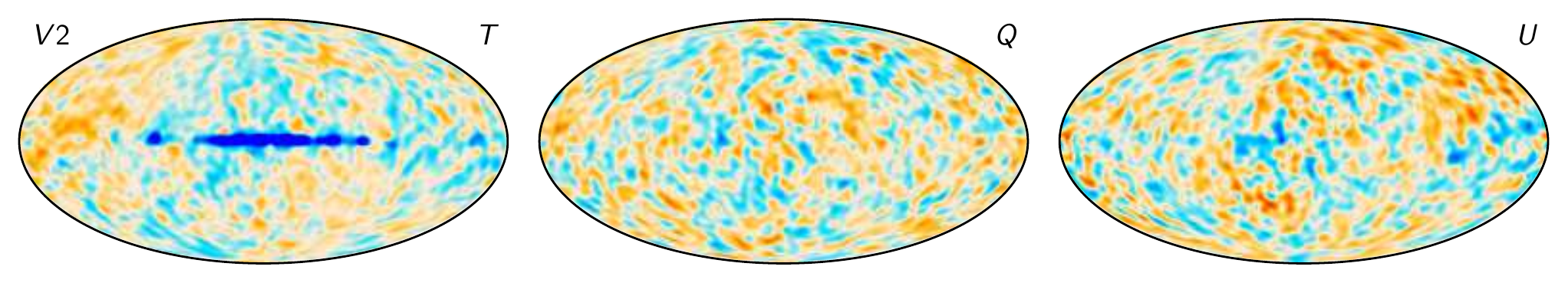}\\
	\includegraphics[width=0.7\textwidth]{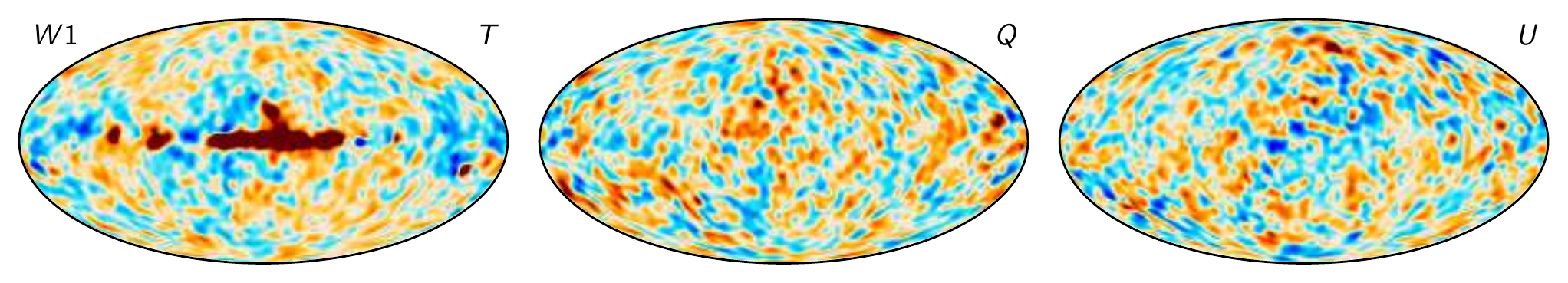}\\
	\includegraphics[width=0.7\textwidth]{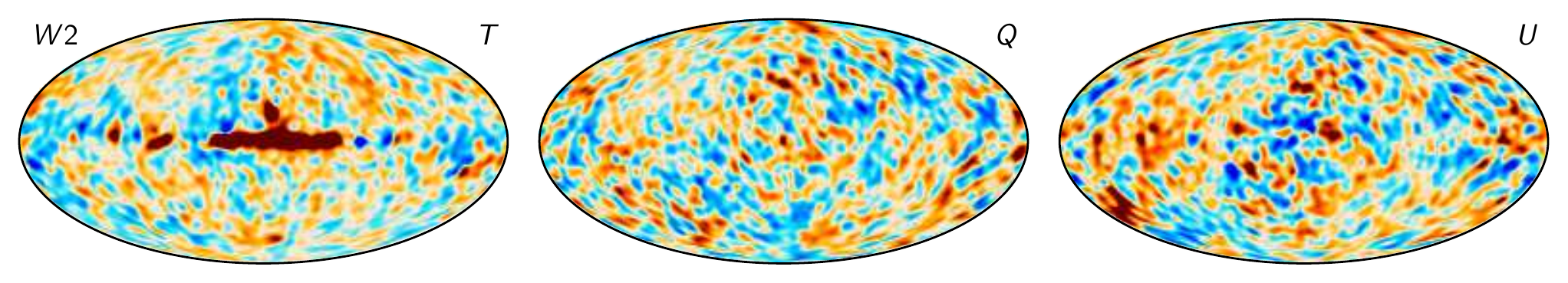}\\
	\includegraphics[width=0.7\textwidth]{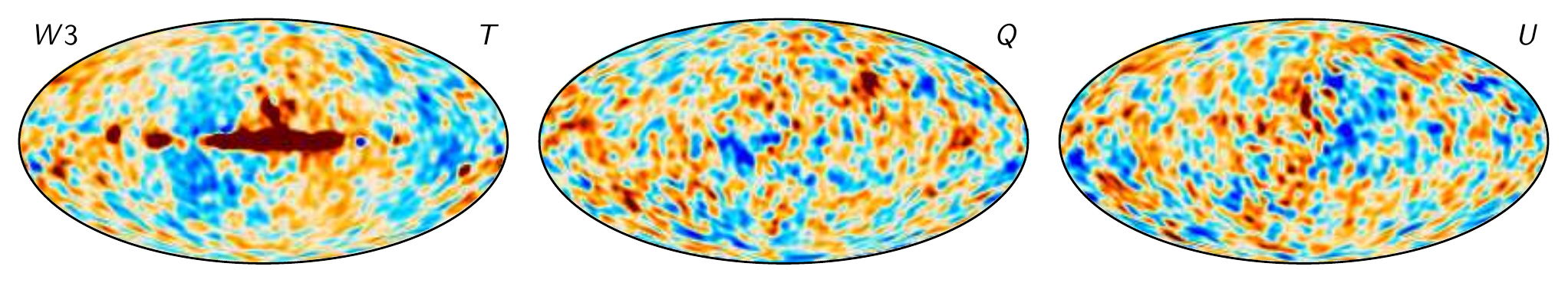}\\
	\includegraphics[width=0.7\textwidth]{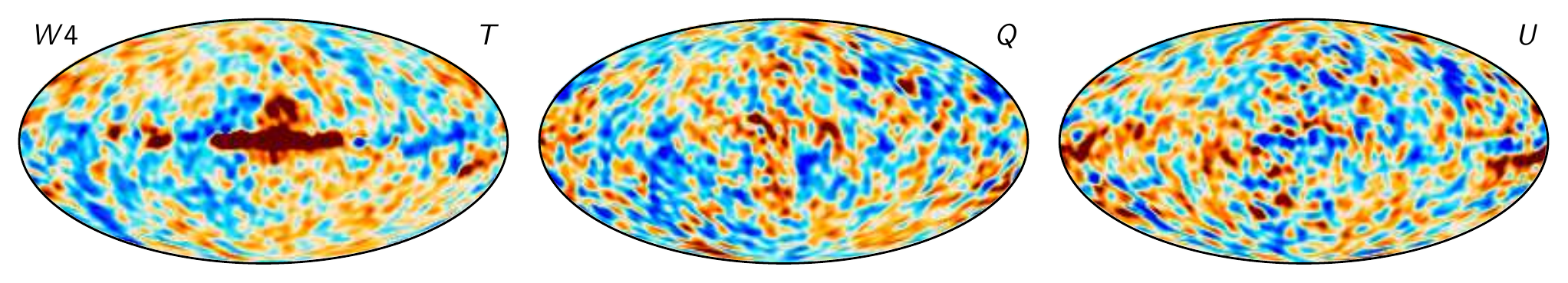}\\
	\includegraphics[width=0.30\textwidth]{figures/cbar_10uK.pdf}
	\caption{Binned TOD-level residuals for each DA, smoothed with a $5^\circ$ FWHM Gaussian beam.}
        \label{fig:todres}
\end{figure*}

\begin{figure*}
	\centering
	\includegraphics[width=0.7\textwidth]{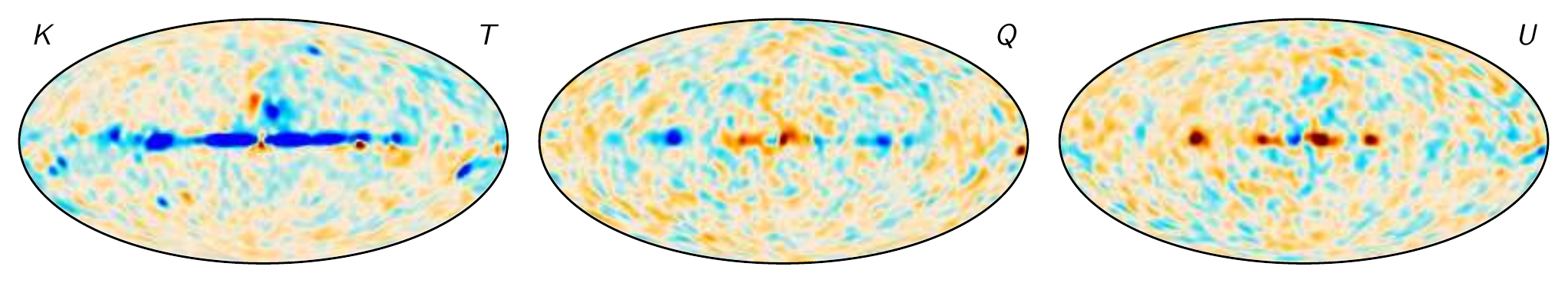}\\
	\includegraphics[width=0.7\textwidth]{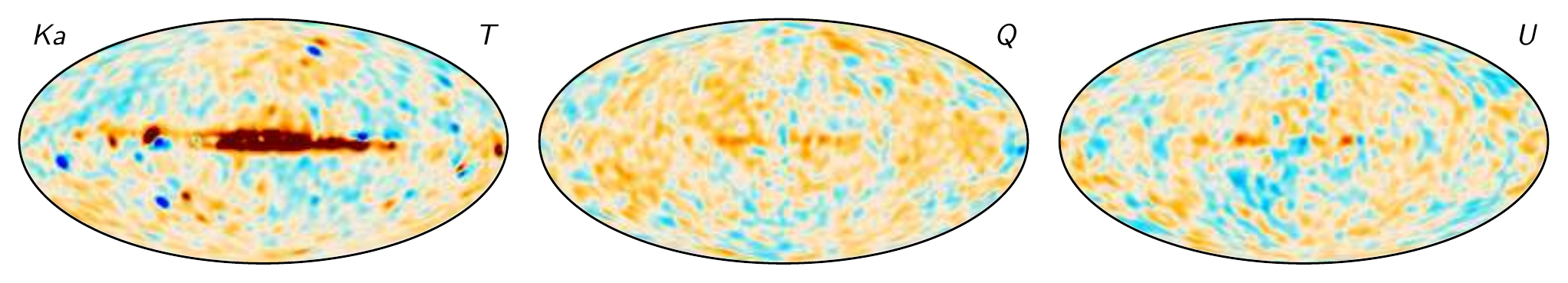}\\
	\includegraphics[width=0.7\textwidth]{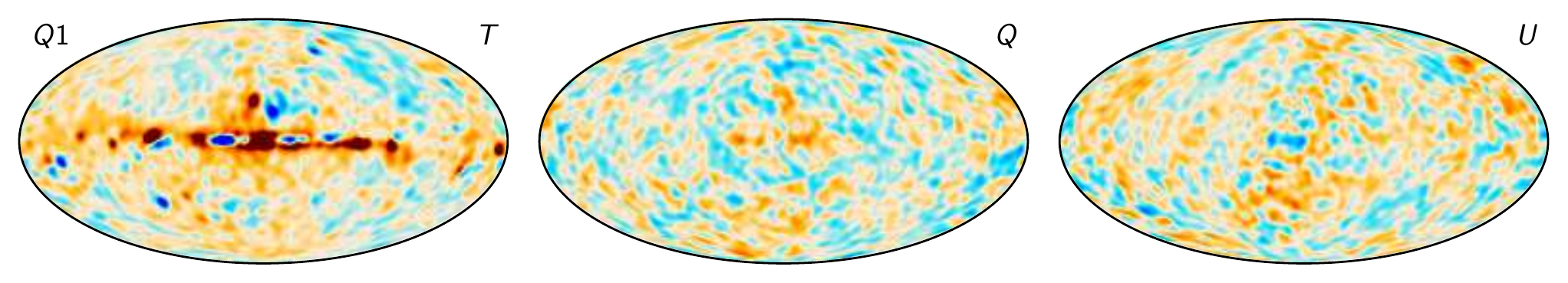}\\
	\includegraphics[width=0.7\textwidth]{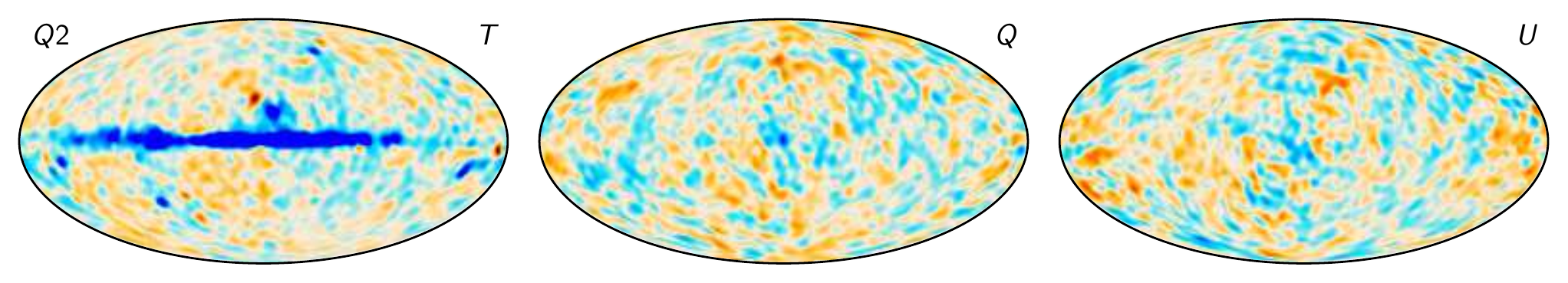}\\
	\includegraphics[width=0.7\textwidth]{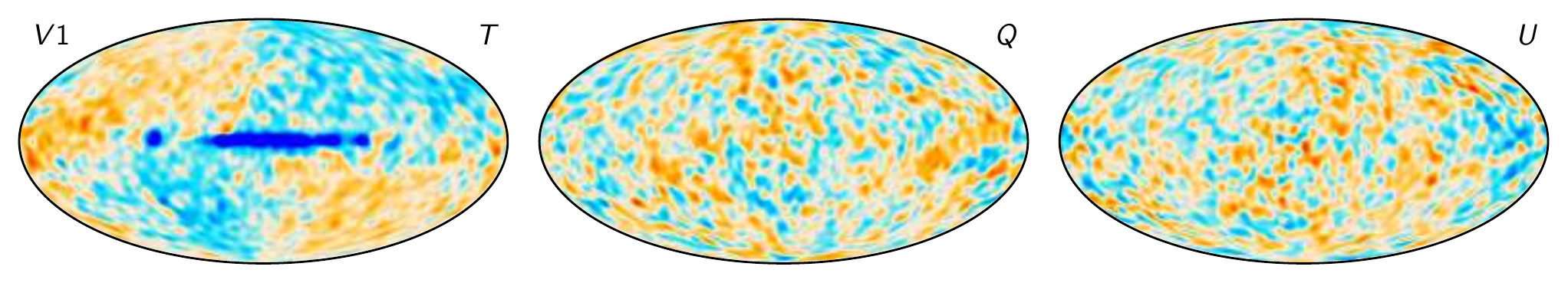}\\
	\includegraphics[width=0.7\textwidth]{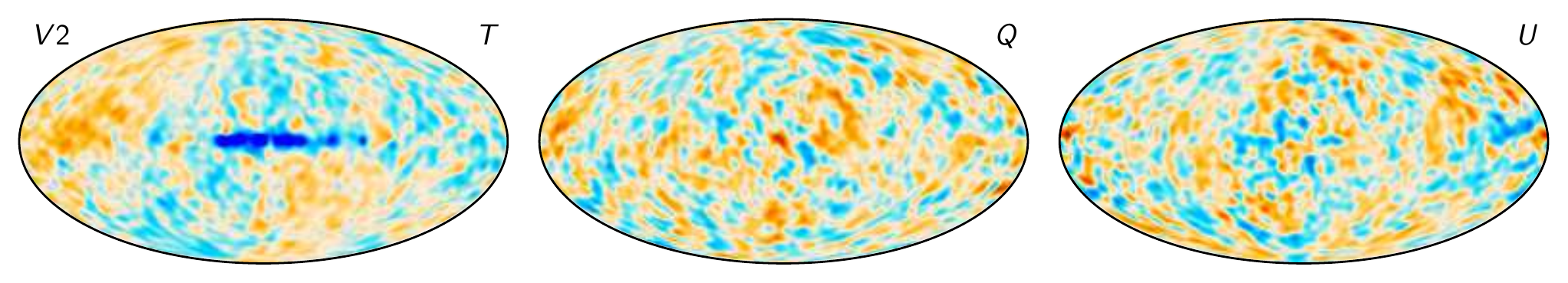}\\
	\includegraphics[width=0.7\textwidth]{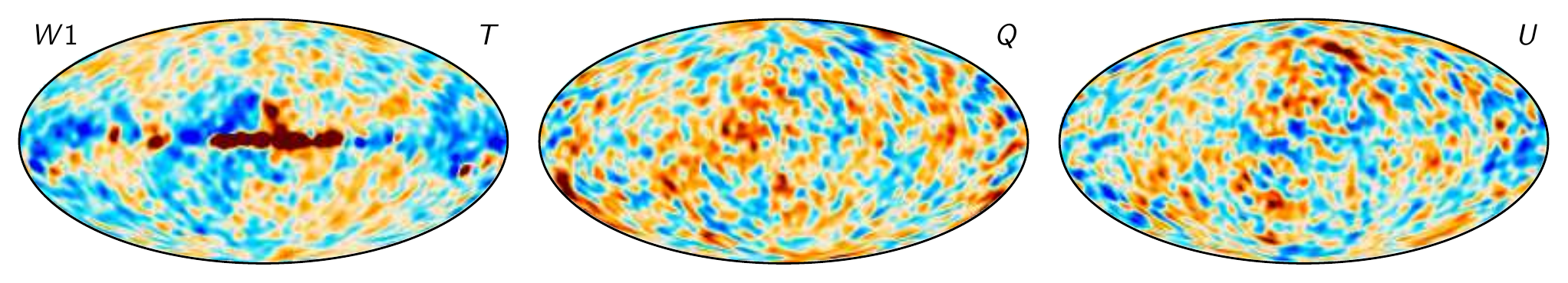}\\
	\includegraphics[width=0.7\textwidth]{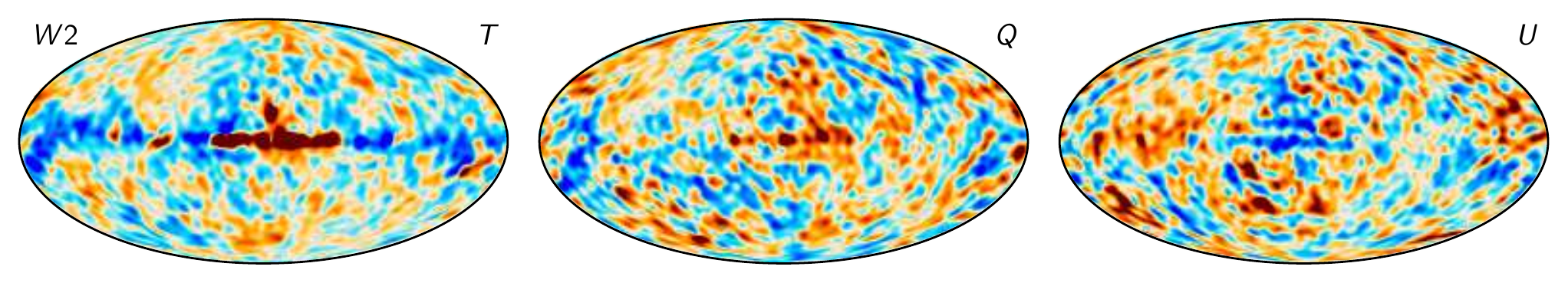}\\
	\includegraphics[width=0.7\textwidth]{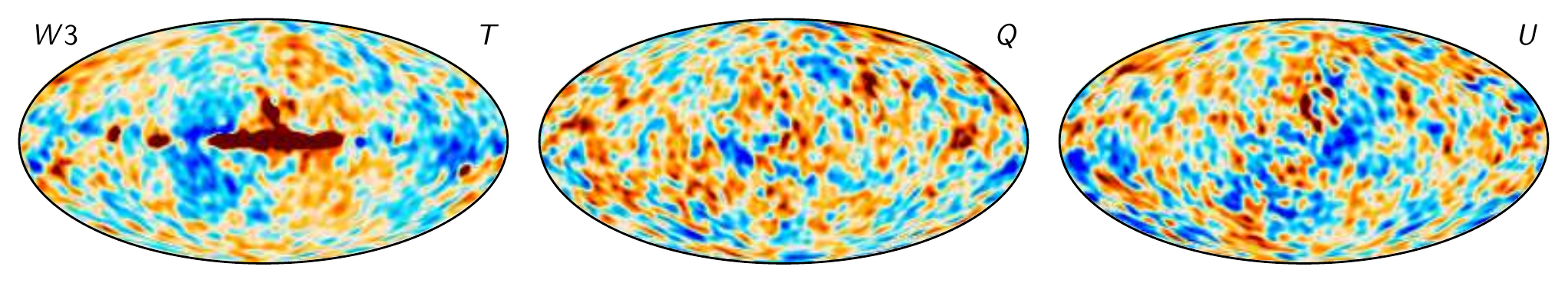}\\
	\includegraphics[width=0.7\textwidth]{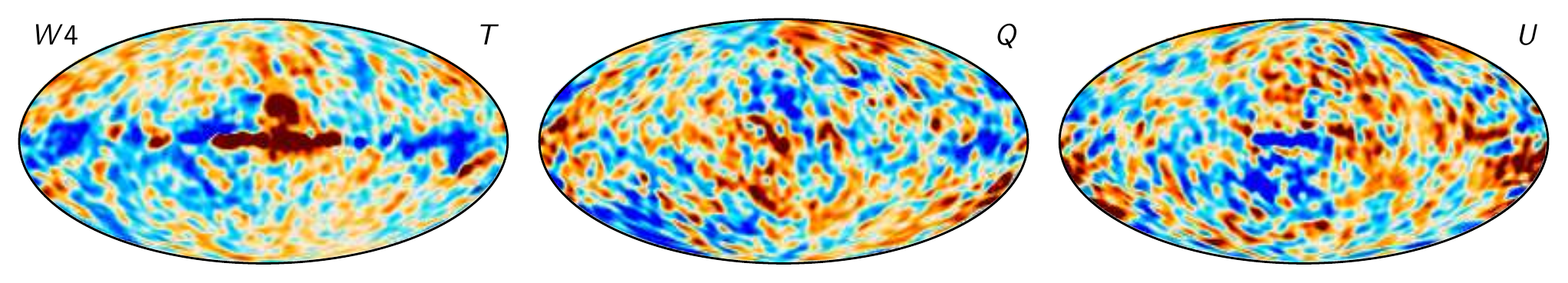}\\
	\includegraphics[width=0.30\textwidth]{figures/cbar_10uK.pdf}
	\caption{Map-level residuals for each DA, smoothed by $5^\circ$.}
	\label{fig:compsep_residual}
\end{figure*}

\noindent\begin{minipage}{\textwidth}
In this appendix, we provide a frequency map survey for all ten DAs. Figure~\ref{fig:skymaps} shows the full DA polarization maps as derived both by \cosmoglobe\ and \WMAPnine, as well as their differences. Figures~\ref{fig:rms} and \ref{fig:std} shows \cosmoglobe\ white noise and posterior rms for each DA and Stokes parameter, as well as the cross-correlation between Stokes $Q$ and $U$. Figure~\ref{fig:sampdiff} shows differences between two Gibbs samples, Figs.~\ref{fig:ncorr} and \ref{fig:todres} show the TOD-level correlated noise and residual, respectively, both obtained by projecting the timestreams into sky maps. Finally, Fig.~\ref{fig:compsep_residual} shows the map-level residual, obtained by subtracting the astrophysical sky model from the corresponding DA map.
\end{minipage}

\noindent\begin{minipage}{\textwidth}
\vspace*{4mm}
\centering
        \includegraphics[width=0.16\linewidth]{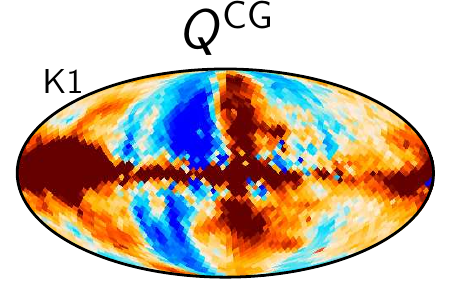}
        \includegraphics[width=0.16\linewidth]{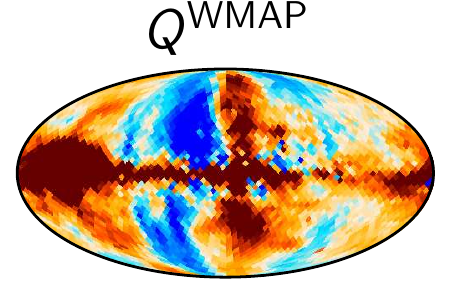}
        \includegraphics[width=0.16\linewidth]{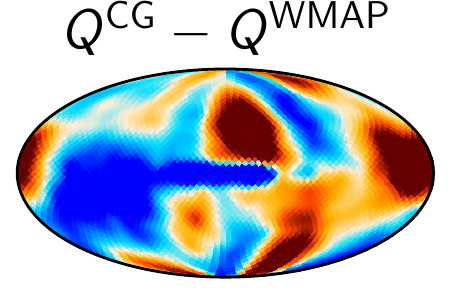}\hspace*{2mm}
        \includegraphics[width=0.16\linewidth]{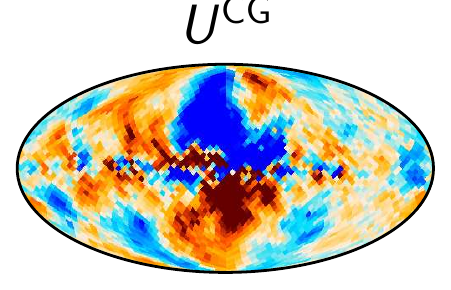}
        \includegraphics[width=0.16\linewidth]{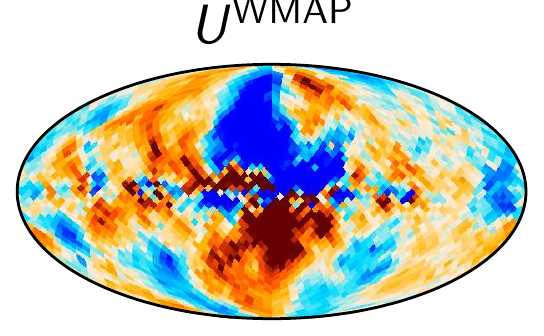}
        \includegraphics[width=0.16\linewidth]{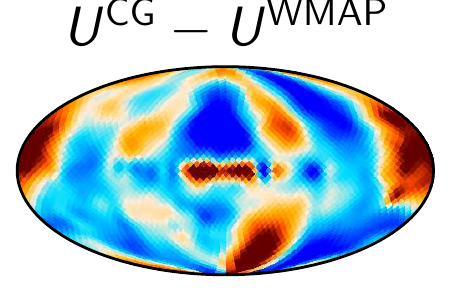}\\
        \includegraphics[width=0.16\linewidth]{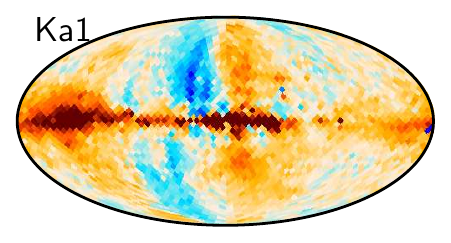}
        \includegraphics[width=0.16\linewidth]{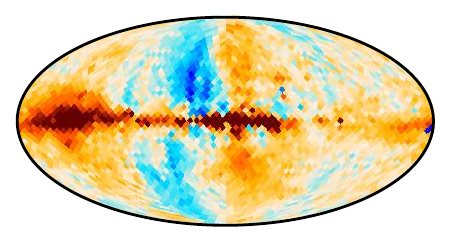}
        \includegraphics[width=0.16\linewidth]{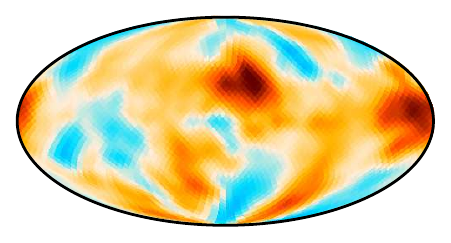}\hspace*{2mm}
        \includegraphics[width=0.16\linewidth]{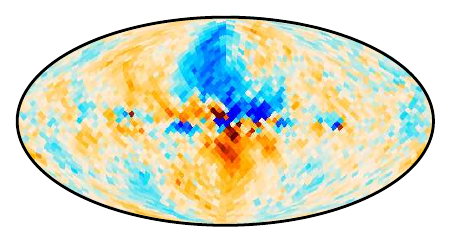}
        \includegraphics[width=0.16\linewidth]{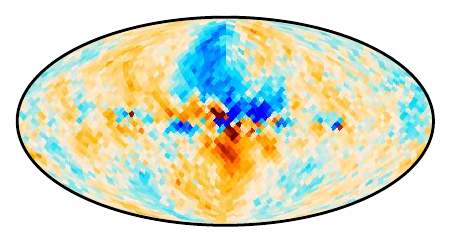}
        \includegraphics[width=0.16\linewidth]{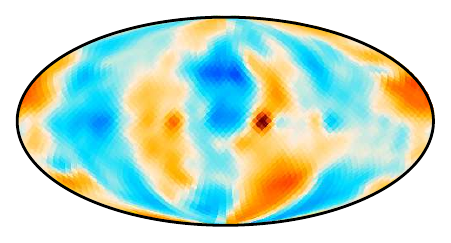}\\
        \includegraphics[width=0.16\linewidth]{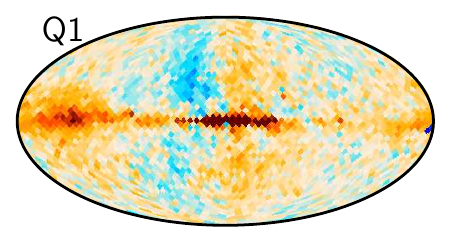}
        \includegraphics[width=0.16\linewidth]{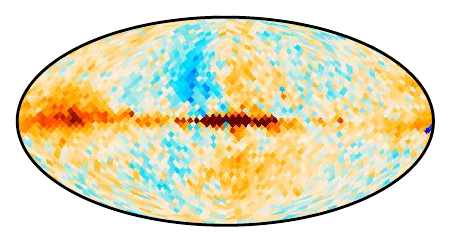}
        \includegraphics[width=0.16\linewidth]{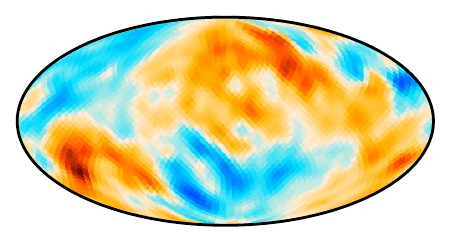}\hspace*{2mm}
        \includegraphics[width=0.16\linewidth]{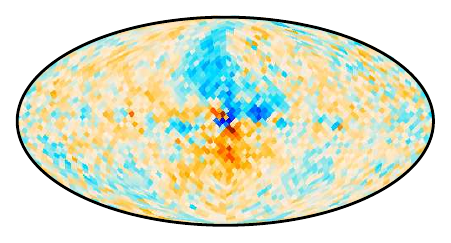}
        \includegraphics[width=0.16\linewidth]{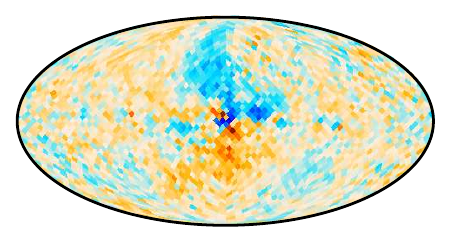}
        \includegraphics[width=0.16\linewidth]{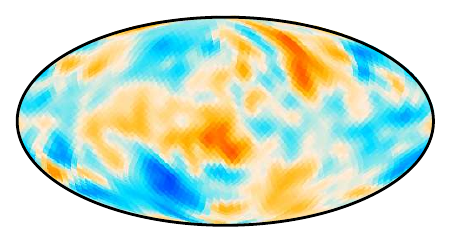}\\
        \includegraphics[width=0.16\linewidth]{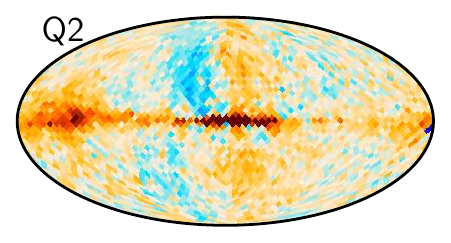}
        \includegraphics[width=0.16\linewidth]{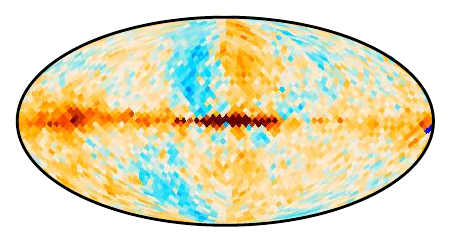}
        \includegraphics[width=0.16\linewidth]{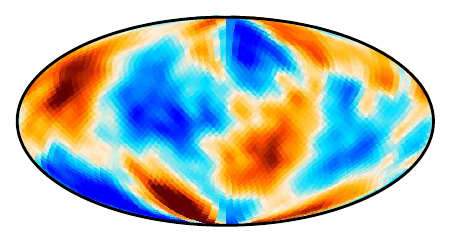}\hspace*{2mm}
        \includegraphics[width=0.16\linewidth]{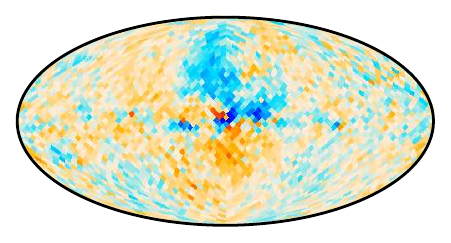}
        \includegraphics[width=0.16\linewidth]{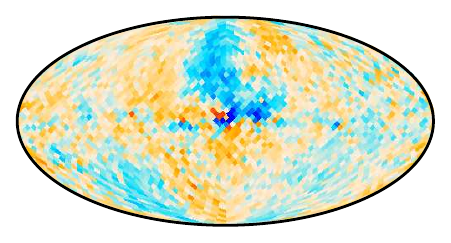}
        \includegraphics[width=0.16\linewidth]{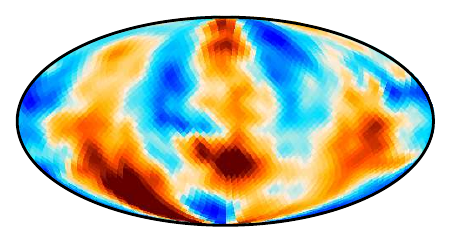}\\
        \includegraphics[width=0.16\linewidth]{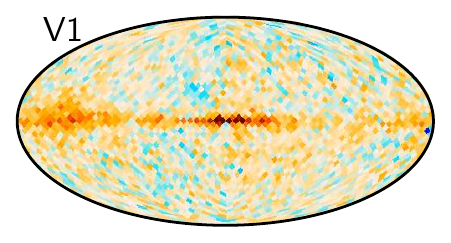}
        \includegraphics[width=0.16\linewidth]{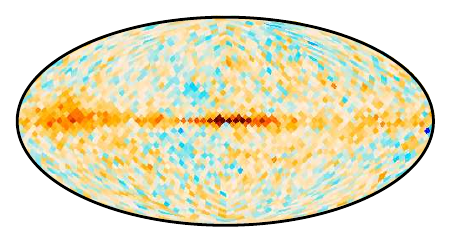}
        \includegraphics[width=0.16\linewidth]{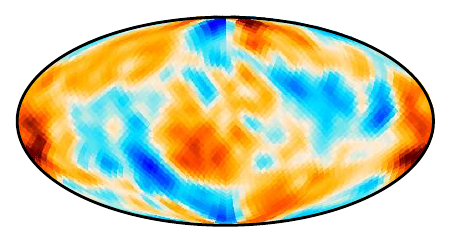}\hspace*{2mm}
        \includegraphics[width=0.16\linewidth]{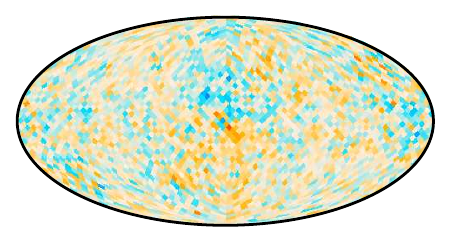}
        \includegraphics[width=0.16\linewidth]{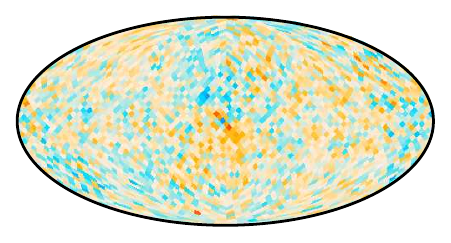}
        \includegraphics[width=0.16\linewidth]{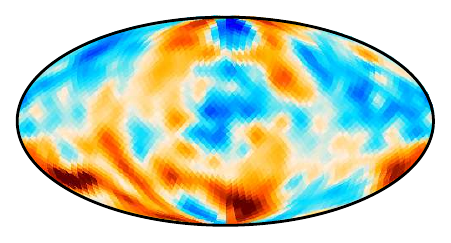}\\
        \includegraphics[width=0.16\linewidth]{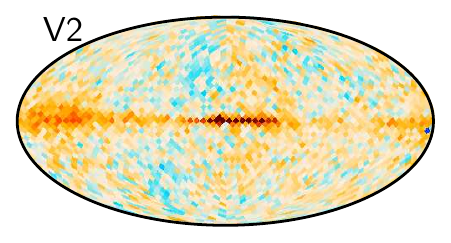}
        \includegraphics[width=0.16\linewidth]{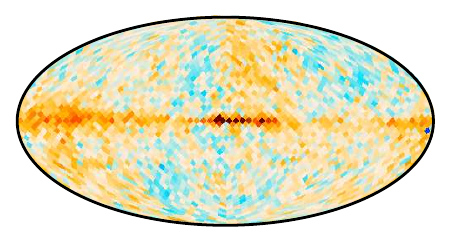}
        \includegraphics[width=0.16\linewidth]{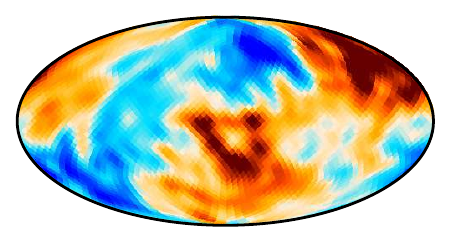}\hspace*{2mm}
        \includegraphics[width=0.16\linewidth]{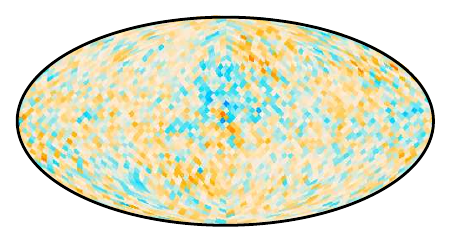}
        \includegraphics[width=0.16\linewidth]{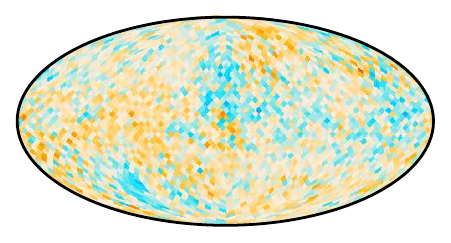}
        \includegraphics[width=0.16\linewidth]{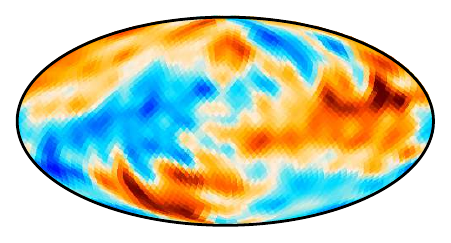}\\
        \includegraphics[width=0.16\linewidth]{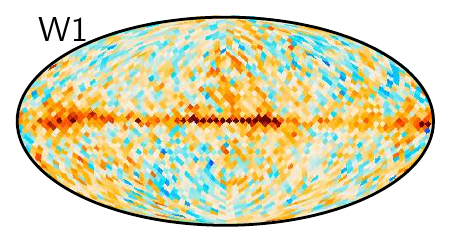}
        \includegraphics[width=0.16\linewidth]{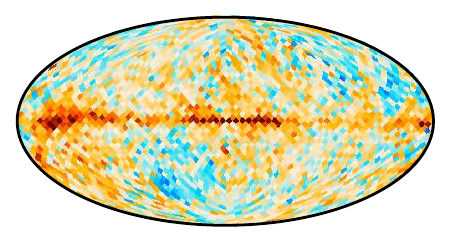}
        \includegraphics[width=0.16\linewidth]{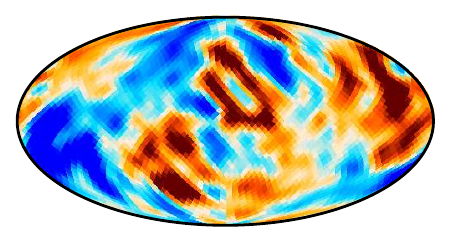}\hspace*{2mm}
        \includegraphics[width=0.16\linewidth]{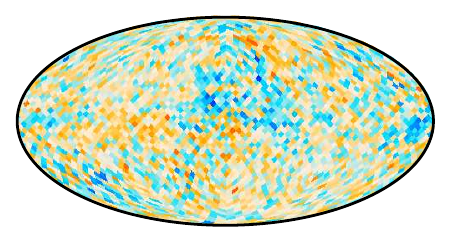}
        \includegraphics[width=0.16\linewidth]{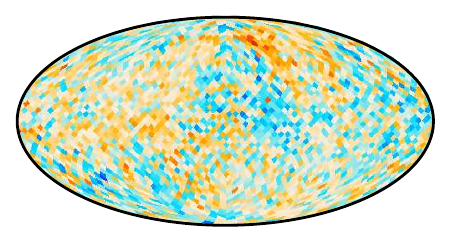}
        \includegraphics[width=0.16\linewidth]{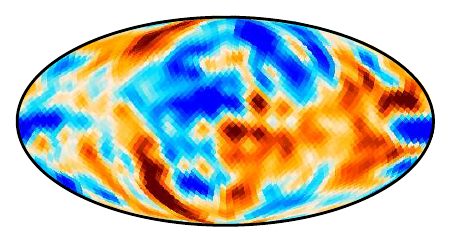}\\
        \includegraphics[width=0.16\linewidth]{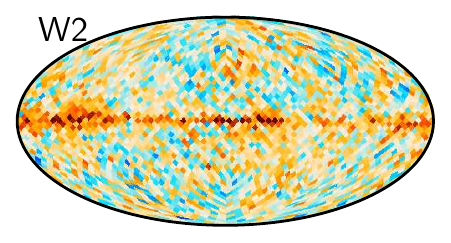}
        \includegraphics[width=0.16\linewidth]{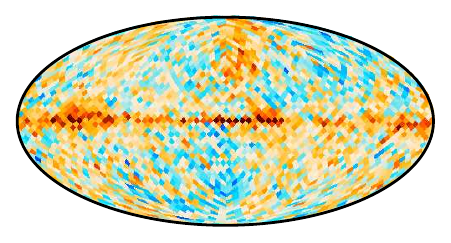}
        \includegraphics[width=0.16\linewidth]{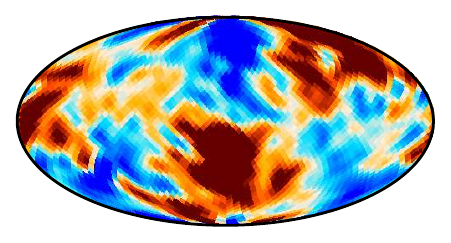}\hspace*{2mm}
        \includegraphics[width=0.16\linewidth]{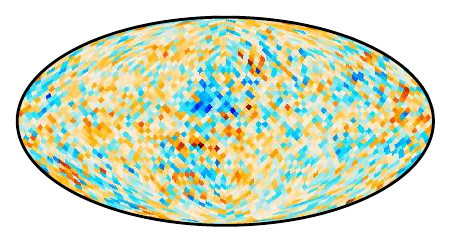}
        \includegraphics[width=0.16\linewidth]{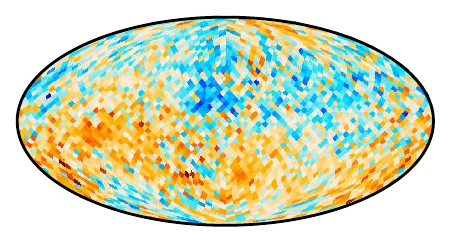}
        \includegraphics[width=0.16\linewidth]{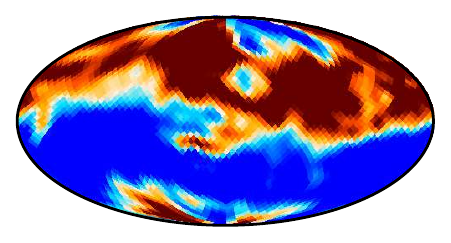}\\
        \includegraphics[width=0.16\linewidth]{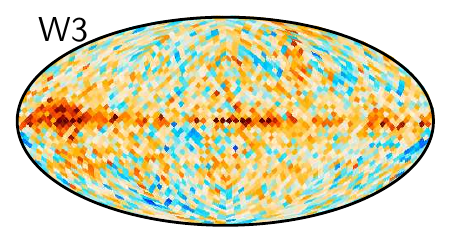}
        \includegraphics[width=0.16\linewidth]{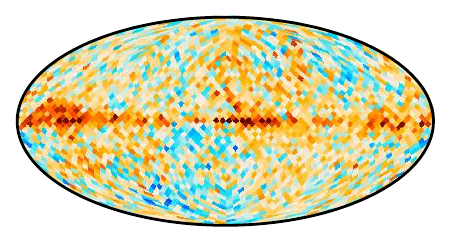}
        \includegraphics[width=0.16\linewidth]{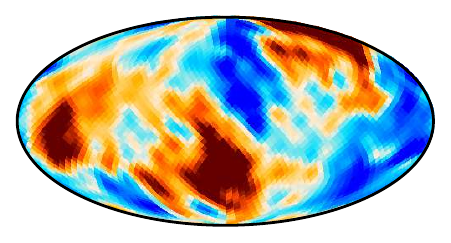}\hspace*{2mm}
        \includegraphics[width=0.16\linewidth]{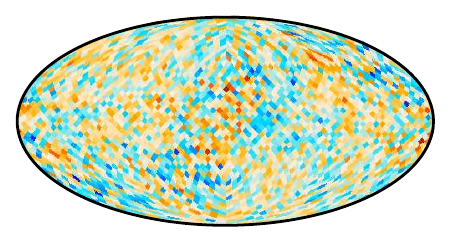}
        \includegraphics[width=0.16\linewidth]{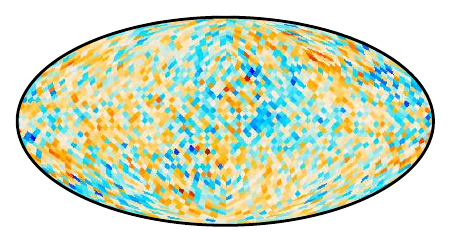}
        \includegraphics[width=0.16\linewidth]{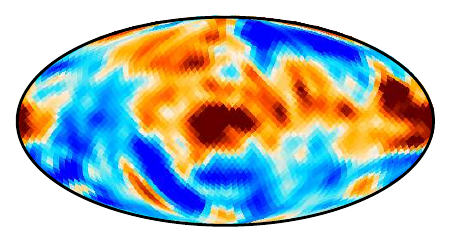}\\
        \includegraphics[width=0.16\linewidth]{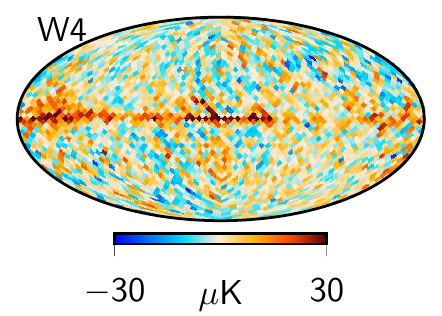}
        \includegraphics[width=0.16\linewidth]{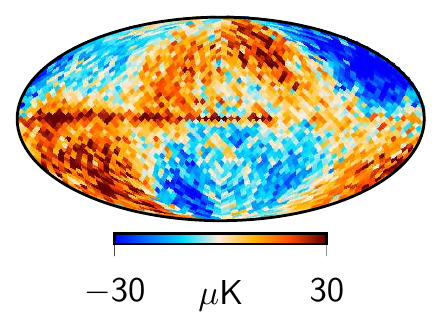}
        \includegraphics[width=0.16\linewidth]{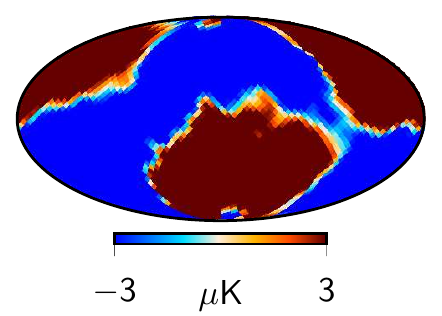}\hspace*{2mm}
        \includegraphics[width=0.16\linewidth]{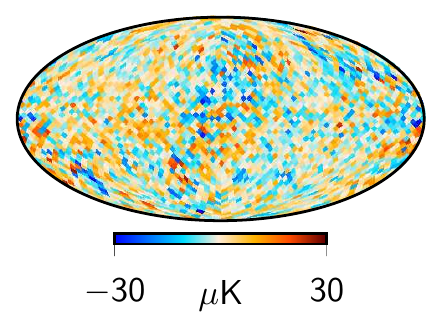}
        \includegraphics[width=0.16\linewidth]{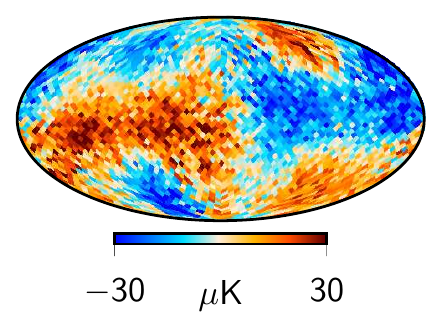}
        \includegraphics[width=0.16\linewidth]{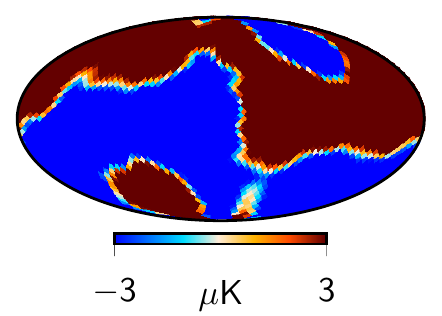}\\                
        \captionof{figure}{Comparison of \cosmoglobe\ and \WMAPnine\ polarization DA maps. Left and right sections show Stokes $Q$ and $U$, respectively, while rows show DAs. Within each section, the left and middle columns show the \cosmoglobe\ and \WMAPnine\ maps, while the right column shows their difference. All full-signal maps are shown at a HEALPix resolution of $N_{\mathrm{side}}=16$, and the difference maps have additionally been smoothed with a $10^{\circ}$ FWHM beam. }
        \label{fig:skymaps}
\end{minipage}


\end{document}